\documentclass[10pt,twocolumn,letterpaper]{article}

\usepackage[pagenumbers]{cvpr} % To force page numbers, e.g. for an arXiv version

\usepackage{graphicx}
\usepackage{amsmath}
\usepackage{amssymb}
\usepackage{booktabs}

\usepackage{mathtools}
\usepackage{amsthm}

\usepackage{multirow}
\usepackage[dvipsnames]{xcolor}

\theoremstyle{plain}

\theoremstyle{definition}

\theoremstyle{remark}

\definecolor{cvprblue}{rgb}{0.21,0.49,0.74}
\usepackage[pagebackref,breaklinks,colorlinks,citecolor=cvprblue]{hyperref}

\usepackage[accsupp]{axessibility}  % Improves PDF readability for those with disabilities.

\usepackage{xcolor, soul}
\sethlcolor{pink}

\usepackage{pgfplots}
    \usepgfplotslibrary{colorbrewer}
    \pgfplotsset{
        cycle list/Dark2,
        cycle multiindex* list={
            mark list*\nextlist
            Dark2\nextlist
        },
    }
\pgfplotsset{compat=1.14}

\definecolor{ForestGreen}{RGB}{34,139,34}
\definecolor{Cerulean}{rgb}{0.0, 0.48, 0.65}
\definecolor{Cerulean_dark}{rgb}{0.0, 0.30, 0.47}
\definecolor{OrangeRed}{RGB}{245,99,0}
\definecolor{goldenpoppy}{rgb}{0.99, 0.76, 0.0}
\definecolor{goldenpoppy2}{rgb}{0.95, 0.72, 0.0}
\definecolor{skyblue}{rgb}{0.53, 0.81, 0.92}
\definecolor{skyblue2}{rgb}{0.55, 0.83, 0.94}
\definecolor{red-violet}{rgb}{0.78, 0.08, 0.52}
\definecolor{darkcerulean}{rgb}{0.03, 0.33, 0.55}
\definecolor{flamingopink}{rgb}{0.99, 0.59, 0.70}
\definecolor{flamingopink_dark}{rgb}{0.75, 0.32, 0.43}
\definecolor{caribbeangreen}{rgb}{0.0, 0.8, 0.6}
\definecolor{caribbeangreen2}{rgb}{0.0, 0.825, 0.625}
\definecolor{darkpastelpurple}{rgb}{0.69, 0.34, 0.84}
\definecolor{darkpastelpurple2}{rgb}{0.68, 0.33, 0.83}
\definecolor{smokyblack}{rgb}{0.1, 0.1, 0.1}
\definecolor{smokyblack2}{rgb}{0.25, 0.25, 0.25}

\usepackage{standalone}

\usepackage{rotating}

\newcommand{\parsection}[1]{\vspace{2mm}\noindent\textbf{\textit{#1}}~ }

\usepackage{booktabs}
\usepackage{array}
\newcolumntype{P}[1]{>{\raggedright\arraybackslash}p{#1}}
\usepackage{makecell}      % easy multi-line cells
\title{Evaluation and Prognostic Validation of Deep Regression\\ Models for WSI-Based Gene-Expression Prediction}

\author{Fredrik K. Gustafsson$^{1,2}$
\and
Constance Boissin$^{1}$
\and
Johan Vallon-Christersson$^{3}$
\and
Mattias Rantalainen$^{*,1}$\vspace{0.5mm}
\and
\normalsize$^{1}$Department of Medical Epidemiology and Biostatistics, Karolinska Institutet, Stockholm, Sweden\vspace{-0.85mm}\\
\normalsize$^{2}$Department of Engineering Science, University of Oxford, Oxford, UK\vspace{-0.85mm}\\
\normalsize$^{3}$Division of Oncology, Department of Clinical Sciences Lund, Lund University, Lund, Sweden\vspace{-0.6mm}\\
\small$^{*}$Corresponding Author, {\tt mattias.rantalainen@ki.se}
}

\begin{document}

\maketitle

\begin{abstract}
    Gene-expression profiling is widely used in research and central to many areas of precision oncology, but remains costly and not universally accessible. Recent advances in computational pathology enable prediction of transcriptomic profiles directly from hematoxylin and eosin (H\&E)-stained whole-slide images (WSIs), although optimal modeling strategies and clinical relevance remain unclear. In this study, we systematically evaluate deep regression models for WSI-based gene-expression prediction across multiple regression formulations and pathology foundation models (PFMs), and assess whether the resulting predicted transcriptomic signals retain prognostic utility. Across four TCGA datasets, we find that direct regression using attention-based multiple instance learning together with PFM feature extractors provides a strong and computationally efficient baseline, with no consistent benefit from separately training multiple models on subsets of genes. We then externally validate the selected configuration on an independent cohort of 997 breast cancer patients, demonstrating robust generalization for clinically relevant gene sets such as PAM50. To assess clinical relevance, we further evaluate predicted gene-expression scores in two independent population-representative breast cancer cohorts comprising 4,172 patients with survival endpoints, where predicted scores retain prognostic value in both the full patient cohort and the ER+ \& HER2- subgroup. Together, these results demonstrate that WSI-based gene-expression prediction can generalize across independent cohorts and recover biologically and clinically meaningful molecular structure, supporting its potential as a scalable approach for transcriptomic phenotyping and risk stratification.
\end{abstract}
\vspace{-12.0mm}

\section*{}

% Intro/background/contributions:
Molecular phenotyping via mRNA gene-expression profiling is an important tool for patient stratification in cancer precision medicine~\cite{van2002gene, van2002agene, mcdermott2011genomics, ren2018whole}. However, molecular profiling remains relatively expensive, time-consuming, and not universally available in routine clinical workflows. In parallel, computational pathology has recently shown that hematoxylin and eosin (H\&E)-stained whole-slide images (WSIs) contain substantial information about the underlying molecular state of a tumor, enabling prediction of gene-expression profiles directly from routine histopathology images~\cite{schmauch2020deep, wang2021predicting, weitz2022transcriptome, mondol2023hist2rna, xie2023spatially, hoang2024deep, jaume2024hest, shulman2024ai}. Such models could potentially be useful both in research settings, by enabling transcriptomic profiling of very large patient cohorts, as well as in clinical settings as a low-cost pre-screening step to identify those patients most likely to benefit from downstream molecular assays.

Despite this promise, several important methodological questions remain unresolved. Gene-expression prediction from WSIs is an extremely \emph{high-dimensional regression problem}, where models must infer continuous values for tens of thousands of genes from weakly supervised slide-level inputs. Many alternative regression formulations from the general machine learning literature~\citep{lathuiliere2019comprehensive, gustafsson2020energy} are plausible in this setting. For example, one may train models to directly regress transcriptomic profiles~\citep{mondol2023hist2rna, hoang2024deep, pizurica2024digital, hoang2025path2omics}, or instead align WSI and gene-expression representations using contrastive objectives before deriving predictions~\citep{xie2023spatially, min2024multimodal}. Similarly, it is unclear whether the full transcriptome should be predicted using a single joint model, or whether improved performance can be achieved by training multiple models on smaller subsets of genes, including one model per gene~\citep{wang2021predicting}. These design choices have received limited systematic evaluation, despite having major implications for both predictive accuracy and computational cost.

A related practical question concerns the choice of patch-level feature extractor. Pathology foundation models (PFMs) have rapidly become a dominant paradigm for representation learning in computational pathology~\citep{chen2024uni, gigapath2024, virchow2024, ding2025multimodal, bilal2025foundation, li2025survey}, and recent benchmarks have shown clear advantages over natural-image pretrained encoders across several downstream tasks~\citep{marza2025thunder, kasireddy2026comprehensive, breen2025comprehensive, bareja2025evaluating, campanella2025clinical, neidlinger2025benchmarking, ma2025pathbench, gustafsson2026benchmarking}. However, their impact on WSI-based gene-expression prediction has not yet been evaluated systematically at this scale.

Beyond model benchmarking, the clinical relevance of WSI-based gene-expression prediction also remains incompletely understood. Prior work has shown that breast cancer proliferation-related signatures can be predicted directly from WSIs and retain prognostic value in an external cohort~\citep{ekholm2024prediction}, highlighting the potential of image-derived transcriptomic signals for clinically meaningful risk assessment. More recently, large-scale pan-cancer studies have demonstrated that transcriptome-wide prediction from WSIs is feasible across many tumor types and can recover biologically meaningful signals~\citep{pizurica2024digital}. In addition, recent work has begun to explore the downstream utility of WSI-derived transcriptomic profiles for survival modelling, for example by constructing prognostic models from predicted pseudo-bulk gene-expression features within TCGA cohorts using cross-validation~\citep{wang2025benchmarking}, or by showing that inferred gene-expression profiles can achieve survival prediction performance comparable to measured transcriptomic data across multiple cancer types, again evaluated within TCGA cross-validation settings~\citep{hoang2025path2omics}. 

These studies represent important advances, but they leave several key questions open. First, there has been limited systematic evaluation of alternative modelling strategies for \emph{transcriptome-wide} WSI-based gene-expression prediction, including whether contrastive learning or multi-model decompositions offer practical advantages over direct regression. Second, the impact of modern PFMs on this task remains insufficiently characterized. Third, while prior work has demonstrated feasibility, and in some cases prognostic value, it remains unclear whether transcriptome-wide WSI-based predictions can robustly recover established gene-expression signatures and retain risk-stratification value in \emph{fully independent external survival cohorts}. Addressing these questions is critical if WSI-based transcriptomic profiling is to move beyond methodological benchmarking and toward practical clinical utility.

To address these gaps, we perform a stepwise evaluation of transcriptome-wide WSI-based gene-expression prediction. We first benchmark alternative deep regression formulations across four TCGA datasets to identify a strong and computationally efficient modelling strategy. We then compare twelve widely used patch-level feature extractors on TCGA-BRCA, including several recent PFMs. Next, we externally validate the best-performing model on SCAN-B-Lund, an independent breast cancer dataset with matched WSIs and gene-expression measurements. Finally, we assess whether predicted gene-expression scores retain clinically meaningful prognostic information in SöS-BC-4 and KS-Solna, two large independent breast cancer survival cohorts. This design allows us to explicitly separate external validation of transcriptomic prediction accuracy from independent validation of downstream prognostic utility.

\textit{Our study makes four main contributions:}
(1) We present a systematic benchmark of deep regression models for WSI-based gene-expression prediction, showing that direct regression with ABMIL provides a strong and computationally efficient baseline, and that training multiple models on gene subsets does not provide a consistent advantage over a single transcriptome-wide model.
(2) We show that pathology-specific feature extractors substantially improve performance for this task, with H-optimus-1 achieving the strongest results on TCGA-BRCA among early pathology-specific models, state-of-the-art PFMs, and vision-language PFMs.
(3) We demonstrate external generalization of the selected gene-expression model on an independent breast cancer cohort of 997 patients, with strong predictive performance for clinically relevant genes and signatures such as PAM50 and an 11-gene proliferation score.
(4) We show that predicted gene-expression scores derived directly from WSIs retain meaningful prognostic information in independent external survival cohorts comprising 4,172 breast cancer patients, indicating that WSI-based transcriptomic prediction can recover clinically relevant molecular structure.

\begin{figure*}[t]
    \centering
    \begin{tikzpicture}
  \centering
  \begin{axis}[
        ybar, axis on top,
        title={},
        height=4.525cm, width=18.0cm, %%%%%%%%%%%%%%%%%%%%%%%%%%%%%%%%%%%%%
        ybar=2.75pt,
        bar width=0.475cm,
        ymajorgrids, tick align=inside,
        enlarge y limits={value=.1,upper},
        ymin=0.0, ymax=0.30,
        set layers,
        ytick={0, 0.1, 0.2, 0.3},
        yticklabels={0, 0.10, 0.20, 0.30},
        axis x line*=bottom,
        axis y line*=left,
        y axis line style={opacity=0},
        tickwidth=0pt,
        % enlarge x limits=true,
        % enlarge x limits=0.185, % (makes the different bar groups closer to eachother!) %%%%%
        % enlarge x limits=0.155, % (makes the different bar groups closer to eachother!) %%%%%
        enlarge x limits=0.25, % (makes the different bar groups closer to eachother!) %%%%%
        % extra y ticks=0.9,
        extra y tick style={
            grid style={
                % black,
                solid,
                line width = 2.5pt,
                /pgfplots/on layer=axis background,
            },
        },
        tick style={
            grid style={
                dashed,
                /pgfplots/on layer=axis background,
            },
        },
        legend style={
            at={(0.5,-0.2)},
            anchor=north,
            legend columns=-1,
            /tikz/every even column/.append style={column sep=0.325cm},
            font=\footnotesize
        },
        ylabel={mean Pearson ($\uparrow$)},
        ylabel style={font=\scriptsize},
        y tick label style={
            /pgf/number format/.cd,
                fixed,
                fixed zerofill,
                precision=2,
            /tikz/.cd
        },
        symbolic x coords={
           TCGA-BRCA,
           TCGA-HNSC,
           TCGA-STAD, 
           TCGA-BLCA},
       xtick=data,
       xticklabel=\empty,
       x tick label style={text width = 2.25cm, align=center, font=\footnotesize},
    ]
    \addplot [fill=darkcerulean, draw=darkcerulean, line width=0mm, error bars/.cd, y dir=both, y explicit, error bar style={thick}, error bar style=black] coordinates {
      (TCGA-BRCA, 0.284548 ) +- (0, 0.00676863)
      (TCGA-HNSC, 0.256569) +- (0, 0.00900385)
      (TCGA-STAD, 0.216453) +- (0, 0.0314102)
      (TCGA-BLCA, 0.253299) +- (0, 0.0296211)};
    \addplot [fill=goldenpoppy, draw=goldenpoppy, line width=0mm, error bars/.cd, y dir=both, y explicit, error bar style={thick}, error bar style=black] coordinates { 
      (TCGA-BRCA, 0.267707) +- (0, 0.00988892)
      (TCGA-HNSC, 0.237809) +- (0, 0.0172336)
      (TCGA-STAD, 0.200889) +- (0, 0.044748)
      (TCGA-BLCA, 0.242186) +- (0, 0.0282367)};
    \addplot [fill=ForestGreen, draw=ForestGreen, line width=0mm, error bars/.cd, y dir=both, y explicit, error bar style={thick}, error bar style=black] coordinates { 
      (TCGA-BRCA, 0.283991) +- (0, 0.00442409)
      (TCGA-HNSC, 0.246884) +- (0, 0.0221961)
      (TCGA-STAD, 0.216922) +- (0, 0.0443081)
      (TCGA-BLCA, 0.249229) +- (0, 0.0256722)};
    \addplot [fill=skyblue2, draw=skyblue2, line width=0mm, error bars/.cd, y dir=both, y explicit, error bar style={thick}, error bar style=black] coordinates { 
      (TCGA-BRCA, 0.157735) +- (0, 0.0112034)
      (TCGA-HNSC, 0.141764) +- (0, 0.0260436)
      (TCGA-STAD, 0.135536) +- (0, 0.0373112)
      (TCGA-BLCA, 0.142848) +- (0, 0.0211071)};
  \end{axis}
  \end{tikzpicture}\vspace{2.0mm}
    \begin{tikzpicture}
  \centering
  \begin{axis}[
        ybar, axis on top,
        title={},
        height=4.525cm, width=18.0cm, %%%%%%%%%%%%%%%%%%%%%%%%%%%%%%%%%%%%%
        ybar=2.75pt,
        bar width=0.475cm,
        ymajorgrids, tick align=inside,
        enlarge y limits={value=.1,upper},
        ymin=0.0, ymax=0.60,
        set layers,
        ytick={0, 0.2, 0.4, 0.6},
        yticklabels={0, 0.20, 0.40, 0.60},
        axis x line*=bottom,
        axis y line*=left,
        y axis line style={opacity=0},
        tickwidth=0pt,
        % enlarge x limits=true,
        % enlarge x limits=0.185, % (makes the different bar groups closer to eachother!) %%%%%
        % enlarge x limits=0.155, % (makes the different bar groups closer to eachother!) %%%%%
        enlarge x limits=0.25, % (makes the different bar groups closer to eachother!) %%%%%
        % extra y ticks=0.9,
        extra y tick style={
            grid style={
                % black,
                solid,
                line width = 2.5pt,
                /pgfplots/on layer=axis background,
            },
        },
        tick style={
            grid style={
                dashed,
                /pgfplots/on layer=axis background,
            },
        },
        legend style={
            at={(0.5,-0.2)},
            anchor=north,
            legend columns=-1,
            /tikz/every even column/.append style={column sep=0.325cm},
            font=\footnotesize
        },
        ylabel={mean Pearson - top 1k genes ($\uparrow$)},
        ylabel style={font=\scriptsize},
        y tick label style={
            /pgf/number format/.cd,
                fixed,
                fixed zerofill,
                precision=2,
            /tikz/.cd
        },
        symbolic x coords={
           TCGA-BRCA,
           TCGA-HNSC,
           TCGA-STAD, 
           TCGA-BLCA},
       xtick=data,
       xticklabel=\empty,
       x tick label style={text width = 2.25cm, align=center, font=\footnotesize},
    ]
    \addplot [fill=darkcerulean, draw=darkcerulean, line width=0mm, error bars/.cd, y dir=both, y explicit, error bar style={thick}, error bar style=black] coordinates {
      (TCGA-BRCA, 0.598019 ) +- (0, 0.0125496)
      (TCGA-HNSC, 0.589236) +- (0, 0.032307)
      (TCGA-STAD, 0.548459) +- (0, 0.0525175)
      (TCGA-BLCA, 0.60033) +- (0, 0.0378758)};
    \addplot [fill=goldenpoppy, draw=goldenpoppy, line width=0mm, error bars/.cd, y dir=both, y explicit, error bar style={thick}, error bar style=black] coordinates { 
      (TCGA-BRCA, 0.57276) +- (0, 0.0108307)
      (TCGA-HNSC, 0.565718) +- (0, 0.027937)
      (TCGA-STAD, 0.533843) +- (0, 0.0846625)
      (TCGA-BLCA, 0.577512) +- (0, 0.03613)};
    \addplot [fill=ForestGreen, draw=ForestGreen, line width=0mm, error bars/.cd, y dir=both, y explicit, error bar style={thick}, error bar style=black] coordinates { 
      (TCGA-BRCA, 0.596739) +- (0, 0.0146391)
      (TCGA-HNSC, 0.576468) +- (0, 0.0332305)
      (TCGA-STAD, 0.54385) +- (0, 0.0664589)
      (TCGA-BLCA, 0.590932) +- (0, 0.0356143)};
    \addplot [fill=skyblue2, draw=skyblue2, line width=0mm, error bars/.cd, y dir=both, y explicit, error bar style={thick}, error bar style=black] coordinates { 
      (TCGA-BRCA, 0.465541) +- (0, 0.0183544)
      (TCGA-HNSC, 0.465364) +- (0, 0.0432175)
      (TCGA-STAD, 0.469073) +- (0, 0.0605809)
      (TCGA-BLCA, 0.464156) +- (0, 0.0282119)};
  \end{axis}
  \end{tikzpicture}\vspace{2.0mm}
    \begin{tikzpicture}
  \centering
  \begin{axis}[
        ybar, axis on top,
        title={},
        height=4.525cm, width=18.0cm, %%%%%%%%%%%%%%%%%%%%%%%%%%%%%%%%%%%%%
        ybar=2.75pt,
        bar width=0.475cm,
        ymajorgrids, tick align=inside,
        enlarge y limits={value=.1,upper},
        ymin=0, ymax=5050,
        set layers,
        ytick={0, 1500, 3000, 4500},
        yticklabels={0, 1.5k, 3.0k, 4.5k},
        axis x line*=bottom,
        axis y line*=left,
        y axis line style={opacity=0},
        tickwidth=0pt,
        % enlarge x limits=true,
        % enlarge x limits=0.185, % (makes the different bar groups closer to eachother!) %%%%%
        % enlarge x limits=0.155, % (makes the different bar groups closer to eachother!) %%%%%
        enlarge x limits=0.25, % (makes the different bar groups closer to eachother!) %%%%%
        % extra y ticks=0.9,
        extra y tick style={
            grid style={
                % black,
                solid,
                line width = 2.5pt,
                /pgfplots/on layer=axis background,
            },
        },
        tick style={
            grid style={
                dashed,
                /pgfplots/on layer=axis background,
            },
        },
        legend style={
            at={(0.5,-0.225)},
            anchor=north,
            legend columns=4,
            /tikz/every even column/.append style={column sep=0.325cm},
            font=\footnotesize
        },
        ylabel={\# genes $\geq 0.4$ Pearson ($\uparrow$)},
        ylabel style={font=\scriptsize},
        symbolic x coords={
           TCGA-BRCA,
           TCGA-HNSC,
           TCGA-STAD, 
           TCGA-BLCA},
       xtick=data,
       x tick label style={text width = 2.25cm, align=center, font={\footnotesize\bfseries}},
       % x tick label style={text width = 2.25cm, align=center, font=\footnotesize},
    ]
    \addplot [fill=darkcerulean, draw=darkcerulean, line width=0mm, error bars/.cd, y dir=both, y explicit, error bar style={thick}, error bar style=black] coordinates {
      (TCGA-BRCA, 4927 ) +- (0, 357.299)
      (TCGA-HNSC, 3987.8) +- (0, 765.179)
      (TCGA-STAD, 3240.2) +- (0, 1342.98)
      (TCGA-BLCA, 4364.6) +- (0, 1188.92)};
    \addplot [fill=goldenpoppy, draw=goldenpoppy, line width=0mm, error bars/.cd, y dir=both, y explicit, error bar style={thick}, error bar style=black] coordinates { 
      (TCGA-BRCA, 4213) +- (0, 353.379)
      (TCGA-HNSC, 3456.4) +- (0, 837.055)
      (TCGA-STAD, 2765.4) +- (0, 1713.97)
      (TCGA-BLCA, 3822.4) +- (0, 1206.4)};
    \addplot [fill=ForestGreen, draw=ForestGreen, line width=0mm, error bars/.cd, y dir=both, y explicit, error bar style={thick}, error bar style=black] coordinates { 
      (TCGA-BRCA, 4886.8) +- (0, 242.807)
      (TCGA-HNSC, 3750.2) +- (0, 835.731)
      (TCGA-STAD, 3184.2) +- (0, 1660.56)
      (TCGA-BLCA, 4183.6) +- (0, 975.421)};
    \addplot [fill=skyblue2, draw=skyblue2, line width=0mm, error bars/.cd, y dir=both, y explicit, error bar style={thick}, error bar style=black] coordinates { 
      (TCGA-BRCA, 1137) +- (0, 247.217)
      (TCGA-HNSC, 1158.6) +- (0, 463.492)
      (TCGA-STAD, 1328) +- (0, 981.967)
      (TCGA-BLCA, 1164.2) +- (0, 451.388)};
    \legend{
            Direct - ABMIL,
            Direct - Patch-Level,
            Contrastive,
            kNN
            }
  \end{axis}
  \end{tikzpicture}\vspace{-1.0mm}
    \caption{\textbf{Benchmarking of deep regression models on TCGA datasets.} Performance comparison of the four regression models across TCGA-BRCA, TCGA-HNSC, TCGA-STAD, and TCGA-BLCA, using UNI as the patch-level feature extractor. \textbf{Top:} mean Pearson correlation across all $N\!=\!20{,}530$ genes. \textbf{Middle:} mean Pearson correlation across the top $1{,}000$ best-predicted genes. \textbf{Bottom:} number of genes (out of $N\!=\!20{,}530$) with Pearson correlation $\geq 0.4$. Higher values indicate better performance for all three metrics. Results are reported as mean $\pm$ standard deviation over 5-fold site-aware cross-validation. Corresponding numerical results are provided in Table~\ref{tab:main_results_brca}~-~\ref{tab:main_results_blca} in the supplementary material, and analogous results using Spearman correlation are shown in Figure~\ref{fig:main_results_spearman}.}
    \label{fig:main_results}
\end{figure*}

\input{figures/fig_main_results_detailed}

\input{figures/fig_main_results_detailed_first800}

\section*{Results}

We present results in a stepwise manner, beginning with benchmarking of regression models on TCGA datasets, followed by external validation of predictive performance, and finally prognostic validation of predicted risk scores.

\subsubsection*{Benchmarking Deep Regression Models}
We first conduct experiments on datasets from the Cancer Genome Atlas (TCGA), for four different cancer types: breast (TCGA-BRCA, $n\!=\!1{,}062$ patients), head-neck (TCGA-HNSC, $n\!=\!450$), stomach (TCGA-STAD, $n\!=\!416$), and bladder (TCGA-BLCA, $n\!=\!386$). For each dataset, we train deep regression models to output gene-level transcription estimates of $N\!=\!20{,}530$ genes (utilizing gene-expression data from UCSC Xena~\cite{goldman2020visualizing}) from the corresponding diagnostic H\&E WSI. Models are trained and evaluated using 5-fold \emph{site-aware} cross-validation~\cite{howard2021impact}. 

We compare four regression models, all based on patch-level feature vectors extracted from the WSI using a frozen UNI~\cite{chen2024uni} model (see \textit{Methods} for details). \textit{Direct - ABMIL} uses attention-based multiple instance learning (ABMIL)~\cite{ilse2018attention, laleh2022benchmarking} to directly output a predicted gene-expression profile $\hat{y}(x) \in \mathbb{R}^N$ for the WSI $x$. \textit{Direct - Patch-Level} removes the trainable ABMIL aggregator and computes $\hat{y}(x) \in \mathbb{R}^N$ as the mean over patch-level predictions. \textit{Contrastive} instead utilizes contrastive learning~\cite{chen2020simple, radford2021learning} to align WSI and gene-expression representations, and computes $\hat{y}(x) \in \mathbb{R}^N$ as a linear combination of the most similar gene-expression profiles from the train set. \textit{kNN} is a simple baseline model (with no trainable parameters) utilizing the k-nearest neighbors algorithm. We also evaluate how the accuracy is affected if, instead of training a single model to regress all $N\!=\!20{,}530$ genes, multiple models are separately trained to regress subsets of genes and then combined at test-time to output a full predicted gene-expression profile $\hat{y}(x) \in \mathbb{R}^N$. The $N\!=\!20{,}530$ genes are grouped into subsets either via sequential chunking, or clustering of correlated genes.

Figure~\ref{fig:main_results} summarizes the performance of the four regression models across the four TCGA datasets, with corresponding numerical results in Table~\ref{tab:main_results_brca}~-~\ref{tab:main_results_blca} in the supplementary material. Across all datasets, \textit{Direct - ABMIL} and \textit{Contrastive} consistently outperform \textit{Direct - Patch-Level} and \textit{kNN}. On TCGA-BRCA, \textit{Direct - ABMIL} achieves a mean Pearson correlation of $0.284\pm0.006$, compared to $0.283\pm0.004$ for \textit{Contrastive}, $0.267\pm0.009$ for \textit{Direct - Patch-Level}, and $0.157\pm0.011$ for \textit{kNN}. It also regresses $4927\pm357$ genes with Pearson correlation of at least $0.4$, compared with $4886\pm242$ for \textit{Contrastive}, $4213\pm353$ for \textit{Direct - Patch-Level}, and $1137\pm247$ for \textit{kNN}. Similar trends are observed on TCGA-HNSC, TCGA-STAD, and TCGA-BLCA. Overall, \textit{Direct - ABMIL} provides the strongest and most consistent performance across datasets, with \textit{Contrastive} performing comparably but not clearly better on any of the evaluated metrics. Table~\ref{tab:main_results_top-genes_abmil}~\&~\ref{tab:main_results_top-genes_contrastive} show the top 20 genes with the highest prediction accuracy across the four datasets for \textit{Direct - ABMIL} and \textit{Contrastive}, respectively.

We next evaluate performance specifically on the PAM50 genes, which are of particular relevance in breast cancer~\cite{parker2009supervised, nielsen2010comparison, wallden2015development}. As shown in Figure~\ref{fig:main_results_pam50} and Table~\ref{tab:main_results_brca_pam50}, the same ranking of models is largely preserved. On TCGA-BRCA, \textit{Direct - ABMIL} achieves a mean Pearson correlation of $0.562\pm0.020$, compared with $0.564\pm0.020$ for \textit{Contrastive}, $0.541\pm0.015$ for \textit{Direct - Patch-Level}, and $0.415\pm0.019$ for \textit{kNN}. Thus, the best-performing models on transcriptome-wide evaluation also perform best on this clinically relevant gene subset. Table~\ref{tab:main_results_brca_pam50_abmil_all_genes}~\&~\ref{tab:main_results_brca_pam50_contrastive_all_genes} list the Pearson correlations for all individual PAM50 genes for \textit{Direct - ABMIL} and \textit{Contrastive}, with corresponding scatter plots for \textit{Direct - ABMIL} shown in Figure~\ref{fig:corr_plots_pam50_1_5}~-~\ref{fig:corr_plots_pam50_36_50}.

We then investigate whether performance can be improved by replacing a single transcriptome-wide model with multiple models trained on subsets of genes. Figure~\ref{fig:main_results_detailed} shows the results when varying the number of \textit{Direct - ABMIL} models and grouping genes either by sequential chunking or clustering. Across the four TCGA datasets, increasing the number of models from one to more than $200$ yields at most modest gains, with no consistent advantage for either grouping strategy. In particular, on TCGA-BRCA the performance remains essentially unchanged, indicating that splitting the regression task does not provide a meaningful benefit in this setting.

To further study this effect, Figure~\ref{fig:main_results_detailed_first800} evaluates performance while progressively increasing the number of genes regressed per model, ranging from the extreme case of one model per gene to a single model covering all $N\!=\!20{,}530$ evaluated genes. To make this analysis computationally feasible, we evaluate models only on the subset of the first $800$ genes, such that the extreme case requires training $800$ models instead of $20{,}530$. We observe that, while the optimal number of genes per model varies slightly across datasets and metrics, it is never optimal to train one model per gene. Instead, performance consistently improves when increasing the number of genes per model from $1$ to $10$, and from $10$ to $20$, after which it either peaks or levels off. 

A similar pattern is observed when training \textit{Direct - ABMIL} models only on the PAM50 subset on TCGA-BRCA (Table~\ref{tab:main_results_brca_pam50}). Increasing the number of genes per model from one ($50$ models, each regressing a single gene) to $50$ (a single model regressing all PAM50 genes) consistently improves performance from a mean Pearson correlation of $0.560\pm0.020$ to $0.576\pm0.020$. Notably, training separate models for each individual PAM50 gene does not outperform the standard \textit{Direct - ABMIL} model trained on all $N\!=\!20{,}530$ genes (mean Pearson $0.562\pm0.020$ on PAM50).

Taken together, these results support three main conclusions. First, \textit{Direct - ABMIL} is a strong and robust choice among the evaluated regression formulations. Second, the trainable ABMIL aggregation step provides a clear benefit over simple patch-level averaging. Third, training a single model to regress all $N\!=\!20{,}530$ genes is a computationally efficient and highly competitive baseline, with no consistent advantage from separately training multiple models on gene subsets. Based on these findings, we select \textit{Direct - ABMIL, with a single model trained on all genes}, as the regression framework for the remainder of the study.

\subsubsection*{Comparison of Pathology Foundation Models}
Next, we compare UNI~\citep{chen2024uni} with eleven other feature extractors in terms of their performance with \textit{Direct - ABMIL} on TCGA-BRCA. The evaluated models include a natural-image baseline (Resnet-IN), two early pathology-specific models (CTransPath~\citep{Wang2023ctranspath}, RetCCL~\citep{wang2023retccl}), six recent PFMs (Prov-GigaPath~\citep{gigapath2024}, UNI2-h~\citep{UNI2h2024}, Virchow~\citep{virchow2024}, Virchow2~\citep{virchow22024}, H-optimus-0~\citep{hoptimus0}, H-optimus-1~\citep{hoptimus1}), and two vision-language PFMs (CONCH~\citep{conch2024}, CONCHv1.5~\citep{ding2025multimodal}).

Table~\ref{table:pfms_rank_info} shows the ranking of all twelve models based on mean Pearson correlation over the PAM50 genes in TCGA-BRCA (5-fold site-aware cross-validation). First, all pathology-specific models substantially outperform the natural-image baseline Resnet-IN. Among the evaluated models, H-optimus-1 achieves the best performance, with a mean Pearson correlation of $0.596 \pm 0.016$, followed by H-optimus-0 ($0.588 \pm 0.013$), UNI2-h ($0.583 \pm 0.016$), and Virchow2 ($0.582 \pm 0.019$). UNI remains competitive ($0.562 \pm 0.026$), but is outperformed by several newer models. The earlier-generation pathology models CTransPath and RetCCL perform substantially worse, while still clearly outperforming Resnet-IN which yields the lowest overall performance ($0.379 \pm 0.035$).

These results highlight the importance of using pathology-specific feature extractors for WSI-based gene-expression prediction. While several recent PFMs perform similarly, \textit{H-optimus-1} gives the strongest results on TCGA-BRCA and is therefore selected as the feature extractor for all subsequent experiments.

\subsubsection*{External Validation of Predictive Accuracy}
We next perform an external evaluation of the selected \textit{H-optimus-1 - Direct - ABMIL} model on SCAN-B-Lund, a cohort of $997$ breast cancer patients with one WSI per patient and matched gene-expression measurements for $19{,}675$ genes. To maximize robustness, we use the five models obtained from the 5-fold site-aware cross-validation on TCGA-BRCA as an ensemble, and apply this ensemble directly to SCAN-B-Lund without any retraining. This provides a direct assessment of cross-cohort generalization.

In contrast to the TCGA cross-validation experiments, where training and evaluation are performed within the same dataset, the external evaluation compares predictions across two cohorts with different gene-expression scales and measurement characteristics. We therefore report both Spearman and Pearson correlations, as Spearman emphasizes preservation of cross-patient rank ordering and is less sensitive to differences in absolute scaling.

In total, $17{,}055$ genes have exactly matching names between the TCGA-BRCA and SCAN-B-Lund gene-expression data. We additionally manually match 11 genes from clinically relevant breast cancer gene sets, resulting in $17{,}066$ evaluated genes in total. Across all these evaluated genes, the model achieves a mean Spearman correlation of $0.234$ and a mean Pearson correlation of $0.221$. Focusing on the best-predicted genes, the mean Spearman correlation is $0.616$ for the top 100 genes and $0.514$ for the top 1{,}000 genes, while the corresponding mean Pearson correlations are $0.632$ and $0.511$. In total, $2{,}116$ genes achieve Spearman correlation $\geq 0.40$, and $1{,}864$ genes achieve Pearson correlation $\geq 0.40$.

\begin{table}[t]
\caption{\textbf{Comparison of PFMs on TCGA-BRCA.} Models are ranked by mean Pearson correlation ($\uparrow$) across the PAM50 genes using \textit{Direct - ABMIL} (5-fold site-aware cross-validation). Model size and pretraining data are also reported for reference.}
\vspace{-2.25mm}
\label{table:pfms_rank_info}
\centering
\resizebox{1.0\linewidth}{!}{%
\begin{tabular}{clccc}
\toprule
Rank & Model Name & mean Pearson ($\uparrow$) 
& Size & Pretraining Data \\
\midrule
1  & H-optimus-1   & 0.596 $\pm$ 0.016 & 1.1B & 1M WSIs \\
2  & H-optimus-0   & 0.588 $\pm$ 0.013 & 1.1B & 500K WSIs \\
3  & UNI2-h        & 0.583 $\pm$ 0.016 & 682M & 350K WSIs \\
4  & Virchow2      & 0.582 $\pm$ 0.019 & 632M & 3.1M WSIs \\
5  & CONCHv1.5     & 0.576 $\pm$ 0.021 & 307M & N/A \\
6  & CONCH         & 0.574 $\pm$ 0.019 & 86M  & \makecell[c]{21K WSIs +\\1.1M img-text pairs} \\
7  & Prov-GigaPath & 0.571 $\pm$ 0.009 & 1.1B & 170K WSIs \\
8  & Virchow       & 0.563 $\pm$ 0.021 & 632M & 1.5M WSIs \\
9  & UNI           & 0.562 $\pm$ 0.026 & 307M & 100K WSIs \\
10 & CTransPath    & 0.517 $\pm$ 0.029 & 22M  & 30K WSIs \\
11 & RetCCL        & 0.450 $\pm$ 0.035 & 25M  & 32K WSIs \\
12 & Resnet-IN     & 0.379 $\pm$ 0.035 & 25M  & 1.3M natural imgs \\
\bottomrule
\end{tabular}
}
\end{table}

\begin{figure}[t]
    \centering
    \includegraphics[clip, trim=0.0cm 0.30cm 0.0cm 1.4cm, width=0.925\linewidth]{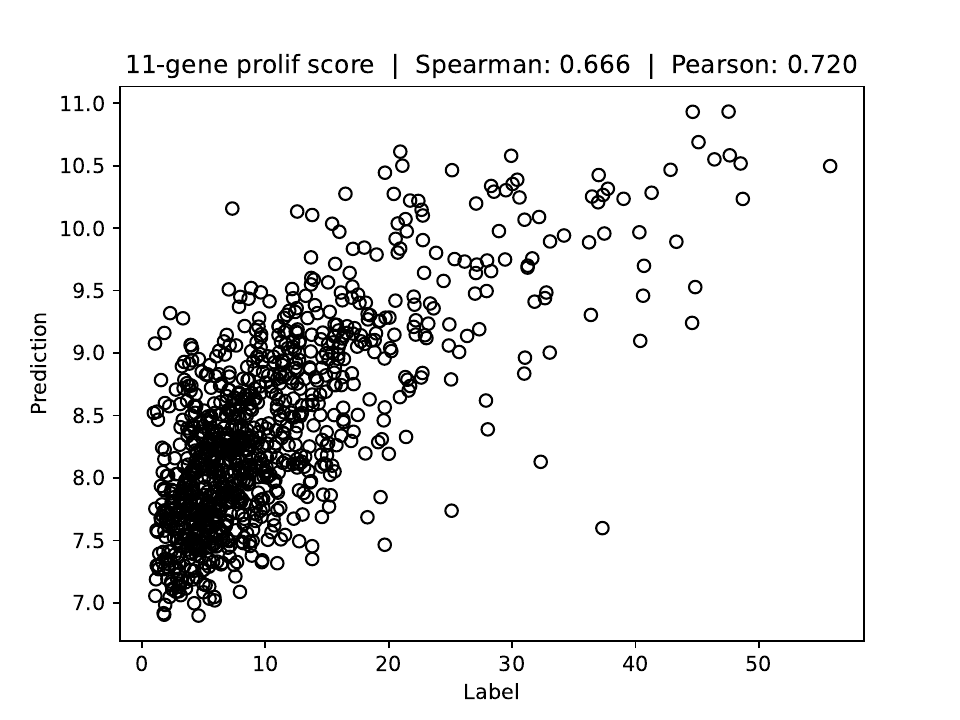}\vspace{-1.0mm}
    \caption{\textbf{Predicted vs observed 11-gene proliferation score on SCAN-B-Lund.} Scatter plot comparing predicted and ground-truth scores, for the \textit{H-optimus-1 - Direct - ABMIL} ensemble trained on TCGA-BRCA and evaluated on SCAN-B-Lund. Each point represents a patient. The predicted score achieves a Spearman correlation of $0.666$ and a Pearson correlation of $0.721$. Note the difference in scale between predicted and observed values.}
    \label{fig:corr_plots_11-gene-prolif-score_scan-b-lund}
\end{figure}

We next evaluate performance on PAM50, which shows strong external performance, with a mean Spearman correlation of $0.473$ and a mean Pearson correlation of $0.443$. Particularly strong results are obtained for an established 11-gene proliferation score~\citep{parker2009supervised, nielsen2010comparison, veta2019predicting, ekholm2024prediction}, defined as the mean gene-expression value of eleven PAM50 genes: \texttt{BIRC5}, \texttt{CCNB1}, \texttt{CDC20}, \texttt{CEP55}, \texttt{MKI67}, \texttt{NDC80}, \texttt{NUF2}, \texttt{PTTG1}, \texttt{RRM2}, \texttt{TYMS}, and \texttt{UBE2C}. A corresponding predicted score is computed by averaging the predicted expression values for these genes. The predicted 11-gene score achieves a Spearman correlation of $0.666$ and a Pearson correlation of $0.721$ on SCAN-B-Lund. While predicted and observed scores are on different absolute scales, they show strong concordance across patients, as illustrated in the scatter plot in Figure~\ref{fig:corr_plots_11-gene-prolif-score_scan-b-lund}. All 11 individual genes are also predicted with relatively high accuracy, with Spearman correlations ranging from $0.562$ (\texttt{TYMS}) to $0.659$ (\texttt{BIRC5}) and Pearson correlations ranging from $0.474$ (\texttt{NUF2}) to $0.692$ (\texttt{CDC20}). Table~\ref{tab:scan-b-lund_pam50_results} lists the Pearson and Spearman correlations for all individual PAM50 genes, with corresponding scatter plots shown in Figure~\ref{fig:corr_plots_pam50_1_5_scan-b-lund}~-~\ref{fig:corr_plots_pam50_36_50_scan-b-lund}.

Overall, these results show that a substantial part of the predictive signal learned on TCGA-BRCA transfers to an independent breast cancer cohort collected under different clinical and technical conditions. In particular, the strong performance on PAM50 and the 11-gene proliferation score indicates that the model captures biologically meaningful and clinically relevant transcriptomic structure. This external validation provides an important foundation for the downstream prognostic analyses.

\subsubsection*{Prognostic Validation of Predicted Risk Scores}
Finally, we evaluate whether predicted gene-expression scores inferred from WSIs retain prognostic information in external survival cohorts with time-to-event outcomes. We perform this evaluation using the selected \textit{H-optimus-1 - Direct - ABMIL} model on two external breast cancer survival cohorts, SöS-BC-4 and KS-Solna, comprising $4{,}172$ patients in total, of whom $289$ experience a progression-free survival (PFS) event. We also evaluate the clinically relevant ER+ \& HER2- subgroup, comprising $3{,}157$ patients with $185$ PFS events. Here, PFS is defined as the time from initial diagnosis to disease recurrence. Death without documented recurrence is not counted as an event and is treated as censoring. Recurrence includes local recurrence, distant metastasis, or detection of contralateral tumors.

\begin{figure*}[t]
\centering
    \begin{subfigure}[t]{0.40\linewidth}
        \centering%
        \includegraphics[clip, trim=0.5cm 1.25cm 0.25cm 0.0cm, width=1.0\linewidth]{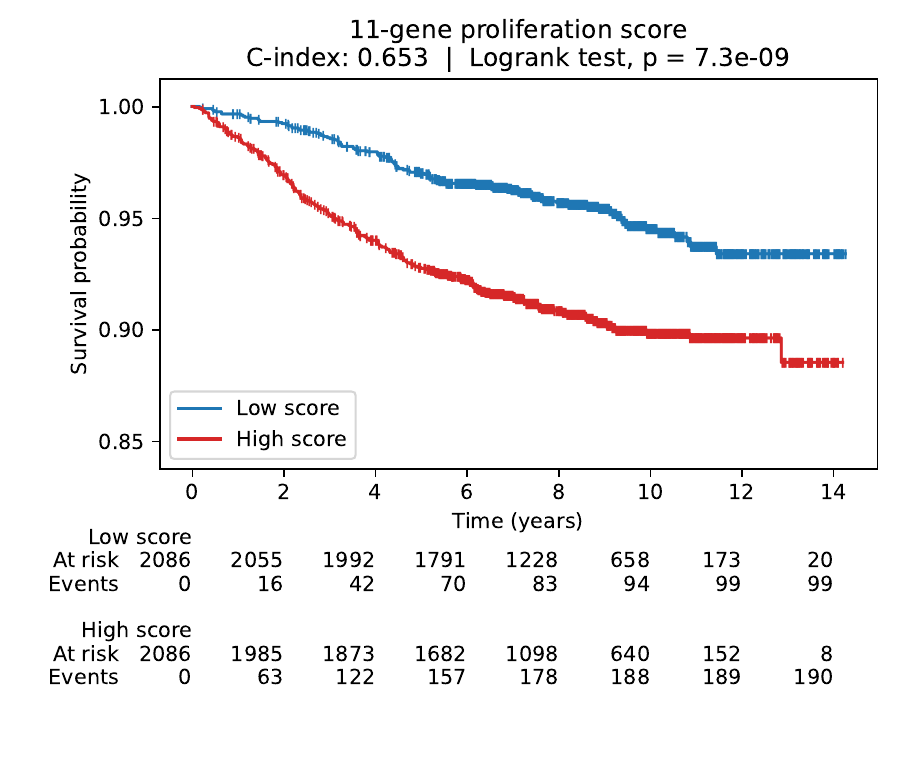}
        \caption{\textbf{11-gene proliferation score -- Full cohort}.}
    \end{subfigure}
    \begin{subfigure}[t]{0.40\linewidth}
        \centering%
        \includegraphics[clip, trim=0.5cm 1.25cm 0.25cm 0.0cm, width=1.0\linewidth]{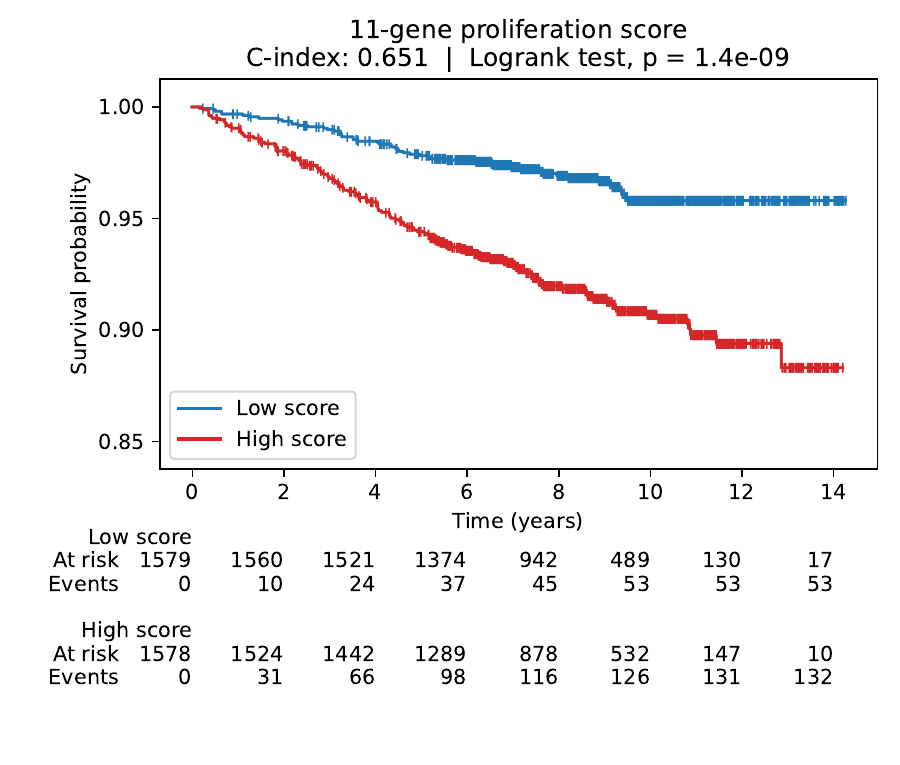}
        \caption{\textbf{11-gene proliferation score -- ER+ \& HER2-}.}
    \end{subfigure}
    \begin{subfigure}[t]{0.40\linewidth}
        \centering%
        \includegraphics[clip, trim=0.5cm 1.25cm 0.25cm 0.0cm, width=1.0\linewidth]{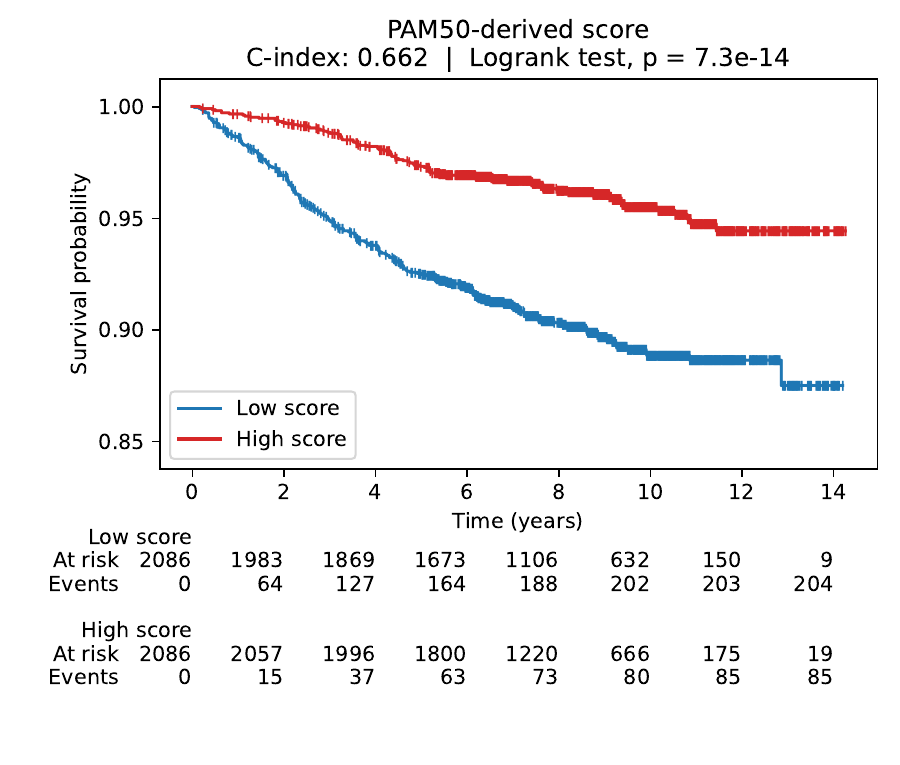}
        \caption{\textbf{PAM50-derived score -- Full cohort}.}
    \end{subfigure}
    \begin{subfigure}[t]{0.40\linewidth}
        \centering%
        \includegraphics[clip, trim=0.5cm 1.25cm 0.25cm 0.0cm, width=1.0\linewidth]{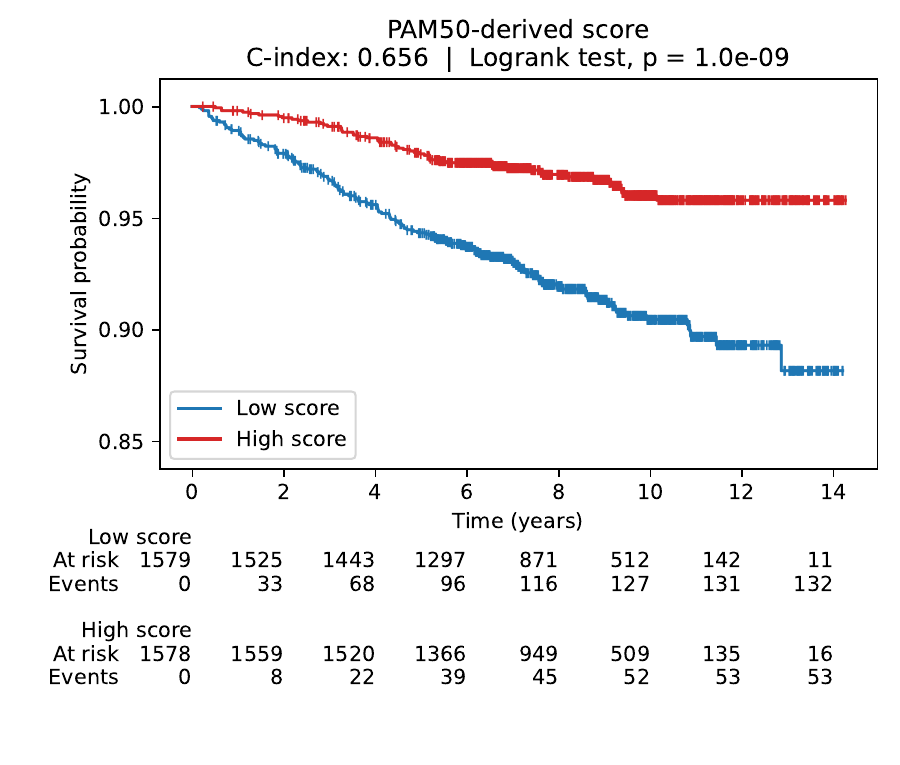}
        \caption{\textbf{PAM50-derived score -- ER+ \& HER2-}.}
    \end{subfigure}\vspace{-1.0mm}
  \caption{\textbf{Two-group Kaplan-Meier risk stratification.} KM survival curves for the 11-gene proliferation score (top) and the PAM50-derived score (bottom), evaluated on the full cohort (left) and the ER+ \& HER2- subgroup (right). Patients are stratified into two groups using the median score. Plots report C-index, log-rank test p-values, and the number of patients at risk and events over time ($0$-$14$ years).}
  \label{fig:km_plots_2groups}
\end{figure*}

\begin{figure*}[t]
\centering
    \begin{subfigure}[t]{0.40\linewidth}
        \centering%
        \includegraphics[clip, trim=0.5cm 0.25cm 1.25cm 0.5cm, width=1.0\linewidth]{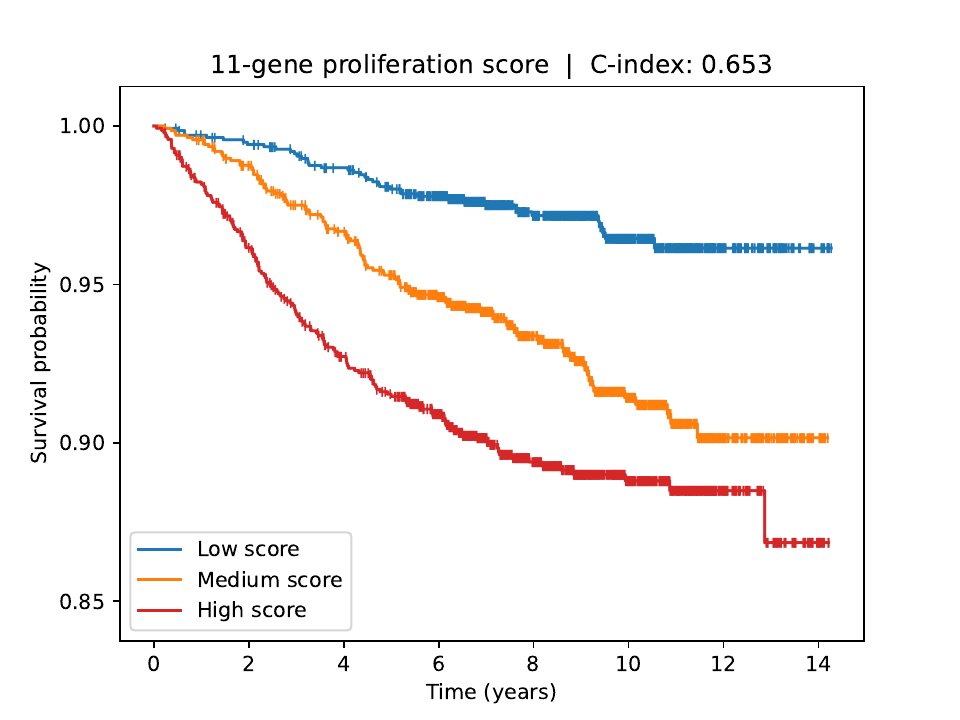}
        \caption{\textbf{11-gene proliferation score -- Full cohort}.}
    \end{subfigure}
    \begin{subfigure}[t]{0.40\linewidth}
        \centering%
        \includegraphics[clip, trim=0.5cm 0.25cm 1.25cm 0.5cm, width=1.0\linewidth]{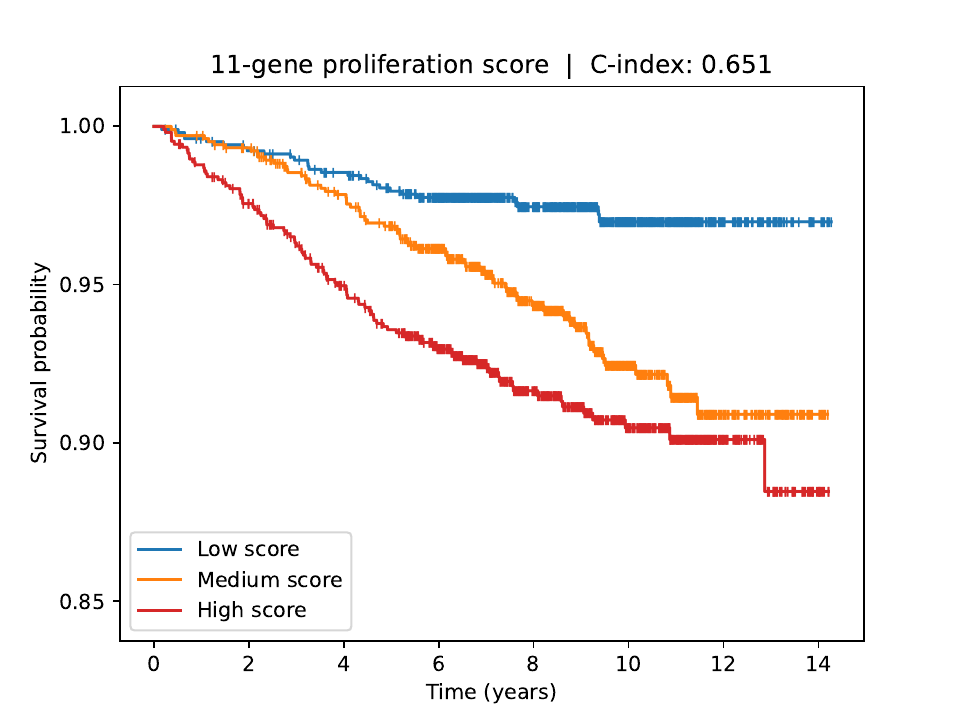}
        \caption{\textbf{11-gene proliferation score -- ER+ \& HER2-}.}
    \end{subfigure}
    \begin{subfigure}[t]{0.40\linewidth}
        \centering%
        \includegraphics[clip, trim=0.5cm 0.25cm 1.25cm 0.5cm, width=1.0\linewidth]{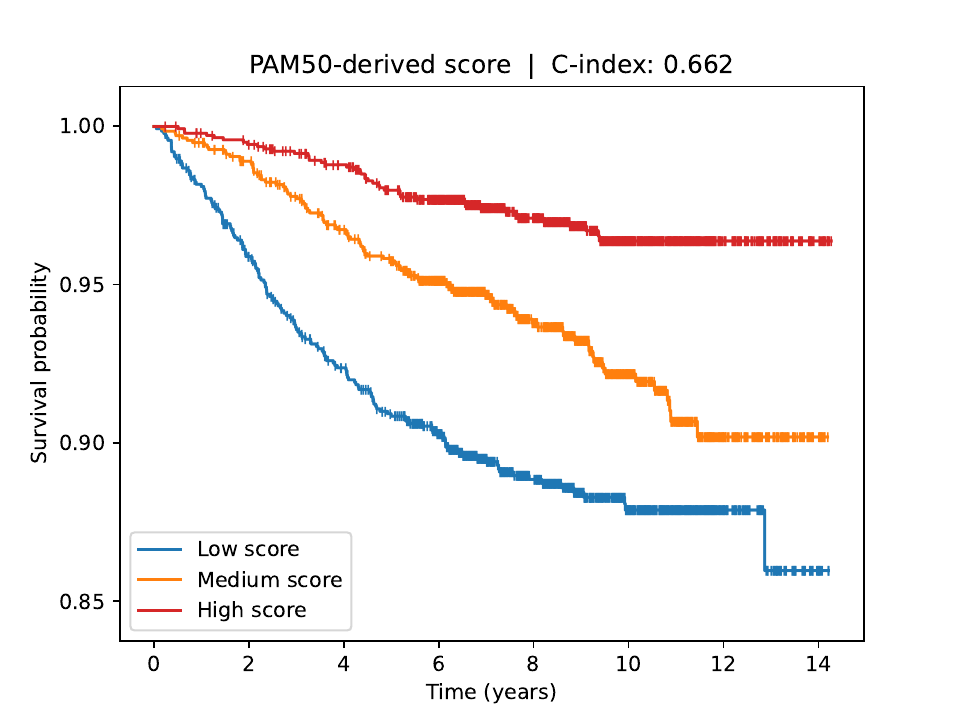}
        \caption{\textbf{PAM50-derived score -- Full cohort}.}
    \end{subfigure}
    \begin{subfigure}[t]{0.40\linewidth}
        \centering%
        \includegraphics[clip, trim=0.5cm 0.25cm 1.25cm 0.5cm, width=1.0\linewidth]{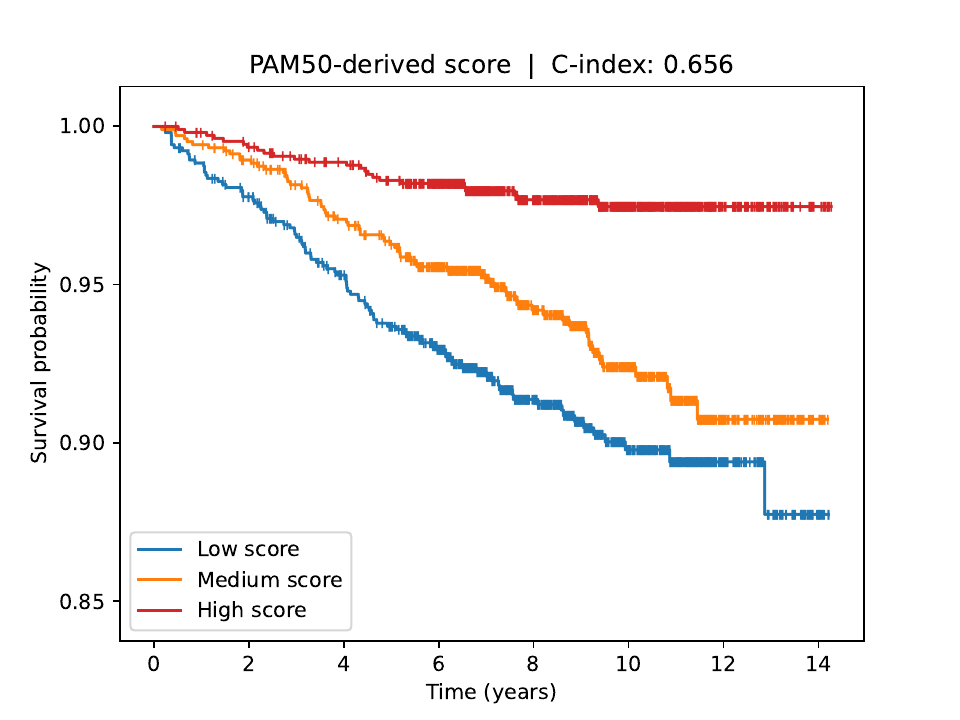}
        \caption{\textbf{PAM50-derived score -- ER+ \& HER2-}.}
    \end{subfigure}\vspace{-1.0mm}
  \caption{\textbf{Three-group Kaplan-Meier risk stratification.} KM survival curves for the 11-gene proliferation score (top) and the PAM50-derived score (bottom), evaluated on the full cohort (left) and the ER+ \& HER2- subgroup (right). Patients are stratified into three groups based on score tertiles. Corresponding plots including the number of patients at risk and events over time ($0$-$14$ years) are in Figure~\ref{fig:km_plots_3groups_counts}.}
  \label{fig:km_plots_3groups}
\end{figure*}

\begin{figure*}[t]
\centering
    \begin{subfigure}[t]{0.40\linewidth}
        \centering%
        \includegraphics[clip, trim=0.5cm 0.25cm 1.25cm 0.5cm, width=1.0\linewidth]{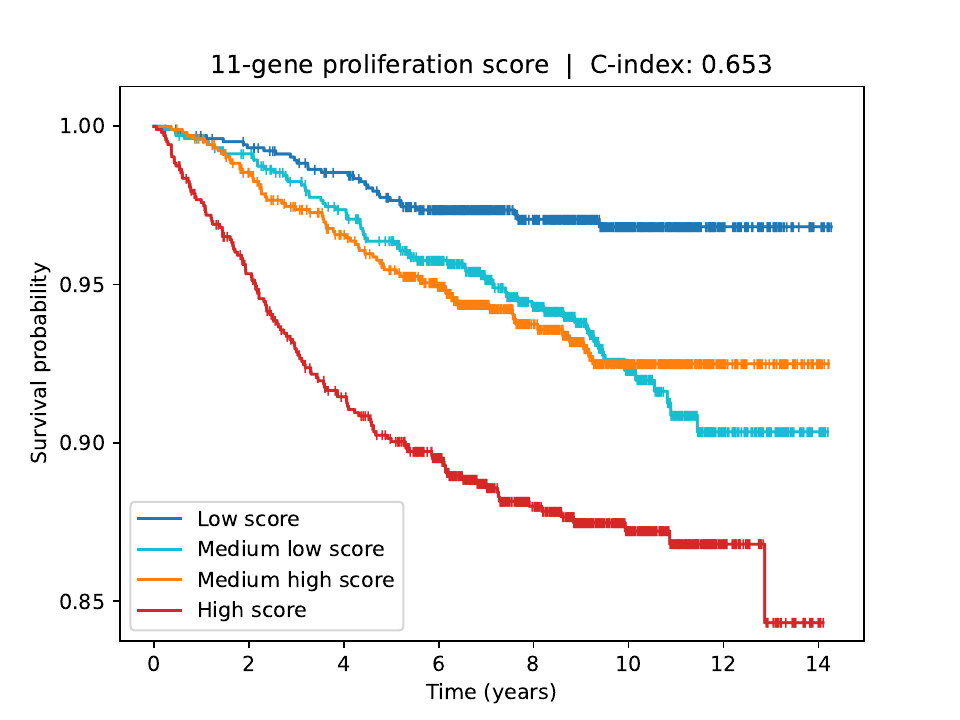}
        \caption{\textbf{11-gene proliferation score -- Full cohort}.}
    \end{subfigure}
    \begin{subfigure}[t]{0.40\linewidth}
        \centering%
        \includegraphics[clip, trim=0.5cm 0.25cm 1.25cm 0.5cm, width=1.0\linewidth]{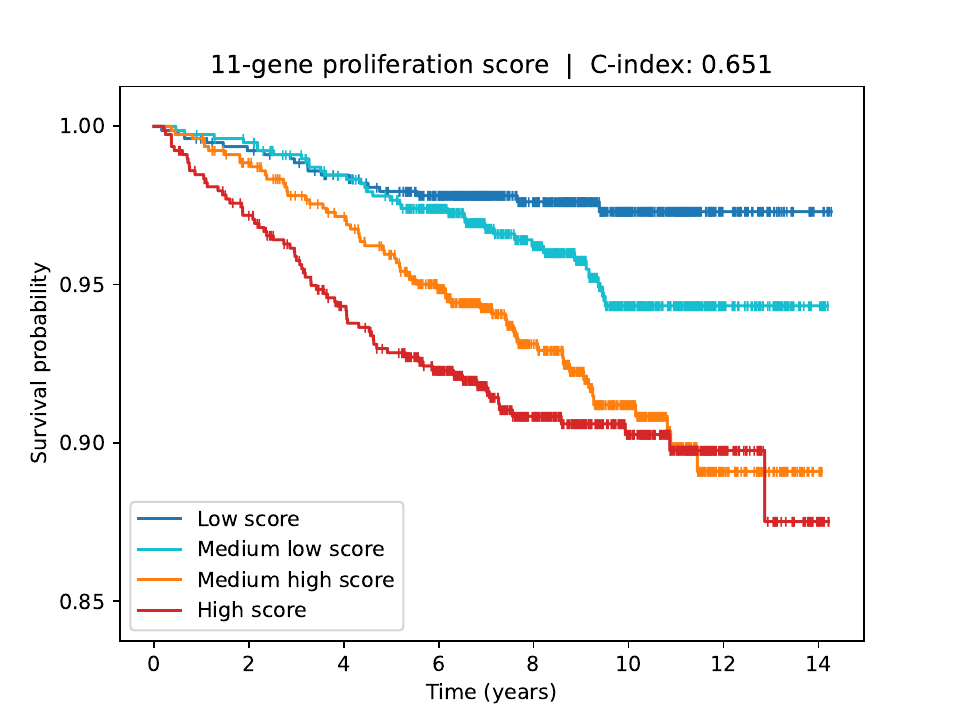}
        \caption{\textbf{11-gene proliferation score -- ER+ \& HER2-}.}
    \end{subfigure}
    \begin{subfigure}[t]{0.40\linewidth}
        \centering%
        \includegraphics[clip, trim=0.5cm 0.25cm 1.25cm 0.5cm, width=1.0\linewidth]{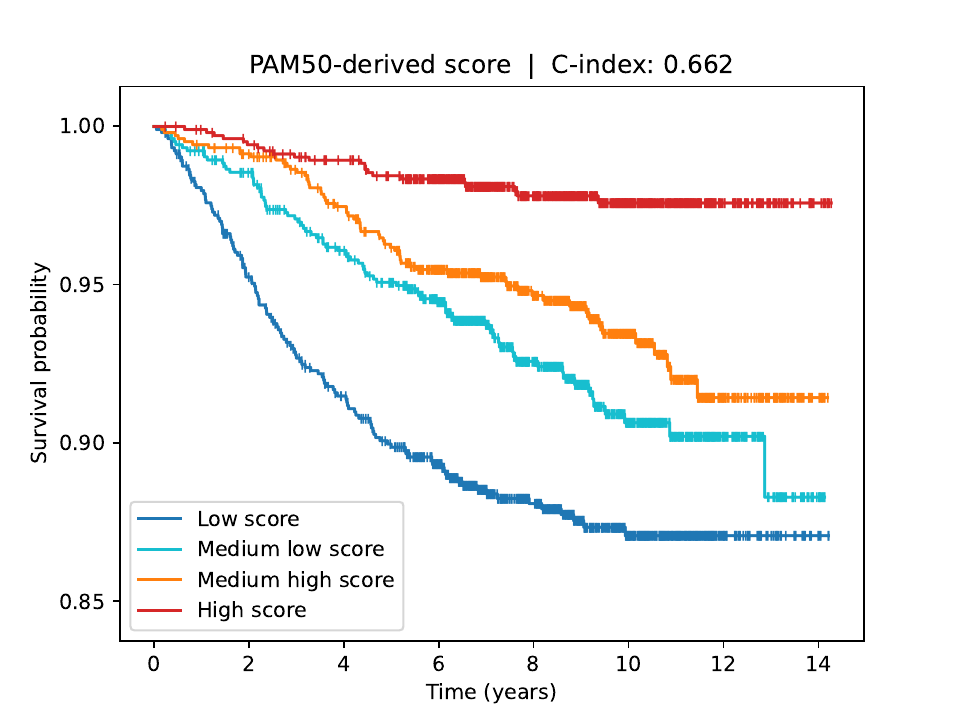}
        \caption{\textbf{PAM50-derived score -- Full cohort}.}
    \end{subfigure}
    \begin{subfigure}[t]{0.40\linewidth}
        \centering%
        \includegraphics[clip, trim=0.5cm 0.25cm 1.25cm 0.5cm, width=1.0\linewidth]{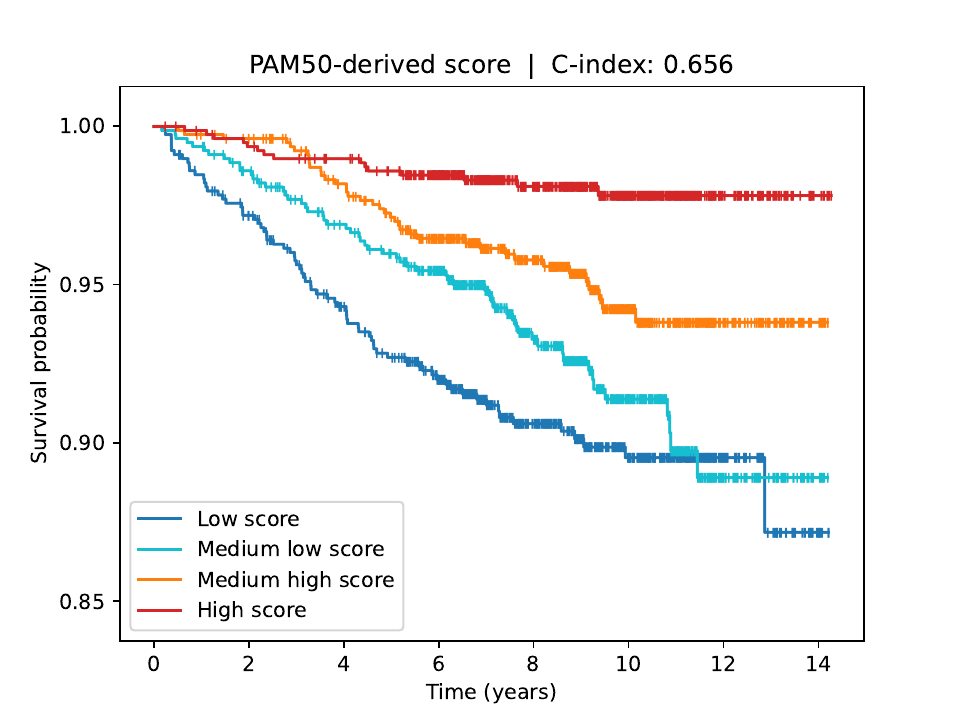}
        \caption{\textbf{PAM50-derived score -- ER+ \& HER2-}.}
    \end{subfigure}\vspace{-1.0mm}
  \caption{\textbf{Four-group Kaplan-Meier risk stratification.} KM survival curves for the 11-gene proliferation score (top) and the PAM50-derived score (bottom), evaluated on the full cohort (left) and the ER+ \& HER2- subgroup (right). Patients are stratified into four groups based on score quartiles. Corresponding plots including the number of patients at risk and events over time ($0$-$14$ years) are in Figure~\ref{fig:km_plots_4groups_counts}.}
  \label{fig:km_plots_4groups}
\end{figure*}

Importantly, the survival cohorts used for prognostic validation (SöS-BC-4, KS-Solna) do not have matched gene-expression data, reflecting a common real-world scenario in which transcriptomic profiling is not routinely available. This setting highlights the potential practical utility of WSI-based gene-expression prediction, as it enables molecular risk stratification in cohorts where direct measurement is infeasible. We emphasize that these analyses are intended to assess prognostic association and risk stratification, rather than to establish clinical utility beyond standard clinicopathological variables.

We first evaluate the established and externally defined 11-gene proliferation score, using the corresponding predicted gene-expression values from the model. Lower values of this score are associated with better prognosis. When computed from WSI-derived predictions, the 11-gene proliferation score achieves a C-index of $0.653$ in the full cohort and $0.651$ in the ER+ \& HER2- subgroup. Two-group Kaplan-Meier (KM) analysis shows clear survival stratification in both settings (Figure~\ref{fig:km_plots_2groups}), with log-rank $p=7.3\times10^{-9}$ in the full cohort and $p=1.4\times10^{-9}$ in the ER+ \& HER2- subgroup. The corresponding three-group Kaplan-Meier curves also show clear separation in both settings (Figure~\ref{fig:km_plots_3groups}~\&~\ref{fig:km_plots_3groups_counts}). With four groups, the score continues to stratify patients, although the separation is less distinct, particularly in the ER+ \& HER2- subgroup (Figure~\ref{fig:km_plots_4groups}~\&~\ref{fig:km_plots_4groups_counts}). Together, these results show that an established proliferation-based prognostic signature can be partially recovered from predicted, rather than directly measured, expression values.

As a secondary and exploratory analysis, we next consider a score derived from the full set of predicted PAM50 genes. SCAN-B-Lund serves here as a score-derivation cohort. Specifically, we define a candidate prognostic score based on PAM50 genes that show both sufficiently strong external prediction accuracy on SCAN-B-Lund (Spearman correlation $\geq 0.4$) and prognostic association with PFS in the same cohort (C-index $\geq 0.65$ when the predicted expression value of each gene is used as a univariate risk score). This yields seven genes for which higher predicted values are associated with better prognosis (\texttt{BCL2}, \texttt{ESR1}, \texttt{FOXA1}, \texttt{MAPT}, \texttt{NAT1}, \texttt{PGR}, \texttt{SLC39A6}) and ten genes for which lower predicted values are associated with better prognosis (\texttt{BIRC5}, \texttt{CCNE1}, \texttt{CDC20}, \texttt{CEP55}, \texttt{EXO1}, \texttt{KIF2C}, \texttt{MYBL2}, \texttt{PTTG1}, \texttt{RRM2}, \texttt{UBE2C}). The final score is defined as the mean normalized predicted value across the seven ``high value = good'' genes minus the corresponding mean across the ten ``low value = good'' genes, using gene-wise normalization based on the TCGA-BRCA training expression data. 

This PAM50-derived score shows similar prognostic performance on the combined SöS-BC-4 and KS-Solna dataset, achieving a C-index of $0.662$ in the full cohort and $0.656$ in the ER+ \& HER2- subgroup. Two-group Kaplan-Meier analysis again shows clear survival stratification in both settings, with log-rank $p=7.3\times10^{-14}$ in the full cohort and $p=1.0\times10^{-9}$ in the ER+ \& HER2- subgroup (Figure~\ref{fig:km_plots_2groups}). Three-group Kaplan-Meier curves also show clear separation for both the full cohort and the ER+ \& HER2- subgroup (Figure~\ref{fig:km_plots_3groups}~\&~\ref{fig:km_plots_3groups_counts}). With four groups, the PAM50-derived score shows particularly clear and approximately monotonic separation in the full cohort, whereas separation in the ER+ \& HER2- subgroup is somewhat less distinct (Figure~\ref{fig:km_plots_4groups}~\&~\ref{fig:km_plots_4groups_counts}). Overall, the risk stratification appears somewhat clearer for the PAM50-derived score than for the 11-gene proliferation score, while separation is generally less pronounced in the ER+ \& HER2- subgroup than in the full cohort. Importantly, SCAN-B-Lund is used only for score construction, whereas all prognostic evaluation of the resulting PAM50-derived score is performed on the independent external cohorts SöS-BC-4 and KS-Solna.

Overall, these results demonstrate that predicted gene-expression scores derived directly from WSIs can retain clinically meaningful prognostic information in independent external cohorts. Importantly, this is observed both for an established proliferation-based score and for a second score derived from predicted PAM50 genes, indicating that the learned WSI-based transcriptomic signal may support multiple clinically relevant forms of risk stratification.

\section*{Discussion}
In this study, we present a systematic evaluation of deep regression models for WSI-based gene-expression prediction, together with external validation and downstream prognostic assessment in breast cancer. Several key insights emerge from our results.

First, among the evaluated regression formulations, \textit{Direct - ABMIL} provides the strongest and most consistent overall performance. It slightly but consistently outperforms the simpler \textit{Direct - Patch-Level} model, indicating that a trainable WSI-level aggregation step is beneficial for this task. \textit{Contrastive} performs competitively, but does not provide a clear and consistent advantage over direct regression. At the same time, the simple \textit{kNN} baseline performs substantially worse across datasets. Taken together, these results suggest that direct transcriptome-wide regression with an ABMIL model constitutes a strong and practical baseline for WSI-based gene-expression prediction.

Second, our results show that the high dimensionality of the transcriptomic output space does not imply that many separate regression models are needed. Across multiple TCGA datasets, separately training multiple models on subsets of genes yields at most modest gains, with no consistent benefit over a single model trained jointly on all $N\!=\!20{,}530$ genes. In particular, it is clearly suboptimal to train one model per gene. Instead, performance improves as more genes are regressed jointly, before eventually saturating. While we do not directly study the mechanism behind this comparatively low performance of single-gene models, one plausible explanation is that jointly regressing larger sets of correlated genes may reduce the impact of noise in individual expression measurements and provide a more stable optimization signal. This is an interesting direction for future work, both for understanding the underlying optimization dynamics and for designing more efficient transcriptomic prediction models. From a practical perspective, a single transcriptome-wide model is therefore appealing not only because it is simpler, but also because it remains highly competitive in predictive accuracy. Accordingly, we recommend starting with such a model and only exploring more complex multi-model decompositions if clear gains are observed on the dataset of interest.

Third, the choice of patch-level feature extractor has a substantial impact on downstream gene-expression prediction performance. All pathology-specific encoders outperform the natural-image baseline Resnet-IN by a large margin, emphasizing the importance of pathology-relevant pretraining for this task. Among the evaluated models, \textit{H-optimus-1} provides the strongest results on TCGA-BRCA. More broadly, the results are consistent with a generational improvement pattern across PFMs, where newer and larger pathology encoders tend to outperform earlier models. At the same time, the absolute differences between several of the top-performing PFMs remain modest, suggesting that once a sufficiently strong pathology-specific representation is available, downstream modeling choices and evaluation setting remain important determinants of final performance. This interpretation is also consistent with the broader pattern observed in recent PFM benchmarking studies, where absolute gaps between top models are often limited. 

Fourth, external evaluation on SCAN-B-Lund shows that a substantial part of the predictive signal learned on TCGA-BRCA transfers to an independent breast cancer cohort collected under different clinical and technical conditions. While the mean correlation across all $17{,}066$ evaluated genes is moderate, a large number of genes remain well predicted externally, and performance is particularly strong for PAM50 and the 11-gene proliferation score. This pattern is encouraging, since it suggests that the most clinically relevant transcriptional programs may be among the most robustly predictable from morphology. The strong correlation of the predicted 11-gene proliferation score on SCAN-B-Lund further indicates that WSI-derived transcriptomic prediction can preserve not only gene-level signal, but also higher-level composite signatures with established biological and clinical meaning.

Finally, the prognostic analyses show that these predicted gene-expression signals are not merely correlated with measured RNA values, but also retain clinically meaningful structure in large independent survival cohorts. This is particularly relevant given that transcriptomic profiling is often unavailable in many clinical settings. While measured gene-expression remains the gold standard, its routine use is limited by cost and turnaround time. In contrast, WSIs are ubiquitously collected as part of standard care. The goal of WSI-based gene-expression prediction is therefore not to replace transcriptomic assays where they are already available, but to extend access to molecular information in settings where such assays are not performed.

Specifically, we find that the predicted 11-gene proliferation score achieves consistent prognostic performance across both the full cohort and the ER+ \& HER2- subgroup, and that a second score derived from predicted PAM50 genes also shows clear prognostic value. The Kaplan-Meier analyses further support this interpretation: both scores produce clear separation in the two-group and three-group settings, while the PAM50-derived score shows somewhat clearer and more monotonic stratification overall. At the same time, survival separation is generally less distinct in the ER+ \& HER2- subgroup than in the full cohort. Importantly, we do not interpret this primarily as evidence for a new prognostic signature. Rather, the key point is that multiple different score constructions based on WSI-derived predicted expression retain meaningful associations with patient outcome. This strengthens the broader conclusion that histopathology-based transcriptomic prediction can recover clinically relevant molecular information, and not only gene-wise statistical correlations. 

Overall, our findings highlight both the promise and the current limitations of WSI-based gene-expression prediction. On the positive side, we identify a strong and computationally efficient modeling strategy, show that recent PFMs materially improve performance, demonstrate external generalization to an independent breast cancer cohort, and provide evidence that predicted transcriptomic signals retain downstream prognostic value. At the same time, transcriptome-wide prediction accuracy remains moderate on average, and external validation shows that only a subset of genes transfer robustly across cohorts. Furthermore, while predicted expression captures meaningful relative structure between patients, the absolute scale and calibration of predictions across datasets remain less well understood. Taken together, these results support WSI-based gene-expression prediction as a promising and scalable approach for molecular phenotyping and risk stratification, while also indicating that further work is needed to improve robustness across datasets, better understand which signals are most reliably recoverable from morphology, and determine how such predictions can best be integrated into downstream clinical use.

% Limitations / future work:
This study has several limitations. First, although we evaluate external generalization and prognostic utility across multiple independent cohorts, all external datasets are breast cancer cohorts from similar healthcare settings within a single country (Sweden), and may therefore not capture the full variability encountered across different institutions, scanners, staining protocols, or international clinical workflows. Second, while our comparison of regression formulations is broad, all experiments are conducted within a frozen-feature pipeline, and we therefore do not assess whether the relative performance of the evaluated methods would change under end-to-end fine-tuning or alternative downstream training strategies. Third, the transcriptomic regression task is evaluated through correlation-based metrics, which are well suited for assessing gene-wise agreement and preservation of patient-level rank ordering, but do not capture the absolute scale of predicted expression values. In particular, predictions may preserve relative ordering between patients while exhibiting systematic shifts or distortions in scale due to differences in staining protocols or whole-slide scanner devices. This aspect is not explicitly evaluated in the present work, nor do we assess gene- or patient-level estimates of prediction uncertainty. Fourth, the prognostic analyses are intentionally focused on demonstrating that predicted gene-expression signals retain clinical meaning, rather than on building an optimized risk model. As such, we do not yet evaluate the added value of these scores beyond established clinicopathological covariates. Finally, the present work focuses on breast cancer, and it therefore remains unclear to what extent these findings generalize to other cancer types, molecular phenotypes, and clinical settings, where the relationship between morphology and gene-expression may differ.

Several directions for future work follow naturally from these findings. One important direction is to evaluate whether the same conclusions hold across a broader range of cancer types and transcriptomic endpoints. A second direction is to investigate whether alternative downstream modeling strategies, including end-to-end fine-tuning or more explicit modeling of gene dependencies, can yield meaningful gains beyond the strong frozen-feature baselines identified here, particularly given that joint prediction of multiple genes consistently outperforms training separate models for individual genes. A third direction is to move beyond gene-wise correlation metrics by explicitly evaluating calibration across cohorts, robustness of absolute expression scales, uncertainty estimation, and performance on additional downstream composite signatures and clinically relevant endpoints. For prognostic applications in particular, it would be valuable to study multivariable models that combine WSI-derived predicted expression with standard clinicopathological variables, and to determine whether these signals provide incremental value beyond existing risk stratification tools.

\emph{The main takeaways from our study are:}
(1) \textit{Direct~-~ABMIL} is a strong and practical baseline for WSI-based gene-expression prediction, consistently outperforming simpler aggregation strategies, while \textit{Contrastive} does not provide a clear or consistent advantage over direct regression.
(2) Training a single model to regress all $N\!=\!20{,}530$ genes is both computationally efficient and highly competitive; training separate models for smaller gene subsets, or even one model per gene, does not provide a consistent benefit. In practice, a single transcriptome-wide model is a strong starting point for new datasets.
(3) Pathology-specific foundation models are important for strong performance, with \textit{H-optimus-1} achieving the best results in our benchmark, although absolute differences between several recent PFMs are modest.
(4) A substantial part of the learned transcriptomic signal transfers externally from TCGA-BRCA to SCAN-B-Lund, with particularly strong predictive performance for clinically relevant gene sets such as PAM50 (mean Pearson correlation of $0.443$).
(5) Predicted gene-expression scores derived directly from WSIs retain meaningful prognostic information in fully independent external survival cohorts (C-index values above $0.65$), indicating that WSI-based transcriptomic prediction captures clinically relevant molecular structure rather than only gene-wise statistical associations.
\section*{Methods}

We conduct a methodological and translational evaluation of WSI-based gene-expression prediction. First, we benchmark alternative regression formulations across four TCGA datasets using site-aware cross-validation. Next, we compare a broad set of patch-level feature extractors on TCGA-BRCA. We then externally validate the selected gene-expression model on SCAN-B-Lund, an independent cohort with matched WSIs and gene-expression measurements. Finally, we evaluate whether predicted gene-expression scores retain prognostic value in the two independent breast cancer survival cohorts SöS-BC-4 and KS-Solna.

\begin{figure*}[t]
    \centering%
    \includegraphics[width=1.0\linewidth]{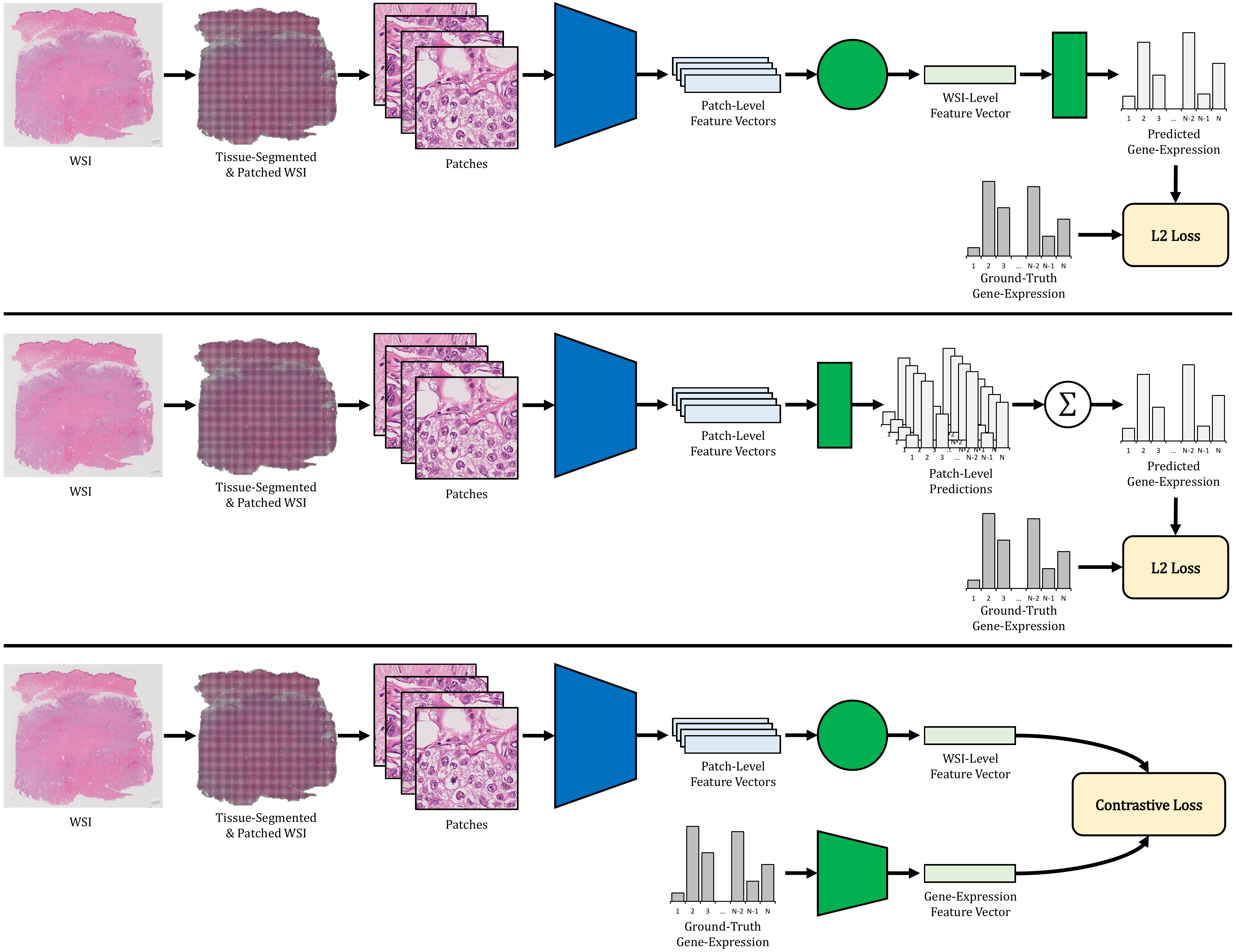}\vspace{-1.0mm}
    \caption{\textbf{Overview of the three main evaluated regression models.} \textbf{Top:} \textit{Direct - ABMIL}. \textbf{Middle:} \textit{Direct - Patch-Level}. \textbf{Bottom:} \textit{Contrastive}. All models use the same WSI preprocessing pipeline. First, the WSI $x$ is tissue-segmented and divided into non-overlapping patches $\tilde{x}_i$ of size $256 \times 256$. Next, a feature vector $p(\tilde{x}_i)$ is extracted for each patch using a pretrained and frozen feature extractor. The models then process these patch-level feature vectors further, outputting a predicted gene-expression profile $\hat{y}(x) \in \mathbb{R}^N$ for all $N=20{,}530$ genes. In the figure, \textcolor{Cerulean}{\textbf{blue}} denotes the pretrained and frozen feature extractor, whereas \textcolor{ForestGreen}{\textbf{green}} denotes trainable model components.}
    \label{fig:model_overview_abmil_patch-level_contrastive}
\end{figure*}

\subsubsection*{Datasets \& Experimental Setup}

\paragraph{TCGA Datasets}
For the initial benchmarking of deep regression models for gene-expression prediction, we use four TCGA datasets: breast invasive carcinoma (TCGA-BRCA), head-neck squamous cell carcinoma (TCGA-HNSC), stomach adenocarcinoma (TCGA-STAD), and urothelial bladder carcinoma (TCGA-BLCA).

We match diagnostic H\&E WSIs with gene-expression data from UCSC Xena~\cite{goldman2020visualizing}. Specifically, we use the \emph{gene expression RNAseq - IlluminaHiSeq} data, containing gene-level transcription estimates for $N=20{,}530$ genes. This results in $1{,}129$ total WSIs with matched gene-expression data for TCGA-BRCA, $464$ WSIs for TCGA-HNSC, $391$ WSIs for TCGA-STAD, and $452$ WSIs for TCGA-BLCA. 

All models are trained and evaluated using 5-fold \emph{site-aware} cross-validation, such that models are never trained and evaluated on samples from the same TCGA collection site~\cite{howard2021impact}. This evaluation protocol is intended to give a fairer estimate of generalization performance than standard random cross-validation.

% (SCAN-B-Lund-survival2025 (survival data updated in 2025))
\paragraph{SCAN-B-Lund}
This dataset is used for external validation of predictive accuracy for the selected gene-expression model, and to construct the PAM50-derived risk score used in the prognostic validation. SCAN-B-Lund is a subset of $1{,}262$ patients enrolled in the prospective SCAN-B study~\citep{vallon2019cross}, diagnosed between 2010 and 2019 in Lund, Sweden~\citep{sharma2024development, sharma2024validation}. Of these, $997$ patients have corresponding gene-expression measurements for $19{,}675$ genes. 

For all external gene-expression and survival evaluations, we use the five models obtained from the 5-fold site-aware cross-validation on TCGA-BRCA as an ensemble, and apply this ensemble directly to the external datasets without any retraining. For each patient and gene, the final predicted expression value is computed as the arithmetic mean of the five fold-specific model predictions. In total, $17{,}055$ genes have exactly matching names between the TCGA-BRCA and SCAN-B-Lund expression data. We additionally manually match 11 genes from clinically relevant breast cancer gene sets, resulting in $17{,}066$ evaluated genes in total.

Clinical follow-up data for SCAN-B-Lund was updated in 2025. The dataset includes $88$ PFS events and has a mean follow-up time of $8.1$ years. PFS is defined as the time from initial diagnosis to disease recurrence, including local recurrence, distant metastasis, or detection of contralateral tumors. Death without documented recurrence is not counted as an event and is treated as censoring. Patients without an event are censored at the date of last follow-up.

\paragraph{External Survival Datasets}
Prognostic validation of predicted risk scores is performed on two independent external breast cancer survival cohorts, SöS-BC-4 and KS-Solna, comprising $4{,}172$ patients in total. The TCGA-BRCA ensemble is also applied in this external evaluation. We evaluate performance both for the full cohort and the clinically relevant subgroup of patients which are oestrogen receptor (ER)-positive and human epidermal growth factor receptor 2 (HER2)-negative (``ER+ \& HER2-''). This ER+ \& HER2- subgroup represents the most common breast cancer subtype and has distinct biological and clinical characteristics, making it particularly relevant for prognostic modeling. The full cohort contains $289$ PFS events, while the ER+ \& HER2- subgroup contains $3{,}157$ patients and $185$ PFS events.

% (SöS-BC-4-survival2025 (survival data updated in 2025))
\parsection{SöS-BC-4}
This is a retrospective observational cohort including patients diagnosed at Södersjukhuset (South General Hospital) in Stockholm, Sweden between 2012 and 2018~\citep{wang2022improved, sharma2024development}. We use a subset of $2{,}315$ patients with available follow-up data, including $144$ PFS events and a mean follow-up time of $7.7$ years. The clinical outcome data was retrieved from the Swedish National Registry for Breast Cancer (NKBC) in 2025.

% (CHIME-Breast-KS-Solna-survival2025 (survival data updated in 2025))
\parsection{KS-Solna}
CHIME breast KS-Solna is a retrospective cohort of primary breast cancer patients treated at the Karolinska University Hospital in Stockholm, diagnosed between 2009 and 2018~\citep{sharma2024validation}. The clinical outcome data was retrieved from NKBC in 2025. We use a subset of $1{,}857$ patients with available follow-up data, including $145$ PFS events and a mean follow-up time of $9.1$ years.

\subsubsection*{Model Overview \& WSI Preprocessing}
\begin{figure}[t]
    \centering%
    \includegraphics[width=1.0\linewidth]{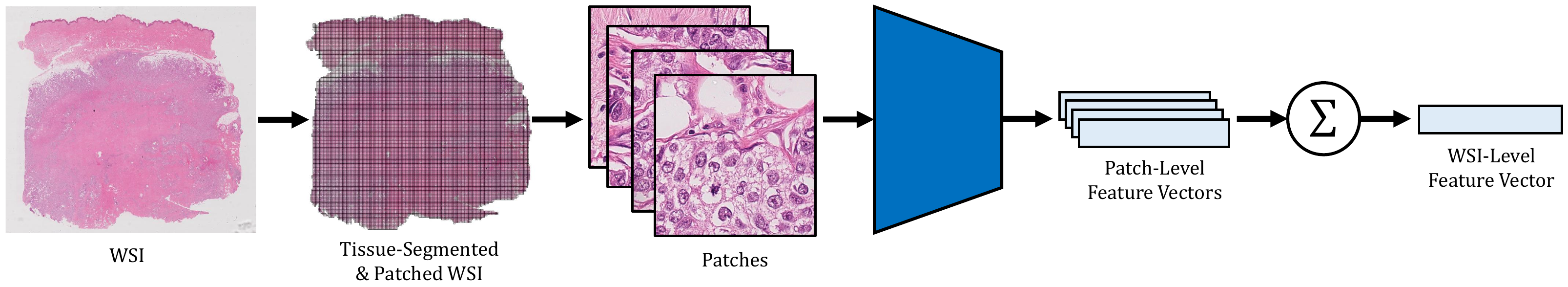}\vspace{-1.0mm}
    \caption{\textbf{Overview of the kNN baseline model}, which contains no trainable model parameters. For the input WSI $x$, a WSI-level feature vector $w(x)$ is directly computed as the mean over the patch-level feature vectors $p(\tilde{x}_i)$. kNN with $k = 100$ is then used to output a predicted gene-expression profile $\hat{y}(x) \in \mathbb{R}^N$.}
    \label{fig:model_overview_knn}
\end{figure}

We evaluate three main different types of regression models (Figure~\ref{fig:model_overview_abmil_patch-level_contrastive}), together with a simple kNN baseline (Figure~\ref{fig:model_overview_knn}). All four models take a WSI $x$ as input and output a predicted gene-expression profile $\hat{y}(x) \in \mathbb{R}^N$. Models thus output predicted values for all $N = 20{,}530$ genes, making this an extremely high-dimensional regression problem.

All four models use the same initial WSI preprocessing steps. First, the input WSI $x$ is tissue-segmented and divided into $P$ non-overlapping patches $\{\tilde{x}_i\}_{i=1}^P$ of size $256 \times 256$ at $20\times$ magnification. The number of extracted tissue patches $P$ varies across WSIs, with a typical range of approximately $100$--$10{,}000$ per slide. Next, a feature vector $p(\tilde{x}_i)$ is extracted for each tissue patch $\tilde{x}_i$ using a pretrained and frozen feature extractor. The four models then process these patch-level feature vectors $\{p(\tilde{x}_i)\}_{i=1}^P$ further, finally outputting a predicted gene-expression profile $\hat{y}(x) \in \mathbb{R}^N$ for the full WSI.

In the benchmarking of regression models on the four TCGA datasets, we implement the WSI preprocessing using CLAM~\citep{lu2021data}. For the PFM comparison on TCGA-BRCA and all external evaluations, we use an in-house preprocessing workflow to enable a unified pipeline across datasets and the broader set of evaluated feature extractors. This workflow consists of four steps: tissue segmentation using Otsu's thresholding~\citep{otsu1979}; extraction of non-overlapping $256 \times 256$ patches at a resolution of $0.4536\,\mu$m/pixel (corresponding to $20\times$ equivalent magnification for a reference slide scanner); removal of blurry patches using the variance of Laplacian (VL) metric~\citep{pech2000diatom}, discarding patches with $\mathrm{VL} < 300$; and color normalization using the Macenko method~\citep{macenko2009method}, adapted for WSI-level color correction following \citet{wang2022improved}.

\subsubsection*{Direct Regression Models}
We evaluate two direct regression models, in which trainable networks are optimized to directly output a predicted gene-expression profile $\hat{y}(x) \in \mathbb{R}^N$ from the patch-level feature vectors. These models are trained by minimizing the squared error $\|y-\hat{y}(x)\|^2$ (L2 loss) between predicted and observed gene-expression profiles (see Figure~\ref{fig:extra_results_losses} for a performance comparison with alternative loss functions).

\paragraph{ABMIL}
The first direct regression model, \textit{Direct - ABMIL} (Figure~\ref{fig:model_overview_abmil_patch-level_contrastive}, top), uses attention-based multiple instance learning (ABMIL)~\cite{ilse2018attention} to aggregate the set of patch-level feature vectors $\{p(\tilde{x}_i)\}_{i=1}^{P}$ into a single WSI-level feature vector $w(x)$. This feature vector $w(x)$ is then fed as input to a small network head of two layers, finally outputting $\hat{y}(x) \in \mathbb{R}^N$. Both the ABMIL aggregator and the network head are trained using the L2 loss.

Our ABMIL implementation is based on CLAM~\cite{lu2021data} without instance-level clustering, and we set the model and training hyperparameters according to UNI (see \textit{Methods - Weakly supervised slide classification} in \citep{chen2024uni}). Specifically, models are trained using the AdamW optimizer~\cite{loshchilov2018decoupled} with a cosine learning rate schedule for a maximum of 20 epochs. For each of the five cross-validation folds, the training portion is split further into train and validation subsets using a random 90\%/10\% split, and early stopping is based on the validation loss.

\parsection{Sequential Chunking} 
Because the regression problem is extremely high-dimensional ($N=20{,}530$ genes), we also conduct experiments where multiple \textit{Direct - ABMIL} models are used to output a full predicted gene-expression profile $\hat{y}(x) \in \mathbb{R}^N$ for the given WSI~$x$. In the \textit{sequential chunking} setup, the $N=20{,}530$ genes are divided into contiguous subsets, and one model is trained per subset. If using five models, for example, we separately train one \textit{Direct - ABMIL} model that regresses the first $20530/5 = 4106$ genes, one model that regresses the next $4106$ genes, and similarly for the three other models. The only modification of the \textit{Direct - ABMIL} model is to change the output dimension of the network head from $20{,}530$ to $4{,}106$. In the extreme case, $N=20{,}530$ models could be trained (at a very high computational cost), each regressing a single gene. 

\parsection{Clustering}
As an alternative decomposition strategy, we also use k-means clustering on the corresponding train split to group the $N=20{,}530$ genes into $k$ clusters of correlated genes. One \textit{Direct - ABMIL} model is then trained per cluster. At test time, predictions from the cluster-specific models are combined to form the full predicted gene-expression profile $\hat{y}(x) \in \mathbb{R}^N$. All gene grouping procedures are performed separately within each cross-validation fold, exclusively using training data to avoid information leakage.

\paragraph{Patch-Level}
The second direct regression model, \textit{Direct - Patch-Level} (Figure~\ref{fig:model_overview_abmil_patch-level_contrastive}, middle), removes the trainable ABMIL WSI-level aggregator. Instead, each patch-level feature vector $p(\tilde{x}_i)$ is directly fed as input to the small network head, outputting a predicted gene-expression profile $\hat{y}(\tilde{x}_i) \in \mathbb{R}^N$ for each tissue patch $\tilde{x}_i$ in the WSI $x$. Then, the mean over these patch-level predictions $\{\hat{y}(\tilde{x}_i)\}_{i=1}^P$ is computed, and output as the final WSI-level predicted gene-expression $\hat{y}(x) \in \mathbb{R}^N$. Compared to \textit{Direct - ABMIL}, this model contains fewer trainable parameters and provides access to patch-level predictions, which potentially could be utilized for spatial analysis.

\subsubsection*{Contrastive Learning-Based Model}
The third evaluated regression model, \textit{Contrastive} (Figure~\ref{fig:model_overview_abmil_patch-level_contrastive}, bottom), is conceptually quite different compared to the direct regression models. Instead of training networks to directly output a predicted gene-expression profile $\hat{y}(x) \in \mathbb{R}^N$ via the L2 loss, the contrastive learning-based model first trains networks to align WSI and gene-expression feature representations using a contrastive loss. Given a WSI $x$, a predicted gene-expression profile $\hat{y}(x) \in \mathbb{R}^N$ is then obtained by computing the similarity between the WSI representation and all gene-expression representations of the train set. 

Our contrastive learning-based model is a relatively straightforward extension of the TANGLE method~\cite{jaume2024transcriptomics} proposed for WSI representation learning, applying it to the gene-expression prediction task. It can also be considered an extension of previous work~\cite{xie2023spatially, min2024multimodal} utilizing contrastive learning for \emph{spatial} gene-expression prediction based on spatial transcriptomics datasets.

\paragraph{Model Architecture \& Training}
We use the TANGLE model~\cite{jaume2024transcriptomics}. As in \textit{Direct - ABMIL}, TANGLE uses an ABMIL model to aggregate patch-level feature vectors $\{p(\tilde{x}_i)\}_{i=1}^P$ into a single WSI-level feature vector $w(x)$. It also consists of a gene-expression encoder model, which takes a gene-expression profile $y \in \mathbb{R}^N$ as input and compresses it into a feature vector $g(y)$, matching the dimension of the WSI-level feature vector $w(x)$. Specifically, the gene-expression encoder is a 3-layer multilayer perceptron (MLP).

The ABMIL WSI-level aggregator and the gene-expression encoder are trained jointly using a symmetric variant of the commonly used contrastive objective~\cite{chen2020simple, radford2021learning}. The model is thus trained to align the WSI-level and gene-expression feature vectors for pairs $(x, y)$ of WSI $x$ and gene-expression profile $y$. Specifically, the model is trained using the AdamW optimizer~\cite{loshchilov2018decoupled} for a maximum of $100$ epochs, with a batch size of $64$. Early stopping is performed based on a smooth rank measure~\cite{garrido2023rankme} of a matrix containing all WSI-level feature vectors $\{w(x_i)\}_{i=1}^M$ from the train set.

\paragraph{Prediction}
We take inspiration from the prediction methods of \citet{xie2023spatially} and \citet{min2024multimodal}, adapted to our setting of \emph{WSI-level} prediction. Given a WSI $x$, we first extract a WSI-level feature vector $w(x)$ using the ABMIL aggregator. Then, we use the gene-expression encoder to obtain feature vectors $\{g(y_i)\}_{i=1}^M$ for all gene-expression profiles $y_i \in \mathbb{R}^N$ of the train set. Next, we compute the cosine similarity $d(w(x), g(y_i))$ between the WSI-level feature vector $w(x)$ and each gene-expression feature vector $g(y_i)$. Finally, a prediction $\hat{y}(x) \in \mathbb{R}^N$ for the WSI $x$ is computed as a weighted sum of the $K=100$ closest gene-expression profiles $\{y_k\}_{k=1}^K$,
\begin{equation}
    \hat{y}(x) = \sum_{k=1}^{K}  \left(\frac{\exp\left\{d(w(x), g(y_k))\right\}}{\sum_{j=1}^{K} \exp\left\{d(w(x), g(y_j))\right\}}\right) y_k.
\end{equation}
Thus, predictions $\hat{y}(x) \in \mathbb{R}^N$ output by the model are always linear combinations of observed gene-expression profiles $\{y_i\}_{i=1}^M$ from the train set.

\subsubsection*{kNN Baseline Model}
As a simple baseline without any trainable model parameters, we also evaluate a kNN-based model. For a given WSI $x$, a WSI-level feature vector $w(x)$ is directly computed as the mean over the patch-level feature vectors $\{p(\tilde{x}_i)\}_{i=1}^P$, as illustrated in Figure~\ref{fig:model_overview_knn}. To output a predicted gene-expression profile $\hat{y}(x) \in \mathbb{R}^N$, we then use \texttt{KNeighborsRegressor} from scikit-learn~\cite{scikit-learn} with $k = 100$. Thus, as for the \textit{Contrastive} model, predictions are always linear combinations of gene-expression profiles $\{y_i\}_{i=1}^M$ from the train set.

\subsubsection*{Patch-Level Feature Extractors}
In the initial benchmarking of regression models on TCGA datasets, we use UNI~\cite{chen2024uni} as the patch-level feature extractor across all four models. UNI is a widely used PFM, pretrained using DINOv2~\cite{oquab2024dinov2} on a pan-cancer dataset (20 major tissue types) of approximately $100$ million tissue patches from more than $100{,}000$ WSIs. It is a vision transformer (ViT-Large)~\cite{dosovitskiy2021an} model and outputs feature vectors of dimension $1024$.

In the PFM comparison on TCGA-BRCA, we evaluate UNI against eleven other feature extractors. These include a natural-image baseline (Resnet-IN), two early pathology-specific models (CTransPath~\citep{Wang2023ctranspath}, RetCCL~\citep{wang2023retccl}), six recent PFMs (Prov-GigaPath~\citep{gigapath2024}, UNI2-h~\citep{UNI2h2024}, Virchow~\citep{virchow2024}, Virchow2~\citep{virchow22024}, H-optimus-0~\citep{hoptimus0}, H-optimus-1~\citep{hoptimus1}), and two vision-language PFMs (CONCH~\citep{conch2024}, CONCHv1.5~\citep{ding2025multimodal}). Table~\ref{table:pfms_rank_info} summarizes the parameter count and pretraining datasets of all evaluated models.

Resnet-IN is a Resnet-50~\citep{he2016deep} pretrained on the ImageNet~\citep{russakovsky2015imagenet} dataset of natural images, serving as a simple reference baseline expected to be significantly outperformed by PFMs. CTransPath and RetCCL were pretrained on approximately $30{,}000$ WSIs, which is substantially smaller than the datasets used to train state-of-the-art PFMs. The six evaluated PFMs were pretrained on datasets ranging from approximately $170{,}000$ WSIs (Prov-GigaPath) to $3.1$ million WSIs (Virchow2). CONCH and CONCHv1.5 were pretrained using paired image-text supervision in addition to WSIs, representing an alternative training paradigm leveraging multimodal datasets.

\subsubsection*{Construction of Predicted Risk Scores}
We evaluate an established 11-gene proliferation score~\citep{parker2009supervised, nielsen2010comparison, veta2019predicting, ekholm2024prediction} defined as the mean gene-expression value of eleven PAM50 genes:
\texttt{BIRC5}, \texttt{CCNB1}, \texttt{CDC20}, \texttt{CEP55}, \texttt{MKI67}, \texttt{NDC80}, \texttt{NUF2}, \texttt{PTTG1}, \texttt{RRM2}, \texttt{TYMS}, and \texttt{UBE2C}. The corresponding predicted score is computed analogously as the mean of the predicted expression values for these genes. Lower values of this score are associated with better prognosis.

We also evaluate a second score derived from the full set of predicted PAM50 genes. To construct this score, we first identify PAM50 genes satisfying two criteria on SCAN-B-Lund: (i) Spearman correlation $\geq 0.4$ for external prediction accuracy, and (ii) C-index $\geq 0.65$ for PFS in survival analysis. For the latter, the predicted expression value of each gene is used as a univariate risk score, and the C-index is computed across all $1{,}262$ patients in SCAN-B-Lund.

This yields seven genes for which higher predicted values are associated with better prognosis:
\texttt{BCL2}, \texttt{ESR1}, \texttt{FOXA1}, \texttt{MAPT}, \texttt{NAT1}, \texttt{PGR}, and \texttt{SLC39A6};
and ten genes for which lower predicted values are associated with better prognosis:
\texttt{BIRC5}, \texttt{CCNE1}, \texttt{CDC20}, \texttt{CEP55}, \texttt{EXO1}, \texttt{KIF2C}, \texttt{MYBL2}, \texttt{PTTG1}, \texttt{RRM2}, and \texttt{UBE2C}. For each gene, we normalize the predicted value using the mean and standard deviation of that gene in the TCGA-BRCA training gene-expression data. We then compute the mean normalized predicted value across the seven ``high value = good'' genes and the mean normalized predicted value across the ten ``low value = good'' genes. The final score is defined as the former minus the latter. Higher values of the final score therefore correspond to lower predicted risk.

Importantly, SCAN-B-Lund is used only to identify the subset of genes included in this score. All prognostic evaluation of the resulting score is performed on the independent external cohorts SöS-BC-4 and KS-Solna.

\subsubsection*{Evaluation Metrics}
For the TCGA gene-expression benchmarking experiments, primary evaluation is performed using Pearson correlation. Specifically, for each gene we compute the Pearson correlation across patients between predicted and observed expression values, and summarize performance using: (i) mean Pearson correlation across all genes, (ii) mean Pearson correlation across the top $1{,}000$ best-predicted genes, and (iii) the number of genes with Pearson correlation $\geq 0.4$. Parallel analyses using Spearman correlation are also performed in supplementary experiments (Figure~\ref{fig:main_results_spearman}). For the external SCAN-B-Lund evaluation, both Pearson and Spearman correlation are reported, as this compares predictions across two cohorts with different gene-expression scales and measurement characteristics. We also report results for PAM50 and the 11-gene proliferation score.

For the prognostic validation, we use the concordance index (C-index)~\citep{harrell1982evaluating} as the primary performance metric, where higher C-index values indicate better alignment between predicted risk ordering and actual patient outcomes. We additionally visualize risk stratification using Kaplan-Meier (KM)~\citep{kaplan1958nonparametric} survival curves. For each predicted score, patients are stratified into risk groups using either a two-group split based on the median, three groups based on the score tertiles, or four groups based on the quartiles. Differences between groups are assessed using the log-rank test. We perform both KM survival analysis and C-index evaluation separately for the full patient cohort and the ER+ \& HER2- subgroup.

\subsubsection*{Ethics Statement}
The study has approval from the Swedish Ethical Review Authority (2017/2106-31, with amendments 2018/1462-32 and 2019–02336). The study was performed in accordance with the Declaration of Helsinki. No additional informed consent was required in accordance with ethical approval in this non-interventional collection and analysis of data from patient records.
\section*{Data Availability}

The whole-slide images for the four utilized TCGA datasets are available at the GDC Data Portal \url{https://portal.gdc.cancer.gov}, while the gene-expression data is available as \emph{gene expression RNAseq - IlluminaHiSeq} at UCSC Xena \url{https://xenabrowser.net/datapages/}. The splits used for 5-fold site-aware cross-validation on TCGA are available at \url{https://github.com/mahmoodlab/SurvPath}. The SöS-BC-4, KS-Solna, and SCAN-B-Lund datasets cannot be made publicly available due to restrictions relating to sensitive patient-related information.
\section*{Code Availability}

The code for this study is based on CLAM and TANGLE, which are available at \url{https://github.com/mahmoodlab/CLAM} and \url{https://github.com/mahmoodlab/TANGLE}. Further implementation details are available from FKG upon reasonable request.
\section*{Acknowledgments}

The project was supported by funding from the Swedish Cancer Society (23 2905 Pj 01 H), VINNOVA (SwAIPP2), Swedish e-science Research Centre (SeRC) (eMPHasis project), Bröstcancerförbundet, Swedish Research Council (2024–06634, 2025-03411), and the AID4BC consortium supported by a WASP/DDLS NEST grant (KAW 2024.0159) from the Knut \& Alice Wallenberg Foundation to SciLifeLab for research in Data-Driven Life Science (DDLS) and the Wallenberg AI, Autonomous Systems and Software Program (WASP). 

The authors acknowledge patients, clinicians, and hospital staff participating in the SCAN-B study; the staff at the central SCAN-B laboratory at the Division of Oncology, Lund University; the Swedish National Breast Cancer Quality Registry (NKBC); the Regional Cancer Center South; and the South Swedish Breast Cancer Group (SSBCG). SCAN-B was funded by the Swedish Cancer Society, the Mrs. Berta Kamprad Foundation, the Lund-Lausanne L2-Bridge/Biltema Foundation, the Mats Paulsson Foundation, and Swedish governmental funding (ALF). 

The results shown here are in whole or part based upon data generated by the TCGA Research Network: \url{https://www.cancer.gov/tcga}. 

\section*{Author Contributions}

FKG was responsible for project conceptualization, software implementation, preparation of figures and tables, and manuscript drafting. CB processed the survival datasets. JVC supported the use of the SCAN-B-Lund dataset. MR supervised the project, contributed to the design of experiments and interpretation of results, and acquired funding. All authors contributed to manuscript revision and finalization.
\section*{Competing Interests}

MR is a co-founder and shareholder of Stratipath AB. All other authors declare no competing interests.
\section*{Declaration of Generative AI Use}

During the preparation of this manuscript, the authors used ChatGPT 5.3 to assist with language editing and drafting. All content produced using this tool was critically reviewed, edited and validated by the authors, who take full responsibility for the final content of the manuscript.

{
\small
\bibliographystyle{plainnat}
\bibliography{references}
}

\clearpage
\appendix
\onecolumn

\renewcommand{\thefigure}{S\arabic{figure}}
\setcounter{figure}{0}

\renewcommand{\thetable}{S\arabic{table}}
\setcounter{table}{0}

\renewcommand{\theequation}{S\arabic{equation}}
\setcounter{equation}{0}

\subsection*{\centering{Evaluation and Prognostic Validation of Deep Regression\\ Models for WSI-Based Gene-Expression Prediction}}
\section*{\centering{Supplementary Material}}

\vspace{6.0mm}

\section{Supplementary Figures}
\label{appendix:figures}

\begin{figure*}[h]
\centering
    \begin{subfigure}[t]{0.29\textwidth}
        \centering%
        \includegraphics[width=1.0\linewidth]{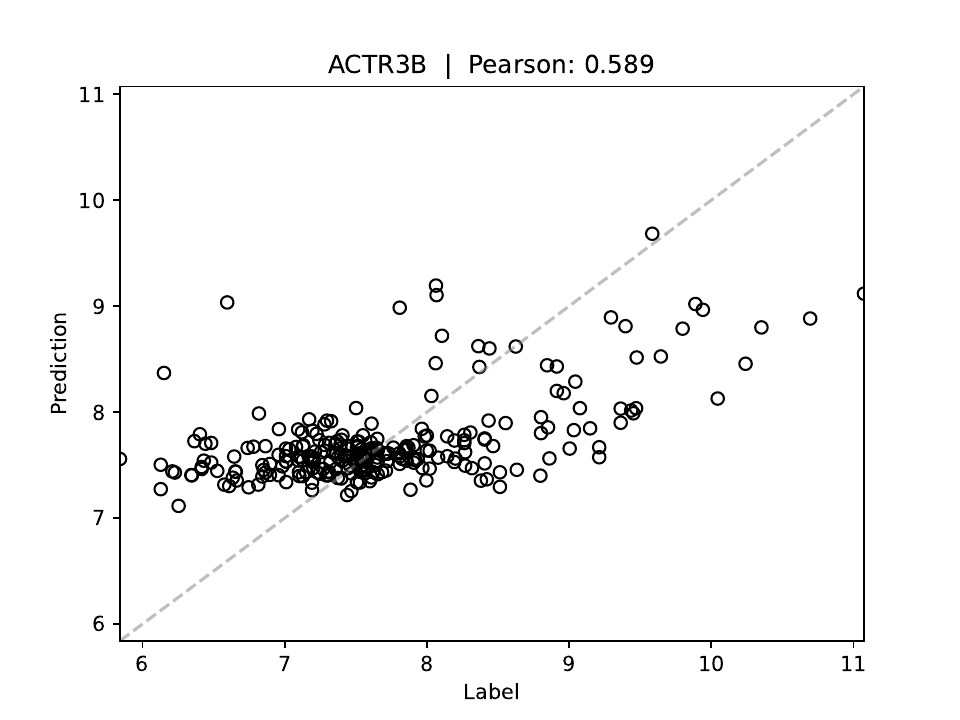}
    \end{subfigure}
    \begin{subfigure}[t]{0.29\textwidth}
        \centering%
        \includegraphics[width=1.0\linewidth]{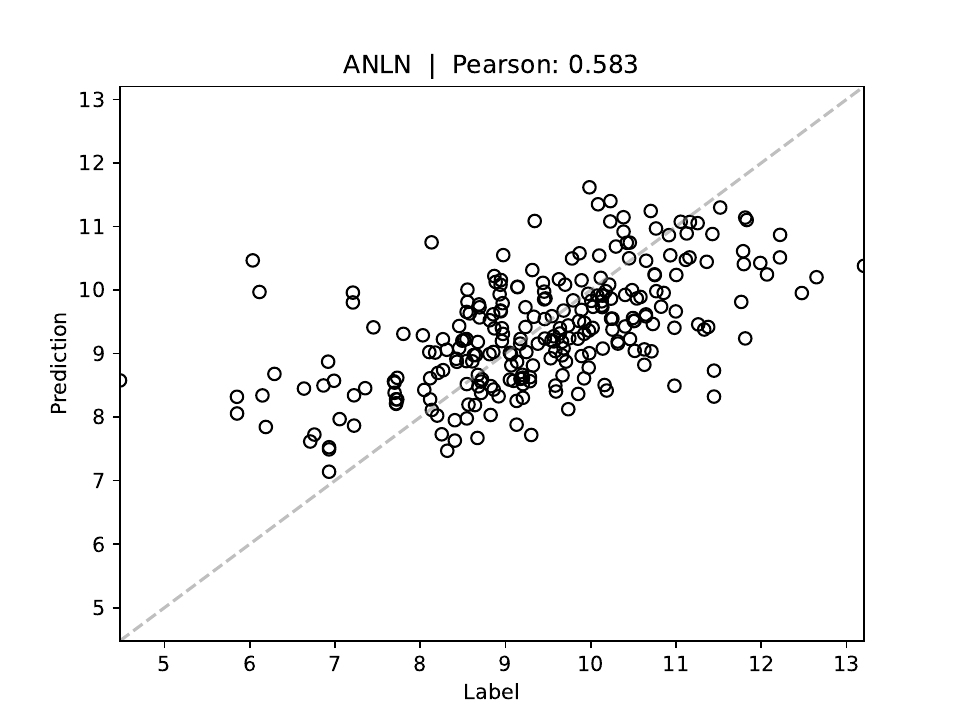}
    \end{subfigure}
    
    \begin{subfigure}[t]{0.29\textwidth}
        \centering%
        \includegraphics[width=1.0\linewidth]{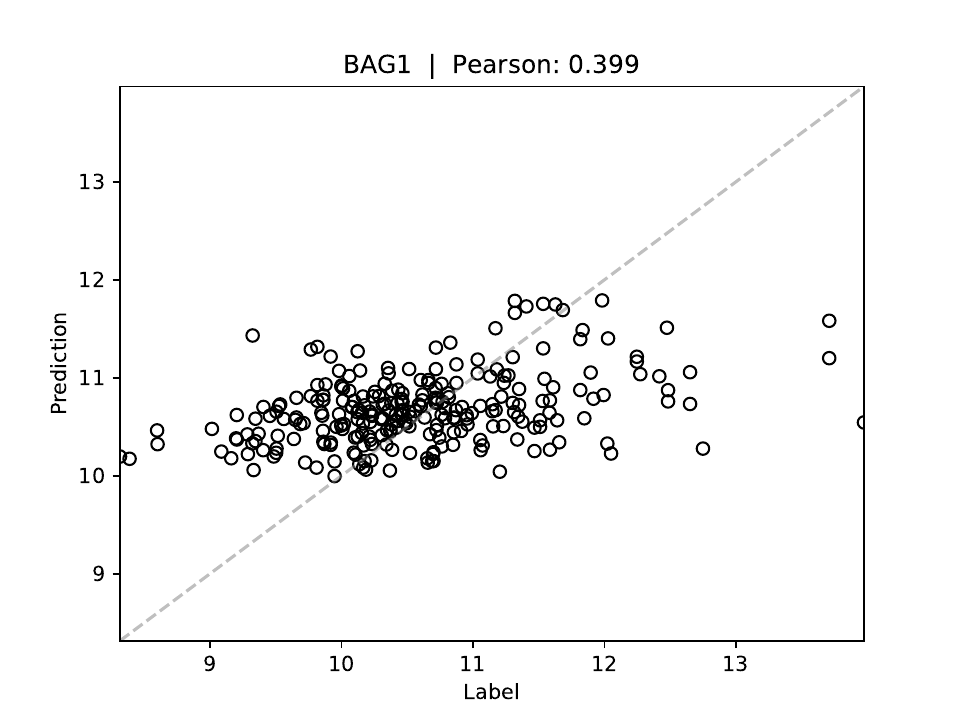}
    \end{subfigure}
    \begin{subfigure}[t]{0.29\textwidth}
        \centering%
        \includegraphics[width=1.0\linewidth]{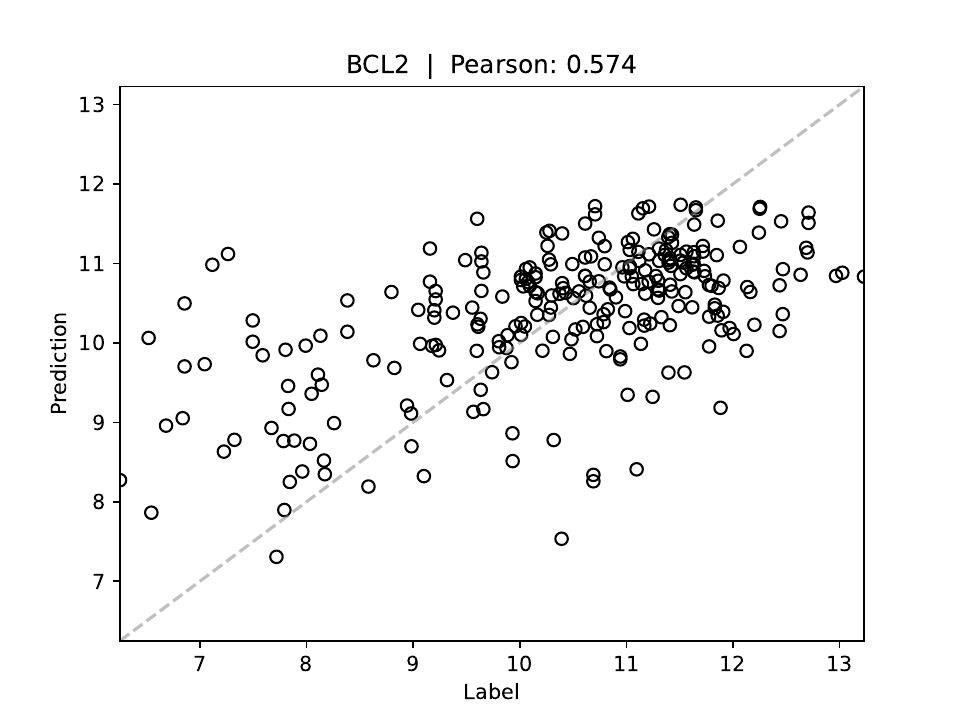}
    \end{subfigure}
    \begin{subfigure}[t]{0.29\textwidth}
        \centering%
        \includegraphics[width=1.0\linewidth]{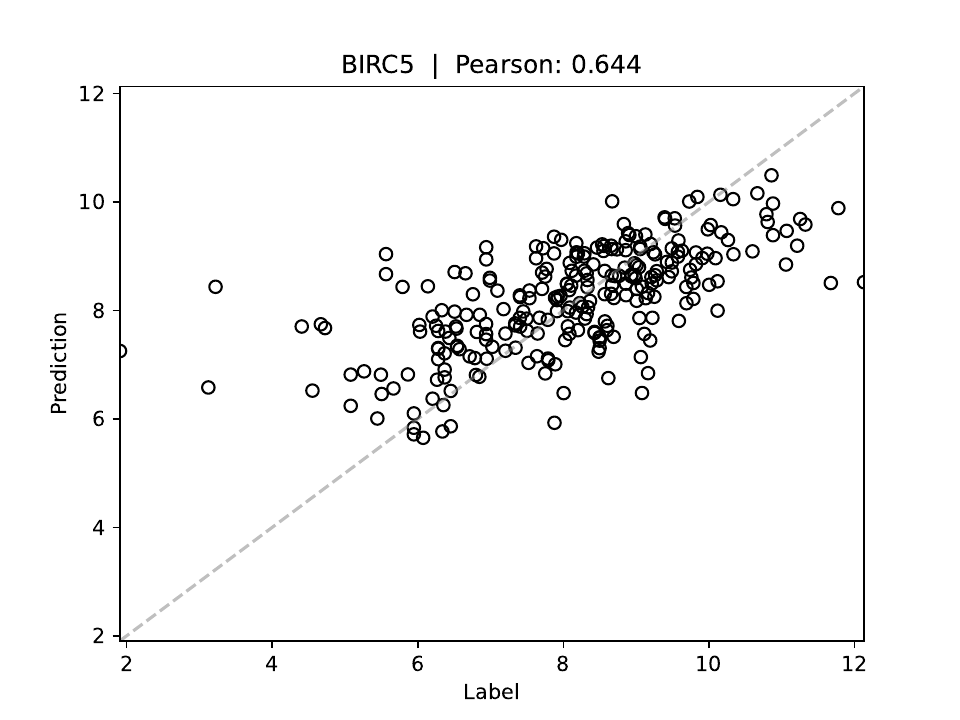}
    \end{subfigure}\vspace{-1.0mm}
    \caption{\textbf{Predicted vs observed gene-expression values for PAM50 gene 1-5 on TCGA-BRCA.} Scatter plots comparing predicted and ground-truth values, for \textit{UNI - Direct - ABMIL} on the TCGA-BRCA dataset, for the test split of the first cross-validation fold.}
  \label{fig:corr_plots_pam50_1_5}
\end{figure*}

\begin{figure*}[h]
\centering
    \begin{subfigure}[t]{0.29\textwidth}
        \centering%
        \includegraphics[width=1.0\linewidth]{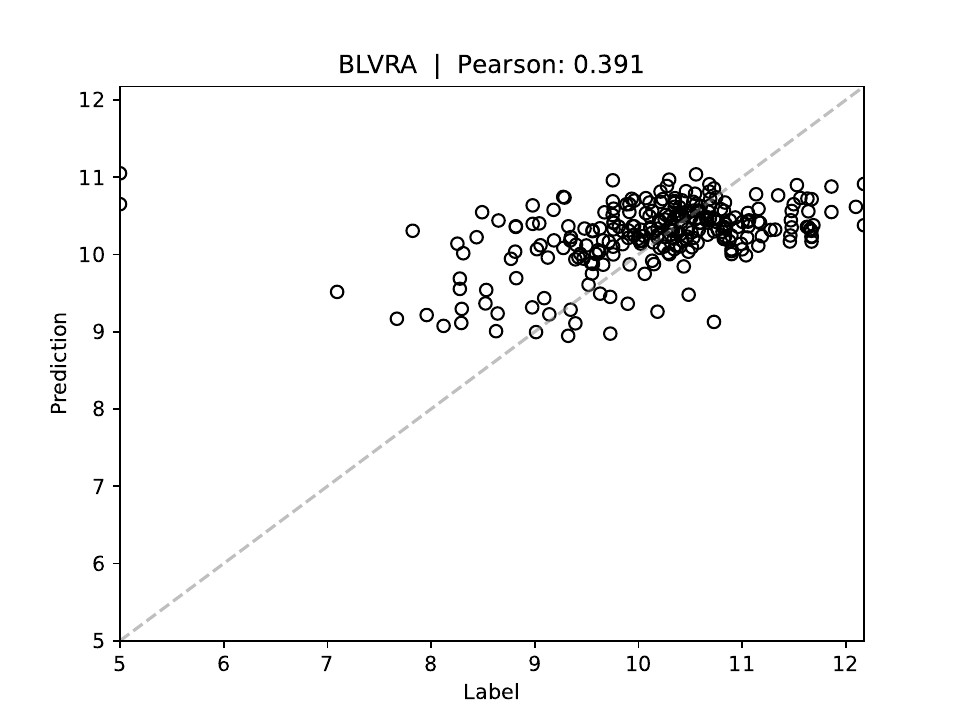}
    \end{subfigure}
    \begin{subfigure}[t]{0.29\textwidth}
        \centering%
        \includegraphics[width=1.0\linewidth]{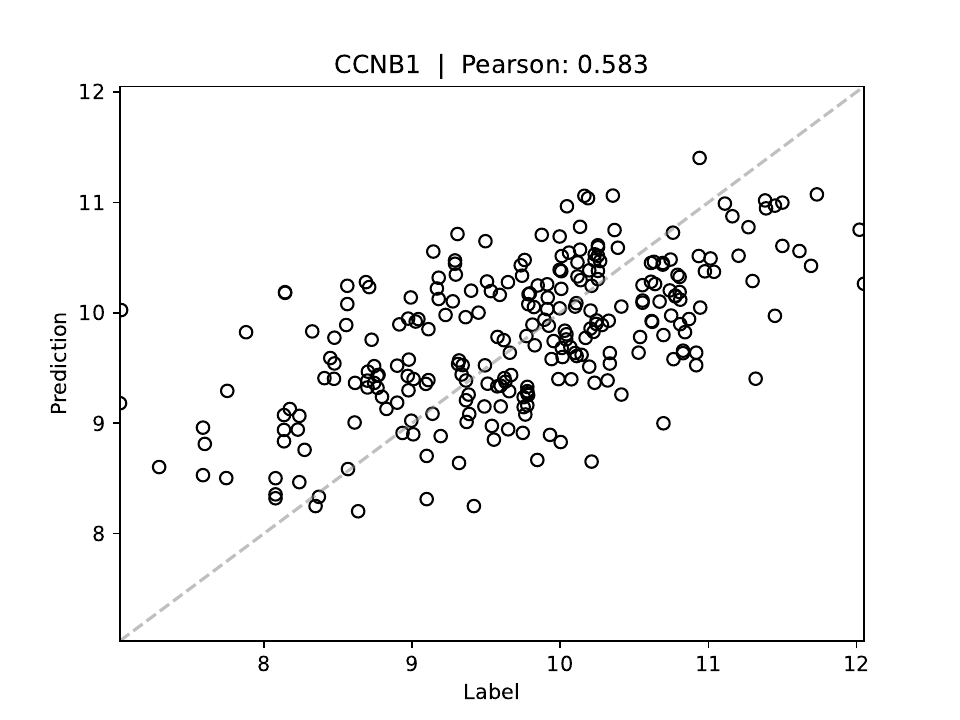}
    \end{subfigure}
    \begin{subfigure}[t]{0.29\textwidth}
        \centering%
        \includegraphics[width=1.0\linewidth]{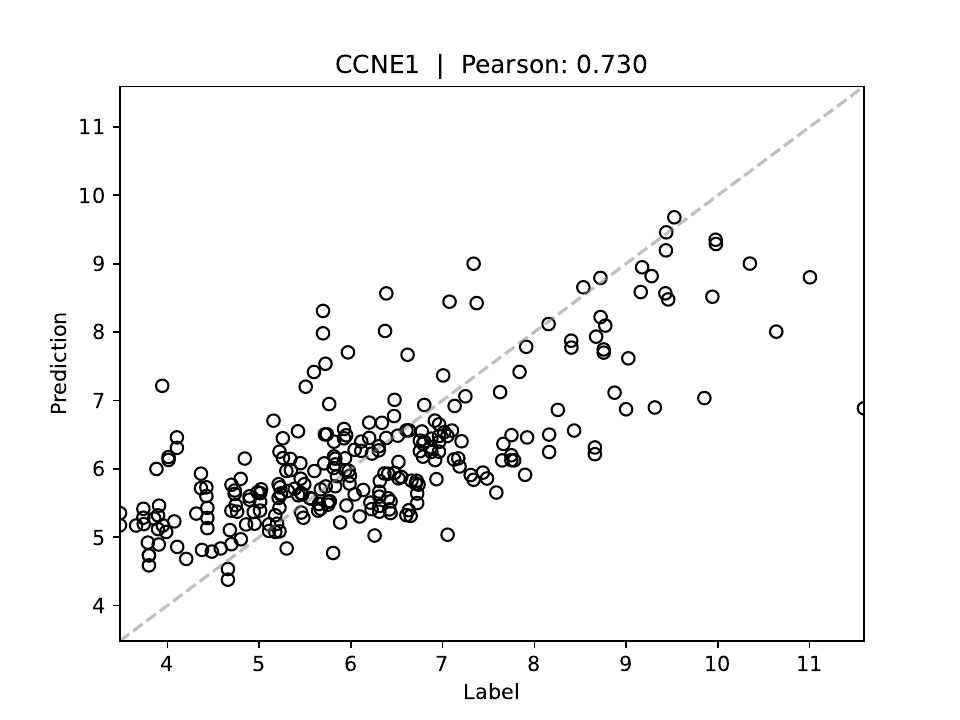}
    \end{subfigure}
    
    \begin{subfigure}[t]{0.29\textwidth}
        \centering%
        \includegraphics[width=1.0\linewidth]{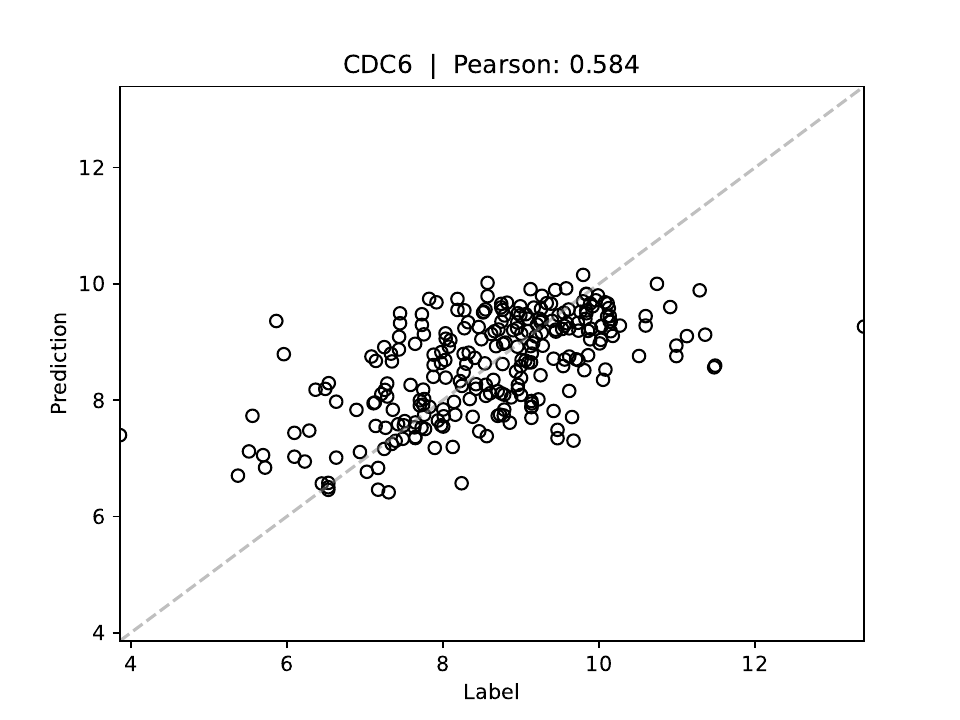}
    \end{subfigure}
    \begin{subfigure}[t]{0.29\textwidth}
        \centering%
        \includegraphics[width=1.0\linewidth]{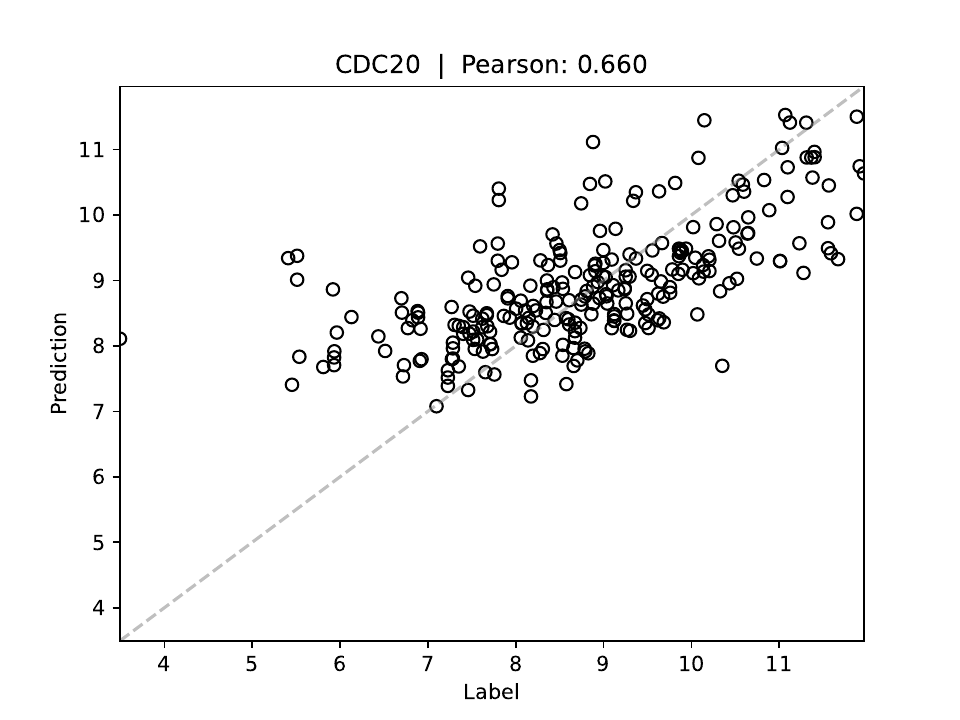}
    \end{subfigure}
    \begin{subfigure}[t]{0.29\textwidth}
        \centering%
        \includegraphics[width=1.0\linewidth]{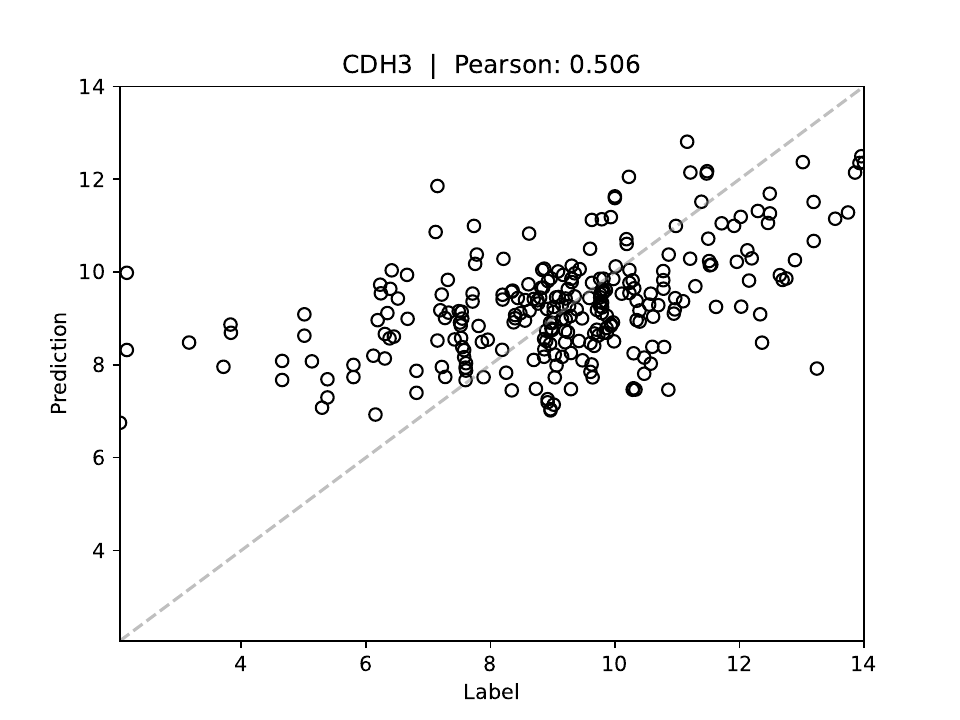}
    \end{subfigure}
    
    \begin{subfigure}[t]{0.29\textwidth}
        \centering%
        \includegraphics[width=1.0\linewidth]{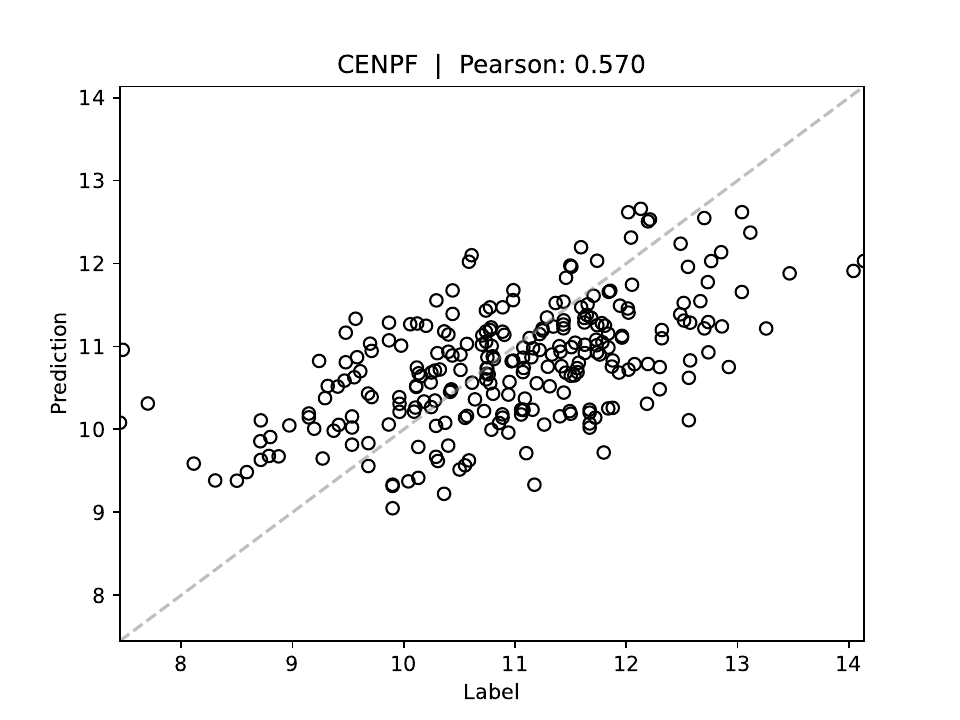}
    \end{subfigure}
    \begin{subfigure}[t]{0.29\textwidth}
        \centering%
        \includegraphics[width=1.0\linewidth]{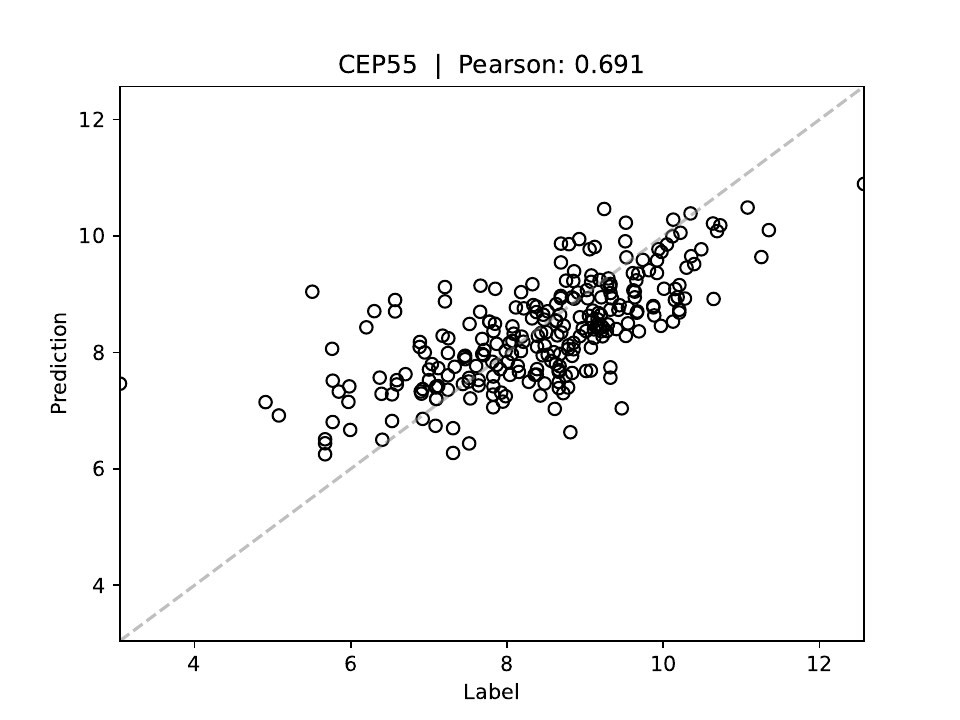}
    \end{subfigure}
    \begin{subfigure}[t]{0.29\textwidth}
        \centering%
        \includegraphics[width=1.0\linewidth]{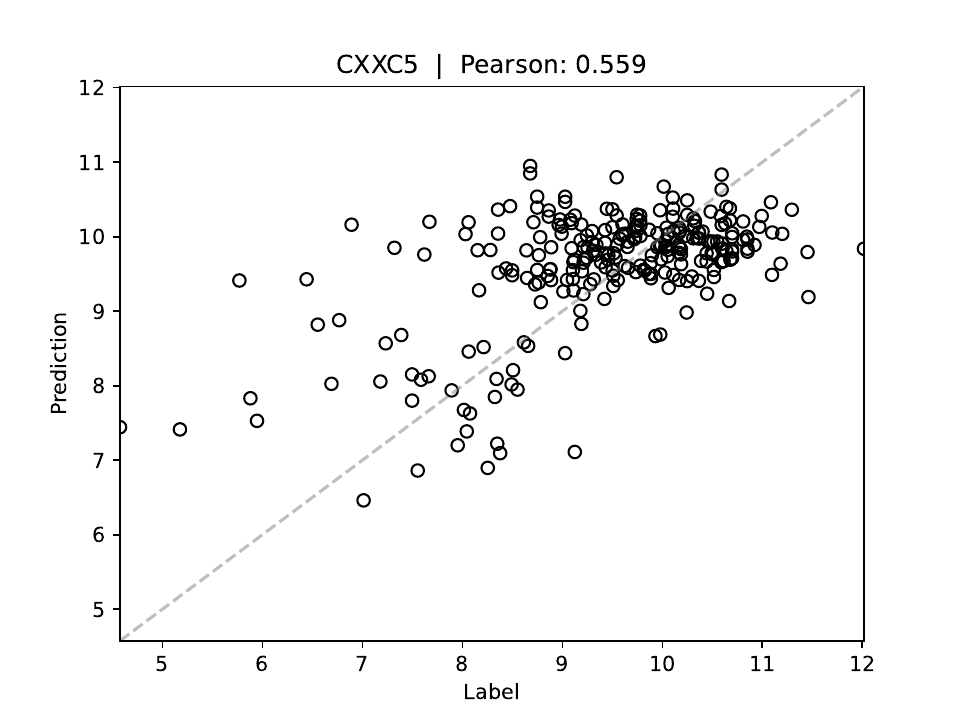}
    \end{subfigure}
    
    \begin{subfigure}[t]{0.29\textwidth}
        \centering%
        \includegraphics[width=1.0\linewidth]{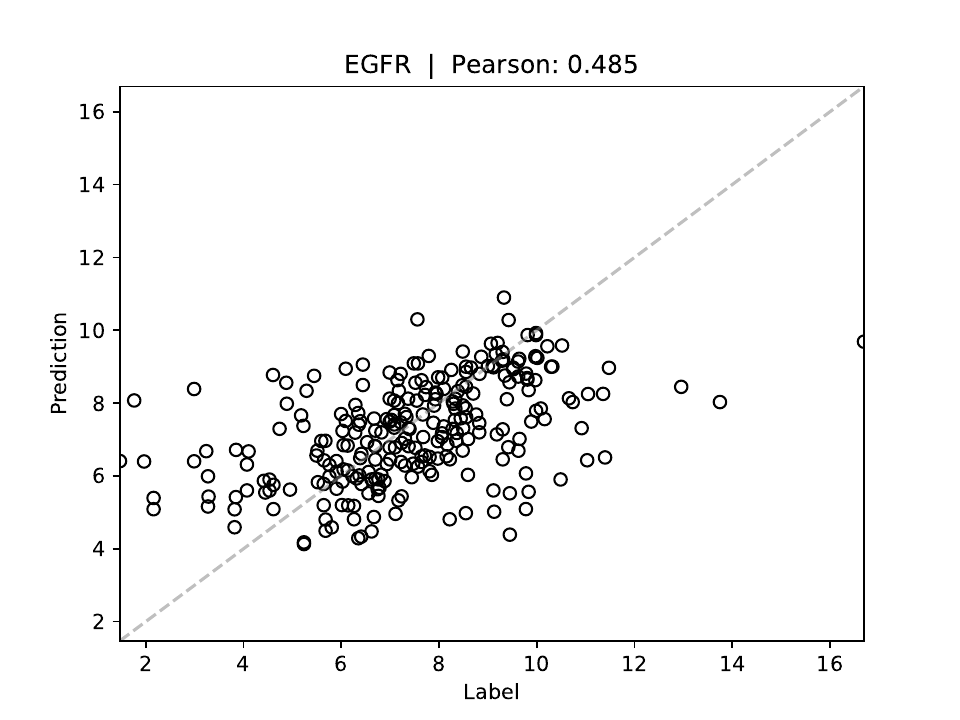}
    \end{subfigure}
    \begin{subfigure}[t]{0.29\textwidth}
        \centering%
        \includegraphics[width=1.0\linewidth]{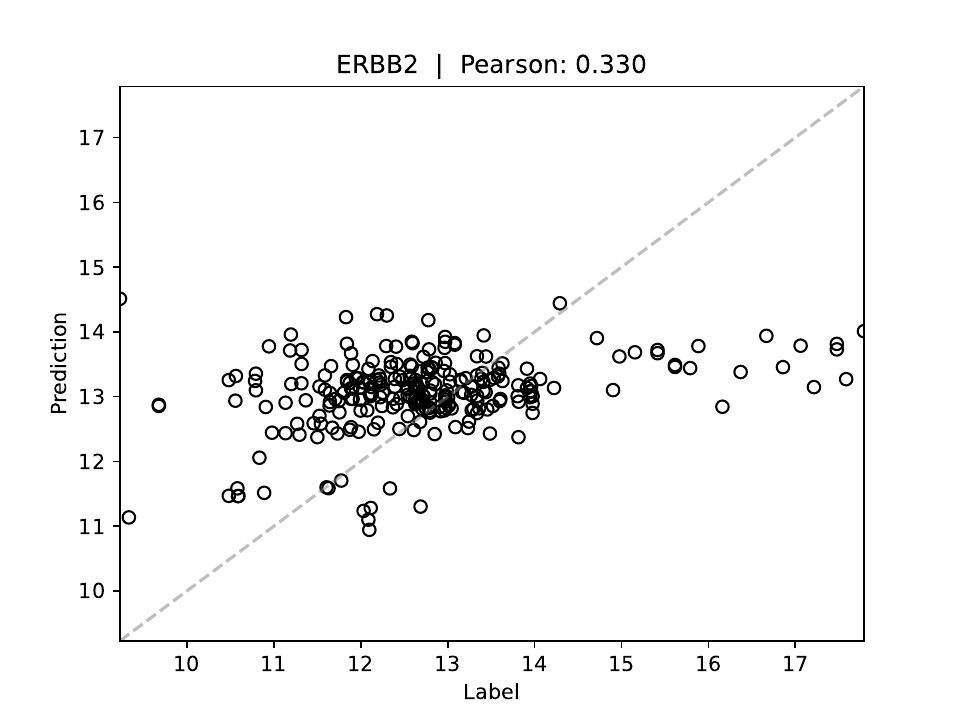}
    \end{subfigure}
    \begin{subfigure}[t]{0.29\textwidth}
        \centering%
        \includegraphics[width=1.0\linewidth]{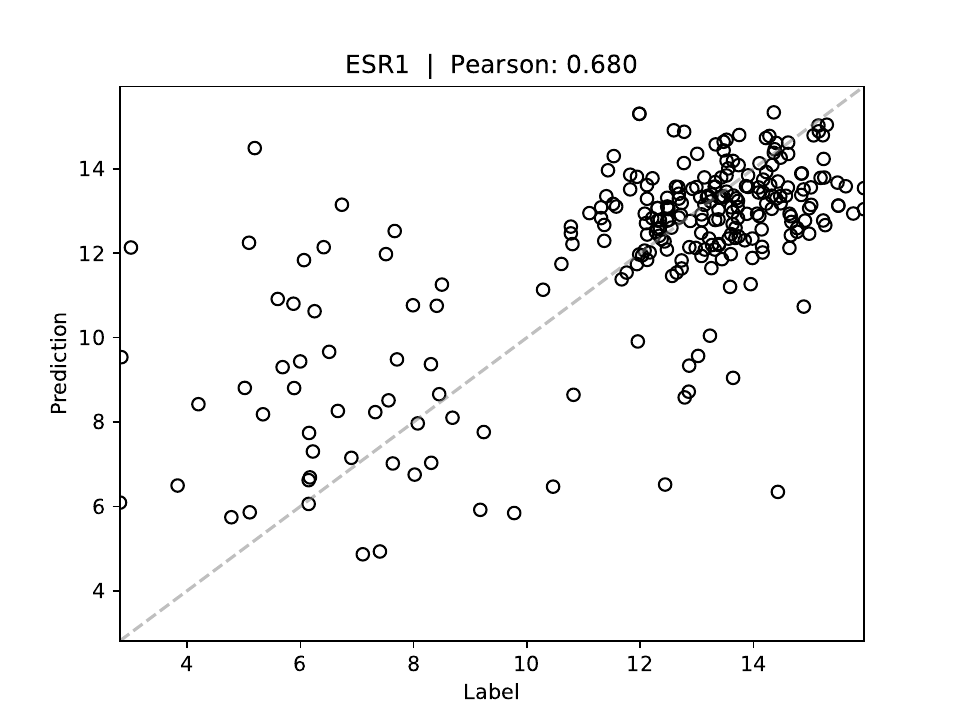}
    \end{subfigure}

    \begin{subfigure}[t]{0.29\textwidth}
        \centering%
        \includegraphics[width=1.0\linewidth]{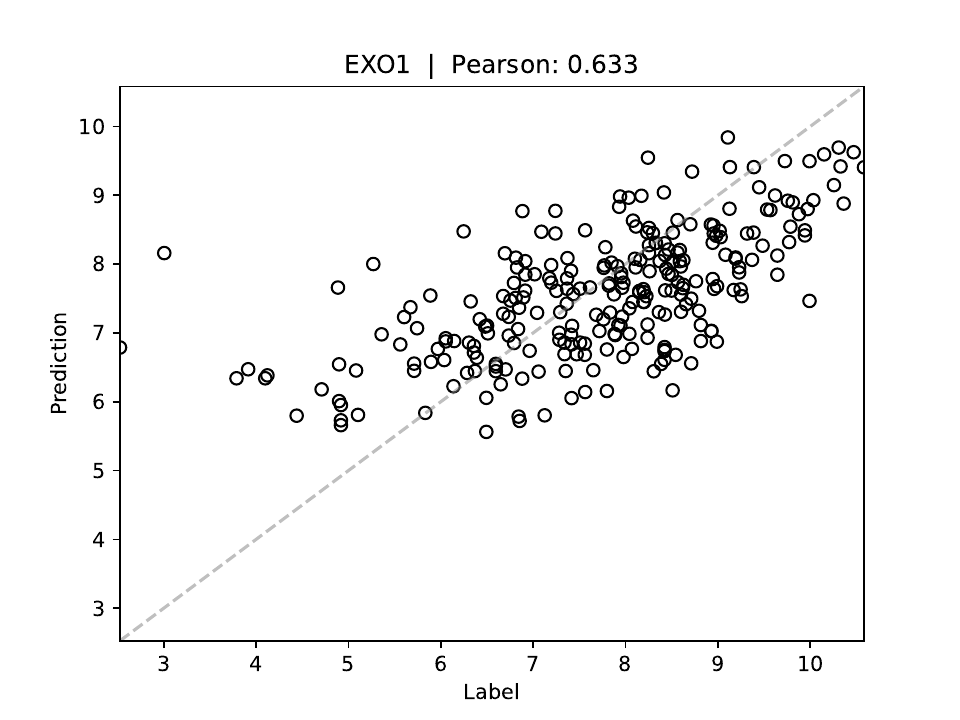}
    \end{subfigure}
    \begin{subfigure}[t]{0.29\textwidth}
        \centering%
        \includegraphics[width=1.0\linewidth]{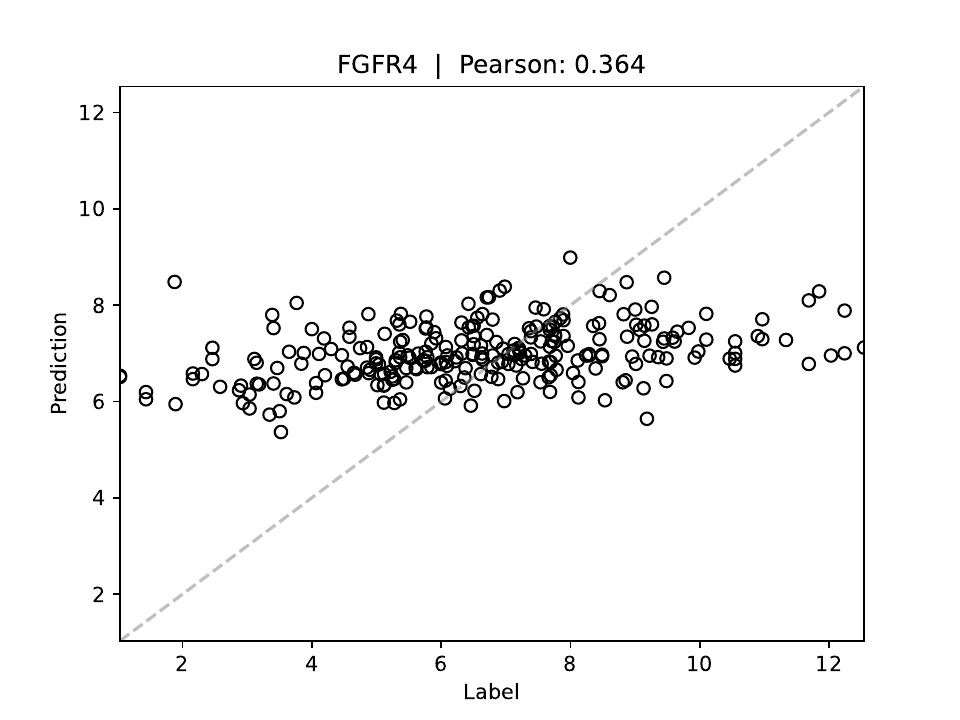}
    \end{subfigure}
    \begin{subfigure}[t]{0.29\textwidth}
        \centering%
        \includegraphics[width=1.0\linewidth]{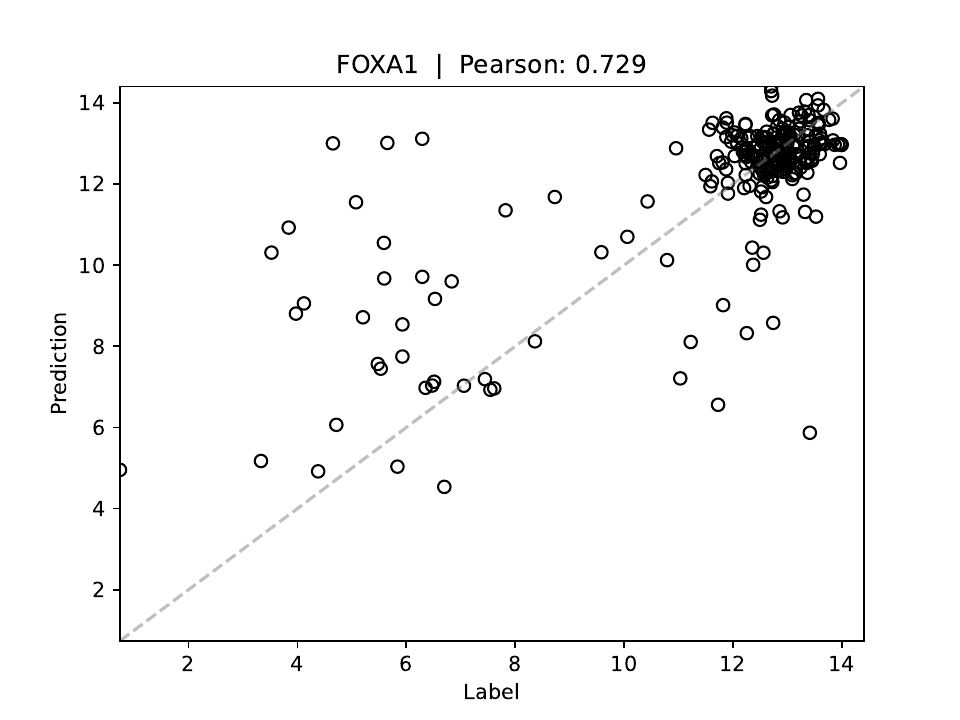}
    \end{subfigure}\vspace{-1.0mm}
    \caption{\textbf{Predicted vs observed gene-expression values for PAM50 gene 6-20 on TCGA-BRCA.} Scatter plots comparing predicted and ground-truth values, for \textit{UNI - Direct - ABMIL} on the TCGA-BRCA dataset, for the test split of the first cross-validation fold.}
  \label{fig:corr_plots_pam50_6_20}
\end{figure*}

\begin{figure*}[h]
\centering
    \begin{subfigure}[t]{0.29\textwidth}
        \centering%
        \includegraphics[width=1.0\linewidth]{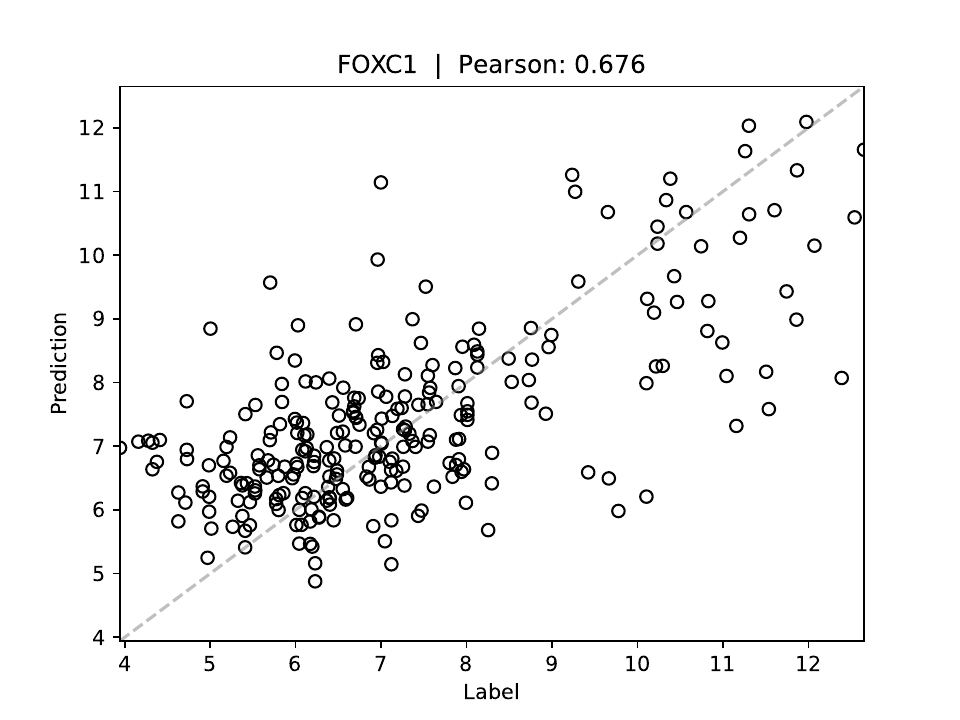}
    \end{subfigure}
    \begin{subfigure}[t]{0.29\textwidth}
        \centering%
        \includegraphics[width=1.0\linewidth]{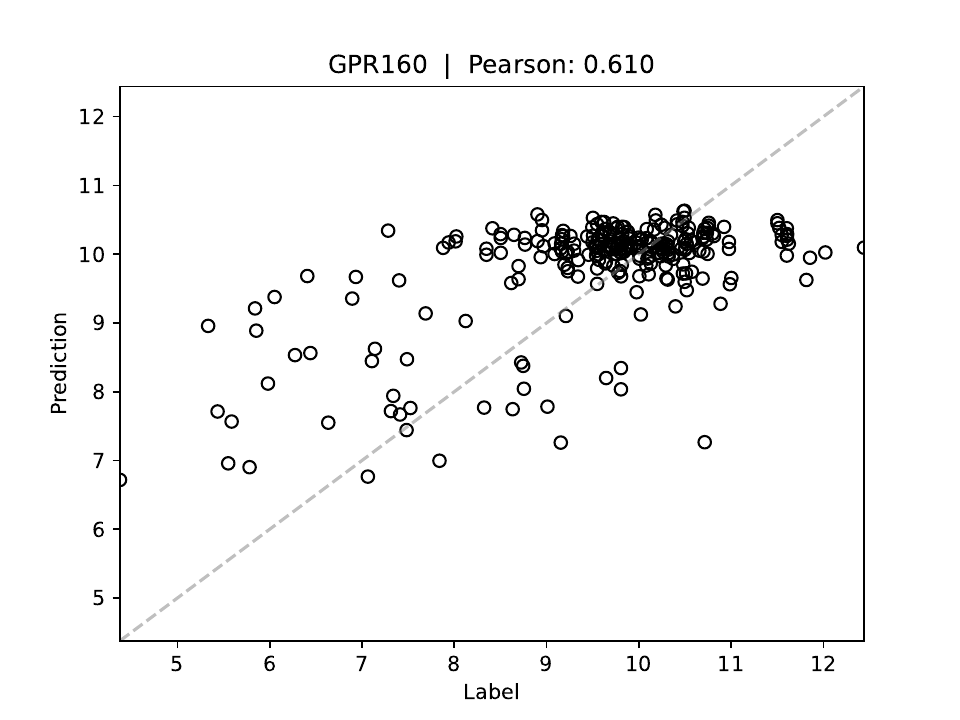}
    \end{subfigure}
    \begin{subfigure}[t]{0.29\textwidth}
        \centering%
        \includegraphics[width=1.0\linewidth]{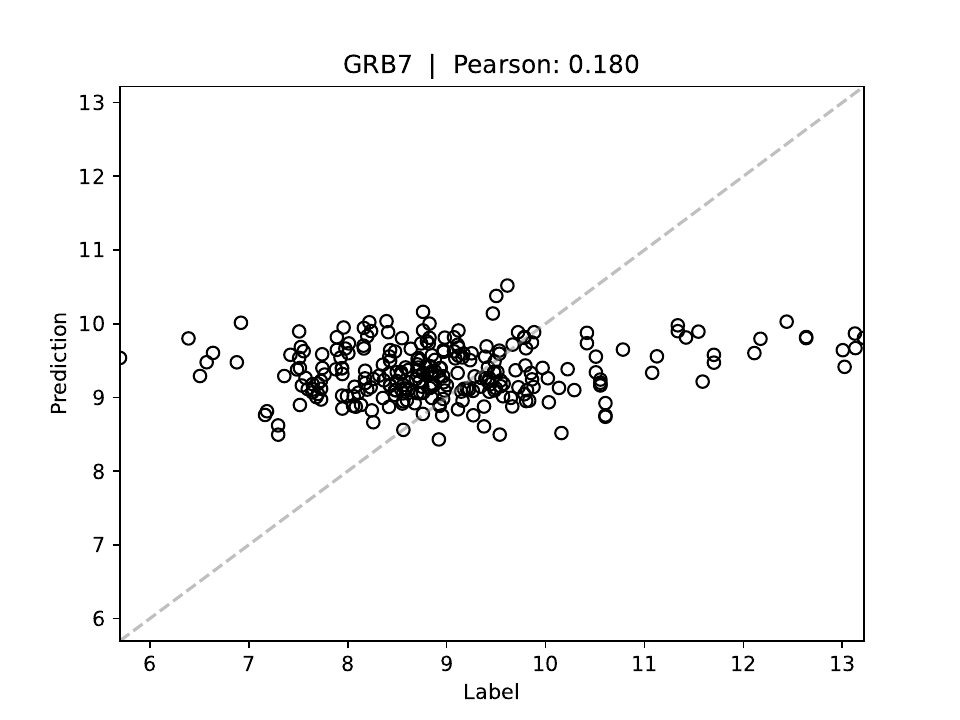}
    \end{subfigure}
    
    \begin{subfigure}[t]{0.29\textwidth}
        \centering%
        \includegraphics[width=1.0\linewidth]{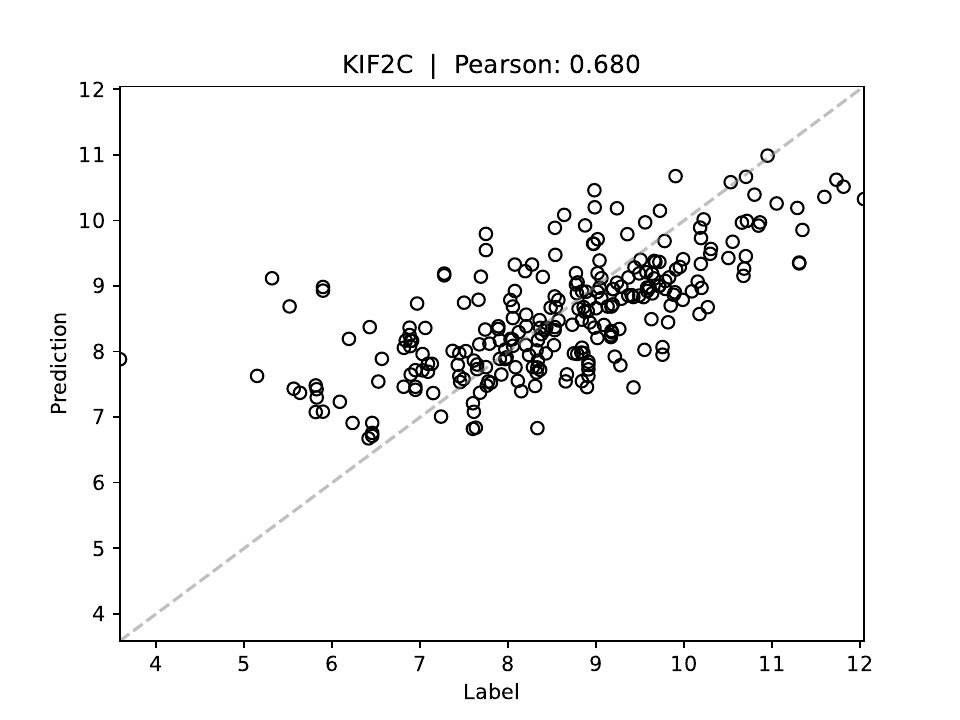}
    \end{subfigure}
    \begin{subfigure}[t]{0.29\textwidth}
        \centering%
        \includegraphics[width=1.0\linewidth]{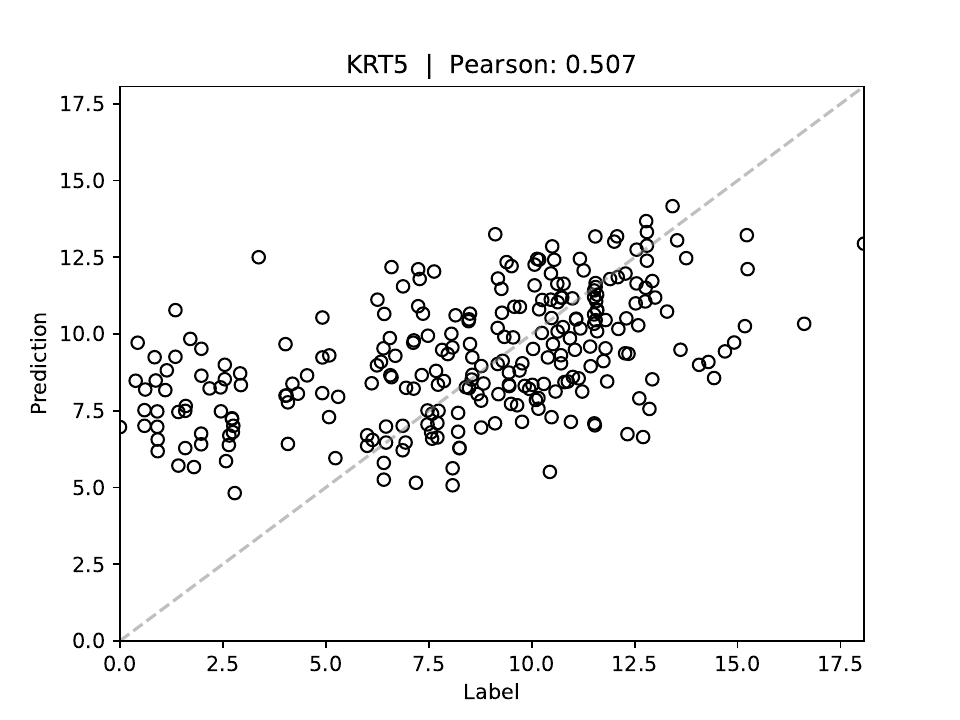}
    \end{subfigure}
    \begin{subfigure}[t]{0.29\textwidth}
        \centering%
        \includegraphics[width=1.0\linewidth]{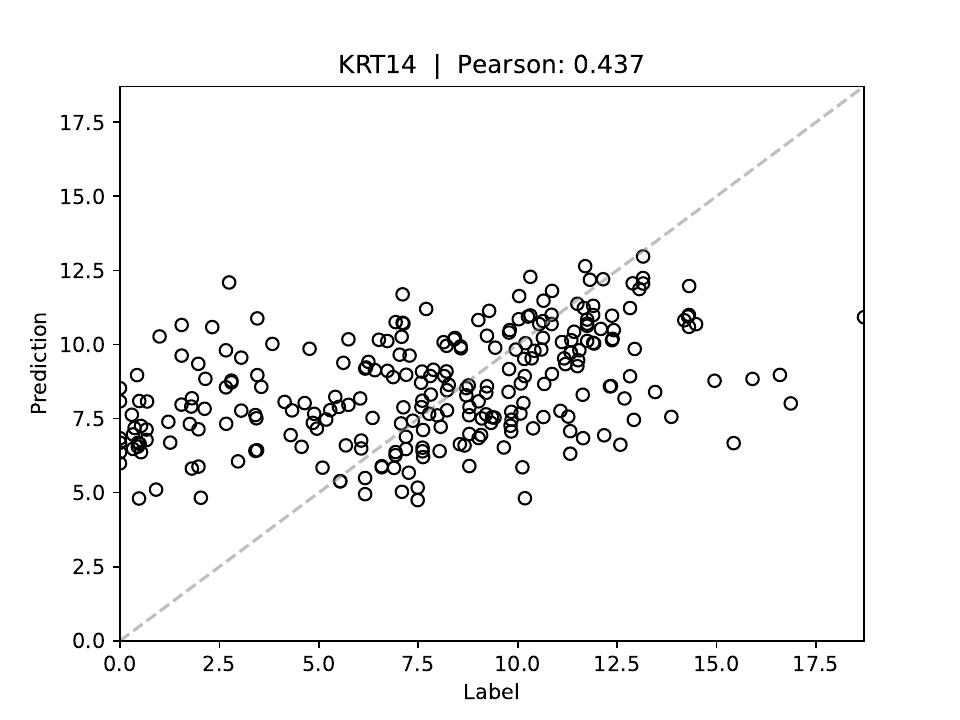}
    \end{subfigure}
    
    \begin{subfigure}[t]{0.29\textwidth}
        \centering%
        \includegraphics[width=1.0\linewidth]{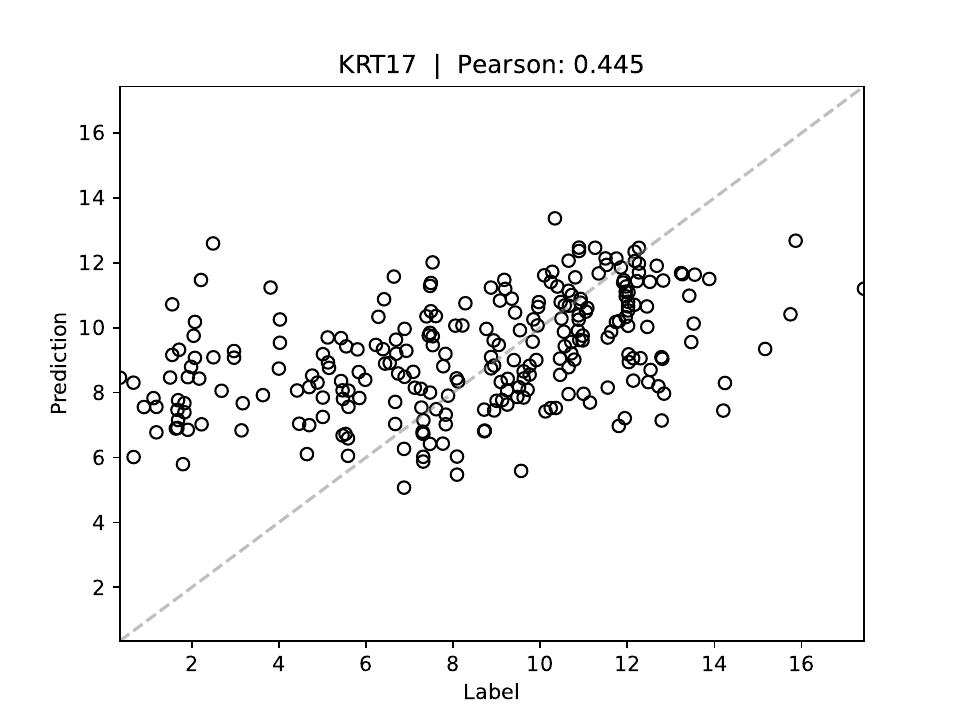}
    \end{subfigure}
    \begin{subfigure}[t]{0.29\textwidth}
        \centering%
        \includegraphics[width=1.0\linewidth]{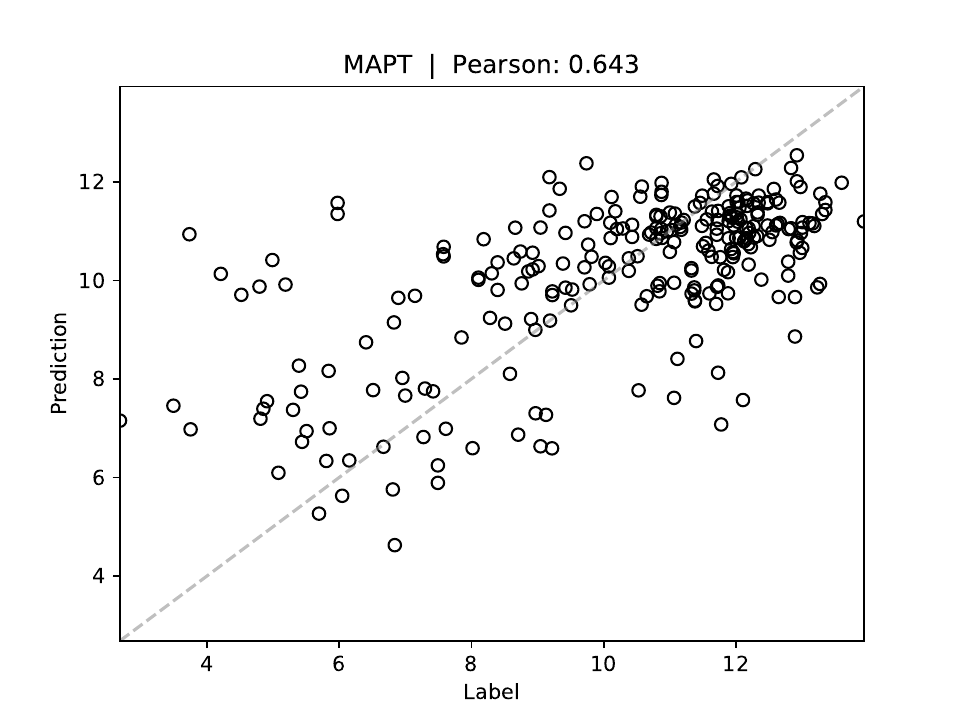}
    \end{subfigure}%\hspace{57.0mm}
    \begin{subfigure}[t]{0.29\textwidth}
        \centering%
        \includegraphics[width=1.0\linewidth]{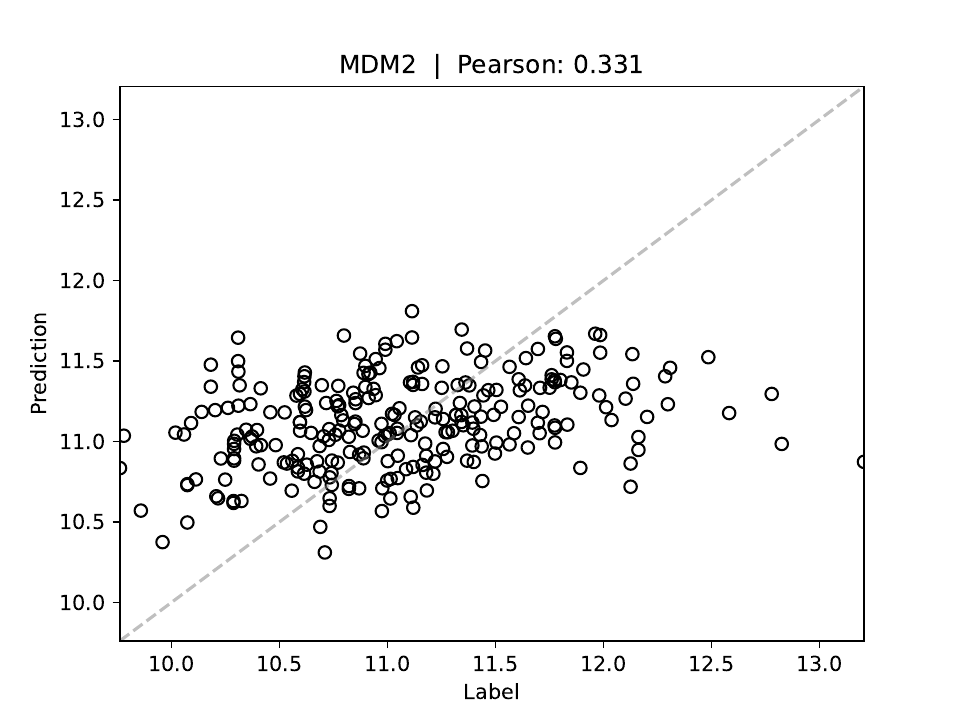}
    \end{subfigure}
    
    \begin{subfigure}[t]{0.29\textwidth}
        \centering%
        \includegraphics[width=1.0\linewidth]{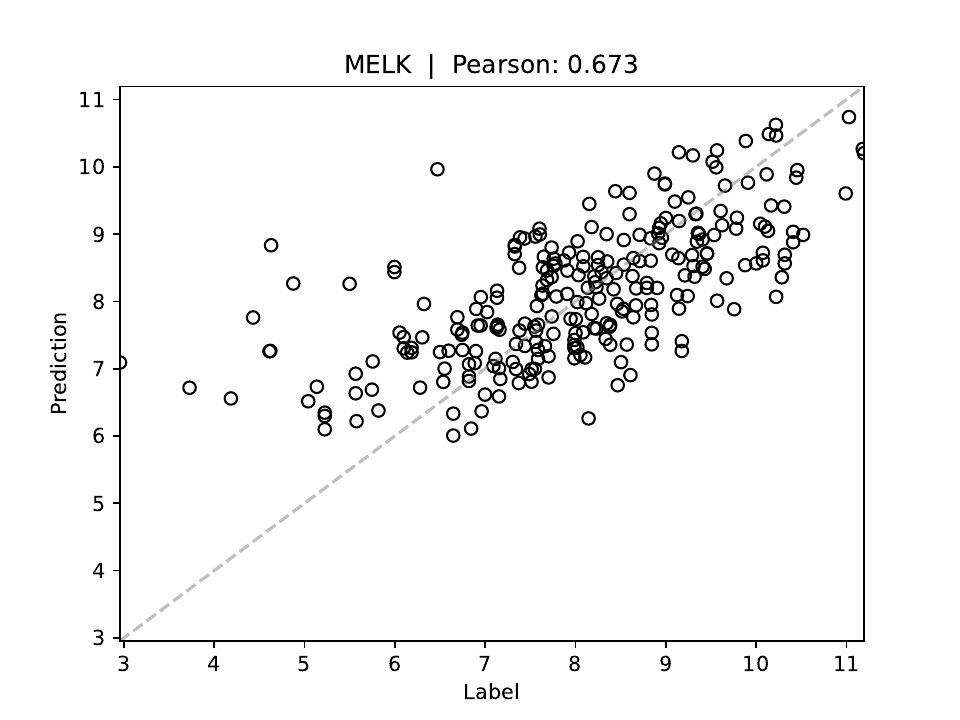}
    \end{subfigure}
    \begin{subfigure}[t]{0.29\textwidth}
        \centering%
        \includegraphics[width=1.0\linewidth]{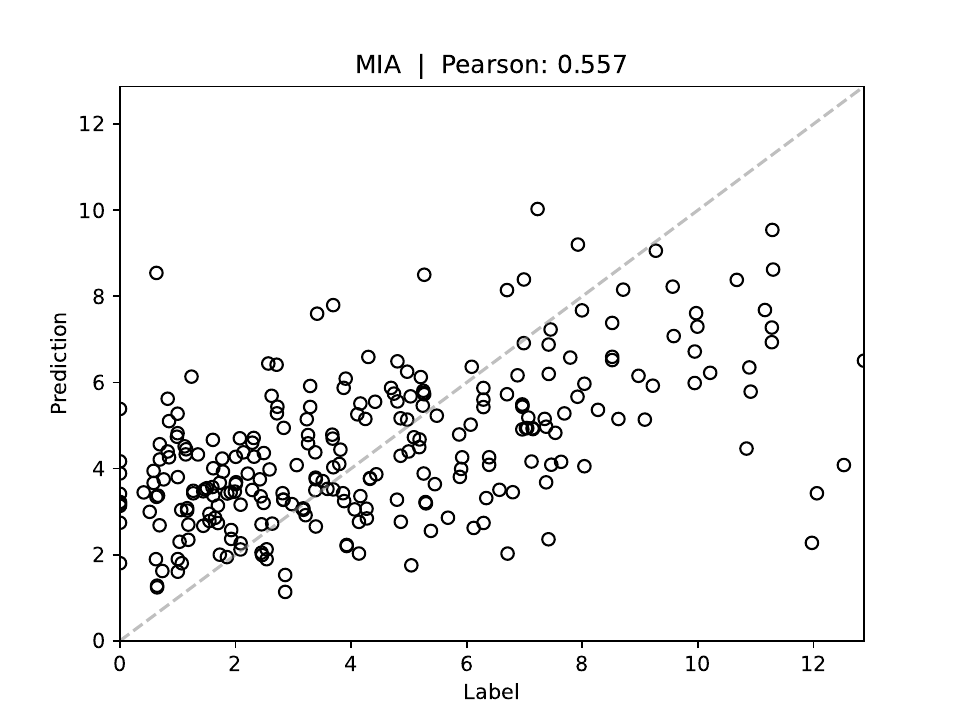}
    \end{subfigure}
    \begin{subfigure}[t]{0.29\textwidth}
        \centering%
        \includegraphics[width=1.0\linewidth]{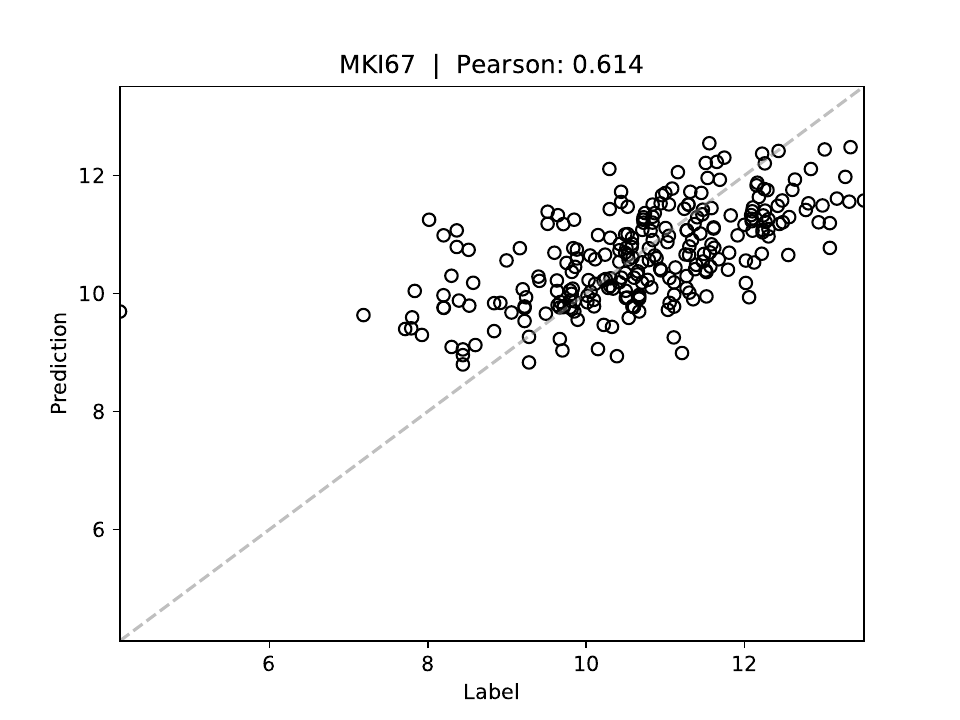}
    \end{subfigure}
    
    \begin{subfigure}[t]{0.29\textwidth}
        \centering%
        \includegraphics[width=1.0\linewidth]{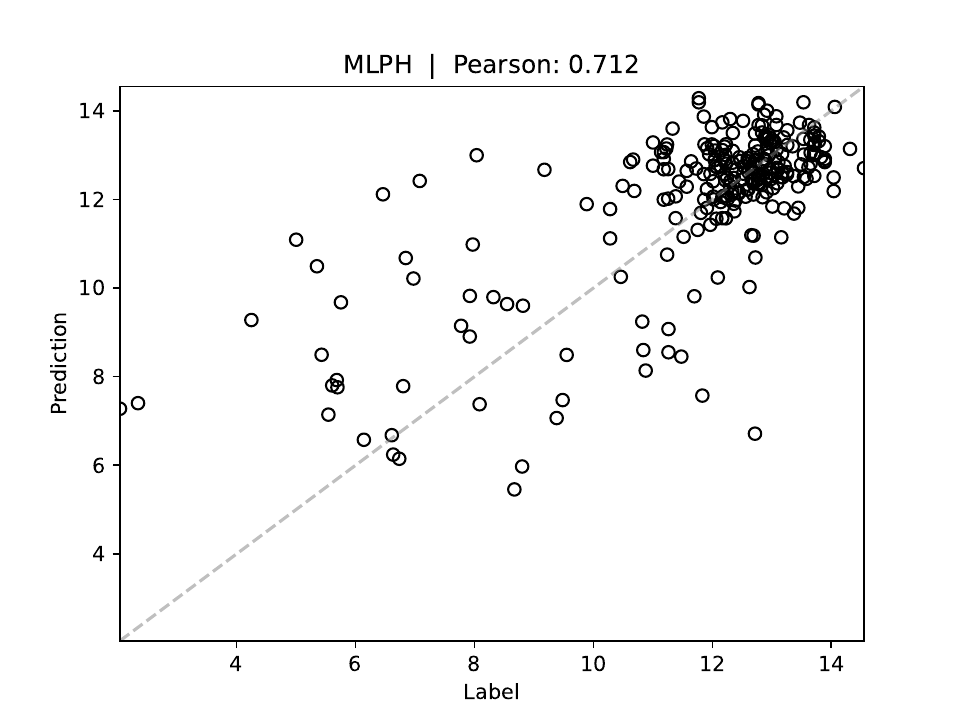}
    \end{subfigure}
    \begin{subfigure}[t]{0.29\textwidth}
        \centering%
        \includegraphics[width=1.0\linewidth]{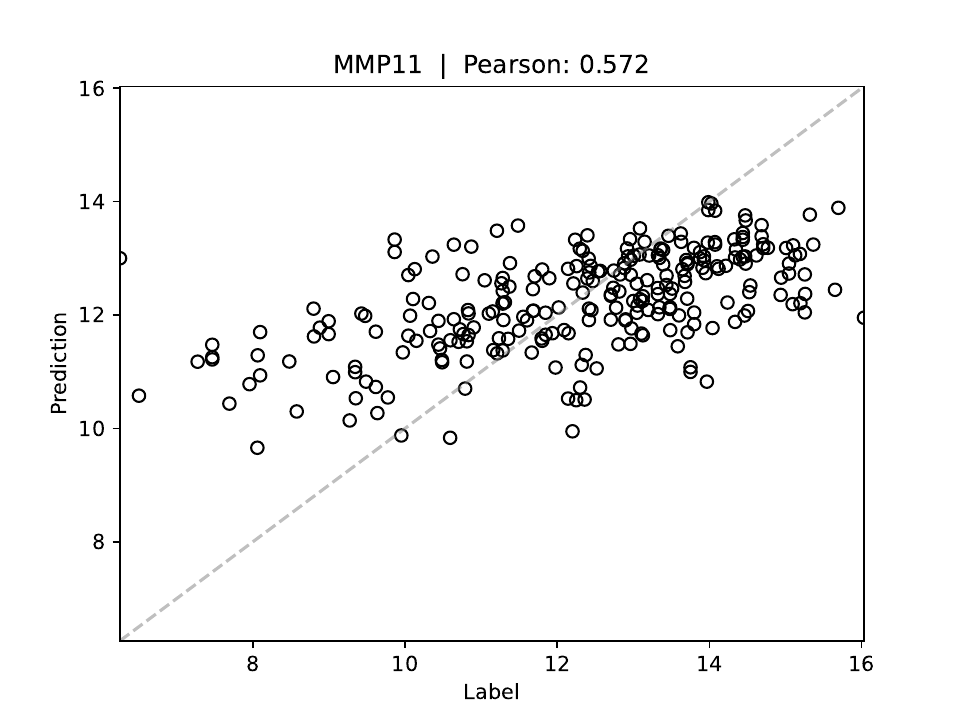}
    \end{subfigure}
    \begin{subfigure}[t]{0.29\textwidth}
        \centering%
        \includegraphics[width=1.0\linewidth]{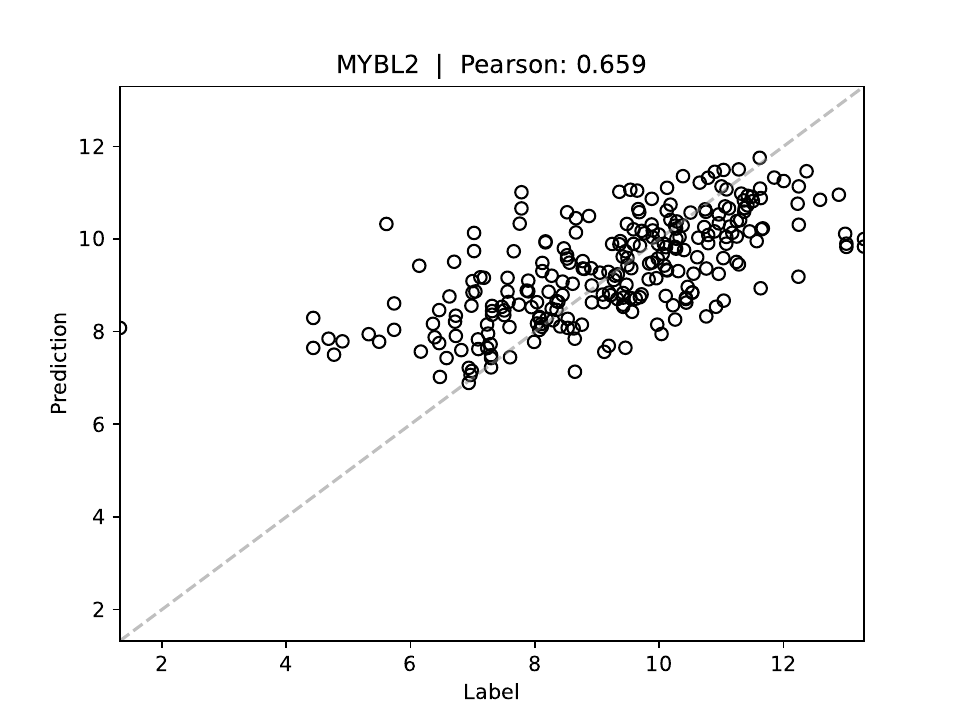}
    \end{subfigure}\vspace{-1.0mm}
    \caption{\textbf{Predicted vs observed gene-expression values for PAM50 gene 21-35 on TCGA-BRCA.} Scatter plots comparing predicted and ground-truth values, for \textit{UNI - Direct - ABMIL} on the TCGA-BRCA dataset, for the test split of the first cross-validation fold.}
  \label{fig:corr_plots_pam50_21_35}
\end{figure*}

\begin{figure*}[h]
\centering
    \begin{subfigure}[t]{0.29\textwidth}
        \centering%
        \includegraphics[width=1.0\linewidth]{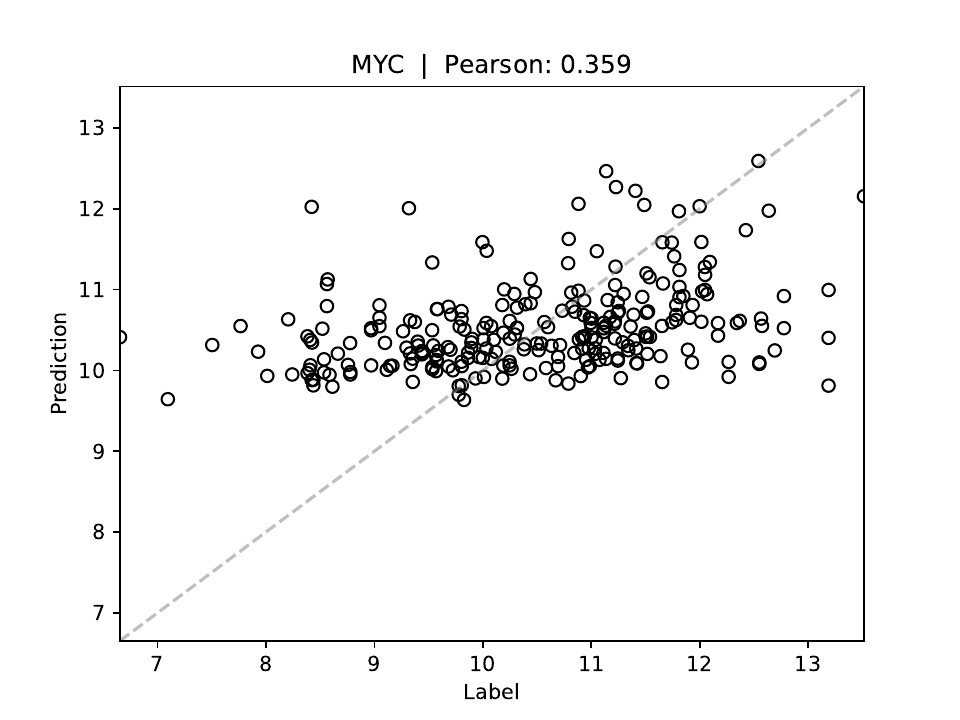}
    \end{subfigure}
    \begin{subfigure}[t]{0.29\textwidth}
        \centering%
        \includegraphics[width=1.0\linewidth]{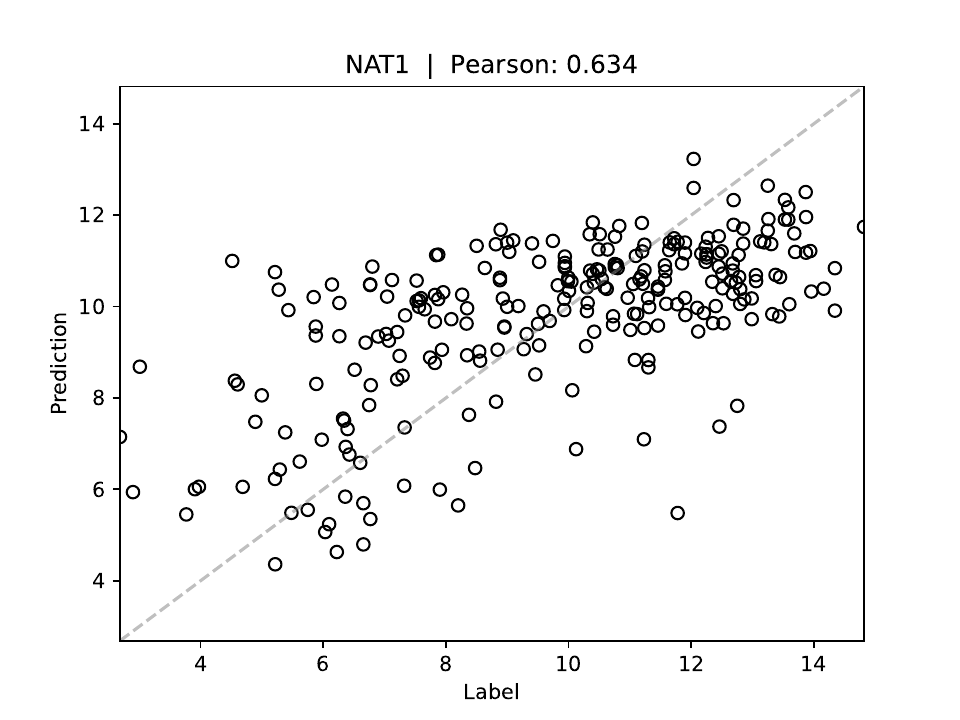}
    \end{subfigure}
    \begin{subfigure}[t]{0.29\textwidth}
        \centering%
        \includegraphics[width=1.0\linewidth]{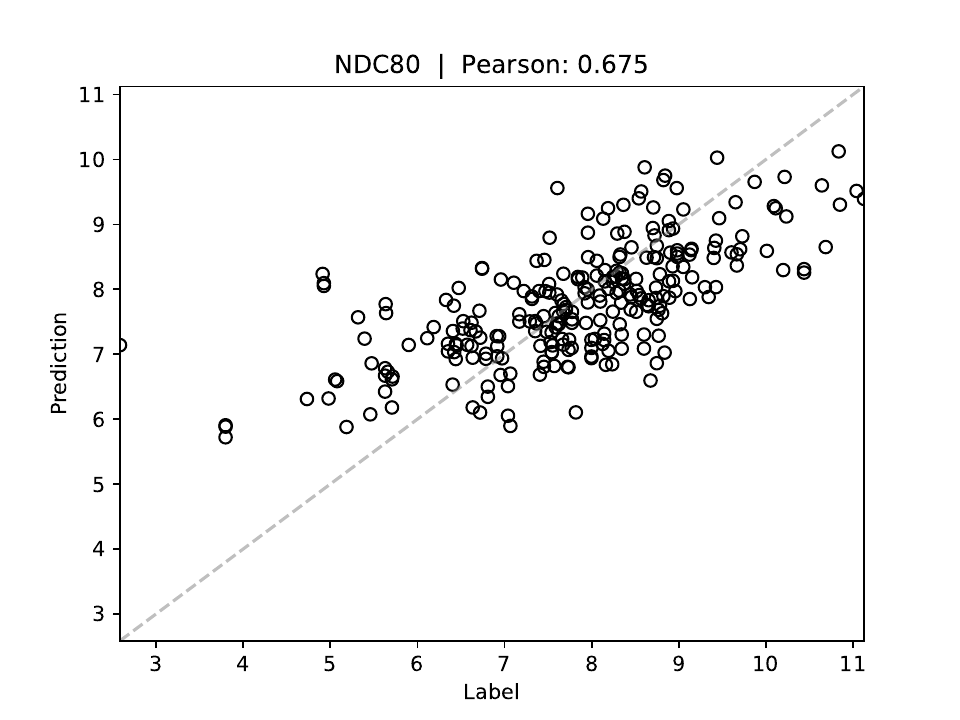}
    \end{subfigure}
    
    \begin{subfigure}[t]{0.29\textwidth}
        \centering%
        \includegraphics[width=1.0\linewidth]{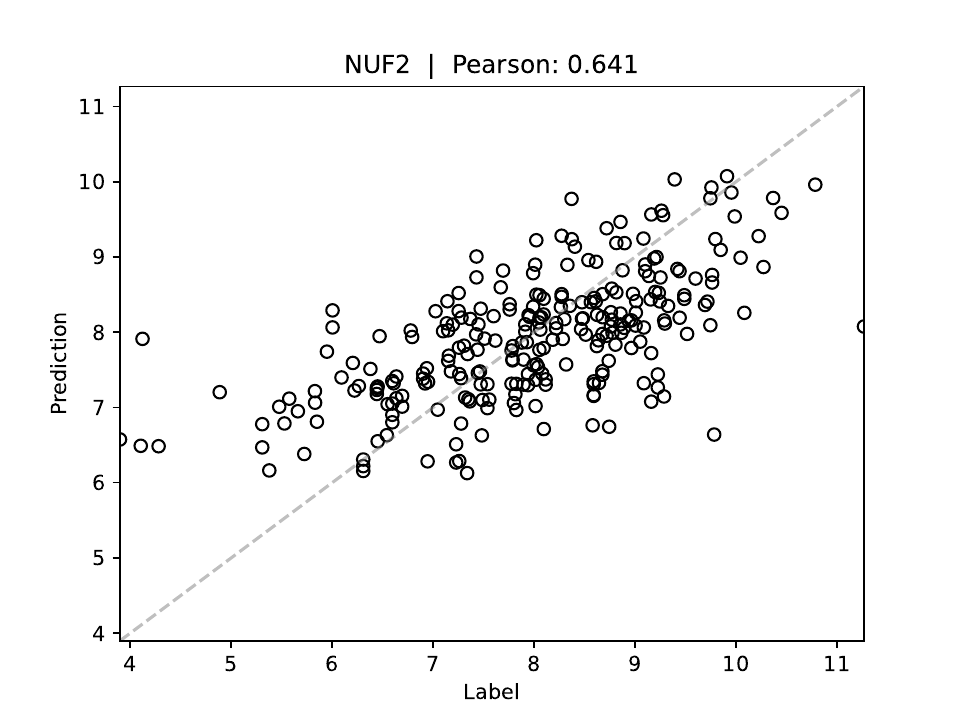}
    \end{subfigure}
    \begin{subfigure}[t]{0.29\textwidth}
        \centering%
        \includegraphics[width=1.0\linewidth]{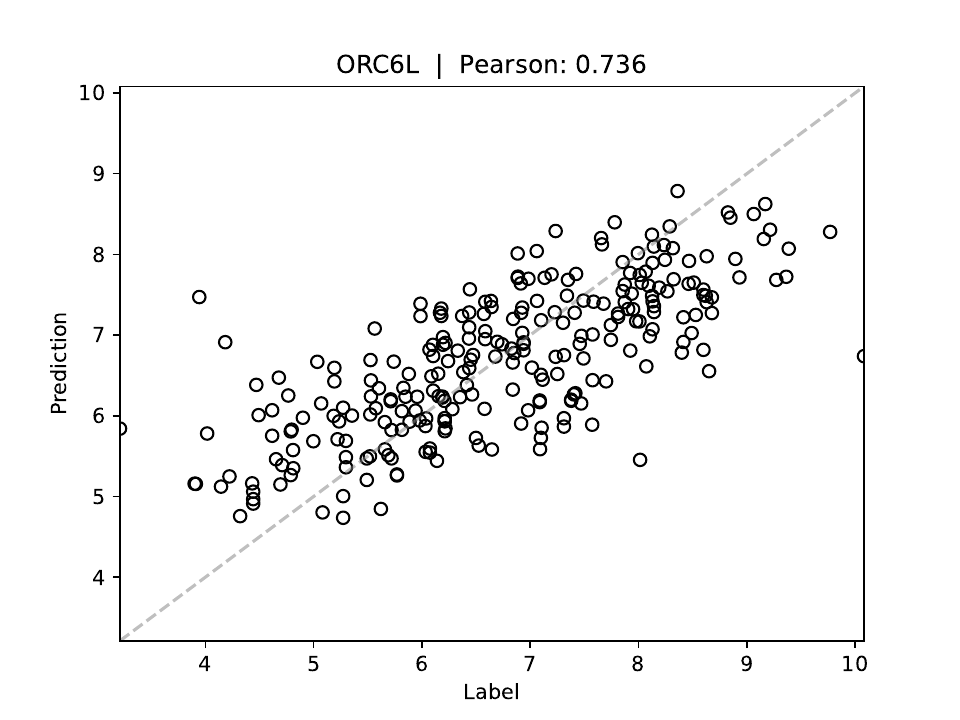}
    \end{subfigure}
    \begin{subfigure}[t]{0.29\textwidth}
        \centering%
        \includegraphics[width=1.0\linewidth]{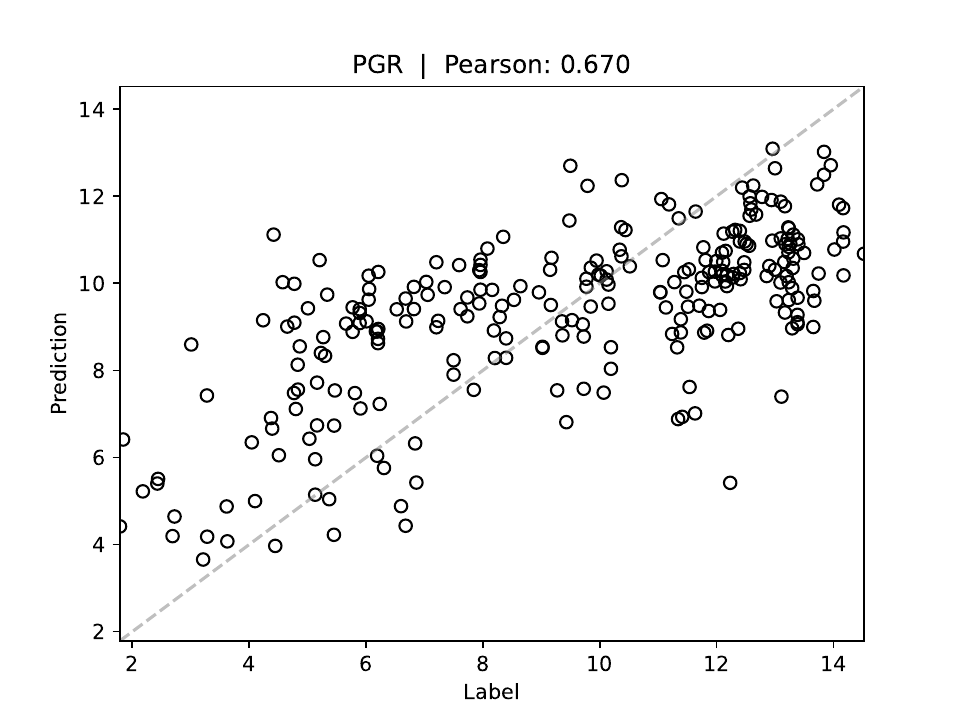}
    \end{subfigure}
    
    \begin{subfigure}[t]{0.29\textwidth}
        \centering%
        \includegraphics[width=1.0\linewidth]{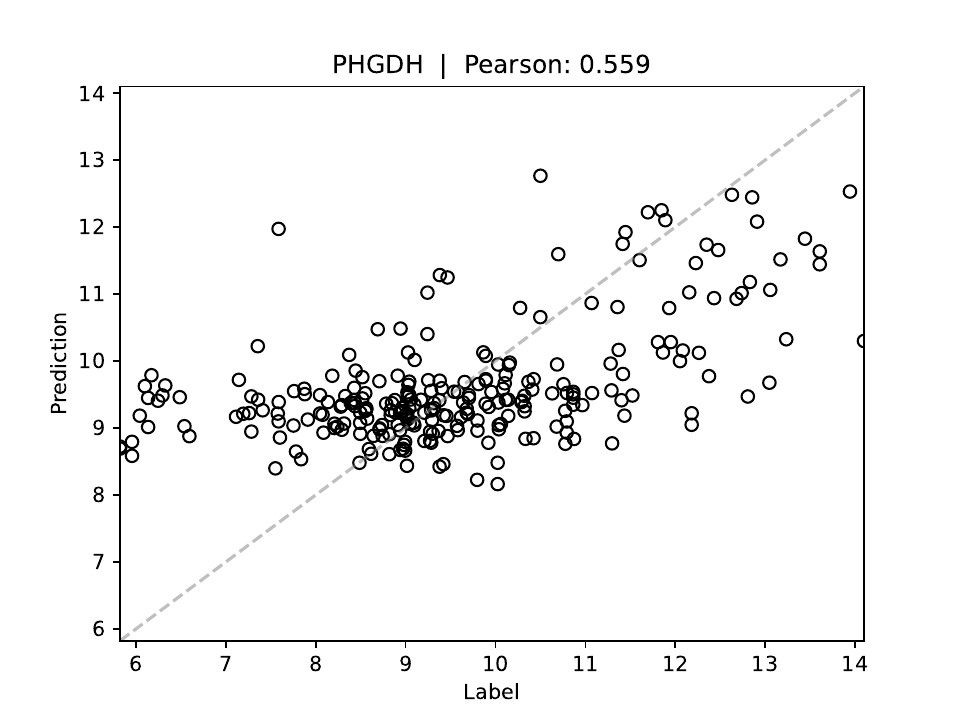}
    \end{subfigure}
    \begin{subfigure}[t]{0.29\textwidth}
        \centering%
        \includegraphics[width=1.0\linewidth]{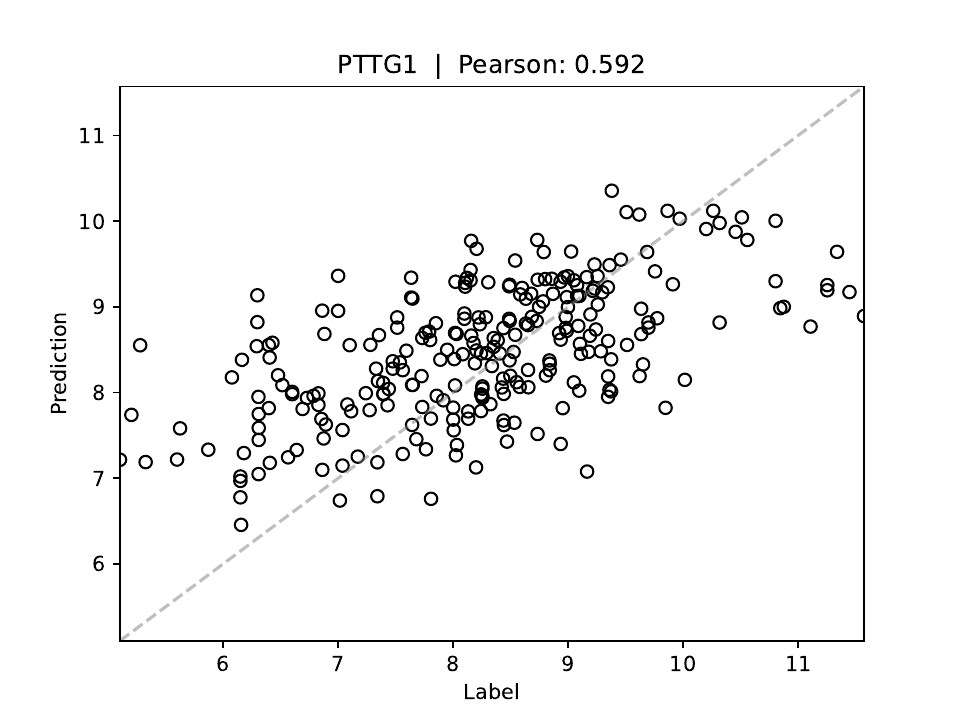}
    \end{subfigure}
    \begin{subfigure}[t]{0.29\textwidth}
        \centering%
        \includegraphics[width=1.0\linewidth]{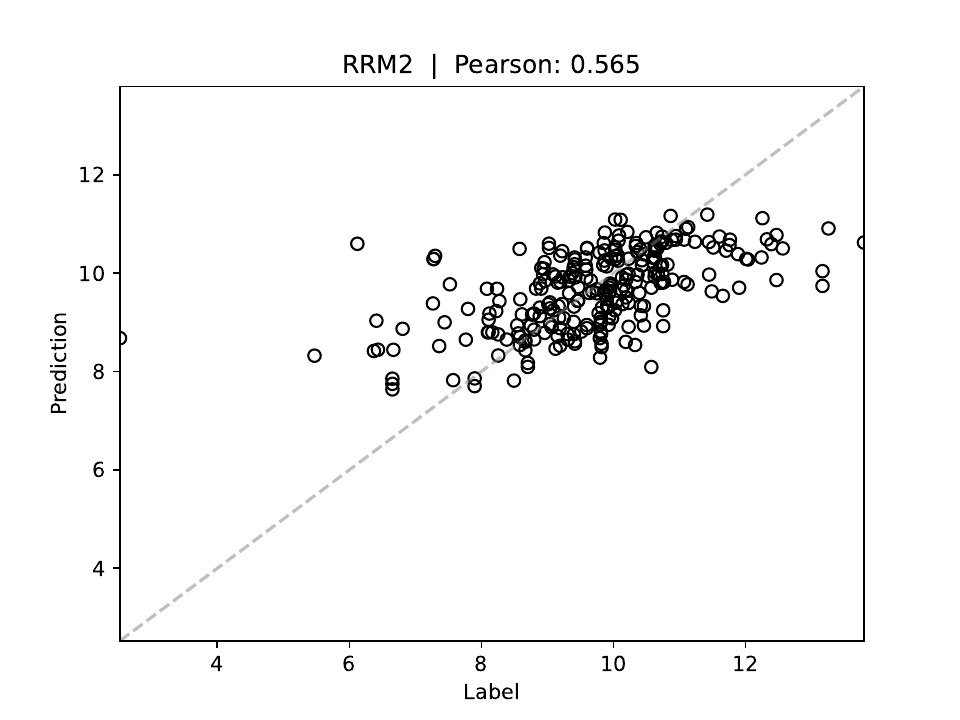}
    \end{subfigure}
    
    \begin{subfigure}[t]{0.29\textwidth}
        \centering%
        \includegraphics[width=1.0\linewidth]{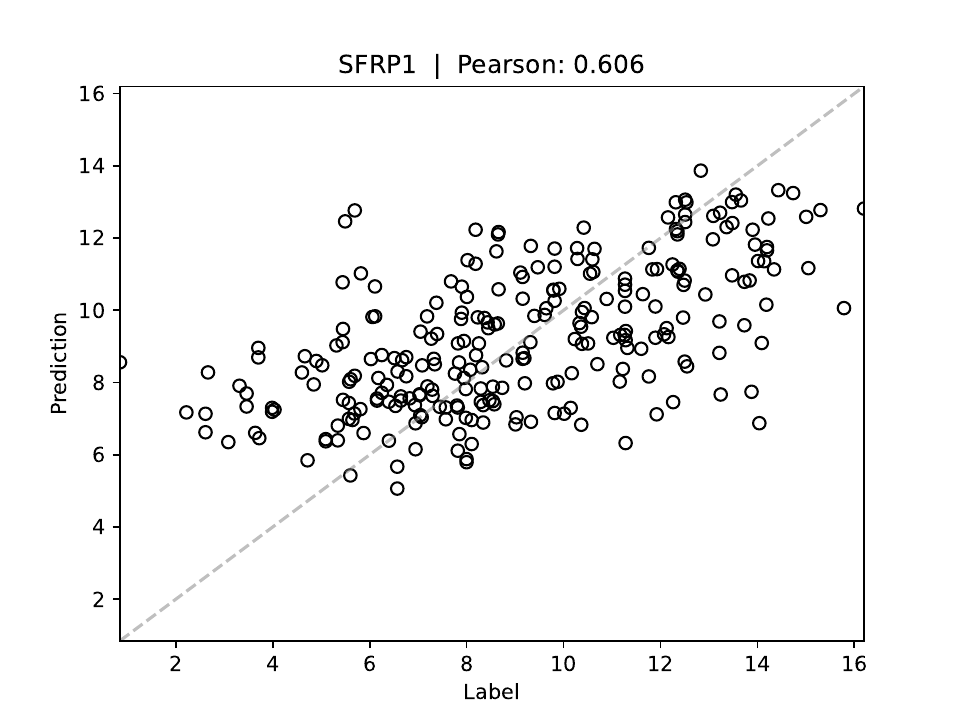}
    \end{subfigure}
    \begin{subfigure}[t]{0.29\textwidth}
        \centering%
        \includegraphics[width=1.0\linewidth]{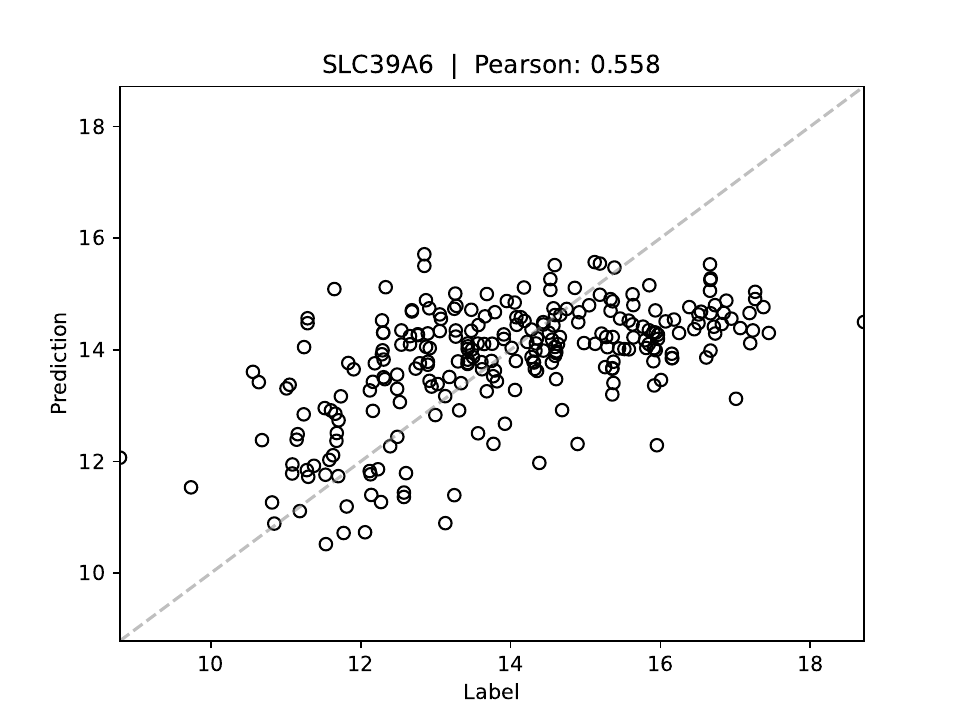}
    \end{subfigure}
    \begin{subfigure}[t]{0.29\textwidth}
        \centering%
        \includegraphics[width=1.0\linewidth]{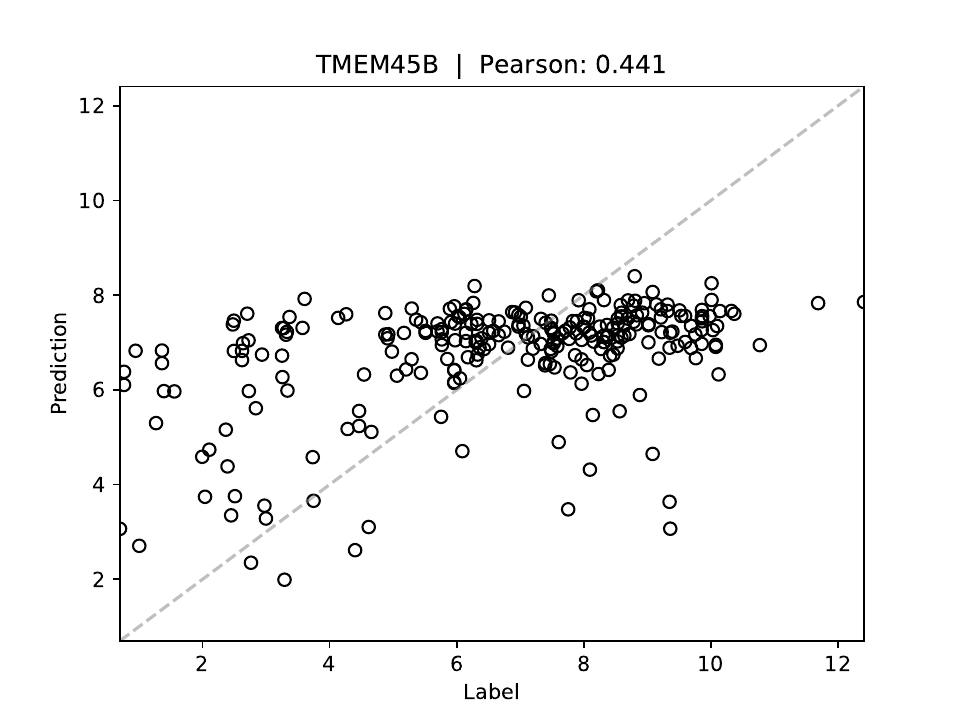}
    \end{subfigure}
    
    \begin{subfigure}[t]{0.29\textwidth}
        \centering%
        \includegraphics[width=1.0\linewidth]{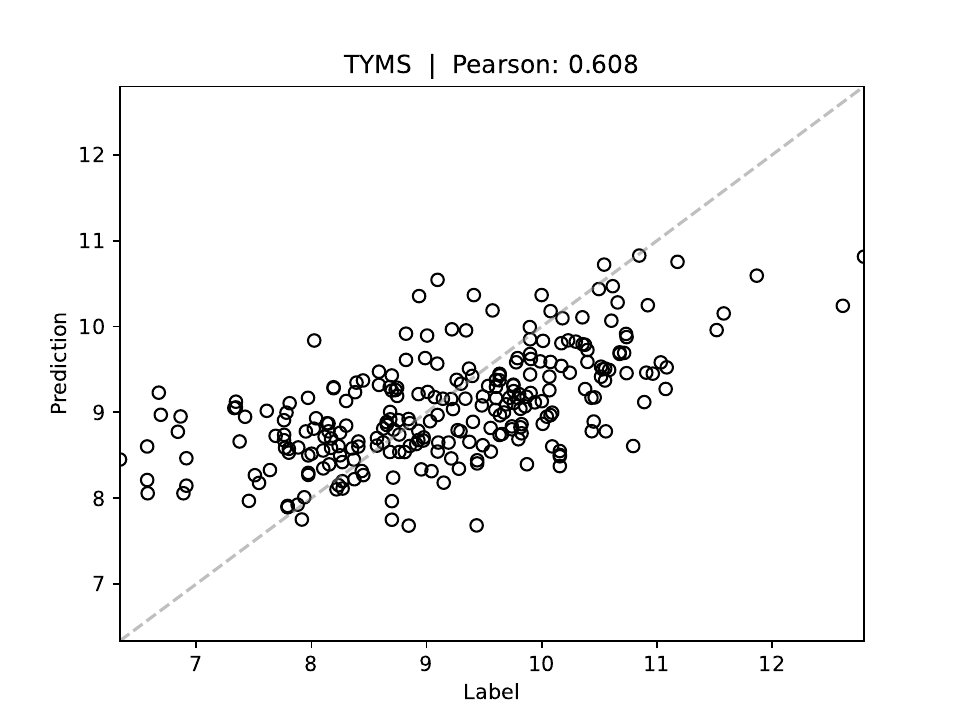}
    \end{subfigure}
    \begin{subfigure}[t]{0.29\textwidth}
        \centering%
        \includegraphics[width=1.0\linewidth]{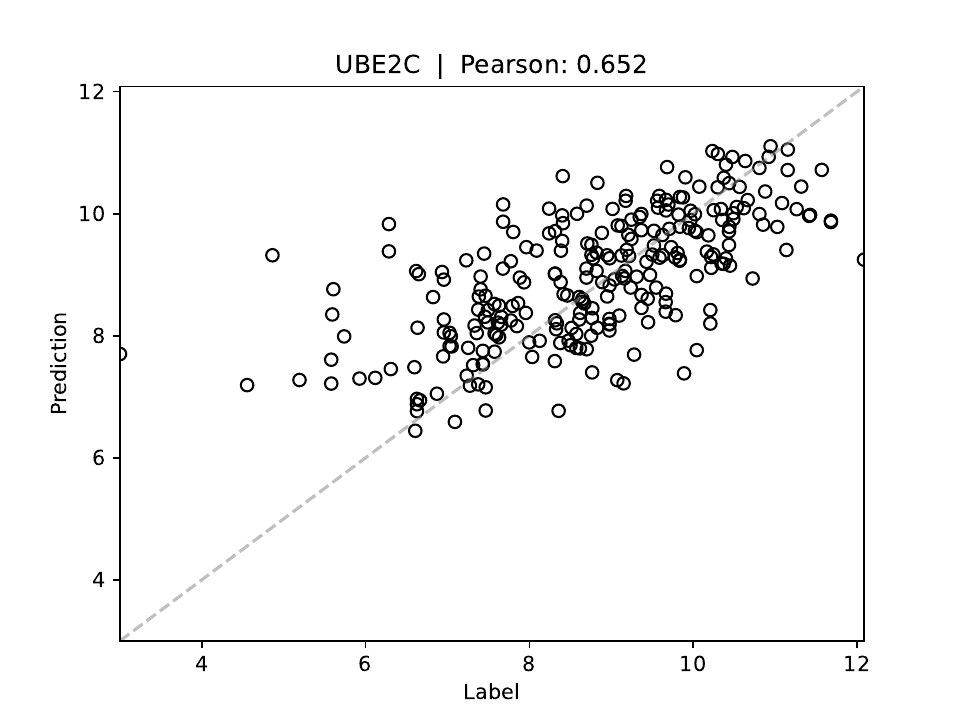}
    \end{subfigure}
    \begin{subfigure}[t]{0.29\textwidth}
        \centering%
        \includegraphics[width=1.0\linewidth]{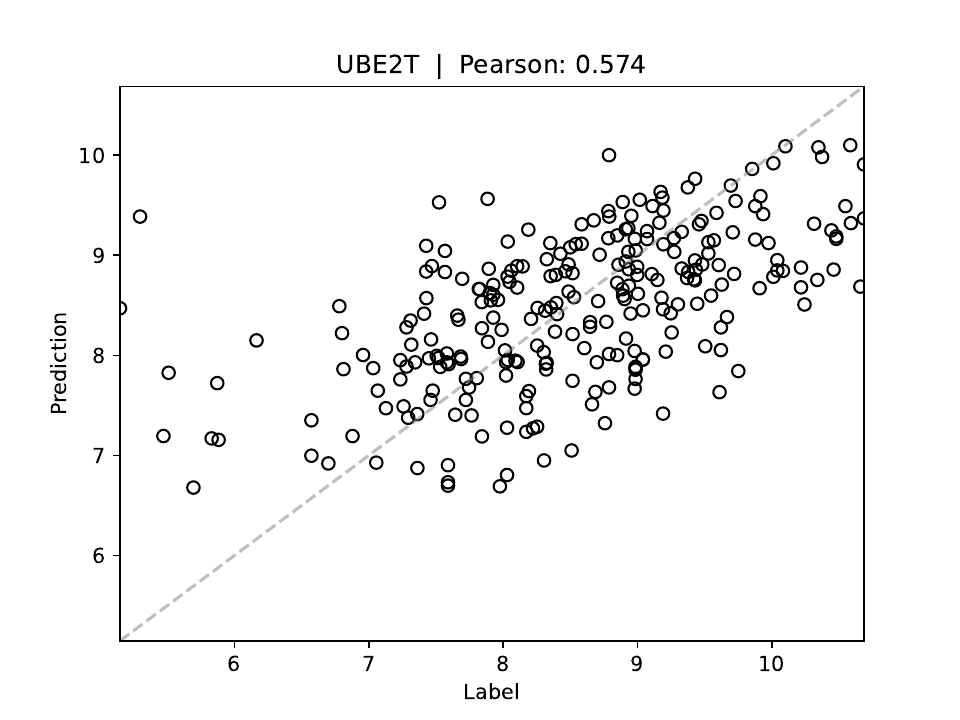}
    \end{subfigure}\vspace{-1.0mm}
    \caption{\textbf{Predicted vs observed gene-expression values for PAM50 gene 36-50 on TCGA-BRCA.} Scatter plots comparing predicted and ground-truth values, for \textit{UNI - Direct - ABMIL} on the TCGA-BRCA dataset, for the test split of the first cross-validation fold.}
  \label{fig:corr_plots_pam50_36_50}
\end{figure*}

\begin{figure*}[h]
\centering
    \begin{subfigure}[t]{0.29\textwidth}
        \centering%
        \includegraphics[width=1.0\linewidth]{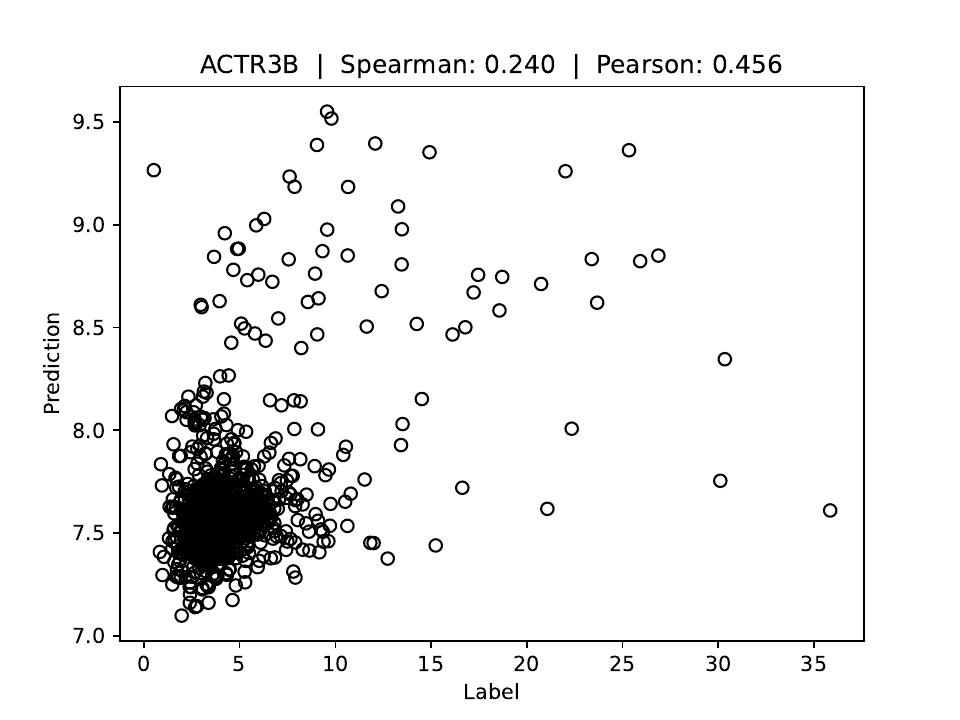}
    \end{subfigure}
    \begin{subfigure}[t]{0.29\textwidth}
        \centering%
        \includegraphics[width=1.0\linewidth]{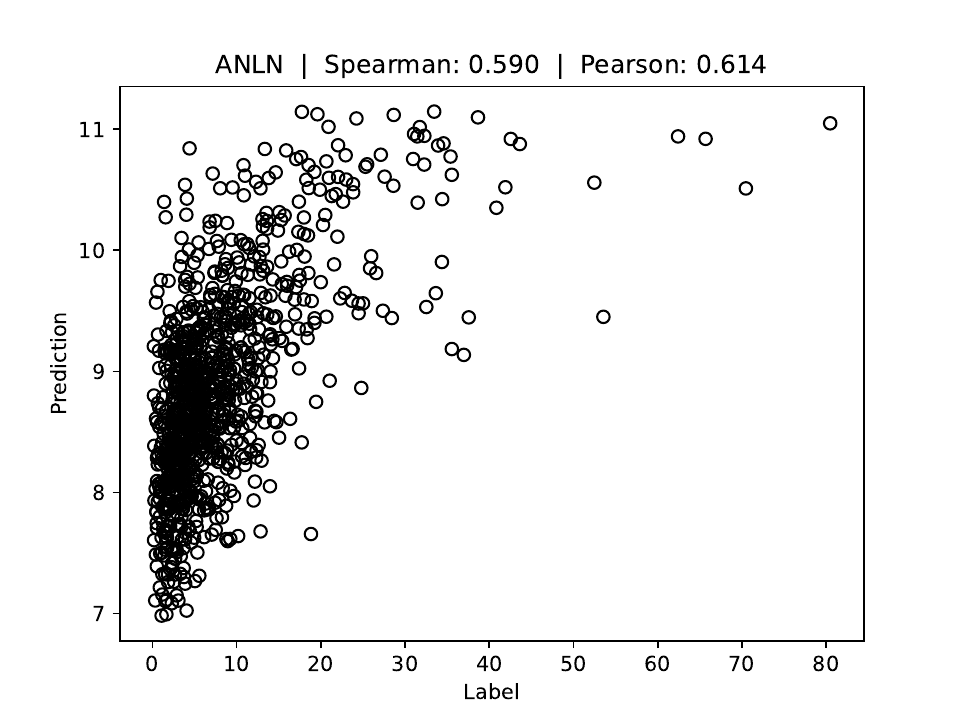}
    \end{subfigure}
    
    \begin{subfigure}[t]{0.29\textwidth}
        \centering%
        \includegraphics[width=1.0\linewidth]{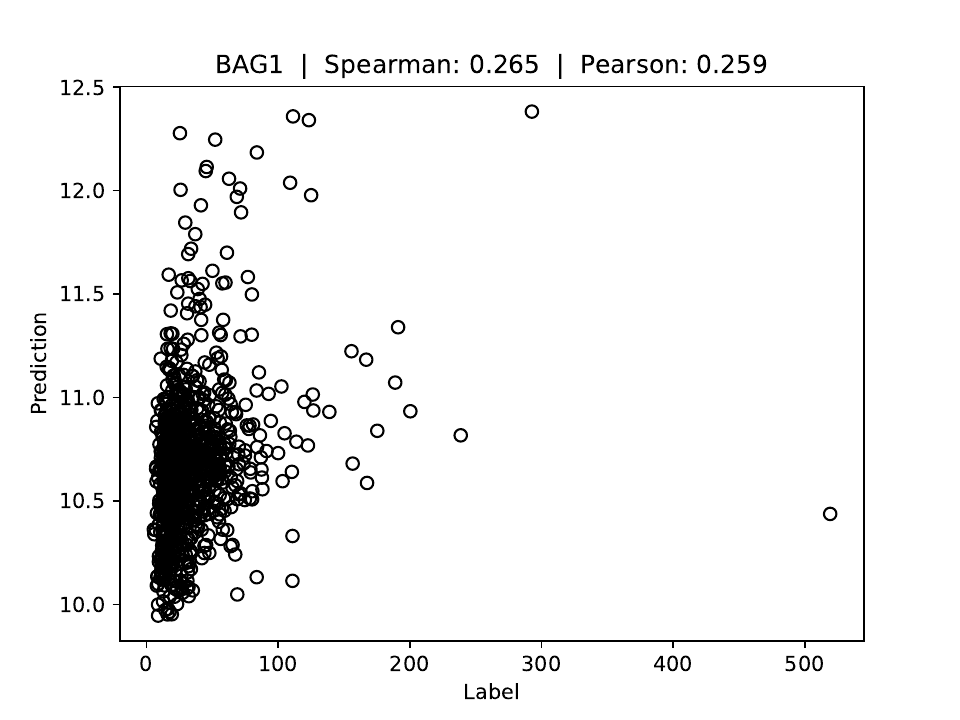}
    \end{subfigure}
    \begin{subfigure}[t]{0.29\textwidth}
        \centering%
        \includegraphics[width=1.0\linewidth]{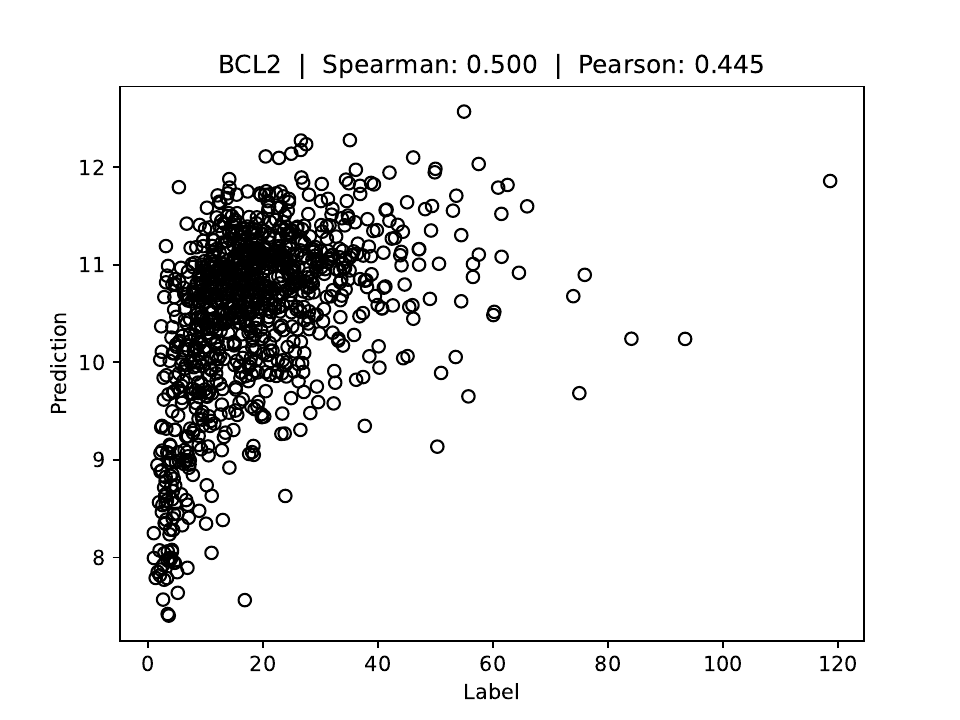}
    \end{subfigure}
    \begin{subfigure}[t]{0.29\textwidth}
        \centering%
        \includegraphics[width=1.0\linewidth]{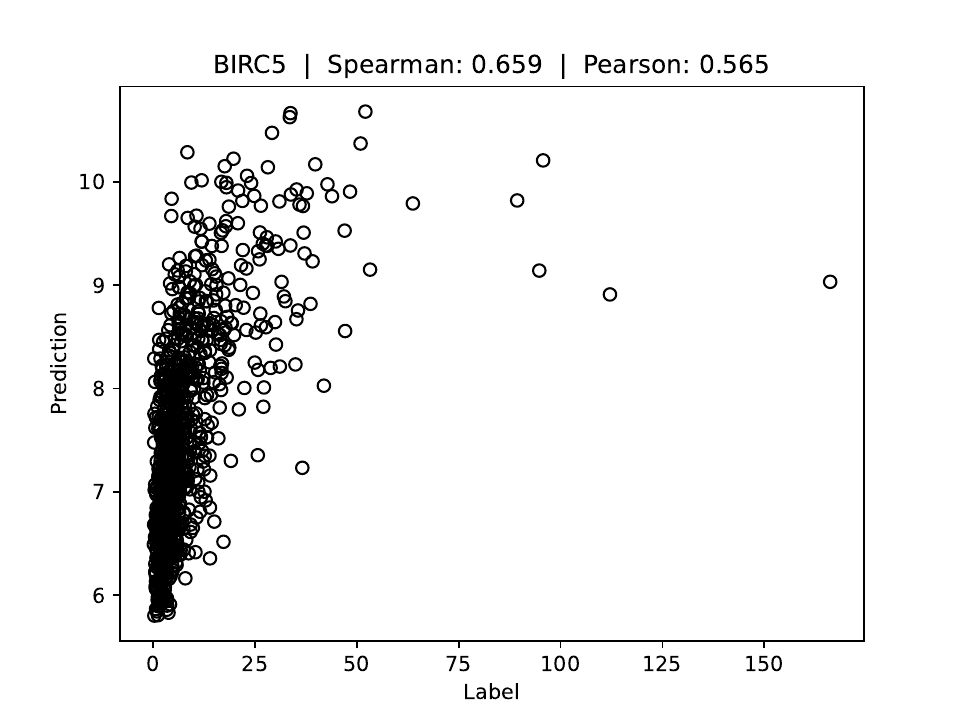}
    \end{subfigure}\vspace{-1.0mm}
    \caption{\textbf{Predicted vs observed gene-expression values for PAM50 gene 1-5 on SCAN-B-Lund.} Scatter plots comparing predicted and ground-truth expression values, for the \textit{H-optimus-1 - Direct - ABMIL} ensemble trained on TCGA-BRCA and evaluated on the external SCAN-B-Lund dataset. Note the difference in scale between predicted and observed gene-expression values (x- and y-axes).}
  \label{fig:corr_plots_pam50_1_5_scan-b-lund}
\end{figure*}

\begin{figure*}[h]
\centering
    \begin{subfigure}[t]{0.29\textwidth}
        \centering%
        \includegraphics[width=1.0\linewidth]{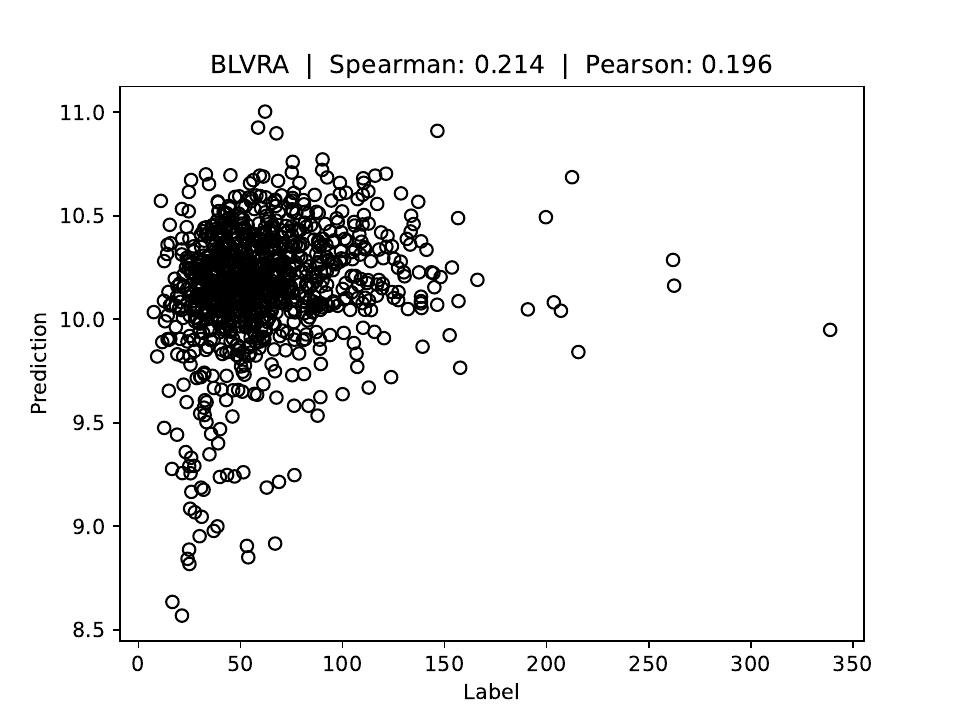}
    \end{subfigure}
    \begin{subfigure}[t]{0.29\textwidth}
        \centering%
        \includegraphics[width=1.0\linewidth]{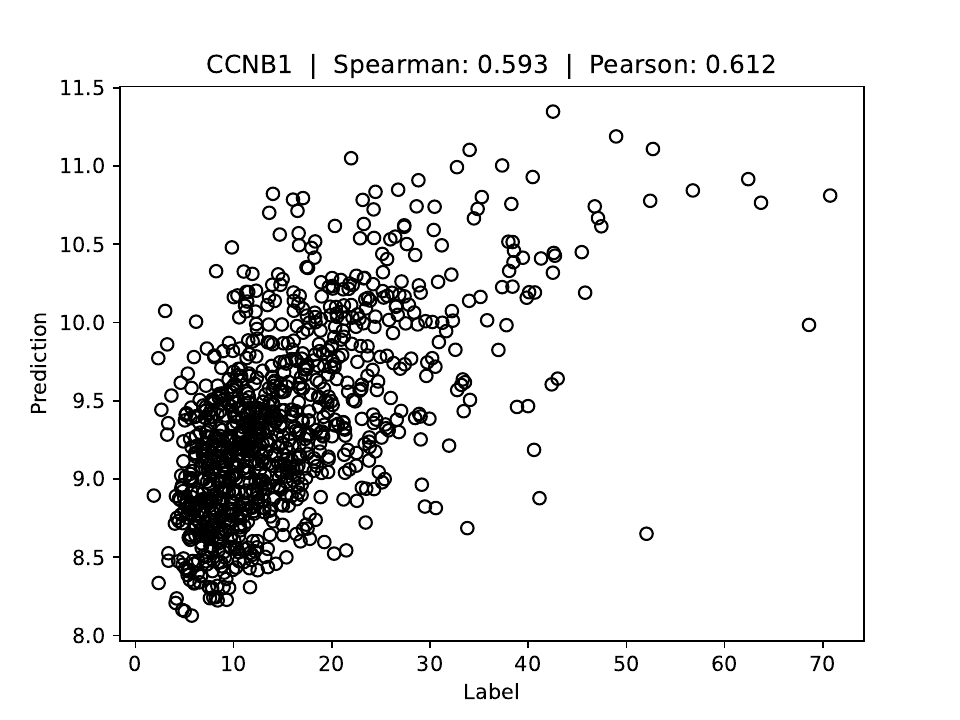}
    \end{subfigure}
    \begin{subfigure}[t]{0.29\textwidth}
        \centering%
        \includegraphics[width=1.0\linewidth]{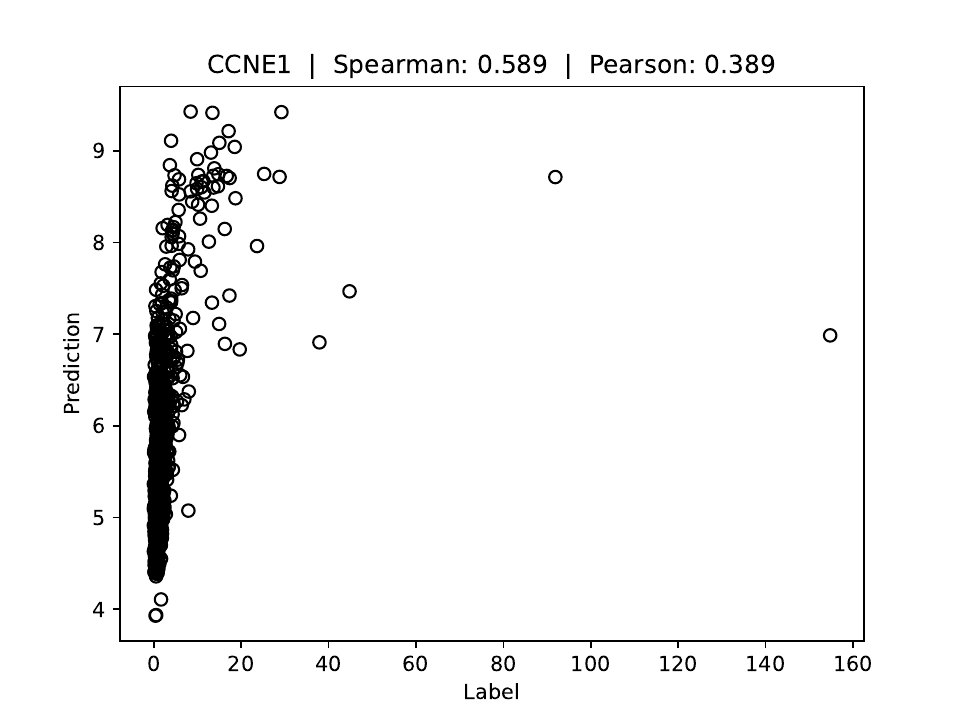}
    \end{subfigure}
    
    \begin{subfigure}[t]{0.29\textwidth}
        \centering%
        \includegraphics[width=1.0\linewidth]{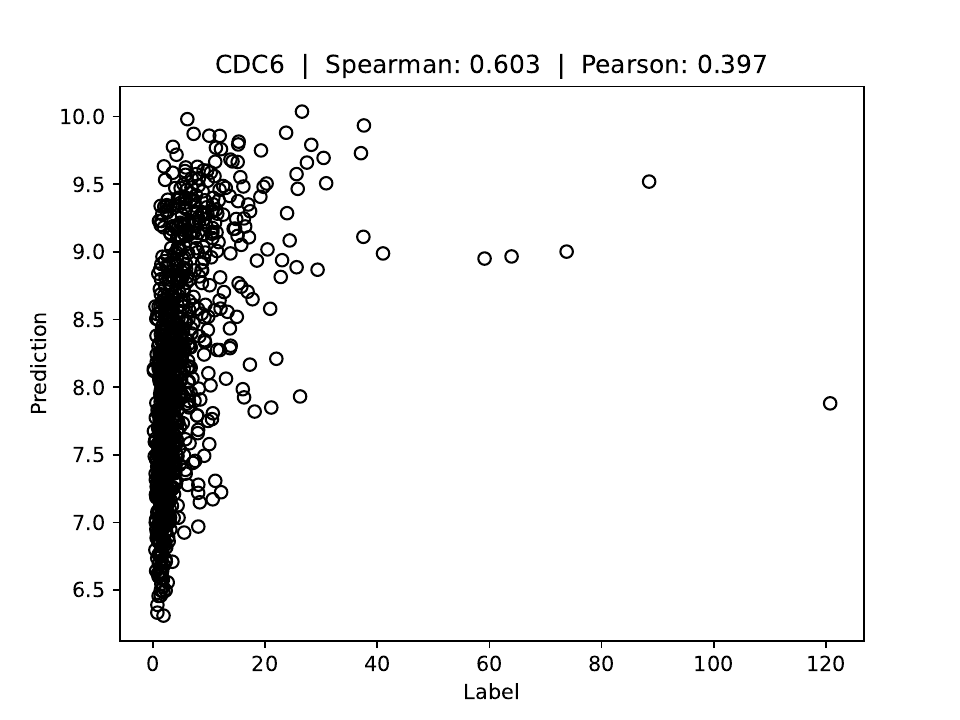}
    \end{subfigure}
    \begin{subfigure}[t]{0.29\textwidth}
        \centering%
        \includegraphics[width=1.0\linewidth]{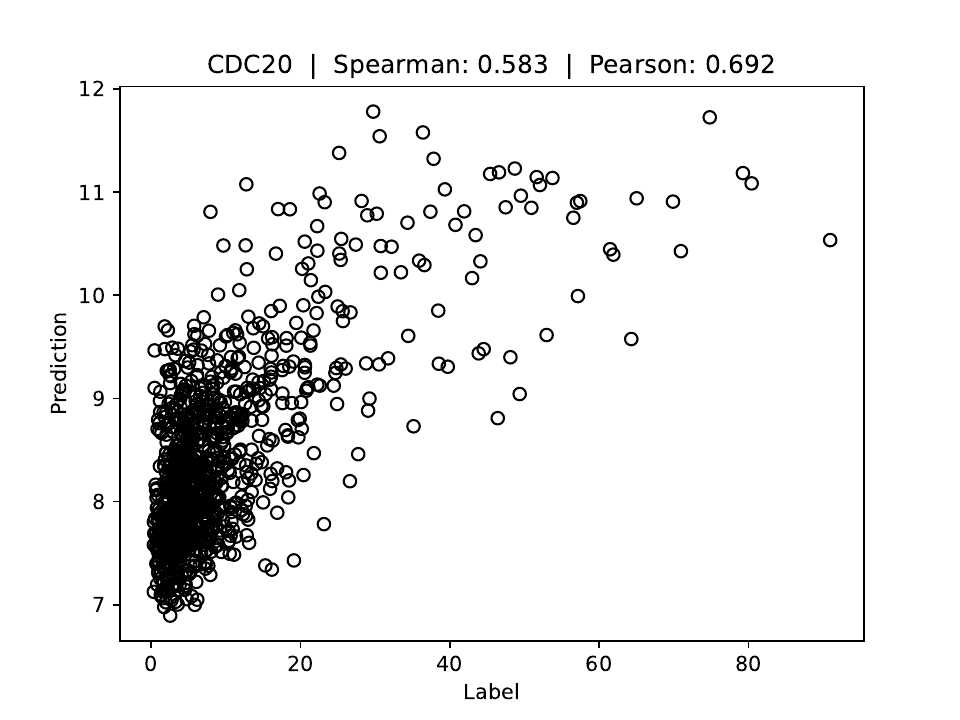}
    \end{subfigure}
    \begin{subfigure}[t]{0.29\textwidth}
        \centering%
        \includegraphics[width=1.0\linewidth]{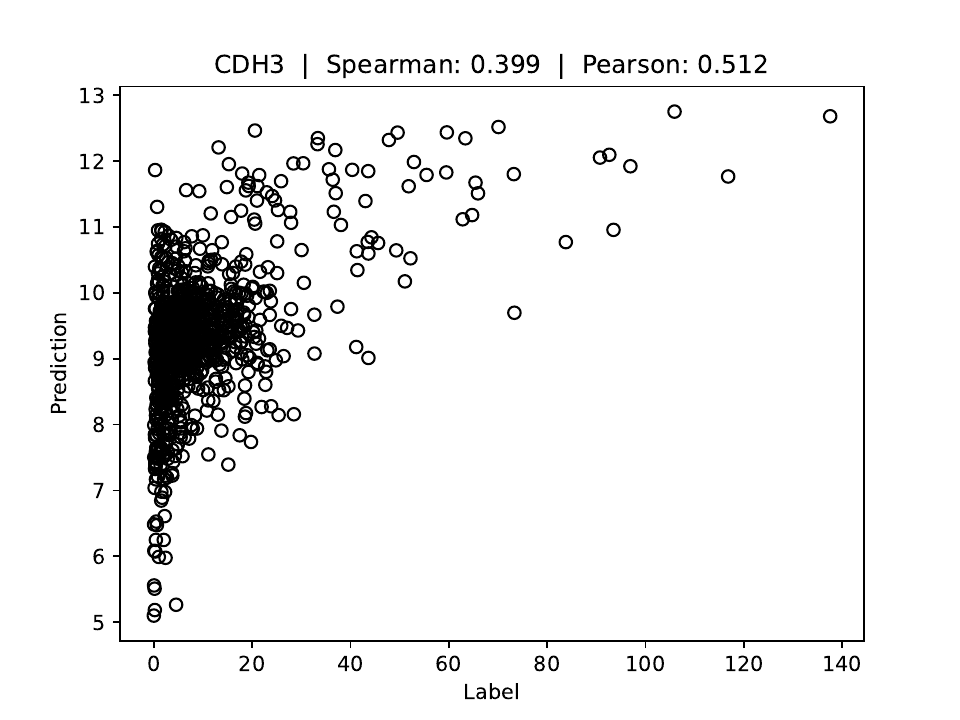}
    \end{subfigure}
    
    \begin{subfigure}[t]{0.29\textwidth}
        \centering%
        \includegraphics[width=1.0\linewidth]{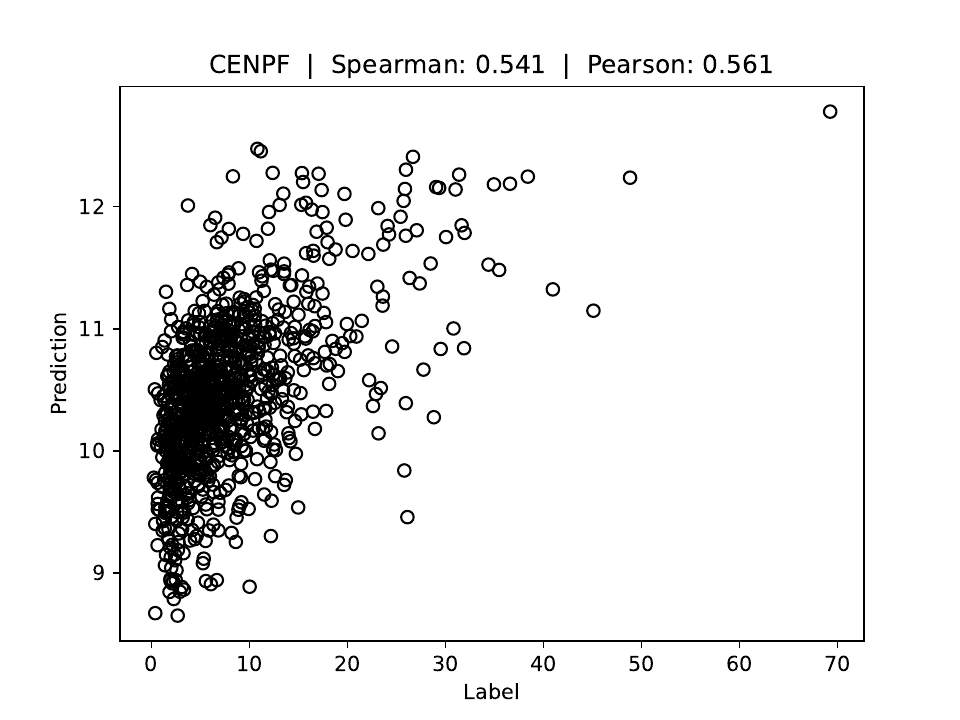}
    \end{subfigure}
    \begin{subfigure}[t]{0.29\textwidth}
        \centering%
        \includegraphics[width=1.0\linewidth]{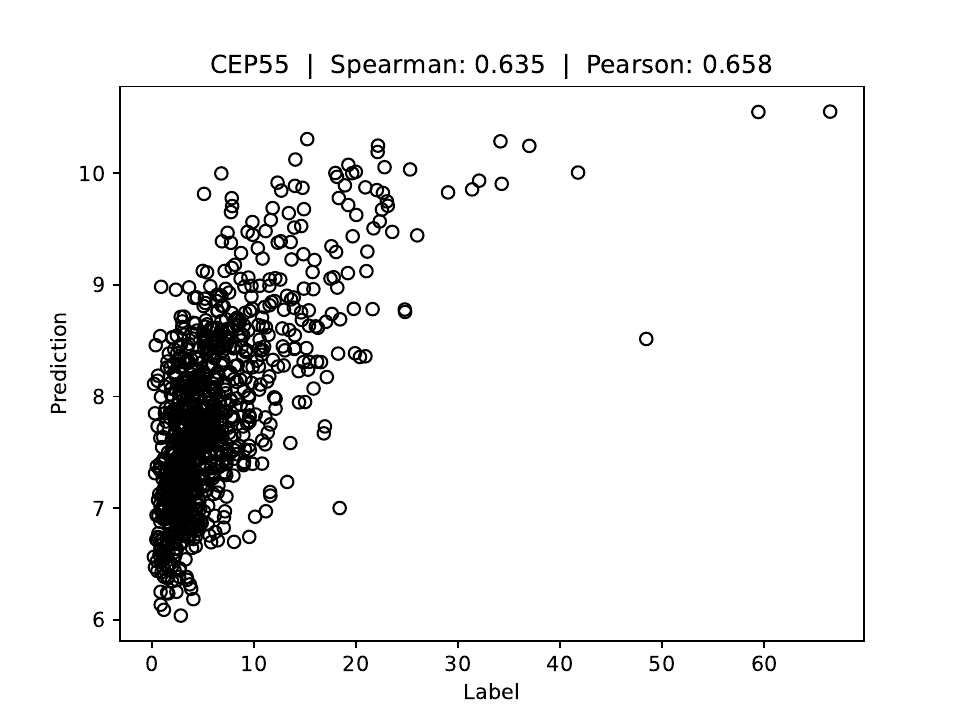}
    \end{subfigure}
    \begin{subfigure}[t]{0.29\textwidth}
        \centering%
        \includegraphics[width=1.0\linewidth]{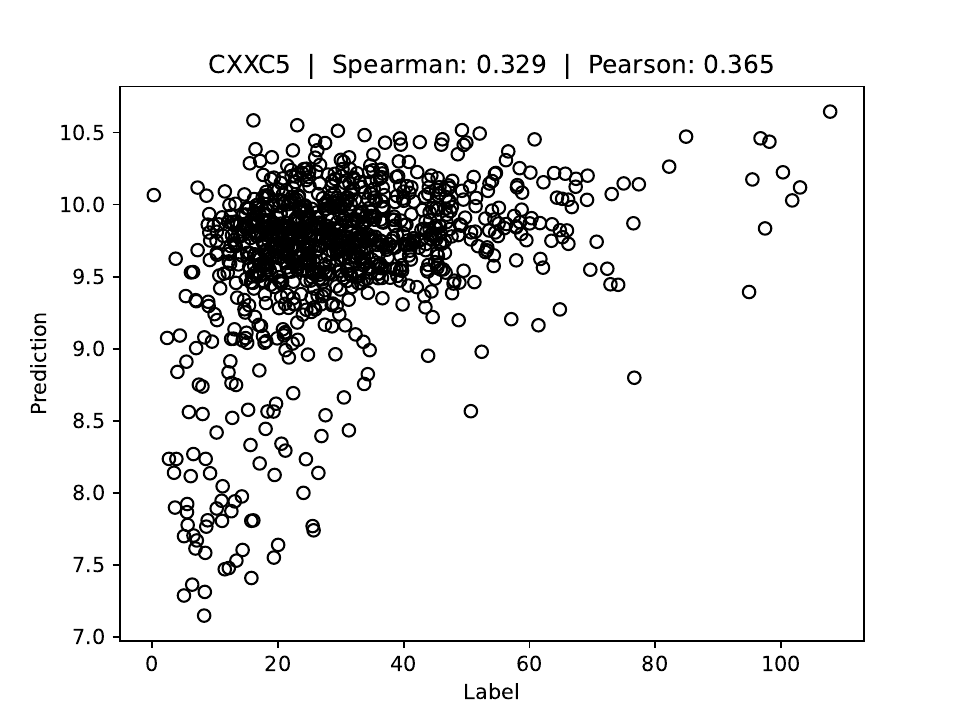}
    \end{subfigure}
    
    \begin{subfigure}[t]{0.29\textwidth}
        \centering%
        \includegraphics[width=1.0\linewidth]{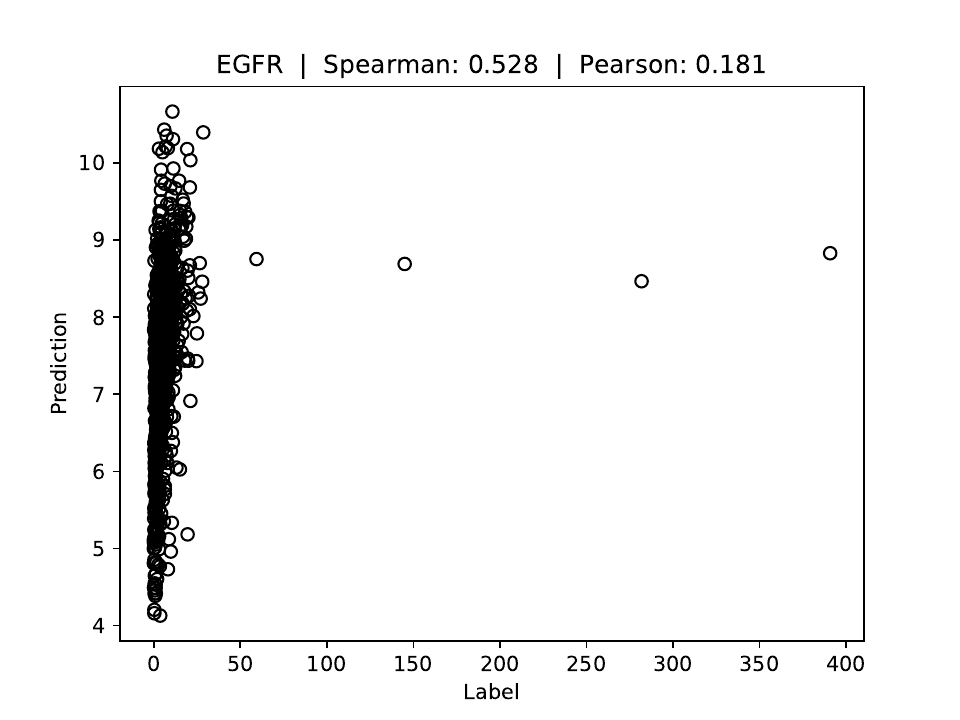}
    \end{subfigure}
    \begin{subfigure}[t]{0.29\textwidth}
        \centering%
        \includegraphics[width=1.0\linewidth]{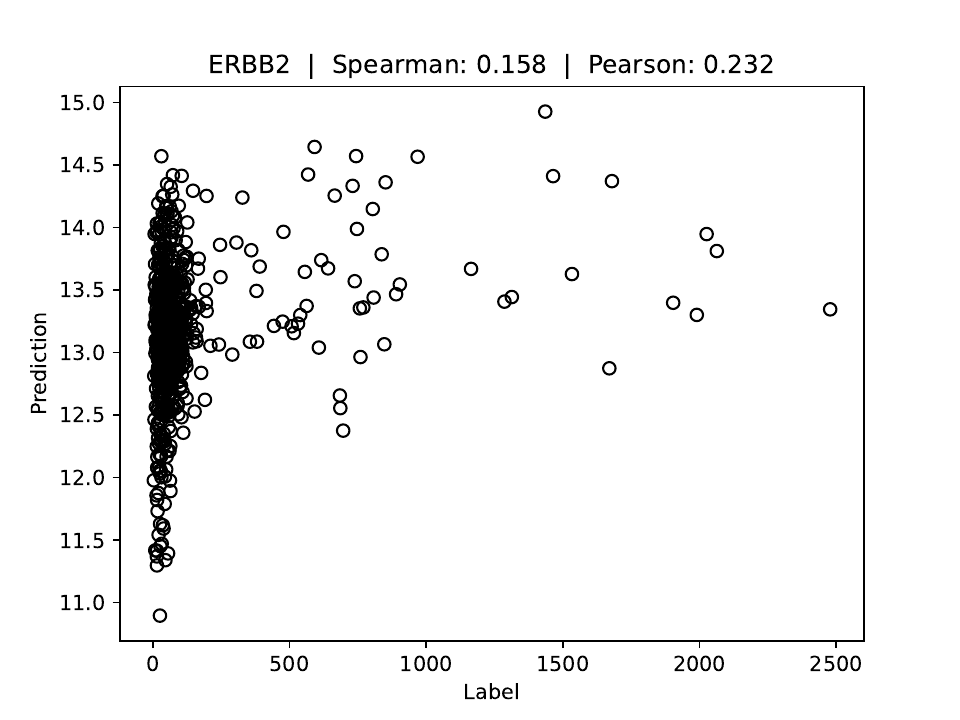}
    \end{subfigure}
    \begin{subfigure}[t]{0.29\textwidth}
        \centering%
        \includegraphics[width=1.0\linewidth]{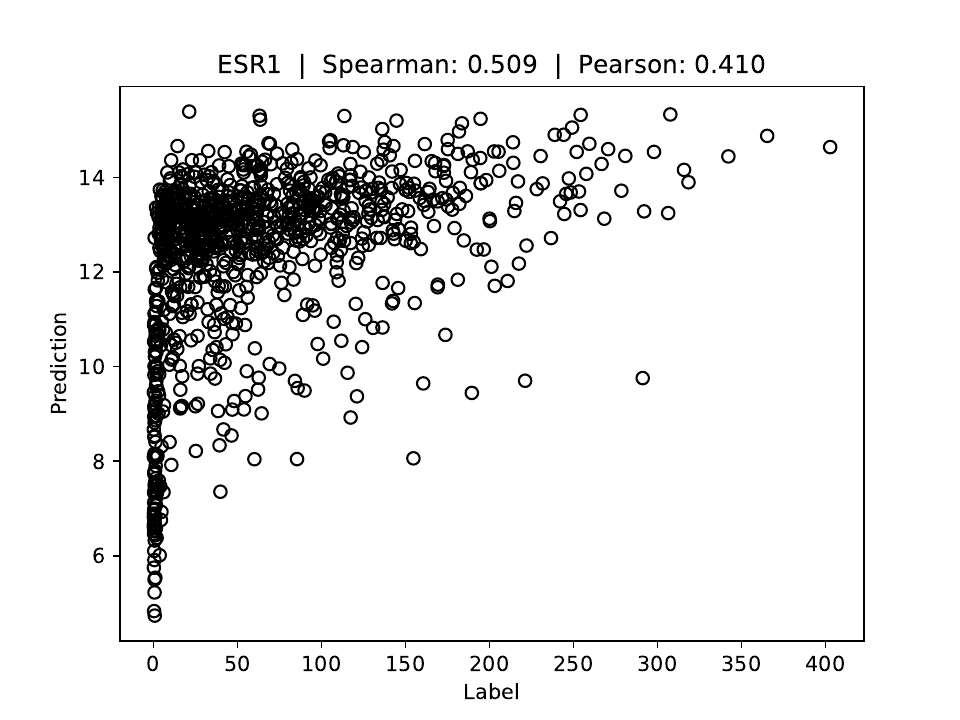}
    \end{subfigure}

    \begin{subfigure}[t]{0.29\textwidth}
        \centering%
        \includegraphics[width=1.0\linewidth]{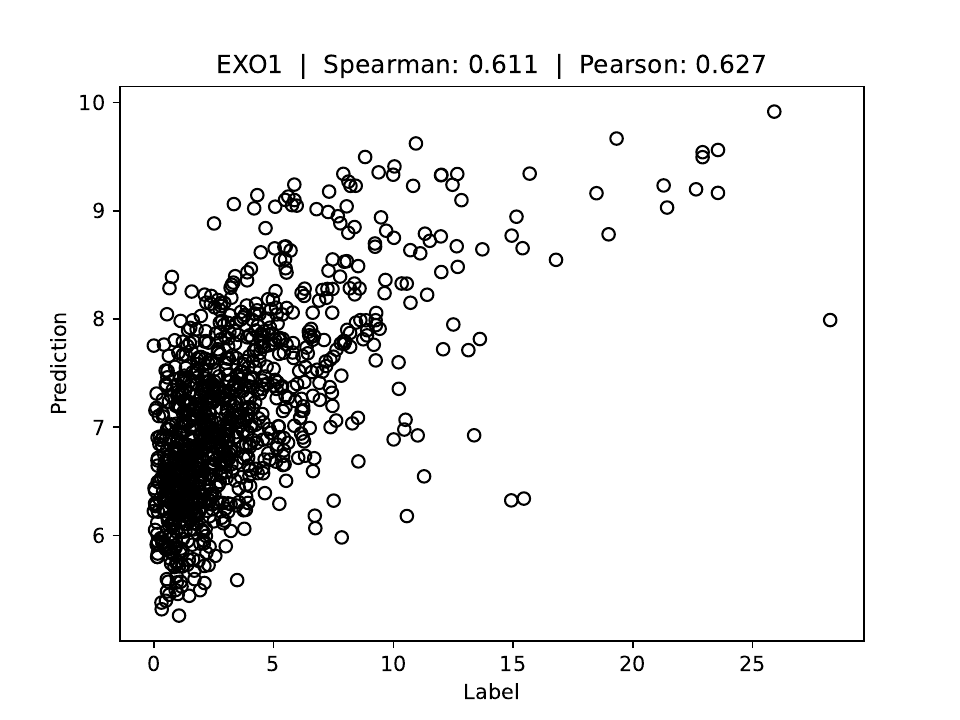}
    \end{subfigure}
    \begin{subfigure}[t]{0.29\textwidth}
        \centering%
        \includegraphics[width=1.0\linewidth]{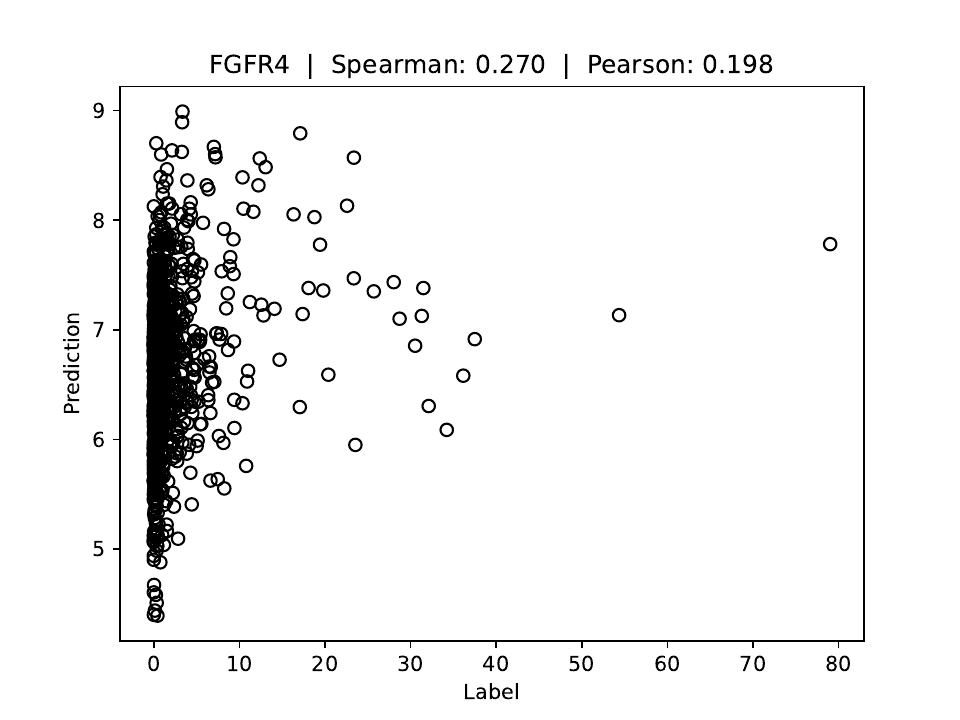}
    \end{subfigure}
    \begin{subfigure}[t]{0.29\textwidth}
        \centering%
        \includegraphics[width=1.0\linewidth]{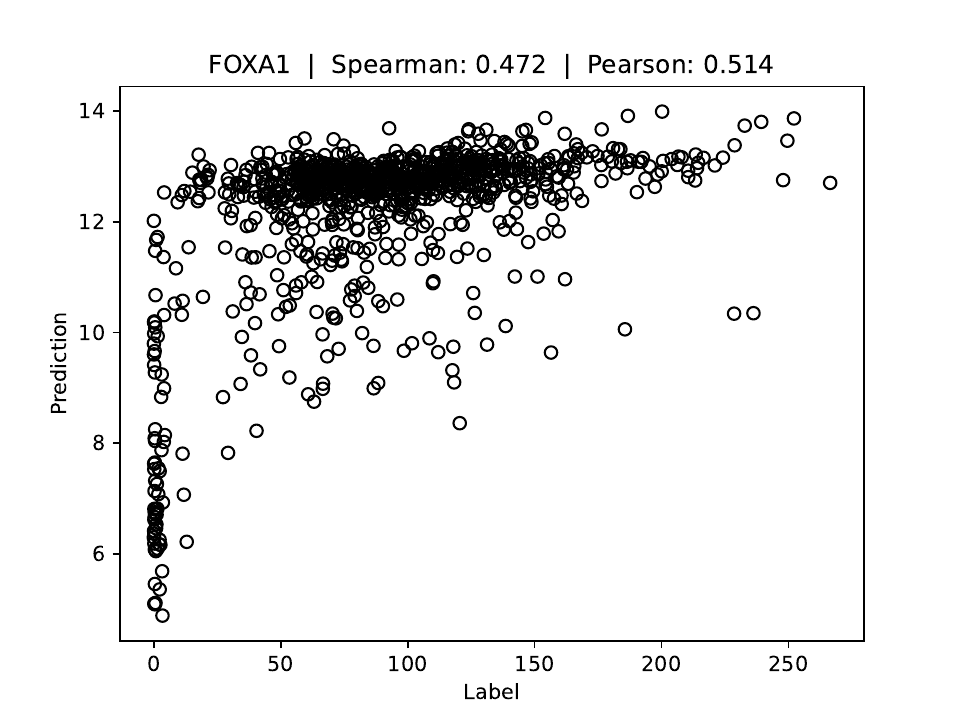}
    \end{subfigure}\vspace{-1.0mm}
    \caption{\textbf{Predicted vs observed gene-expression values for PAM50 gene 6-20 on SCAN-B-Lund.} Scatter plots comparing predicted and ground-truth expression values, for the \textit{H-optimus-1 - Direct - ABMIL} ensemble trained on TCGA-BRCA and evaluated on the external SCAN-B-Lund dataset. Note the difference in scale between predicted and observed gene-expression values (x- and y-axes).}
  \label{fig:corr_plots_pam50_6_20_scan-b-lund}
\end{figure*}

\begin{figure*}[h]
\centering
    \begin{subfigure}[t]{0.29\textwidth}
        \centering%
        \includegraphics[width=1.0\linewidth]{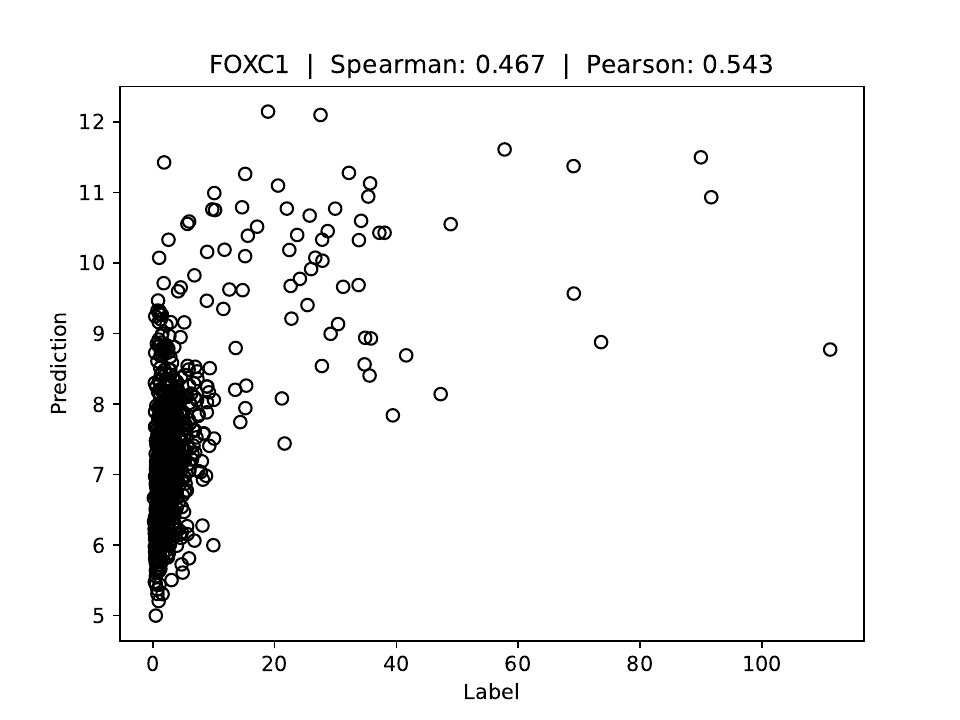}
    \end{subfigure}
    \begin{subfigure}[t]{0.29\textwidth}
        \centering%
        \includegraphics[width=1.0\linewidth]{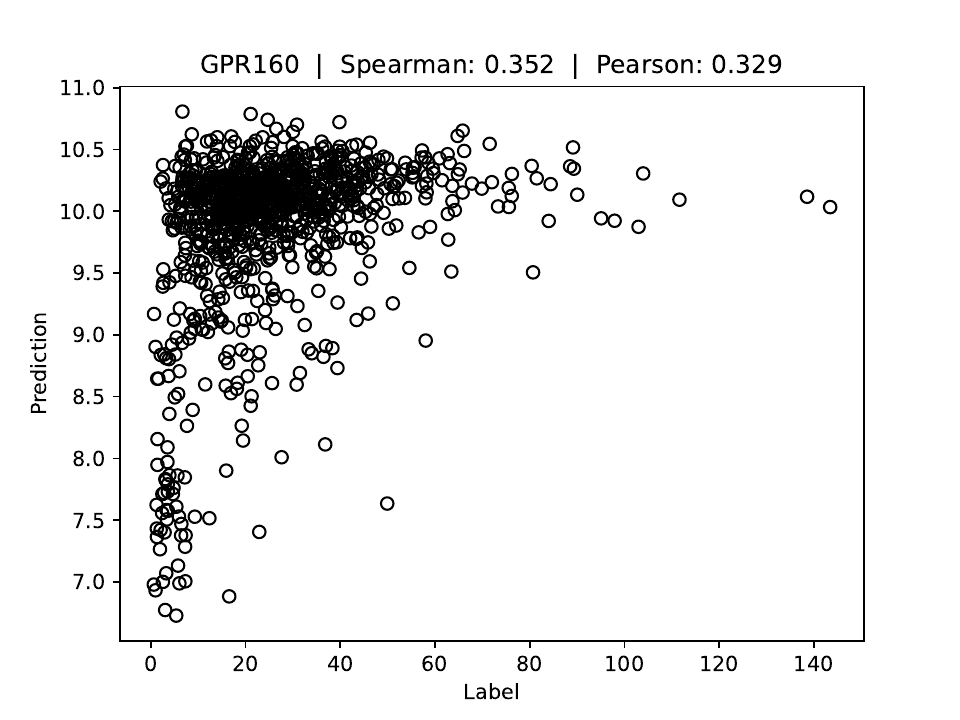}
    \end{subfigure}
    \begin{subfigure}[t]{0.29\textwidth}
        \centering%
        \includegraphics[width=1.0\linewidth]{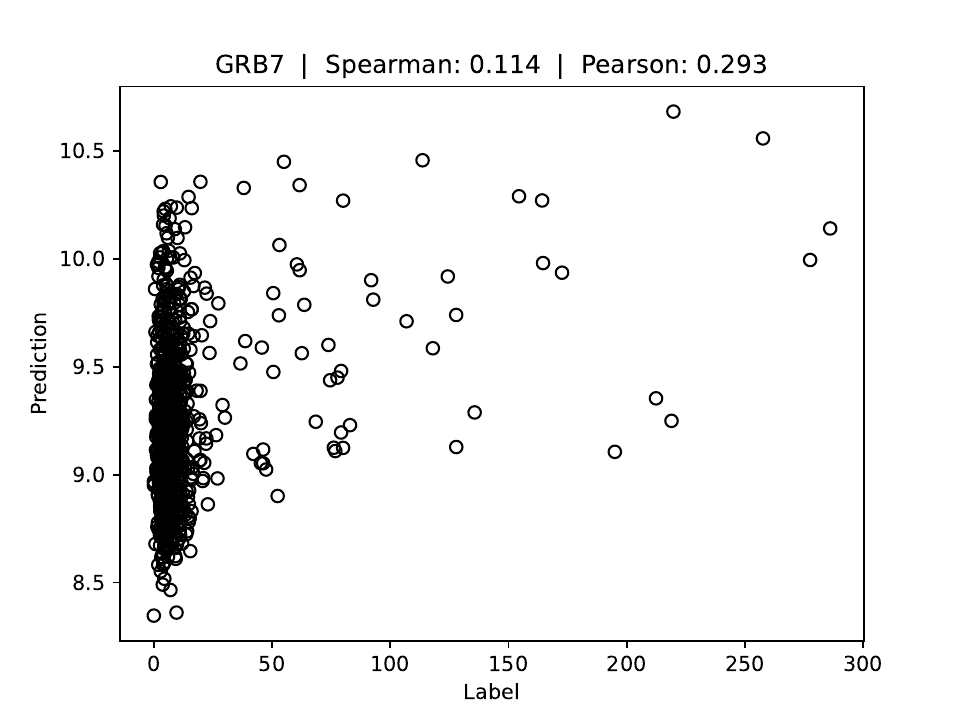}
    \end{subfigure}
    
    \begin{subfigure}[t]{0.29\textwidth}
        \centering%
        \includegraphics[width=1.0\linewidth]{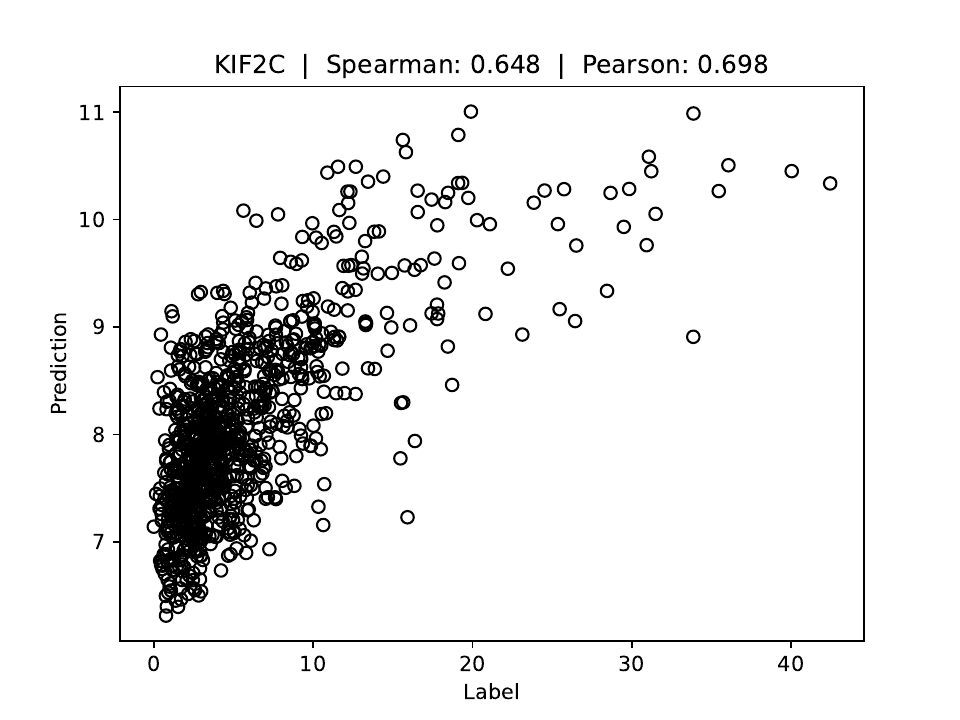}
    \end{subfigure}
    \begin{subfigure}[t]{0.29\textwidth}
        \centering%
        \includegraphics[width=1.0\linewidth]{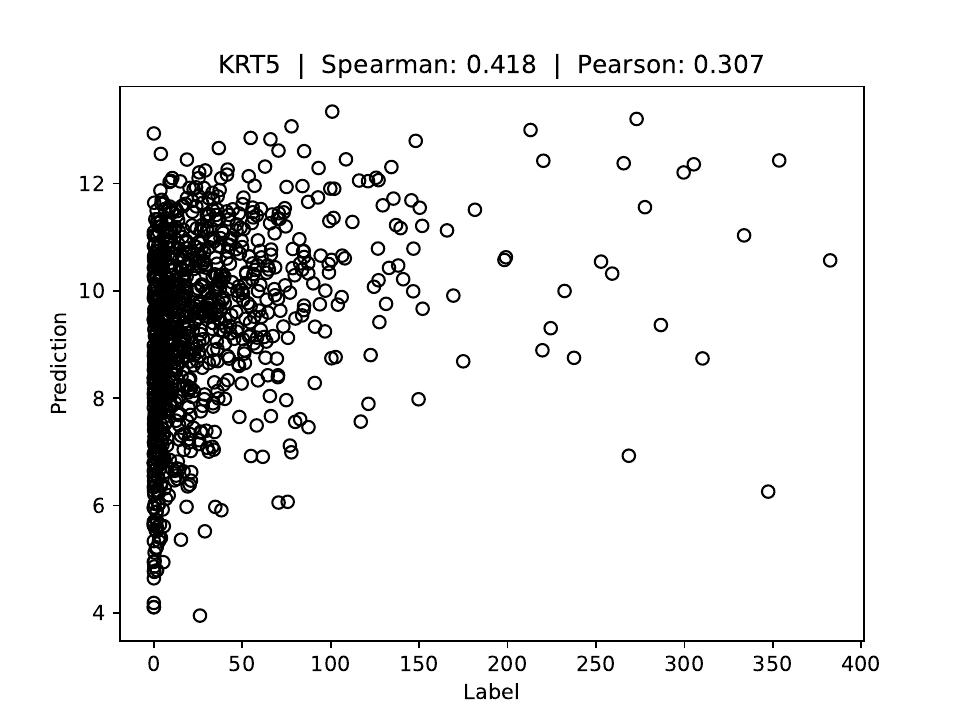}
    \end{subfigure}
    \begin{subfigure}[t]{0.29\textwidth}
        \centering%
        \includegraphics[width=1.0\linewidth]{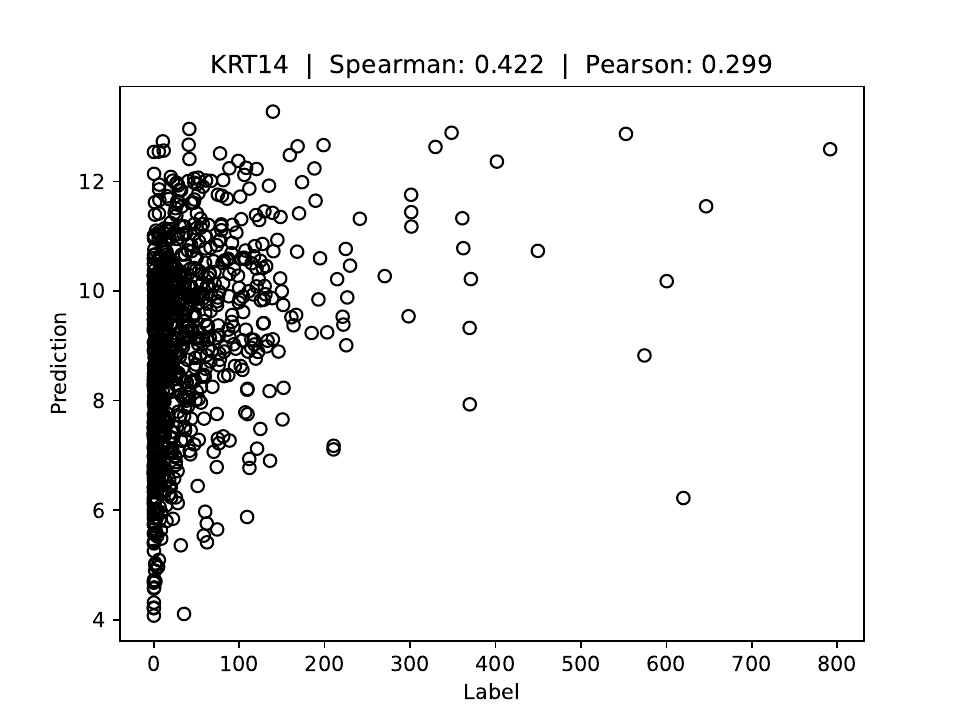}
    \end{subfigure}
    
    \begin{subfigure}[t]{0.29\textwidth}
        \centering%
        \includegraphics[width=1.0\linewidth]{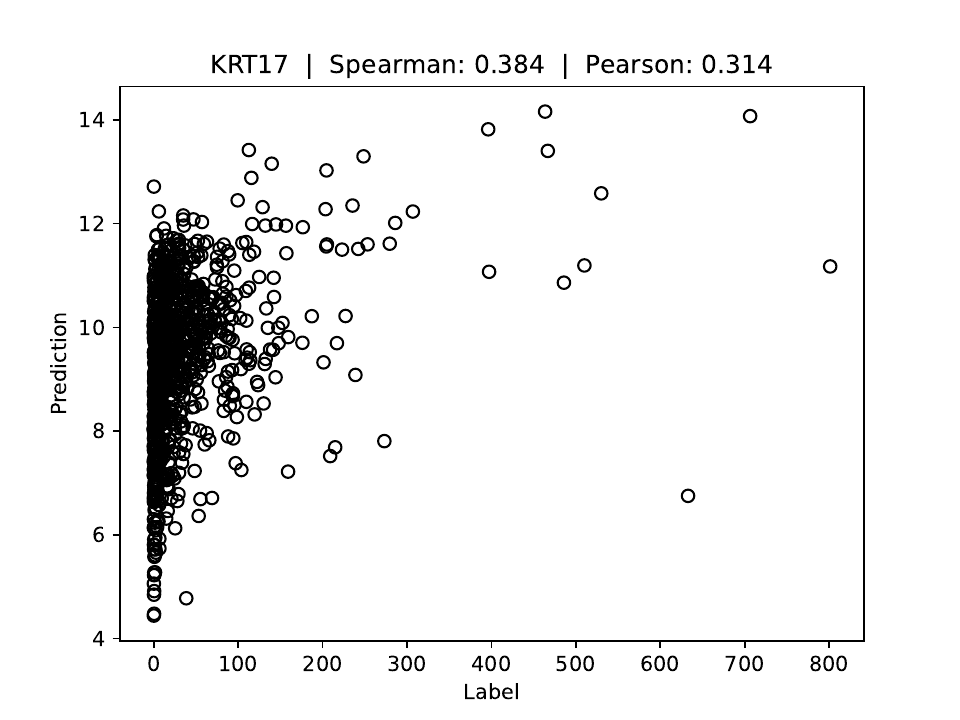}
    \end{subfigure}
    \begin{subfigure}[t]{0.29\textwidth}
        \centering%
        \includegraphics[width=1.0\linewidth]{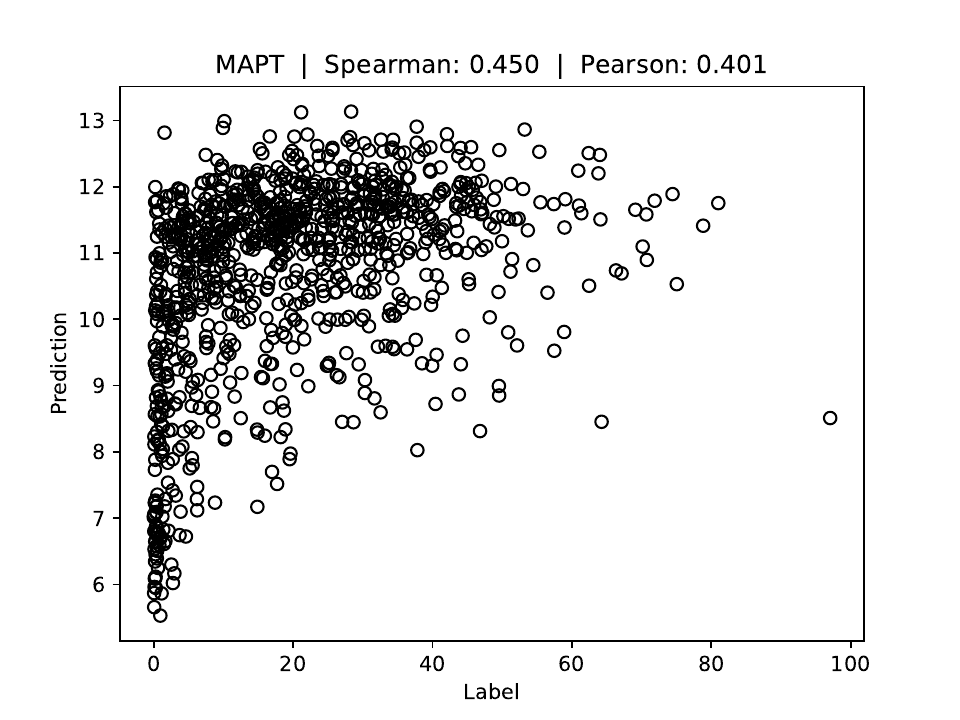}
    \end{subfigure}%\hspace{57.0mm}
    \begin{subfigure}[t]{0.29\textwidth}
        \centering%
        \includegraphics[width=1.0\linewidth]{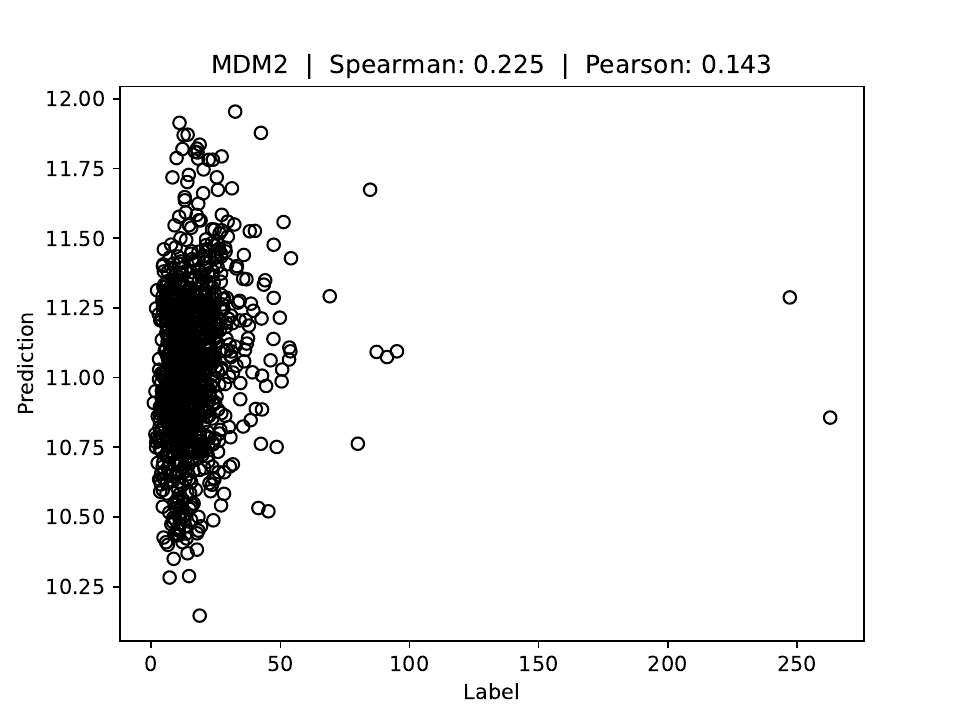}
    \end{subfigure}
    
    \begin{subfigure}[t]{0.29\textwidth}
        \centering%
        \includegraphics[width=1.0\linewidth]{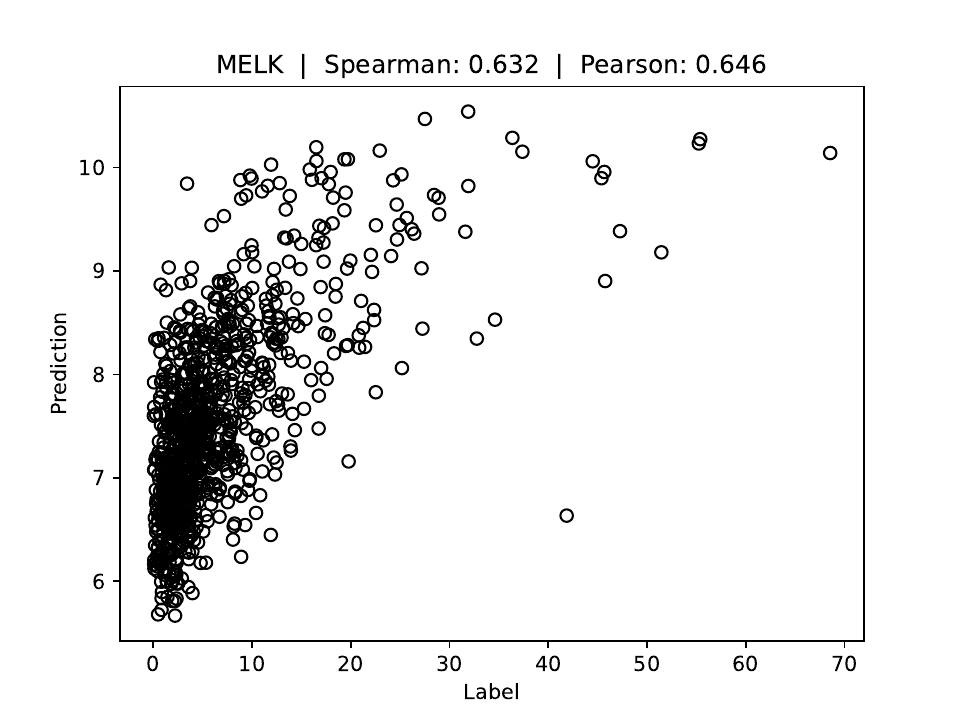}
    \end{subfigure}
    \begin{subfigure}[t]{0.29\textwidth}
        \centering%
        \includegraphics[width=1.0\linewidth]{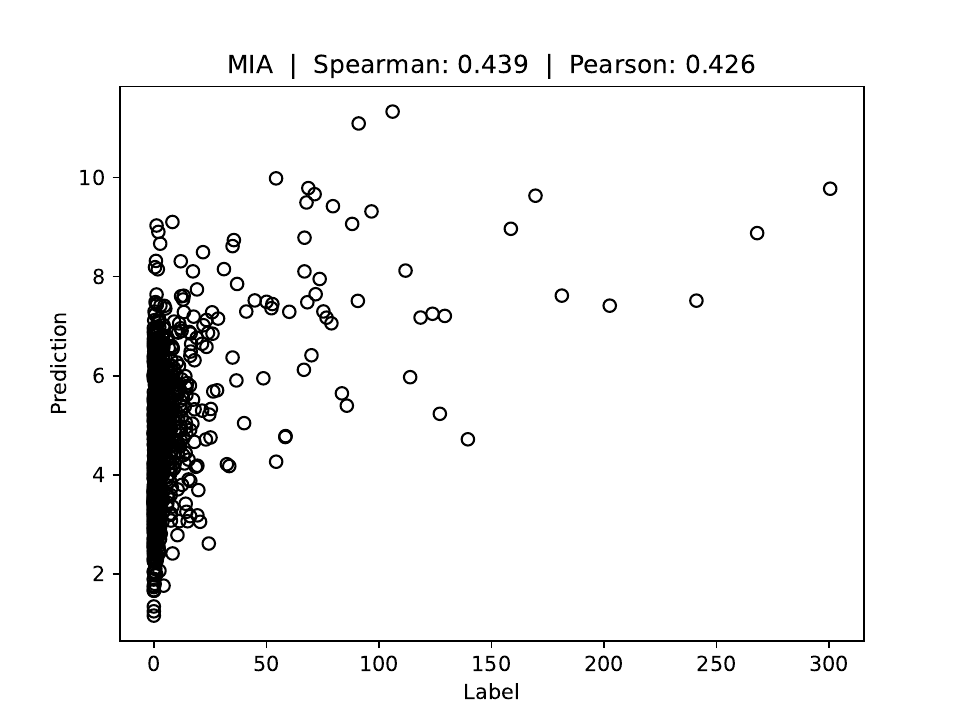}
    \end{subfigure}
    \begin{subfigure}[t]{0.29\textwidth}
        \centering%
        \includegraphics[width=1.0\linewidth]{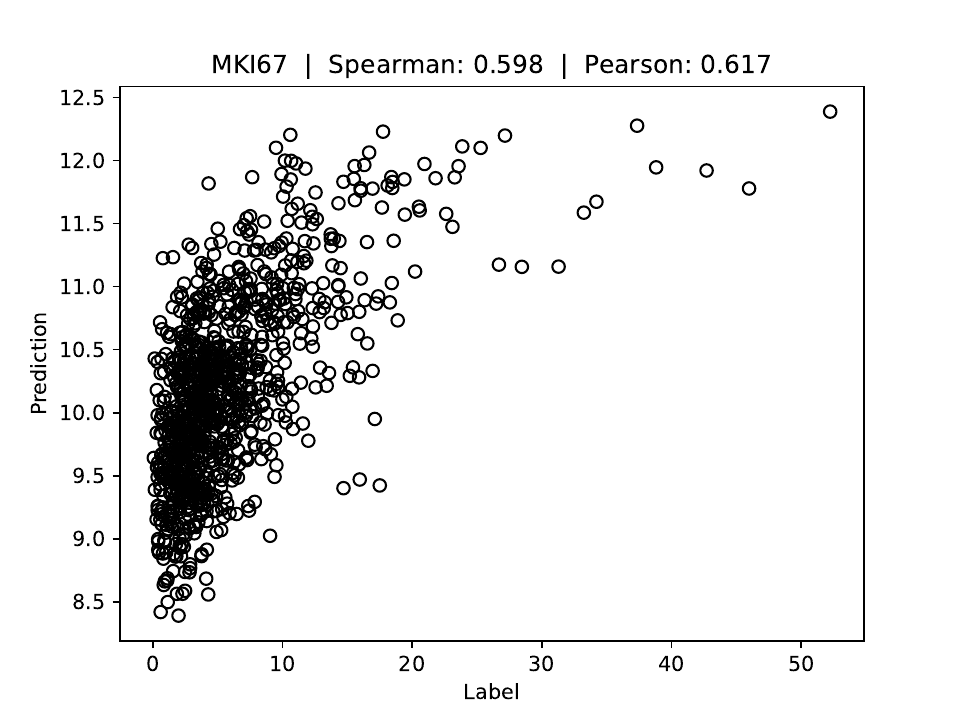}
    \end{subfigure}
    
    \begin{subfigure}[t]{0.29\textwidth}
        \centering%
        \includegraphics[width=1.0\linewidth]{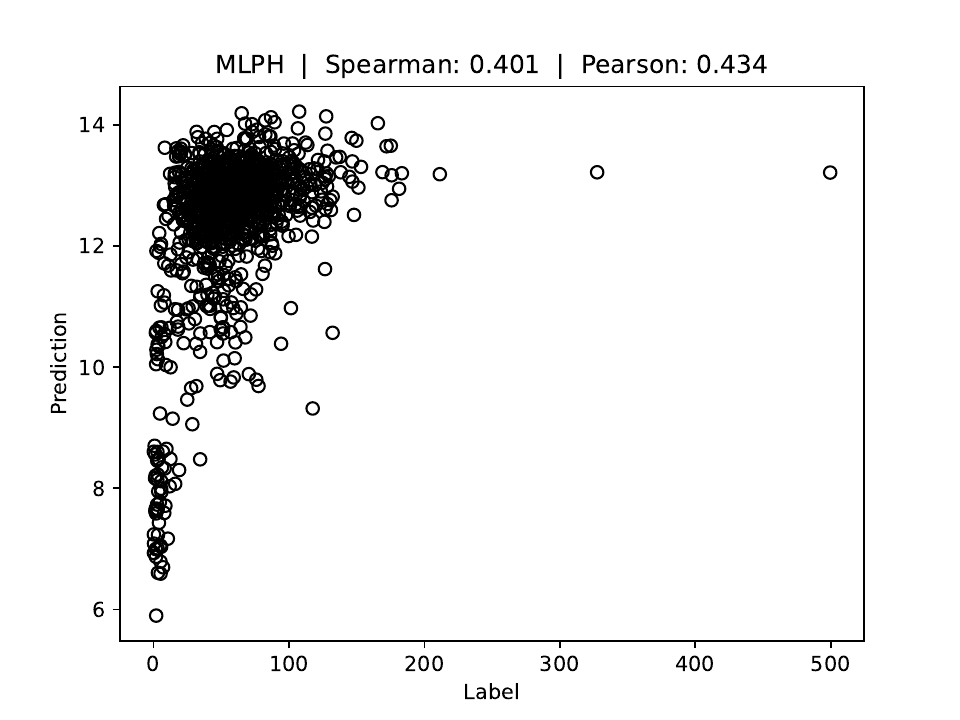}
    \end{subfigure}
    \begin{subfigure}[t]{0.29\textwidth}
        \centering%
        \includegraphics[width=1.0\linewidth]{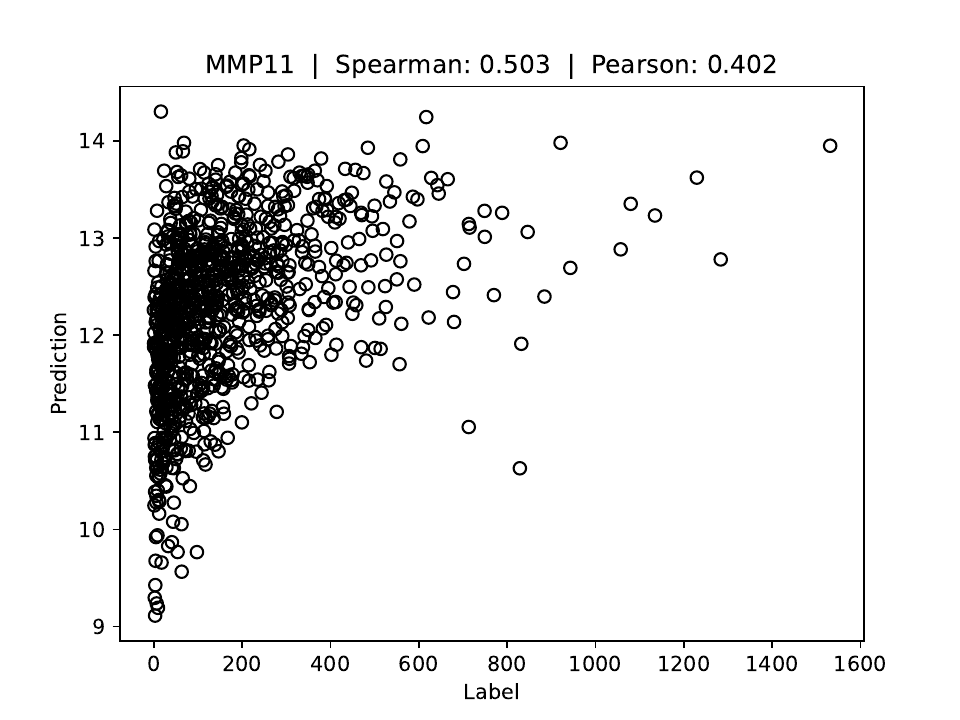}
    \end{subfigure}
    \begin{subfigure}[t]{0.29\textwidth}
        \centering%
        \includegraphics[width=1.0\linewidth]{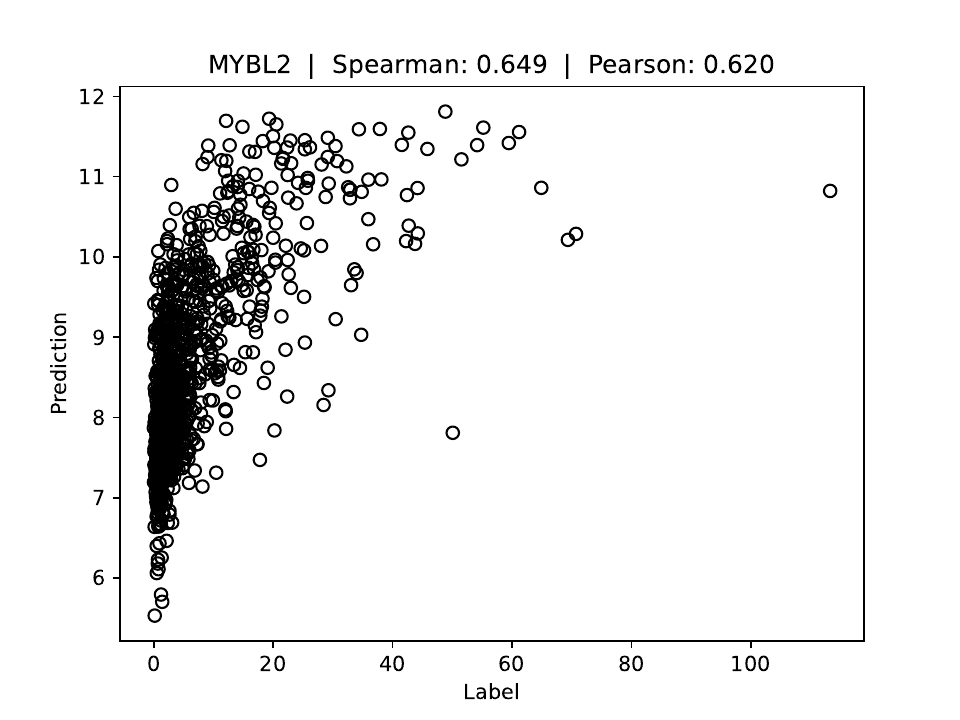}
    \end{subfigure}\vspace{-1.0mm}
    \caption{\textbf{Predicted vs observed gene-expression values for PAM50 gene 21-35 on SCAN-B-Lund.} Scatter plots comparing predicted and ground-truth expression values, for the \textit{H-optimus-1 - Direct - ABMIL} ensemble trained on TCGA-BRCA and evaluated on the external SCAN-B-Lund dataset. Note the difference in scale between predicted and observed gene-expression values (x- and y-axes).}
  \label{fig:corr_plots_pam50_21_35_scan-b-lund}
\end{figure*}

\begin{figure*}[h]
\centering
    \begin{subfigure}[t]{0.29\textwidth}
        \centering%
        \includegraphics[width=1.0\linewidth]{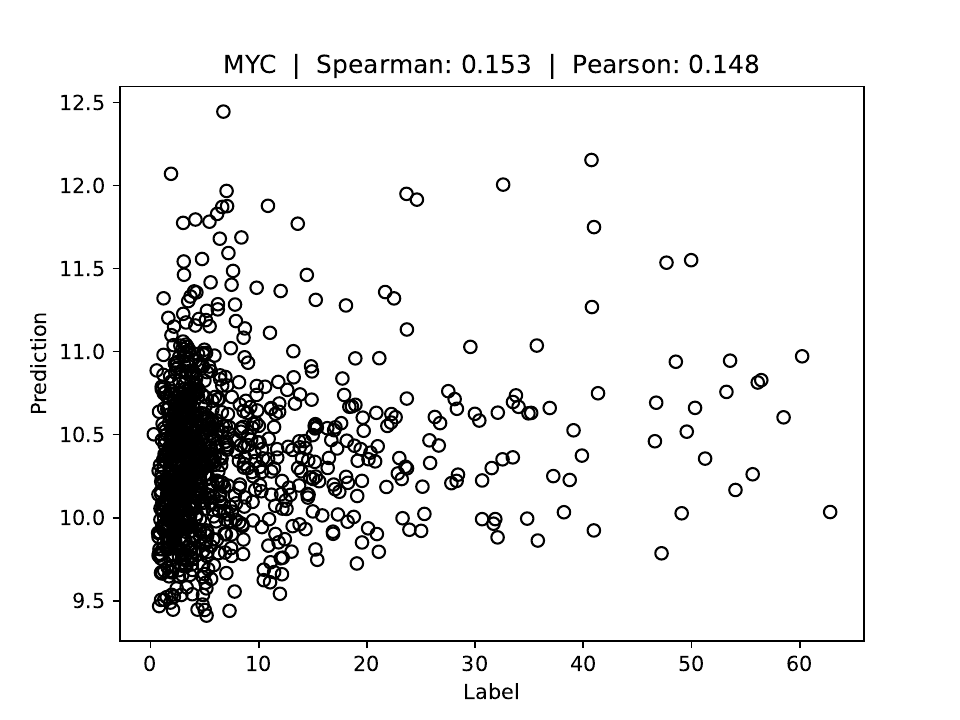}
    \end{subfigure}
    \begin{subfigure}[t]{0.29\textwidth}
        \centering%
        \includegraphics[width=1.0\linewidth]{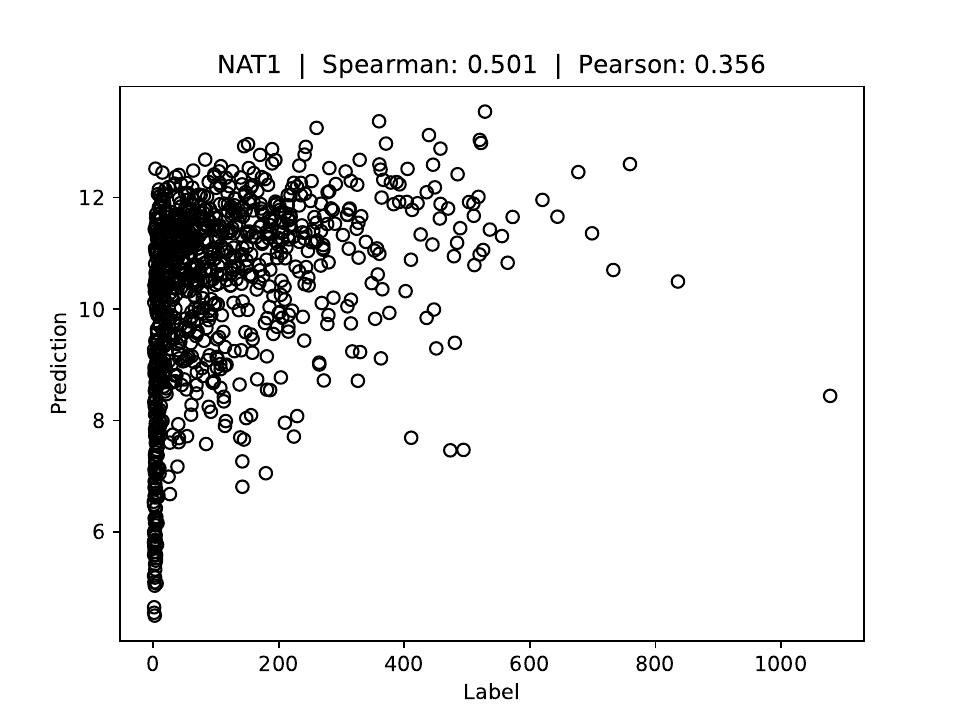}
    \end{subfigure}
    \begin{subfigure}[t]{0.29\textwidth}
        \centering%
        \includegraphics[width=1.0\linewidth]{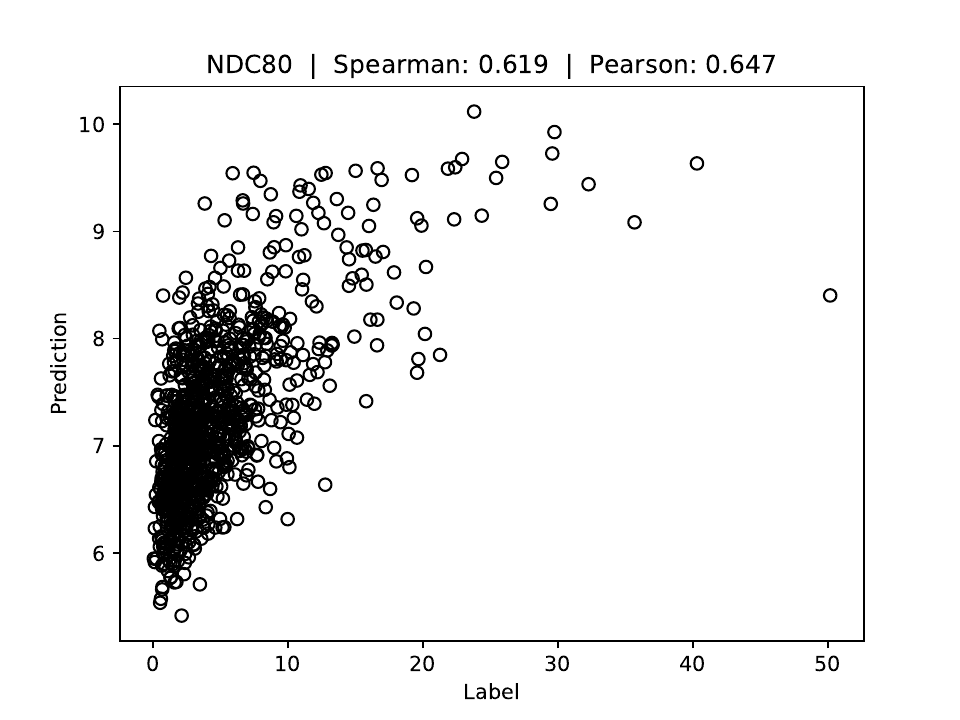}
    \end{subfigure}
    
    \begin{subfigure}[t]{0.29\textwidth}
        \centering%
        \includegraphics[width=1.0\linewidth]{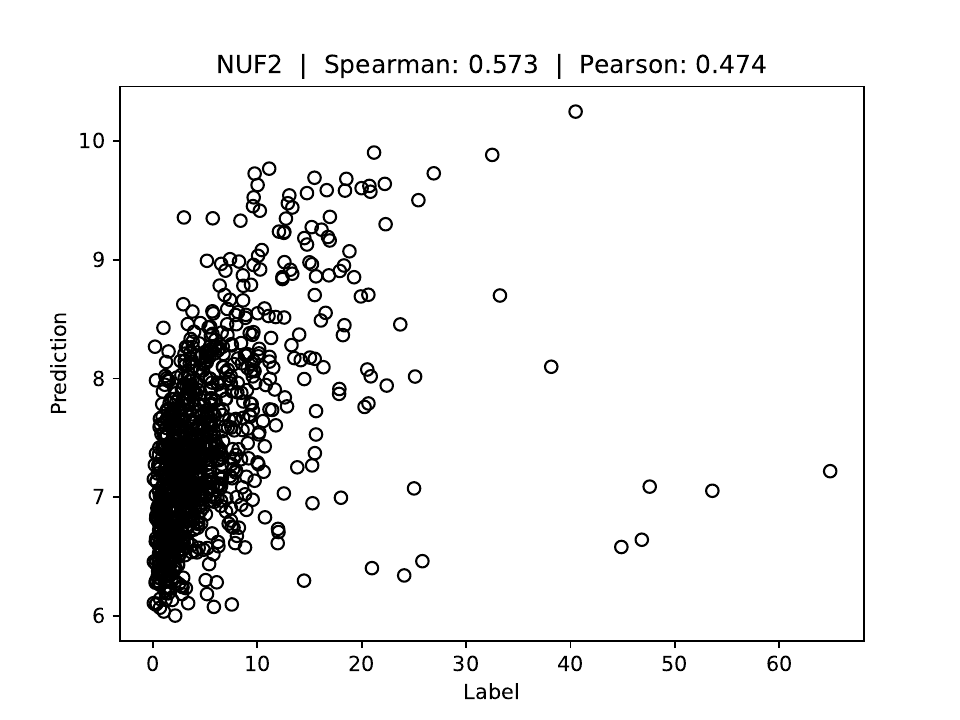}
    \end{subfigure}
    \begin{subfigure}[t]{0.29\textwidth}
        \centering%
        \includegraphics[width=1.0\linewidth]{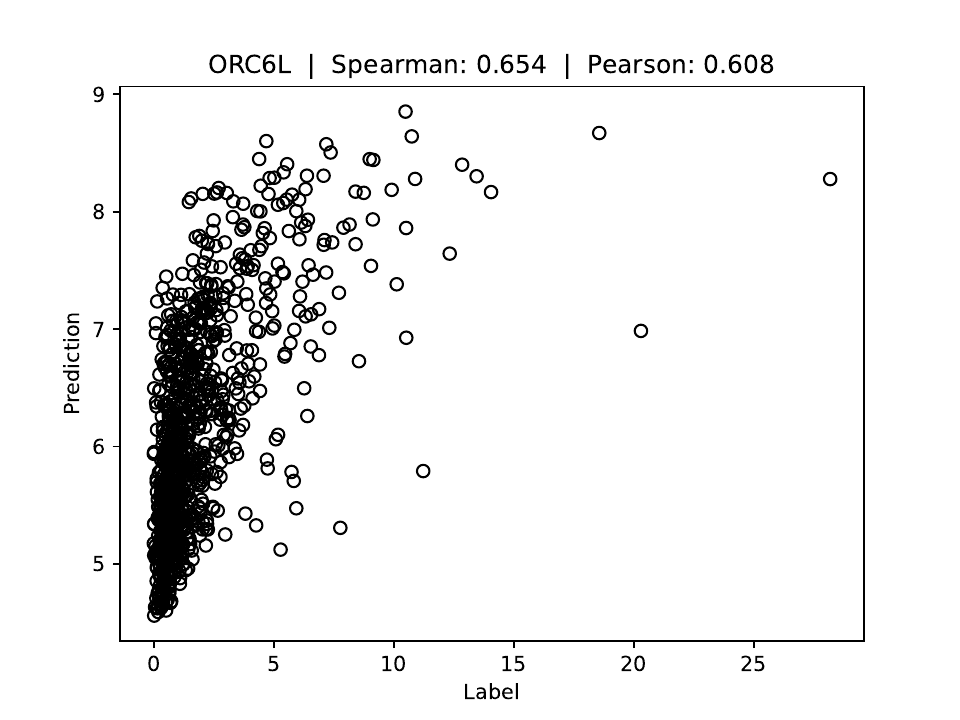}
    \end{subfigure}
    \begin{subfigure}[t]{0.29\textwidth}
        \centering%
        \includegraphics[width=1.0\linewidth]{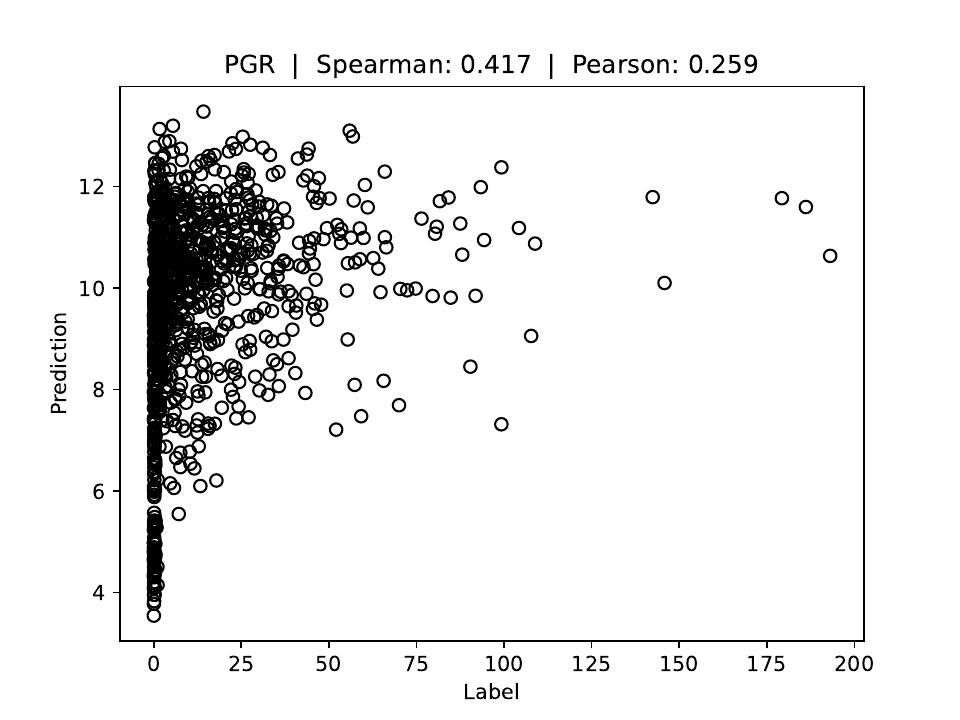}
    \end{subfigure}
    
    \begin{subfigure}[t]{0.29\textwidth}
        \centering%
        \includegraphics[width=1.0\linewidth]{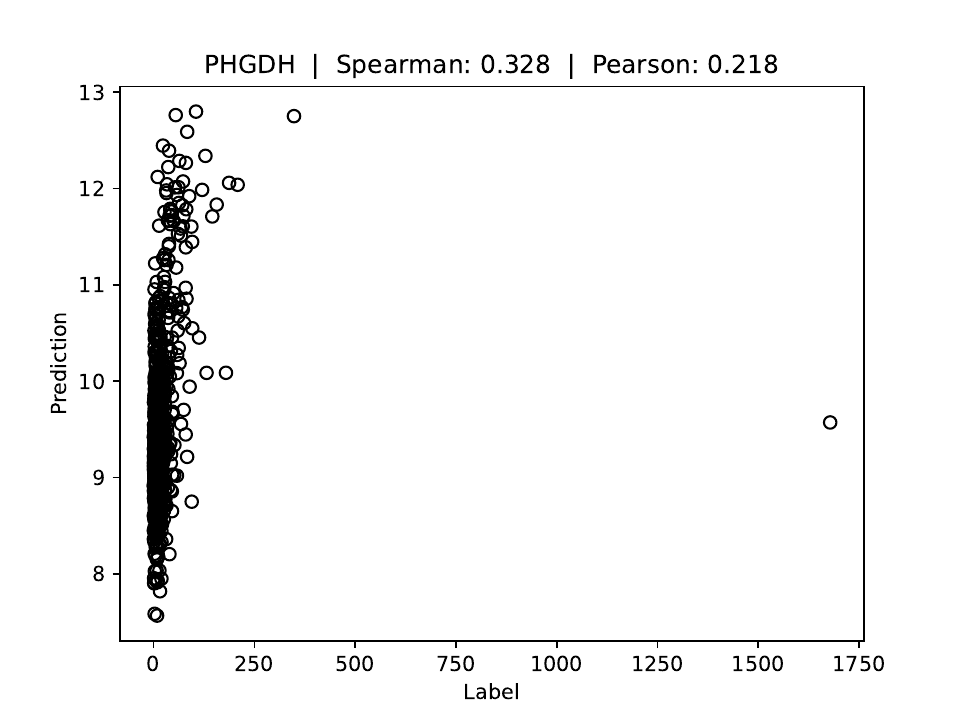}
    \end{subfigure}
    \begin{subfigure}[t]{0.29\textwidth}
        \centering%
        \includegraphics[width=1.0\linewidth]{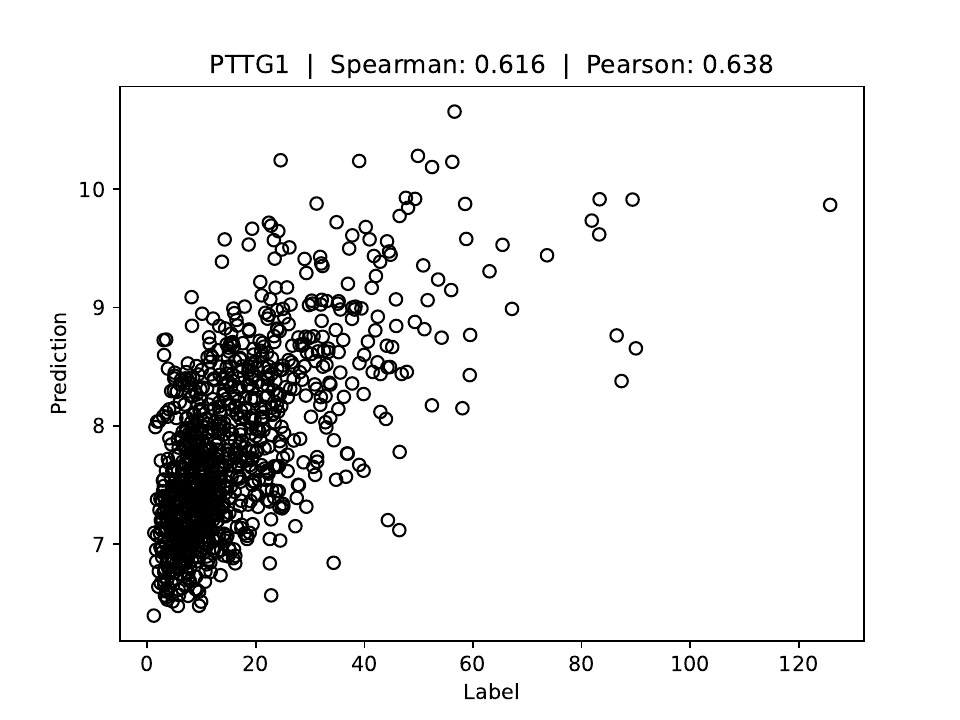}
    \end{subfigure}
    \begin{subfigure}[t]{0.29\textwidth}
        \centering%
        \includegraphics[width=1.0\linewidth]{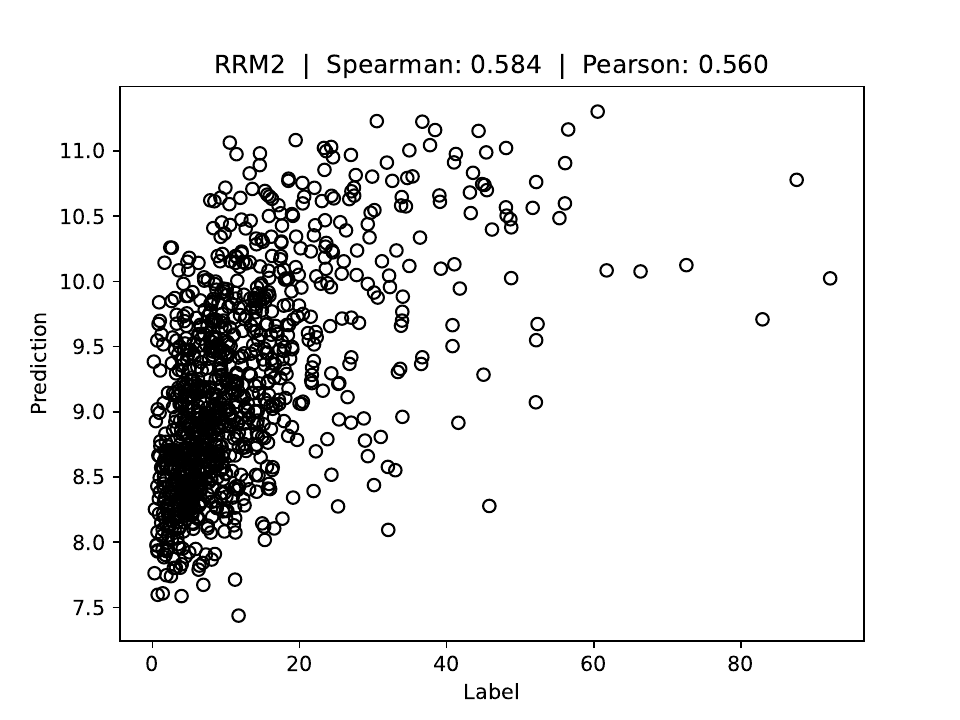}
    \end{subfigure}
    
    \begin{subfigure}[t]{0.29\textwidth}
        \centering%
        \includegraphics[width=1.0\linewidth]{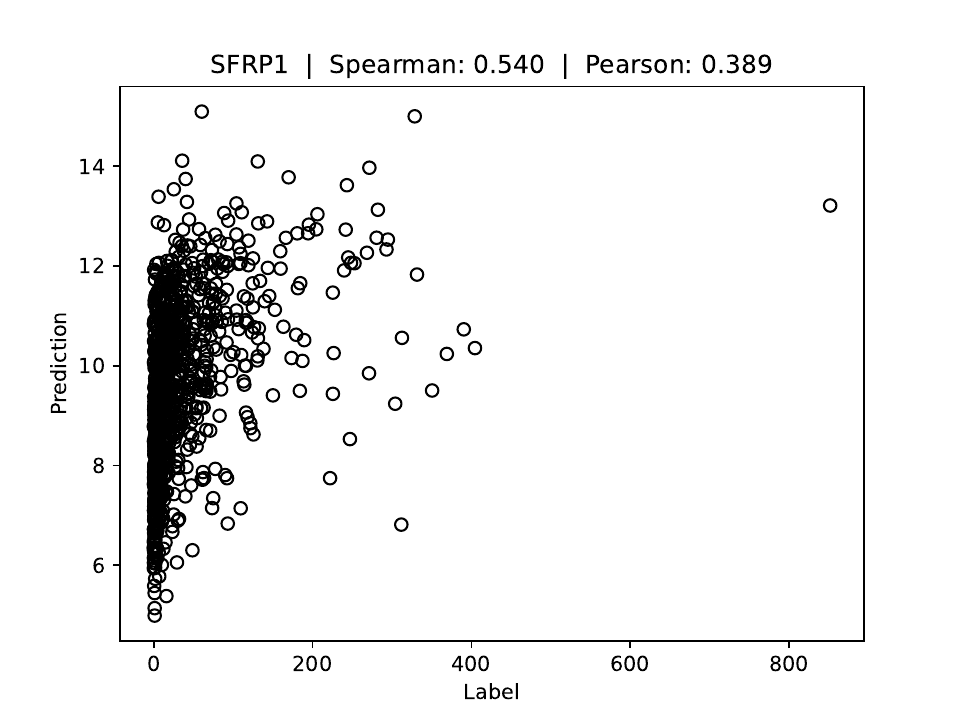}
    \end{subfigure}
    \begin{subfigure}[t]{0.29\textwidth}
        \centering%
        \includegraphics[width=1.0\linewidth]{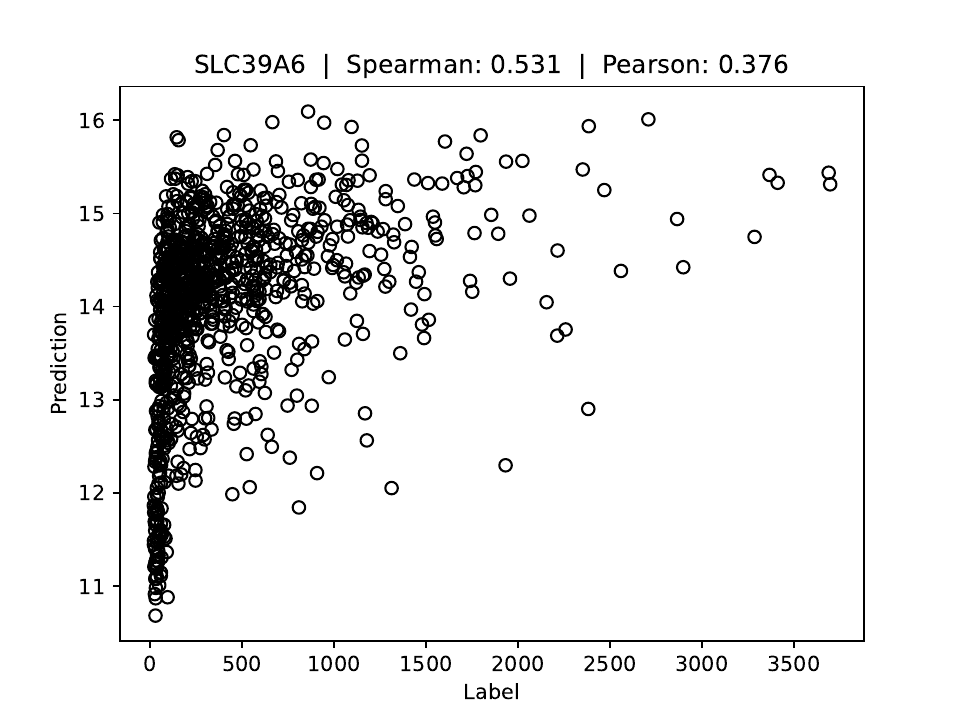}
    \end{subfigure}
    \begin{subfigure}[t]{0.29\textwidth}
        \centering%
        \includegraphics[width=1.0\linewidth]{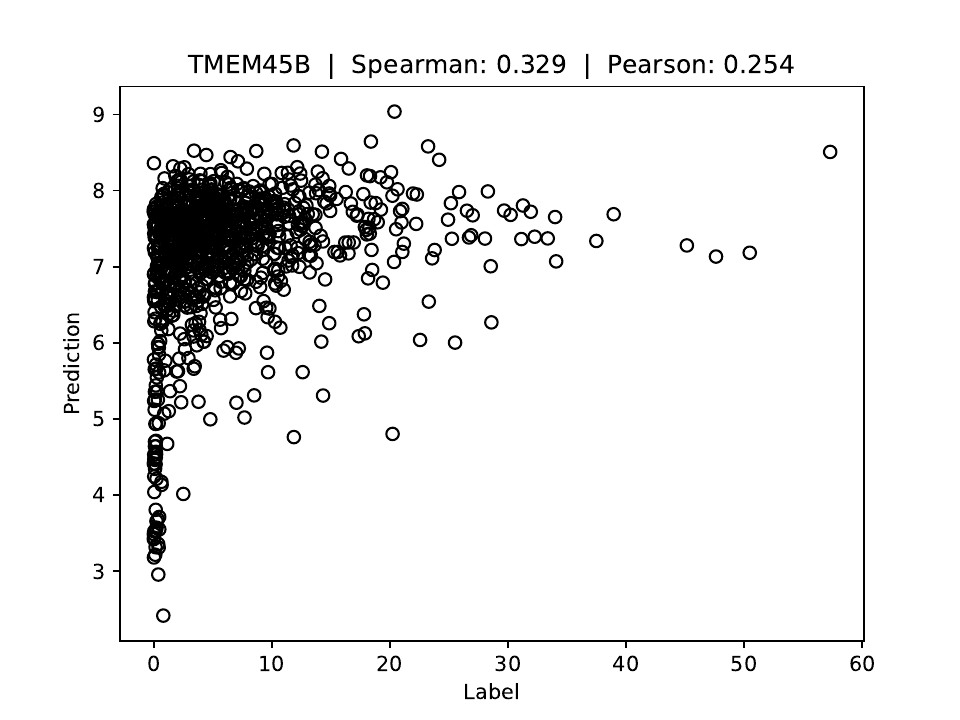}
    \end{subfigure}
    
    \begin{subfigure}[t]{0.29\textwidth}
        \centering%
        \includegraphics[width=1.0\linewidth]{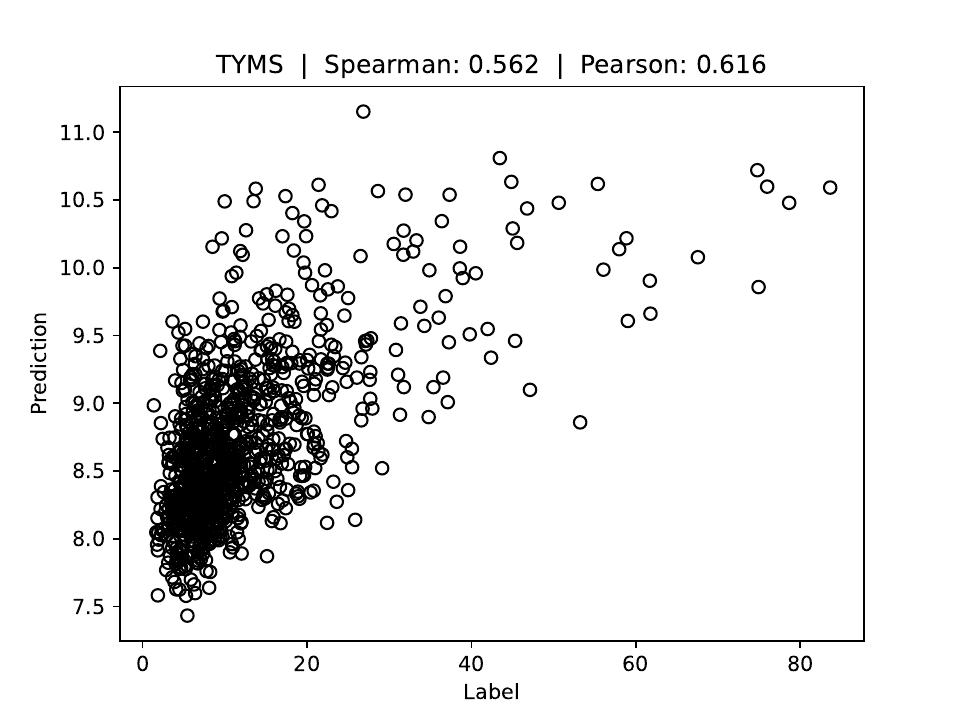}
    \end{subfigure}
    \begin{subfigure}[t]{0.29\textwidth}
        \centering%
        \includegraphics[width=1.0\linewidth]{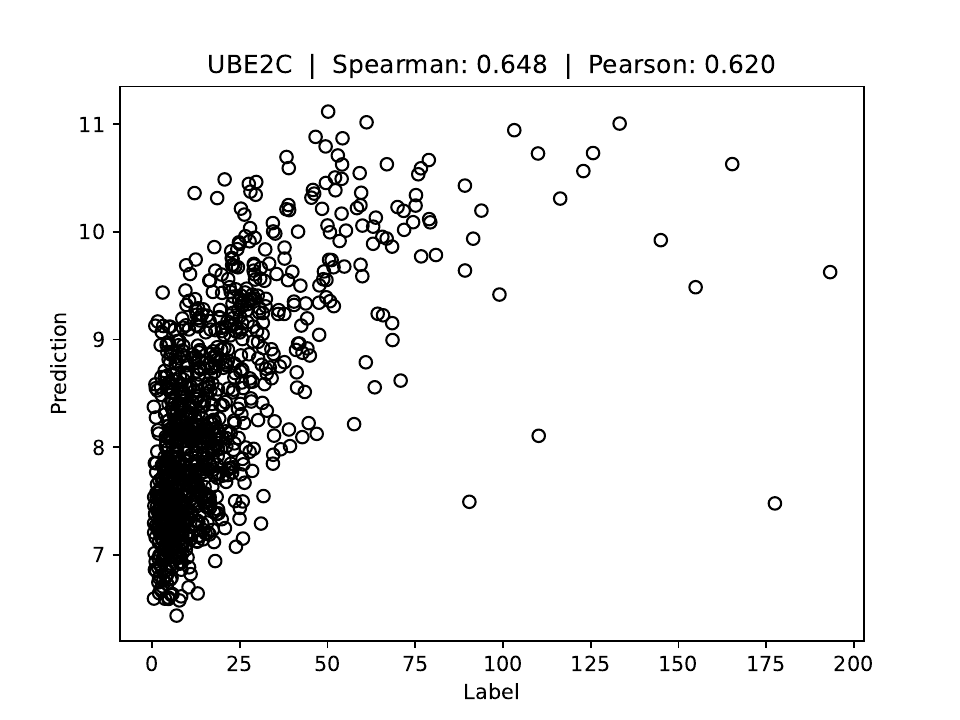}
    \end{subfigure}
    \begin{subfigure}[t]{0.29\textwidth}
        \centering%
        \includegraphics[width=1.0\linewidth]{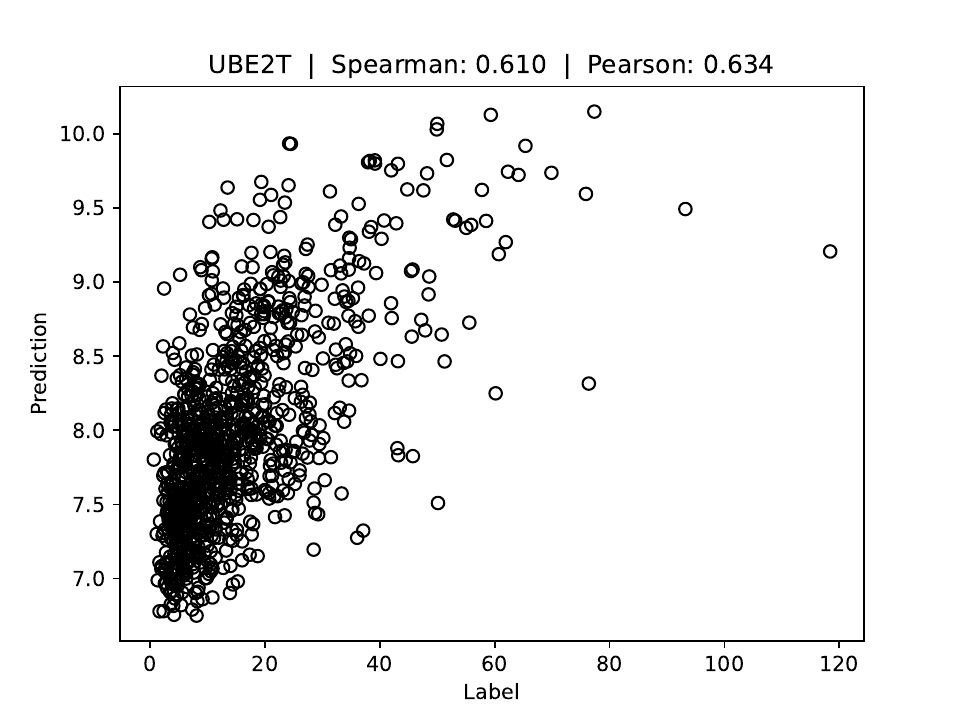}
    \end{subfigure}\vspace{-1.0mm}
    \caption{\textbf{Predicted vs observed gene-expression values for PAM50 gene 36-50 on SCAN-B-Lund.} Scatter plots comparing predicted and ground-truth expression values, for the \textit{H-optimus-1 - Direct - ABMIL} ensemble trained on TCGA-BRCA and evaluated on the external SCAN-B-Lund dataset. Note the difference in scale between predicted and observed gene-expression values (x- and y-axes).}
  \label{fig:corr_plots_pam50_36_50_scan-b-lund}
\end{figure*}

\begin{figure*}[t]
\centering
    \begin{subfigure}[t]{0.495\textwidth}
        \centering%
        \includegraphics[clip, trim=0.35cm 1.25cm 0.25cm 0.0cm, width=0.95\linewidth]{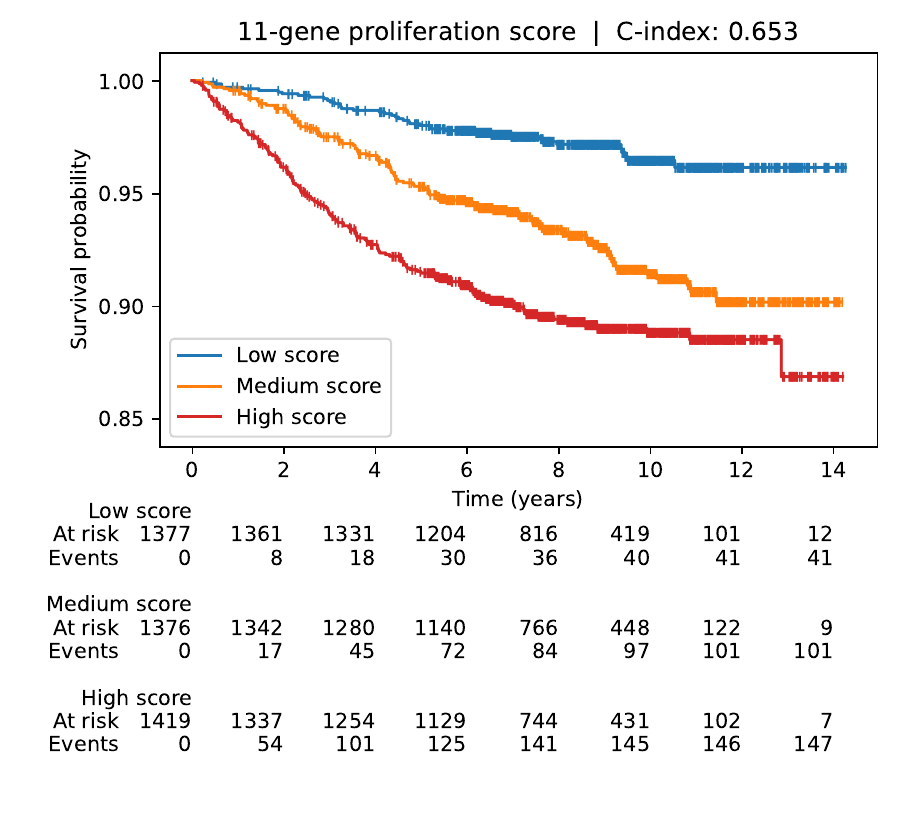}
        \caption{\textbf{11-gene proliferation score -- Full cohort}.}
    \end{subfigure}
    \begin{subfigure}[t]{0.495\textwidth}
        \centering%
        \includegraphics[clip, trim=0.35cm 1.25cm 0.25cm 0.0cm, width=0.95\linewidth]{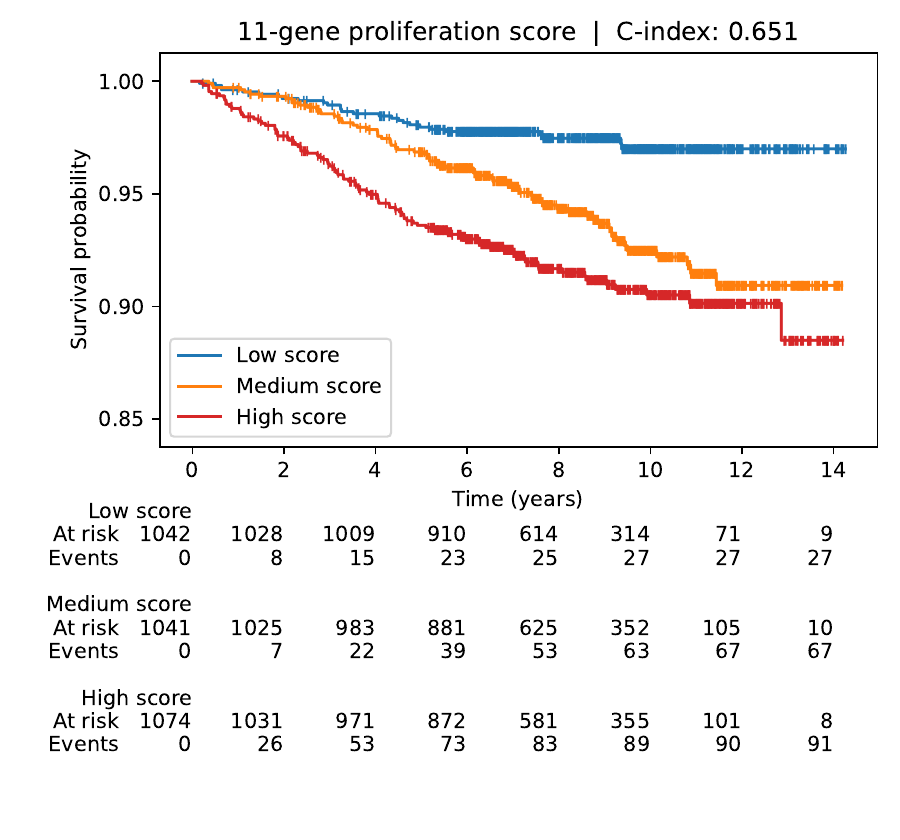}
        \caption{\textbf{11-gene proliferation score -- ER+ \& HER2-}.}
    \end{subfigure}
    \begin{subfigure}[t]{0.495\textwidth}
        \centering%
        \includegraphics[clip, trim=0.35cm 1.25cm 0.25cm 0.0cm, width=0.95\linewidth]{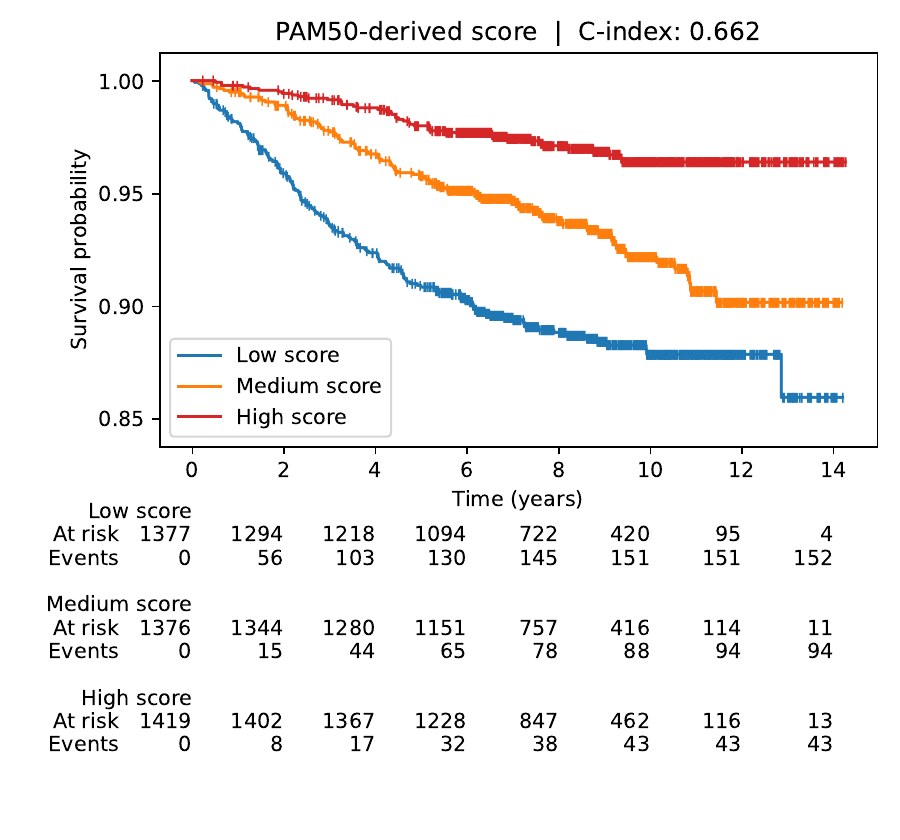}
        \caption{\textbf{PAM50-derived score -- Full cohort}.}
    \end{subfigure}
    \begin{subfigure}[t]{0.495\textwidth}
        \centering%
        \includegraphics[clip, trim=0.35cm 1.25cm 0.25cm 0.0cm, width=0.95\linewidth]{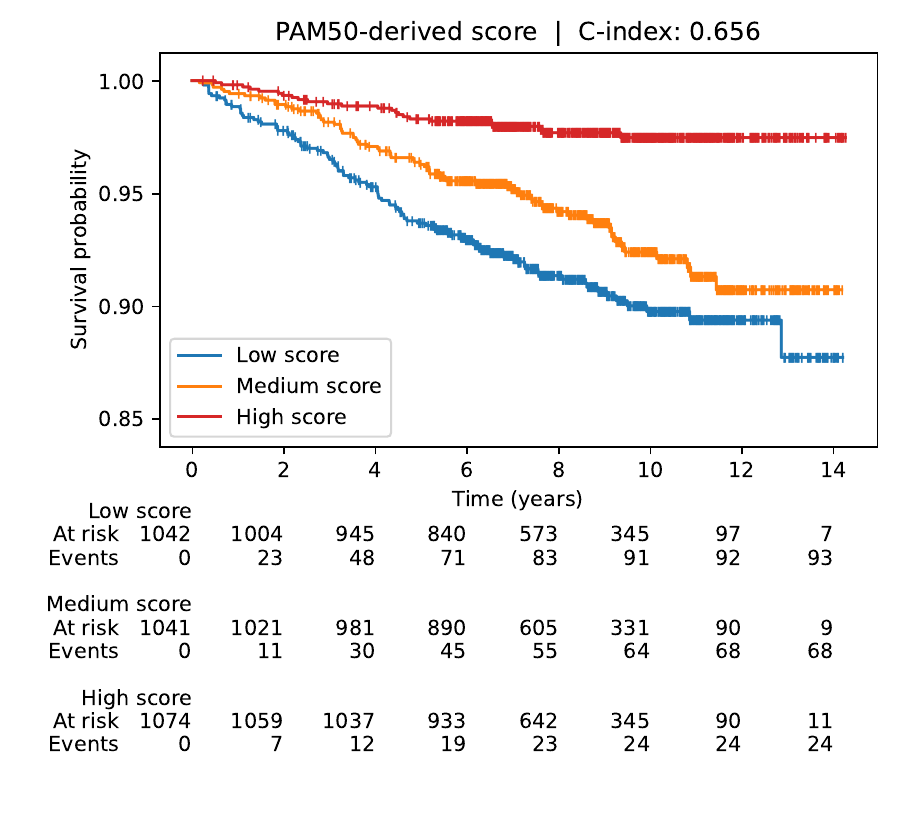}
        \caption{\textbf{PAM50-derived score -- ER+ \& HER2-}.}
    \end{subfigure}\vspace{-1.0mm}
    \caption{\textbf{Three-group Kaplan-Meier risk stratification with at-risk and event counts}. KM survival curves corresponding to Figure~\ref{fig:km_plots_3groups}, including the number of patients at risk and the number of events over time ($0$-$14$ years) for each of the three risk groups.}
  \label{fig:km_plots_3groups_counts}
\end{figure*}

\begin{figure*}[t]
\centering
    \begin{subfigure}[t]{0.495\textwidth}
        \centering%
        \includegraphics[clip, trim=0.35cm 1.25cm 0.25cm 0.0cm, width=0.95\linewidth]{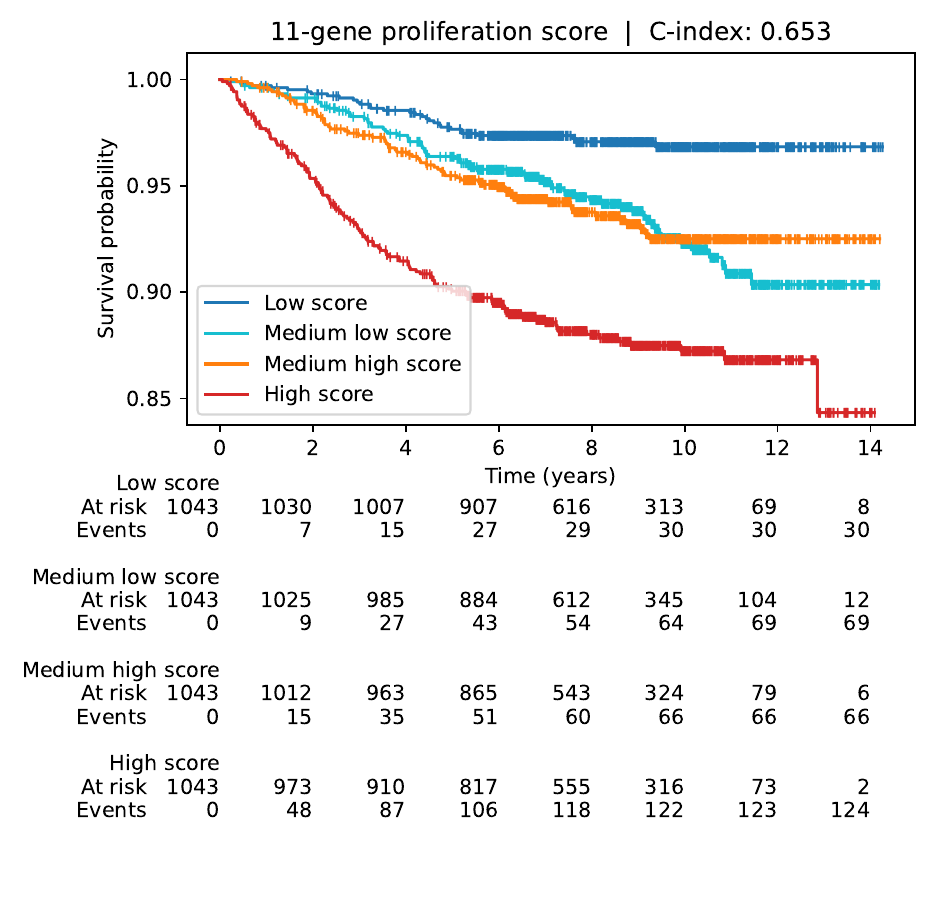}
        \caption{\textbf{11-gene proliferation score -- Full cohort}.}
    \end{subfigure}
    \begin{subfigure}[t]{0.495\textwidth}
        \centering%
        \includegraphics[clip, trim=0.35cm 1.25cm 0.25cm 0.0cm, width=0.95\linewidth]{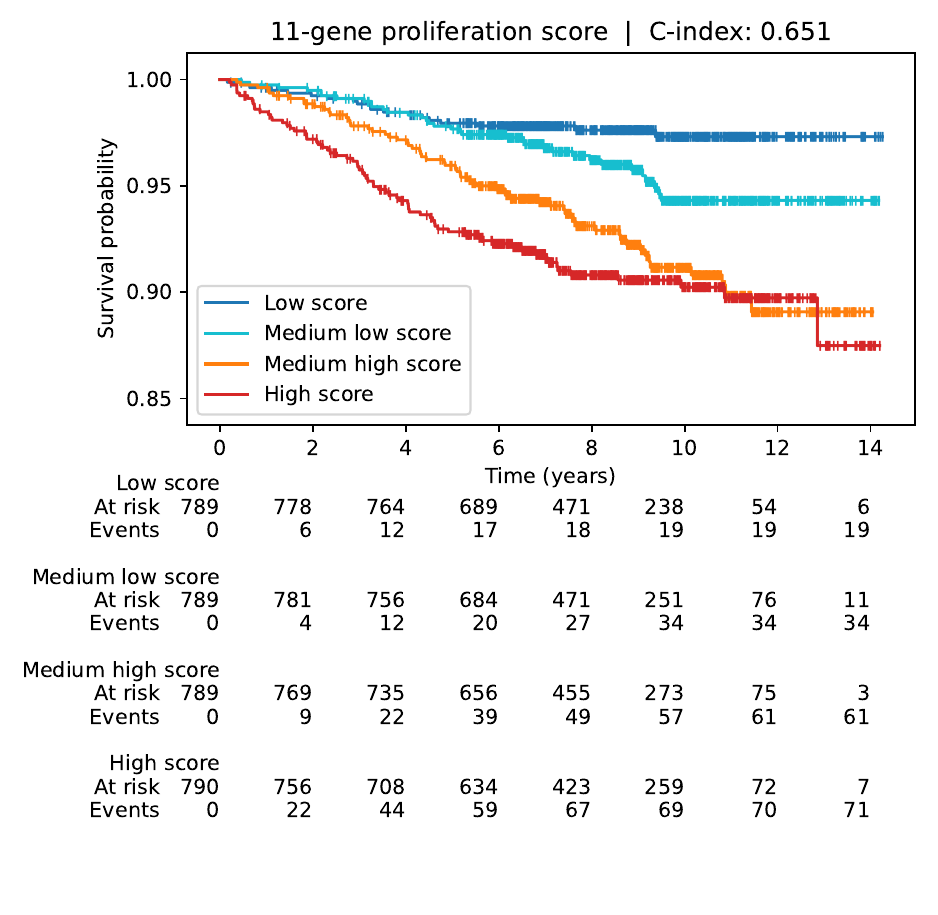}
        \caption{\textbf{11-gene proliferation score -- ER+ \& HER2-}.}
    \end{subfigure}
    \begin{subfigure}[t]{0.495\textwidth}
        \centering%
        \includegraphics[clip, trim=0.35cm 1.25cm 0.25cm 0.0cm, width=0.95\linewidth]{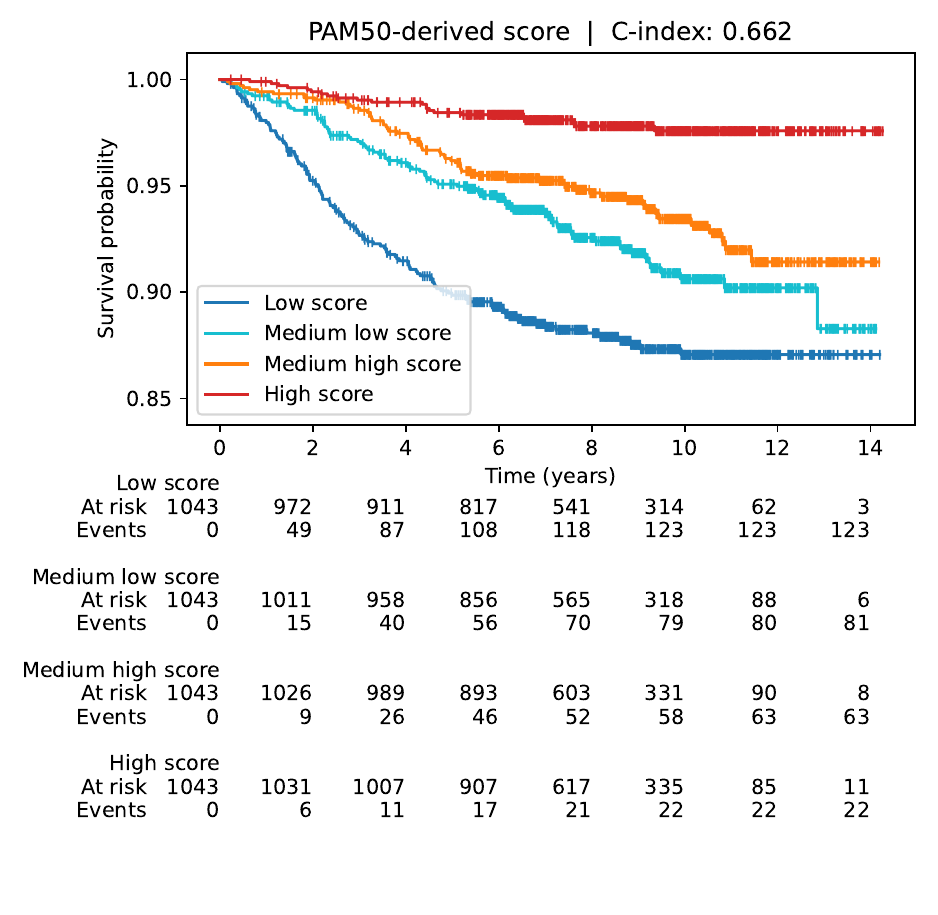}
        \caption{\textbf{PAM50-derived score -- Full cohort}.}
    \end{subfigure}
    \begin{subfigure}[t]{0.495\textwidth}
        \centering%
        \includegraphics[clip, trim=0.35cm 1.25cm 0.25cm 0.0cm, width=0.95\linewidth]{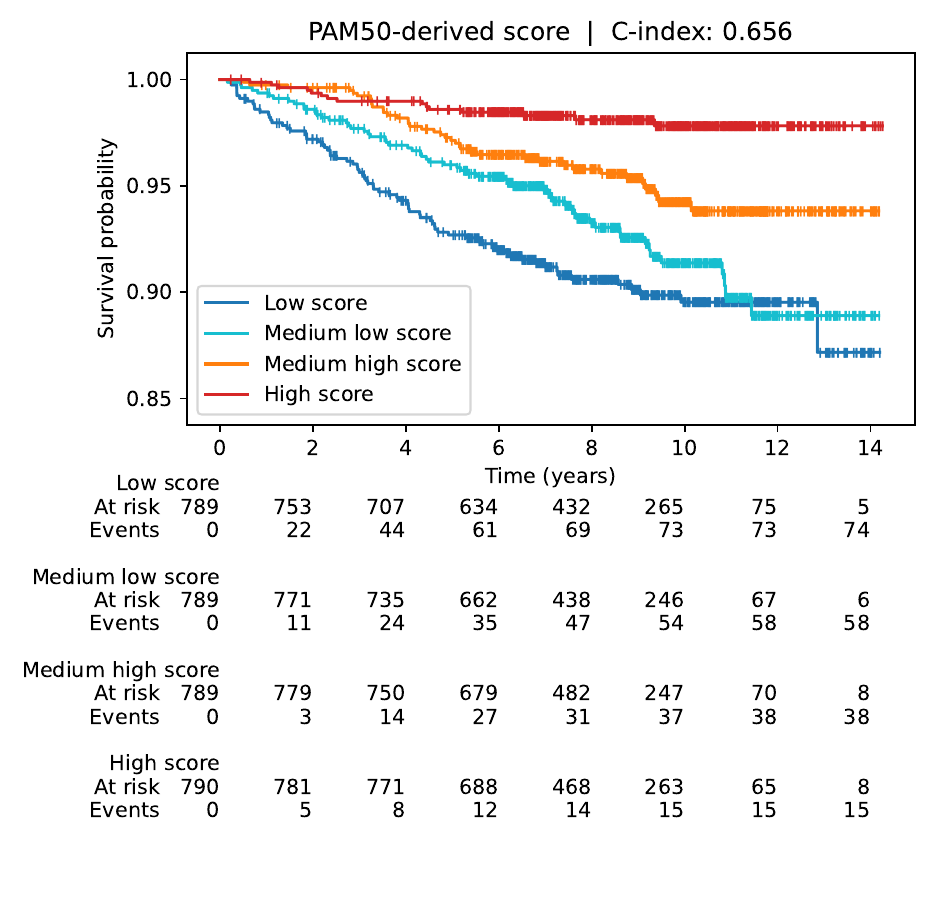}
        \caption{\textbf{PAM50-derived score -- ER+ \& HER2-}.}
    \end{subfigure}\vspace{-1.0mm}
    \caption{\textbf{Four-group Kaplan-Meier risk stratification with at-risk and event counts}. KM survival curves corresponding to Figure~\ref{fig:km_plots_4groups}, including the number of patients at risk and the number of events over time ($0$-$14$ years) for each of the four risk groups.}
  \label{fig:km_plots_4groups_counts}
\end{figure*}

This section contains Figure~\ref{fig:main_results_spearman} - \ref{fig:km_plots_4groups_counts}.

\begin{figure*}[h]
    \centering
    \begin{tikzpicture}
  \centering
  \begin{axis}[
        ybar, axis on top,
        title={},
        height=4.525cm, width=18.0cm, %%%%%%%%%%%%%%%%%%%%%%%%%%%%%%%%%%%%%
        ybar=2.75pt,
        bar width=0.475cm,
        ymajorgrids, tick align=inside,
        enlarge y limits={value=.1,upper},
        ymin=0.0, ymax=0.30,
        set layers,
        ytick={0, 0.1, 0.2, 0.3},
        yticklabels={0, 0.10, 0.20, 0.30},
        axis x line*=bottom,
        axis y line*=left,
        y axis line style={opacity=0},
        tickwidth=0pt,
        % enlarge x limits=true,
        % enlarge x limits=0.185, % (makes the different bar groups closer to eachother!) %%%%%
        % enlarge x limits=0.155, % (makes the different bar groups closer to eachother!) %%%%%
        enlarge x limits=0.25, % (makes the different bar groups closer to eachother!) %%%%%
        % extra y ticks=0.9,
        extra y tick style={
            grid style={
                % black,
                solid,
                line width = 2.5pt,
                /pgfplots/on layer=axis background,
            },
        },
        tick style={
            grid style={
                dashed,
                /pgfplots/on layer=axis background,
            },
        },
        legend style={
            at={(0.5,-0.2)},
            anchor=north,
            legend columns=-1,
            /tikz/every even column/.append style={column sep=0.325cm},
            font=\footnotesize
        },
        ylabel={mean Spearman ($\uparrow$)},
        ylabel style={font=\scriptsize},
        y tick label style={
            /pgf/number format/.cd,
                fixed,
                fixed zerofill,
                precision=2,
            /tikz/.cd
        },
        symbolic x coords={
           TCGA-BRCA,
           TCGA-HNSC,
           TCGA-STAD, 
           TCGA-BLCA},
       xtick=data,
       xticklabel=\empty,
       x tick label style={text width = 2.25cm, align=center, font=\footnotesize},
    ]
    \addplot [fill=darkcerulean, draw=darkcerulean, line width=0mm, error bars/.cd, y dir=both, y explicit, error bar style={thick}, error bar style=black] coordinates {
      (TCGA-BRCA, 0.276704) +- (0, 0.00515587)
      (TCGA-HNSC, 0.249034) +- (0, 0.0104545)
      (TCGA-STAD, 0.211412) +- (0, 0.0268514)
      (TCGA-BLCA, 0.246421) +- (0, 0.0275692)};
    \addplot [fill=goldenpoppy, draw=goldenpoppy, line width=0mm, error bars/.cd, y dir=both, y explicit, error bar style={thick}, error bar style=black] coordinates { 
      (TCGA-BRCA, 0.262753) +- (0, 0.00884314)
      (TCGA-HNSC, 0.23401) +- (0, 0.0149405)
      (TCGA-STAD, 0.198086) +- (0, 0.0443216)
      (TCGA-BLCA, 0.239056) +- (0, 0.0278722)};
    \addplot [fill=ForestGreen, draw=ForestGreen, line width=0mm, error bars/.cd, y dir=both, y explicit, error bar style={thick}, error bar style=black] coordinates { 
      (TCGA-BRCA, 0.278233) +- (0, 0.00495665)
      (TCGA-HNSC, 0.245035) +- (0, 0.0193272)
      (TCGA-STAD, 0.21548) +- (0, 0.0419687)
      (TCGA-BLCA, 0.247848) +- (0, 0.0237725)};
    \addplot [fill=skyblue2, draw=skyblue2, line width=0mm, error bars/.cd, y dir=both, y explicit, error bar style={thick}, error bar style=black] coordinates { 
      (TCGA-BRCA, 0.151575) +- (0, 0.011684)
      (TCGA-HNSC, 0.142326) +- (0, 0.0246204)
      (TCGA-STAD, 0.135455) +- (0, 0.0363681)
      (TCGA-BLCA, 0.142091) +- (0, 0.0193079)};
  \end{axis}
  \end{tikzpicture}\vspace{2.0mm}
    \begin{tikzpicture}
  \centering
  \begin{axis}[
        ybar, axis on top,
        title={},
        height=4.525cm, width=18.0cm, %%%%%%%%%%%%%%%%%%%%%%%%%%%%%%%%%%%%%
        ybar=2.75pt,
        bar width=0.475cm,
        ymajorgrids, tick align=inside,
        enlarge y limits={value=.1,upper},
        ymin=0.0, ymax=0.60,
        set layers,
        ytick={0, 0.2, 0.4, 0.6},
        yticklabels={0, 0.20, 0.40, 0.60},
        axis x line*=bottom,
        axis y line*=left,
        y axis line style={opacity=0},
        tickwidth=0pt,
        % enlarge x limits=true,
        % enlarge x limits=0.185, % (makes the different bar groups closer to eachother!) %%%%%
        % enlarge x limits=0.155, % (makes the different bar groups closer to eachother!) %%%%%
        enlarge x limits=0.25, % (makes the different bar groups closer to eachother!) %%%%%
        % extra y ticks=0.9,
        extra y tick style={
            grid style={
                % black,
                solid,
                line width = 2.5pt,
                /pgfplots/on layer=axis background,
            },
        },
        tick style={
            grid style={
                dashed,
                /pgfplots/on layer=axis background,
            },
        },
        legend style={
            at={(0.5,-0.2)},
            anchor=north,
            legend columns=-1,
            /tikz/every even column/.append style={column sep=0.325cm},
            font=\footnotesize
        },
        ylabel={mean Spearman - top 1k genes ($\uparrow$)},
        ylabel style={font=\scriptsize},
        y tick label style={
            /pgf/number format/.cd,
                fixed,
                fixed zerofill,
                precision=2,
            /tikz/.cd
        },
        symbolic x coords={
           TCGA-BRCA,
           TCGA-HNSC,
           TCGA-STAD, 
           TCGA-BLCA},
       xtick=data,
       xticklabel=\empty,
       x tick label style={text width = 2.25cm, align=center, font=\footnotesize},
    ]
    \addplot [fill=darkcerulean, draw=darkcerulean, line width=0mm, error bars/.cd, y dir=both, y explicit, error bar style={thick}, error bar style=black] coordinates {
      (TCGA-BRCA, 0.577347) +- (0, 0.00870563)
      (TCGA-HNSC, 0.582472) +- (0, 0.0314083)
      (TCGA-STAD, 0.544475) +- (0, 0.049921)
      (TCGA-BLCA, 0.589089) +- (0, 0.0434532)};
    \addplot [fill=goldenpoppy, draw=goldenpoppy, line width=0mm, error bars/.cd, y dir=both, y explicit, error bar style={thick}, error bar style=black] coordinates { 
      (TCGA-BRCA, 0.560283) +- (0, 0.00778584)
      (TCGA-HNSC, 0.565656) +- (0, 0.0258026)
      (TCGA-STAD, 0.535415) +- (0, 0.0822393)
      (TCGA-BLCA, 0.571624) +- (0, 0.0461009)};
    \addplot [fill=ForestGreen, draw=ForestGreen, line width=0mm, error bars/.cd, y dir=both, y explicit, error bar style={thick}, error bar style=black] coordinates { 
      (TCGA-BRCA, 0.576493) +- (0, 0.00846886)
      (TCGA-HNSC, 0.574388) +- (0, 0.0312932)
      (TCGA-STAD, 0.543024) +- (0, 0.0687386)
      (TCGA-BLCA, 0.585774) +- (0, 0.0395914)};
    \addplot [fill=skyblue2, draw=skyblue2, line width=0mm, error bars/.cd, y dir=both, y explicit, error bar style={thick}, error bar style=black] coordinates { 
      (TCGA-BRCA, 0.452247) +- (0, 0.0153078)
      (TCGA-HNSC, 0.471308) +- (0, 0.04202)
      (TCGA-STAD, 0.476998) +- (0, 0.0613084)
      (TCGA-BLCA, 0.455372) +- (0, 0.032232)};
  \end{axis}
  \end{tikzpicture}\vspace{2.0mm}
    \begin{tikzpicture}
  \centering
  \begin{axis}[
        ybar, axis on top,
        title={},
        height=4.525cm, width=18.0cm, %%%%%%%%%%%%%%%%%%%%%%%%%%%%%%%%%%%%%
        ybar=2.75pt,
        bar width=0.475cm,
        ymajorgrids, tick align=inside,
        enlarge y limits={value=.1,upper},
        ymin=0, ymax=5050,
        set layers,
        ytick={0, 1500, 3000, 4500},
        yticklabels={0, 1.5k, 3.0k, 4.5k},
        axis x line*=bottom,
        axis y line*=left,
        y axis line style={opacity=0},
        tickwidth=0pt,
        % enlarge x limits=true,
        % enlarge x limits=0.185, % (makes the different bar groups closer to eachother!) %%%%%
        % enlarge x limits=0.155, % (makes the different bar groups closer to eachother!) %%%%%
        enlarge x limits=0.25, % (makes the different bar groups closer to eachother!) %%%%%
        % extra y ticks=0.9,
        extra y tick style={
            grid style={
                % black,
                solid,
                line width = 2.5pt,
                /pgfplots/on layer=axis background,
            },
        },
        tick style={
            grid style={
                dashed,
                /pgfplots/on layer=axis background,
            },
        },
        legend style={
            at={(0.5,-0.25)},
            anchor=north,
            legend columns=4,
            /tikz/every even column/.append style={column sep=0.325cm},
            font=\footnotesize
        },
        ylabel={\# genes $\geq 0.4$ Spearman ($\uparrow$)},
        ylabel style={font=\scriptsize},
        symbolic x coords={
           TCGA-BRCA,
           TCGA-HNSC,
           TCGA-STAD, 
           TCGA-BLCA},
       xtick=data,
       x tick label style={text width = 2.25cm, align=center, font={\footnotesize\bfseries}},
       % x tick label style={text width = 2.25cm, align=center, font=\footnotesize},
    ]
    \addplot [fill=darkcerulean, draw=darkcerulean, line width=0mm, error bars/.cd, y dir=both, y explicit, error bar style={thick}, error bar style=black] coordinates {
      (TCGA-BRCA, 4424.4) +- (0, 288.354)
      (TCGA-HNSC, 3710.2) +- (0, 770.687)
      (TCGA-STAD, 3062.6) +- (0, 1167.27)
      (TCGA-BLCA, 4107) +- (0, 1180.19)};
    \addplot [fill=goldenpoppy, draw=goldenpoppy, line width=0mm, error bars/.cd, y dir=both, y explicit, error bar style={thick}, error bar style=black] coordinates { 
      (TCGA-BRCA, 3894.8) +- (0, 263.533)
      (TCGA-HNSC, 3284.4) +- (0, 782.431)
      (TCGA-STAD, 2704.8) +- (0, 1609.12)
      (TCGA-BLCA, 3727) +- (0, 1304.28)};
    \addplot [fill=ForestGreen, draw=ForestGreen, line width=0mm, error bars/.cd, y dir=both, y explicit, error bar style={thick}, error bar style=black] coordinates { 
      (TCGA-BRCA, 4501.4) +- (0, 171.096)
      (TCGA-HNSC, 3611.6) +- (0, 734.985)
      (TCGA-STAD, 3099) +- (0, 1550.73)
      (TCGA-BLCA, 4123.8) +- (0, 1019.43)};
    \addplot [fill=skyblue2, draw=skyblue2, line width=0mm, error bars/.cd, y dir=both, y explicit, error bar style={thick}, error bar style=black] coordinates { 
      (TCGA-BRCA, 924.4) +- (0, 193.692)
      (TCGA-HNSC, 1200.6) +- (0, 446.165)
      (TCGA-STAD, 1382.6) +- (0, 936.065)
      (TCGA-BLCA, 1060.6) +- (0, 446.393)};
    \legend{
            Direct - ABMIL,
            Direct - Patch-Level,
            Contrastive,
            kNN
            }
  \end{axis}
  \end{tikzpicture}\vspace{-1.0mm}
    \caption{\textbf{Benchmarking of deep regression models on TCGA datasets (Spearman).} The same model performance comparison as in Figure~\ref{fig:main_results}, but using Spearman correlation metrics instead of Pearson. Higher is better for all three metrics. The ranking of the four regression models is virtually identical as in Figure~\ref{fig:main_results}.}
    \label{fig:main_results_spearman}
\end{figure*}

\begin{figure*}[h]
    \centering
    \begin{tikzpicture}
  \centering
  \begin{axis}[
        ybar, axis on top,
        title={},
        height=4.525cm, width=18.0cm, %%%%%%%%%%%%%%%%%%%%%%%%%%%%%%%%%%%%%
        ybar=2.75pt,
        bar width=0.475cm,
        ymajorgrids, tick align=inside,
        enlarge y limits={value=.1,upper},
        ymin=0, ymax=0.6,
        set layers,
        ytick={0, 0.2, 0.4, 0.6},
        yticklabels={0, 0.20, 0.40, 0.60},
        axis x line*=bottom,
        axis y line*=left,
        y axis line style={opacity=0},
        tickwidth=0pt,
        % enlarge x limits=true,
        % enlarge x limits=0.185, % (makes the different bar groups closer to eachother!) %%%%%
        % enlarge x limits=0.155, % (makes the different bar groups closer to eachother!) %%%%%
        enlarge x limits=0.25, % (makes the different bar groups closer to eachother!) %%%%%
        % extra y ticks=0.9,
        extra y tick style={
            grid style={
                % black,
                solid,
                line width = 2.5pt,
                /pgfplots/on layer=axis background,
            },
        },
        tick style={
            grid style={
                dashed,
                /pgfplots/on layer=axis background,
            },
        },
        legend style={
            at={(0.5,-0.25)},
            anchor=north,
            legend columns=4,
            /tikz/every even column/.append style={column sep=0.325cm},
            font=\footnotesize
        },
        ylabel={mean Pearson ($\uparrow$)},
        ylabel style={font=\scriptsize},
        symbolic x coords={
           TCGA-BRCA,
           TCGA-HNSC,
           TCGA-STAD, 
           TCGA-BLCA},
       xtick=data,
       x tick label style={text width = 2.25cm, align=center, font={\footnotesize\bfseries}},
       % x tick label style={text width = 2.25cm, align=center, font=\footnotesize},
    ]
    \addplot [fill=darkcerulean, draw=darkcerulean, line width=0mm, error bars/.cd, y dir=both, y explicit, error bar style={thick}, error bar style=black] coordinates {
      (TCGA-BRCA, 0.562243) +- (0, 0.0208197)
      (TCGA-HNSC, 0.330339) +- (0, 0.0220682)
      (TCGA-STAD, 0.358467) +- (0, 0.0677657)
      (TCGA-BLCA, 0.395625) +- (0, 0.0410757)};
    \addplot [fill=goldenpoppy, draw=goldenpoppy, line width=0mm, error bars/.cd, y dir=both, y explicit, error bar style={thick}, error bar style=black] coordinates { 
      (TCGA-BRCA, 0.541048) +- (0, 0.0154627)
      (TCGA-HNSC, 0.308577) +- (0, 0.0357018)
      (TCGA-STAD, 0.326961) +- (0, 0.081407)
      (TCGA-BLCA, 0.361964) +- (0, 0.041211)};
    \addplot [fill=ForestGreen, draw=ForestGreen, line width=0mm, error bars/.cd, y dir=both, y explicit, error bar style={thick}, error bar style=black] coordinates { 
      (TCGA-BRCA, 0.563735) +- (0, 0.0205112)
      (TCGA-HNSC, 0.316051) +- (0, 0.0255539)
      (TCGA-STAD, 0.349776) +- (0, 0.0739082)
      (TCGA-BLCA, 0.378201) +- (0, 0.0563399)};
    \addplot [fill=skyblue2, draw=skyblue2, line width=0mm, error bars/.cd, y dir=both, y explicit, error bar style={thick}, error bar style=black] coordinates { 
      (TCGA-BRCA, 0.414828) +- (0, 0.0194548)
      (TCGA-HNSC, 0.189866) +- (0, 0.0308896)
      (TCGA-STAD, 0.261739) +- (0, 0.076039)
      (TCGA-BLCA, 0.197647) +- (0, 0.0495538)};
    \legend{
            Direct - ABMIL,
            Direct - Patch-Level,
            Contrastive,
            kNN
            }
  \end{axis}
  \end{tikzpicture}\vspace{-1.0mm}
    \caption{\textbf{Benchmarking of deep regression models on TCGA datasets (PAM50 genes).} The same model performance comparison as in Figure~\ref{fig:main_results}, but models are evaluated only on a subset of 50 genes (PAM50) with demonstrated prognostic value for breast cancer~\cite{nielsen2010comparison, wallden2015development}. The performance on TCGA-BRCA is strong compared to the three other (non-breast) TCGA datasets, which is reasonable.}
    \label{fig:main_results_pam50}
\end{figure*}

\clearpage
\begin{figure*}[h]
    \centering
    \begin{tikzpicture}
  \centering
  \begin{axis}[
        ybar, axis on top,
        title={},
        height=4.525cm, width=18.0cm, %%%%%%%%%%%%%%%%%%%%%%%%%%%%%%%%%%%%%
        ybar=2.75pt,
        bar width=0.475cm,
        ymajorgrids, tick align=inside,
        enlarge y limits={value=.1,upper},
        ymin=0.0, ymax=0.30,
        set layers,
        axis x line*=bottom,
        axis y line*=left,
        y axis line style={opacity=0},
        tickwidth=0pt,
        % enlarge x limits=true,
        % enlarge x limits=0.185, % (makes the different bar groups closer to eachother!) %%%%%
        % enlarge x limits=0.155, % (makes the different bar groups closer to eachother!) %%%%%
        enlarge x limits=0.25, % (makes the different bar groups closer to eachother!) %%%%%
        % extra y ticks=0.9,
        extra y tick style={
            grid style={
                % black,
                solid,
                line width = 2.5pt,
                /pgfplots/on layer=axis background,
            },
        },
        tick style={
            grid style={
                dashed,
                /pgfplots/on layer=axis background,
            },
        },
        legend style={
            at={(0.5,-0.2)},
            anchor=north,
            legend columns=-1,
            /tikz/every even column/.append style={column sep=0.325cm},
            font=\footnotesize
        },
        ylabel={mean Pearson ($\uparrow$)},
        ylabel style={font=\scriptsize},
        y tick label style={
            /pgf/number format/.cd,
                fixed,
                % fixed zerofill,
                precision=2,
            /tikz/.cd
        },
        symbolic x coords={
           TCGA-BRCA,
           TCGA-HNSC,
           TCGA-STAD, 
           TCGA-BLCA},
       xtick=data,
       xticklabel=\empty,
       x tick label style={text width = 2.25cm, align=center, font=\footnotesize},
    ]
    \addplot [fill=darkcerulean, draw=darkcerulean, line width=0mm, error bars/.cd, y dir=both, y explicit, error bar style={thick}, error bar style=black] coordinates {
      (TCGA-BRCA, 0.284548 ) +- (0, 0.00676863)
      (TCGA-HNSC, 0.256569) +- (0, 0.00900385)
      (TCGA-STAD, 0.216453) +- (0, 0.0314102)
      (TCGA-BLCA, 0.253299) +- (0, 0.0296211)};
    \addplot [fill=red-violet, draw=red-violet, line width=0mm, error bars/.cd, y dir=both, y explicit, error bar style={thick}, error bar style=black] coordinates { 
      (TCGA-BRCA, 0.251903) +- (0, 0.0133063)
      (TCGA-HNSC, 0.204926) +- (0, 0.0316067)
      (TCGA-STAD, 0.192518) +- (0, 0.0501626)
      (TCGA-BLCA, 0.229752) +- (0, 0.0411661)};
    \addplot [fill=flamingopink, draw=flamingopink, line width=0mm, error bars/.cd, y dir=both, y explicit, error bar style={thick}, error bar style=black] coordinates { 
      (TCGA-BRCA, 0.268581) +- (0, 0.0106736)
      (TCGA-HNSC, 0.248558) +- (0, 0.0132508)
      (TCGA-STAD, 0.20765) +- (0, 0.0344402)
      (TCGA-BLCA, 0.256112) +- (0, 0.0301302)};
    \addplot [fill=OrangeRed, draw=OrangeRed, line width=0mm, error bars/.cd, y dir=both, y explicit, error bar style={thick}, error bar style=black] coordinates { 
      (TCGA-BRCA, 0.28164) +- (0, 0.00967983)
      (TCGA-HNSC, 0.252767) +- (0, 0.0118318)
      (TCGA-STAD, 0.221846) +- (0, 0.0358783)
      (TCGA-BLCA, 0.255076) +- (0, 0.027621)};
  \end{axis}
  \end{tikzpicture}\vspace{3.0mm}
    \begin{tikzpicture}
  \centering
  \begin{axis}[
        ybar, axis on top,
        title={},
        height=4.525cm, width=18.0cm, %%%%%%%%%%%%%%%%%%%%%%%%%%%%%%%%%%%%%
        ybar=2.75pt,
        bar width=0.475cm,
        ymajorgrids, tick align=inside,
        enlarge y limits={value=.1,upper},
        ymin=0.0, ymax=0.60,
        set layers,
        axis x line*=bottom,
        axis y line*=left,
        y axis line style={opacity=0},
        tickwidth=0pt,
        % enlarge x limits=true,
        % enlarge x limits=0.185, % (makes the different bar groups closer to eachother!) %%%%%
        % enlarge x limits=0.155, % (makes the different bar groups closer to eachother!) %%%%%
        enlarge x limits=0.25, % (makes the different bar groups closer to eachother!) %%%%%
        % extra y ticks=0.9,
        extra y tick style={
            grid style={
                % black,
                solid,
                line width = 2.5pt,
                /pgfplots/on layer=axis background,
            },
        },
        tick style={
            grid style={
                dashed,
                /pgfplots/on layer=axis background,
            },
        },
        legend style={
            at={(0.5,-0.2)},
            anchor=north,
            legend columns=-1,
            /tikz/every even column/.append style={column sep=0.325cm},
            font=\footnotesize
        },
        ylabel={mean Pearson - top 1k genes ($\uparrow$)},
        ylabel style={font=\scriptsize},
        y tick label style={
            /pgf/number format/.cd,
                fixed,
                % fixed zerofill,
                precision=2,
            /tikz/.cd
        },
        symbolic x coords={
           TCGA-BRCA,
           TCGA-HNSC,
           TCGA-STAD, 
           TCGA-BLCA},
       xtick=data,
       xticklabel=\empty,
       x tick label style={text width = 2.25cm, align=center, font=\footnotesize},
    ]
    \addplot [fill=darkcerulean, draw=darkcerulean, line width=0mm, error bars/.cd, y dir=both, y explicit, error bar style={thick}, error bar style=black] coordinates {
      (TCGA-BRCA, 0.598019 ) +- (0, 0.0125496)
      (TCGA-HNSC, 0.589236) +- (0, 0.032307)
      (TCGA-STAD, 0.548459) +- (0, 0.0525175)
      (TCGA-BLCA, 0.60033) +- (0, 0.0378758)};
    \addplot [fill=red-violet, draw=red-violet, line width=0mm, error bars/.cd, y dir=both, y explicit, error bar style={thick}, error bar style=black] coordinates { 
      (TCGA-BRCA, 0.579001) +- (0, 0.0231828)
      (TCGA-HNSC, 0.556876) +- (0, 0.052583)
      (TCGA-STAD, 0.555951) +- (0, 0.0877791)
      (TCGA-BLCA, 0.582352) +- (0, 0.0405786)};
    \addplot [fill=flamingopink, draw=flamingopink, line width=0mm, error bars/.cd, y dir=both, y explicit, error bar style={thick}, error bar style=black] coordinates { 
      (TCGA-BRCA, 0.598249) +- (0, 0.0133318)
      (TCGA-HNSC, 0.585707) +- (0, 0.0278542)
      (TCGA-STAD, 0.549142) +- (0, 0.0572022)
      (TCGA-BLCA, 0.603652) +- (0, 0.0376991)};
    \addplot [fill=OrangeRed, draw=OrangeRed, line width=0mm, error bars/.cd, y dir=both, y explicit, error bar style={thick}, error bar style=black] coordinates { 
      (TCGA-BRCA, 0.597209) +- (0, 0.0133384)
      (TCGA-HNSC, 0.585948) +- (0, 0.0259985)
      (TCGA-STAD, 0.560038) +- (0, 0.057932)
      (TCGA-BLCA, 0.601764) +- (0, 0.0375224)};
  \end{axis}
  \end{tikzpicture}\vspace{2.0mm}
    \begin{tikzpicture}
  \centering
  \begin{axis}[
        ybar, axis on top,
        title={},
        height=4.525cm, width=18.0cm, %%%%%%%%%%%%%%%%%%%%%%%%%%%%%%%%%%%%%
        ybar=2.75pt,
        bar width=0.475cm,
        ymajorgrids, tick align=inside,
        enlarge y limits={value=.1,upper},
        ymin=0, ymax=5550,
        set layers,
        ytick={0, 1500, 3000, 4500},
        axis x line*=bottom,
        axis y line*=left,
        y axis line style={opacity=0},
        tickwidth=0pt,
        % enlarge x limits=true,
        % enlarge x limits=0.185, % (makes the different bar groups closer to eachother!) %%%%%
        % enlarge x limits=0.155, % (makes the different bar groups closer to eachother!) %%%%%
        enlarge x limits=0.25, % (makes the different bar groups closer to eachother!) %%%%%
        % extra y ticks=0.9,
        extra y tick style={
            grid style={
                % black,
                solid,
                line width = 2.5pt,
                /pgfplots/on layer=axis background,
            },
        },
        tick style={
            grid style={
                dashed,
                /pgfplots/on layer=axis background,
            },
        },
        legend style={
            at={(0.5,-0.25)},
            anchor=north,
            legend columns=-1,
            /tikz/every even column/.append style={column sep=0.325cm},
            font=\footnotesize
        },
        ylabel={\# genes $\geq 0.4$ Pearson ($\uparrow$)},
        ylabel style={font=\scriptsize},
        symbolic x coords={
           TCGA-BRCA,
           TCGA-HNSC,
           TCGA-STAD, 
           TCGA-BLCA},
       xtick=data,
       x tick label style={text width = 2.25cm, align=center, font={\footnotesize\bfseries}},
       % x tick label style={text width = 2.25cm, align=center, font=\footnotesize},
    ]
    \addplot [fill=darkcerulean, draw=darkcerulean, line width=0mm, error bars/.cd, y dir=both, y explicit, error bar style={thick}, error bar style=black] coordinates {
      (TCGA-BRCA, 4927 ) +- (0, 357.299)
      (TCGA-HNSC, 3987.8) +- (0, 765.179)
      (TCGA-STAD, 3240.2) +- (0, 1342.98)
      (TCGA-BLCA, 4364.6) +- (0, 1188.92)};
    \addplot [fill=red-violet, draw=red-violet, line width=0mm, error bars/.cd, y dir=both, y explicit, error bar style={thick}, error bar style=black] coordinates { 
      (TCGA-BRCA, 3897.6) +- (0, 606.933)
      (TCGA-HNSC, 2842.8) +- (0, 1078.47)
      (TCGA-STAD, 2979.2) +- (0, 1804.83)
      (TCGA-BLCA, 3816.8) +- (0, 1291.7)};
    \addplot [fill=flamingopink, draw=flamingopink, line width=0mm, error bars/.cd, y dir=both, y explicit, error bar style={thick}, error bar style=black] coordinates { 
      (TCGA-BRCA, 4645) +- (0, 417.769)
      (TCGA-HNSC, 3881.2) +- (0, 742.523)
      (TCGA-STAD, 3104.2) +- (0, 1386.35)
      (TCGA-BLCA, 4626.6) +- (0, 1440.98)};
    \addplot [fill=OrangeRed, draw=OrangeRed, line width=0mm, error bars/.cd, y dir=both, y explicit, error bar style={thick}, error bar style=black] coordinates { 
      (TCGA-BRCA, 4906.4) +- (0, 441.478)
      (TCGA-HNSC, 3938.6) +- (0, 630.541)
      (TCGA-STAD, 3537.4) +- (0, 1525.5)
      (TCGA-BLCA, 4429.4) +- (0, 1181.22)};
    \legend{
            % Contrastive,
            L2,
            Gaussian,
            L1,
            smoothL1
            }
  \end{axis}
  \end{tikzpicture}\vspace{-1.0mm}
    \caption{\textbf{Performance comparison of \textit{UNI - Direct - ABMIL} on TCGA datasets when trained with different loss functions.} Comparison of L2 (as used in the main paper), Gaussian, L1, and smoothL1 regression loss functions. Same metrics as in Figure~\ref{fig:main_results}. Gaussian is outperformed by the three other losses. L2, L1, and smoothL1 achieve similar performance overall.}
    \label{fig:extra_results_losses}
\end{figure*}

\clearpage
\clearpage
\clearpage

\clearpage
\clearpage
\clearpage

\clearpage
\section{Supplementary Tables}
\label{appendix:tables}

This section contains Table~\ref{tab:main_results_brca} - \ref{tab:scan-b-lund_pam50_results}.

\begin{table}[h]
	\caption{\textbf{Benchmarking results on TCGA-BRCA.} Numerical results corresponding to Figure~\ref{fig:main_results} for the TCGA-BRCA dataset. Results are reported as mean $\pm$ standard deviation over 5-fold site-aware cross-validation. \textbf{Bold} indicates the best mean value for each metric.}\vspace{-2.25mm}    
    \label{tab:main_results_brca}
    \centering
	\resizebox{0.80\textwidth}{!}{%
		% \begin{tabular}{lccc}
% \toprule
%  &mean Pearson ($\uparrow$) &mean Pearson - top 1k genes ($\uparrow$) &\# genes $\geq 0.4$ Pearson ($\uparrow$)\\
% \midrule
% UNI - Direct - ABMIL           &\textbf{0.284}$\pm$0.006    &\textbf{0.598}$\pm$0.012    &\textbf{4927}$\pm$357\\
% UNI - Direct - Patch-Level     &0.267$\pm$0.009    &0.572$\pm$0.010    &4213$\pm$353\\
% UNI - Contrastive     &0.283$\pm$0.004    &0.596$\pm$0.014    &4886$\pm$242\\
% UNI - kNN             &0.157$\pm$0.011    &0.465$\pm$0.018    &1137$\pm$247\\
% \midrule
% Resnet-IN - Direct - ABMIL           &0.151$\pm$0.027    &0.460$\pm$0.029    &1129$\pm$420\\
% Resnet-IN - Direct - Patch-Level     &0.128$\pm$0.016    &0.417$\pm$0.014    &574$\pm$128\\
% Resnet-IN - Contrastive     &\textbf{0.217}$\pm$0.029    &\textbf{0.515}$\pm$0.042    &\textbf{2519}$\pm$941\\
% Resnet-IN - kNN             &0.103$\pm$0.014    &0.376$\pm$0.018    &281$\pm$94\\
% \bottomrule
% \end{tabular}
\begin{tabular}{lccc}
\toprule
 &mean Pearson ($\uparrow$) &mean Pearson - top 1k genes ($\uparrow$) &\# genes $\geq 0.4$ Pearson ($\uparrow$)\\
\midrule
UNI - Direct - ABMIL           &\textbf{0.284}$\pm$0.006    &\textbf{0.598}$\pm$0.012    &\textbf{4927}$\pm$357\\
UNI - Direct - Patch-Level     &0.267$\pm$0.009    &0.572$\pm$0.010    &4213$\pm$353\\
UNI - Contrastive     &0.283$\pm$0.004    &0.596$\pm$0.014    &4886$\pm$242\\
UNI - kNN             &0.157$\pm$0.011    &0.465$\pm$0.018    &1137$\pm$247\\
\bottomrule
\end{tabular}
	}
\end{table}

\begin{table}[h]
	\caption{\textbf{Benchmarking results on TCGA-HNSC.} Numerical results corresponding to Figure~\ref{fig:main_results} for the TCGA-HNSC dataset. Results are reported as mean $\pm$ standard deviation over 5-fold site-aware cross-validation. \textbf{Bold} indicates the best mean value for each metric.}\vspace{-2.25mm}  
    \label{tab:main_results_hnsc}
    \centering
	\resizebox{0.80\textwidth}{!}{%
		% \begin{tabular}{lccc}
% \toprule
%  &mean Pearson ($\uparrow$) &mean Pearson - top 1k genes ($\uparrow$) &\# genes $\geq 0.4$ Pearson ($\uparrow$)\\
% \midrule
% UNI - Direct - ABMIL           &0\textbf{.256}$\pm$0.009    &\textbf{0.589}$\pm$0.032    &\textbf{3987}$\pm$765\\
% UNI - Direct - Patch-Level     &0.237$\pm$0.017    &0.565$\pm$0.027    &3456$\pm$837\\
% UNI - Contrastive     &0.246$\pm$0.022    &0.576$\pm$0.033    &3750$\pm$835\\
% UNI - kNN             &0.141$\pm$0.026    &0.465$\pm$0.043    &1158$\pm$463\\
% \midrule
% Resnet-IN - Direct - ABMIL           &0.143$\pm$0.015    &0.493$\pm$0.048    &1494$\pm$671\\
% Resnet-IN - Direct - Patch-Level     &0.115$\pm$0.026    &0.419$\pm$0.068    &767$\pm$656\\
% Resnet-IN - Contrastive     &\textbf{0.183}$\pm$0.022    &\textbf{0.533}$\pm$0.047    &\textbf{2264}$\pm$806\\
% Resnet-IN - kNN             &0.087$\pm$0.031    &0.382$\pm$0.056    &416$\pm$294\\
% \bottomrule
% \end{tabular}
\begin{tabular}{lccc}
\toprule
 &mean Pearson ($\uparrow$) &mean Pearson - top 1k genes ($\uparrow$) &\# genes $\geq 0.4$ Pearson ($\uparrow$)\\
\midrule
UNI - Direct - ABMIL           &0\textbf{.256}$\pm$0.009    &\textbf{0.589}$\pm$0.032    &\textbf{3987}$\pm$765\\
UNI - Direct - Patch-Level     &0.237$\pm$0.017    &0.565$\pm$0.027    &3456$\pm$837\\
UNI - Contrastive     &0.246$\pm$0.022    &0.576$\pm$0.033    &3750$\pm$835\\
UNI - kNN             &0.141$\pm$0.026    &0.465$\pm$0.043    &1158$\pm$463\\
\bottomrule
\end{tabular}
	}
\end{table}

\begin{table}[h]
	\caption{\textbf{Benchmarking results on TCGA-STAD.} Numerical results corresponding to Figure~\ref{fig:main_results} for the TCGA-STAD dataset. Results are reported as mean $\pm$ standard deviation over 5-fold site-aware cross-validation. \textbf{Bold} indicates the best mean value for each metric.}\vspace{-2.25mm}  
    \label{tab:main_results_stad}
    \centering
	\resizebox{0.80\textwidth}{!}{%
		% \begin{tabular}{lccc}
% \toprule
%  &mean Pearson ($\uparrow$) &mean Pearson - top 1k genes ($\uparrow$) &\# genes $\geq 0.4$ Pearson ($\uparrow$)\\
% \midrule
% UNI - Direct - ABMIL           &\textbf{0.216}$\pm$0.031    &\textbf{0.548}$\pm$0.052    &\textbf{3240}$\pm$1342\\
% UNI - Direct - Patch-Level     &0.200$\pm$0.044    &0.533$\pm$0.084    &2765$\pm$1713\\
% UNI - Contrastive     &\textbf{0.216}$\pm$0.044    &0.543$\pm$0.066    &3184$\pm$1660\\
% UNI - kNN             &0.135$\pm$0.037    &0.469$\pm$0.060    &1328$\pm$981\\
% \midrule
% Resnet-IN - Direct - ABMIL           &0.110$\pm$0.038    &0.431$\pm$0.101    &1132$\pm$1119\\
% Resnet-IN - Direct - Patch-Level     &0.115$\pm$0.033    &0.438$\pm$0.077    &1068$\pm$839\\
% Resnet-IN - Contrastive     &\textbf{0.124}$\pm$0.035    &\textbf{0.444}$\pm$0.069    &\textbf{1181}$\pm$1238\\
% Resnet-IN - kNN             &0.098$\pm$0.022    &0.421$\pm$0.066    &794$\pm$573\\
% \bottomrule
% \end{tabular}
\begin{tabular}{lccc}
\toprule
 &mean Pearson ($\uparrow$) &mean Pearson - top 1k genes ($\uparrow$) &\# genes $\geq 0.4$ Pearson ($\uparrow$)\\
\midrule
UNI - Direct - ABMIL           &\textbf{0.216}$\pm$0.031    &\textbf{0.548}$\pm$0.052    &\textbf{3240}$\pm$1342\\
UNI - Direct - Patch-Level     &0.200$\pm$0.044    &0.533$\pm$0.084    &2765$\pm$1713\\
UNI - Contrastive     &\textbf{0.216}$\pm$0.044    &0.543$\pm$0.066    &3184$\pm$1660\\
UNI - kNN             &0.135$\pm$0.037    &0.469$\pm$0.060    &1328$\pm$981\\
\bottomrule
\end{tabular}
	}
\end{table}

\begin{table}[h]
	\caption{\textbf{Benchmarking results on TCGA-BLCA.} Numerical results corresponding to Figure~\ref{fig:main_results} for the TCGA-BLCA dataset. Results are reported as mean $\pm$ standard deviation over 5-fold site-aware cross-validation. \textbf{Bold} indicates the best mean value for each metric.}\vspace{-2.25mm}  
    \label{tab:main_results_blca}
    \centering
	\resizebox{0.80\textwidth}{!}{%
		% \begin{tabular}{lccc}
% \toprule
%  &mean Pearson ($\uparrow$) &mean Pearson - top 1k genes ($\uparrow$) &\# genes $\geq 0.4$ Pearson ($\uparrow$)\\
% \midrule
% UNI - Direct - ABMIL           &\textbf{0.253}$\pm$0.029    &\textbf{0.600}$\pm$0.037    &\textbf{4364}$\pm$1188\\
% UNI - Direct - Patch-Level     &0.242$\pm$0.028    &0.577$\pm$0.036    &3822$\pm$1206\\
% UNI - Contrastive     &0.249$\pm$0.025    &0.590$\pm$0.035    &4183$\pm$975\\
% UNI - kNN             &0.142$\pm$0.021    &0.464$\pm$0.028    &1164$\pm$451\\
% \midrule
% Resnet-IN - Direct - ABMIL           &0.093$\pm$0.026    &0.373$\pm$0.062    &366$\pm$491\\
% Resnet-IN - Direct - Patch-Level     &0.087$\pm$0.024    &0.389$\pm$0.060    &479$\pm$373\\
% Resnet-IN - Contrastive     &\textbf{0.184}$\pm$0.032    &\textbf{0.505}$\pm$0.061    &\textbf{2143}$\pm$1077\\
% Resnet-IN - kNN             &0.088$\pm$0.016    &0.380$\pm$0.049    &347$\pm$289\\
% \bottomrule
% \end{tabular}
\begin{tabular}{lccc}
\toprule
 &mean Pearson ($\uparrow$) &mean Pearson - top 1k genes ($\uparrow$) &\# genes $\geq 0.4$ Pearson ($\uparrow$)\\
\midrule
UNI - Direct - ABMIL           &\textbf{0.253}$\pm$0.029    &\textbf{0.600}$\pm$0.037    &\textbf{4364}$\pm$1188\\
UNI - Direct - Patch-Level     &0.242$\pm$0.028    &0.577$\pm$0.036    &3822$\pm$1206\\
UNI - Contrastive     &0.249$\pm$0.025    &0.590$\pm$0.035    &4183$\pm$975\\
UNI - kNN             &0.142$\pm$0.021    &0.464$\pm$0.028    &1164$\pm$451\\
\bottomrule
\end{tabular}
	}
\end{table}

\begin{table}[h]
	\caption{\textbf{Benchmarking results on TCGA-BRCA (PAM50).} Results for the TCGA-BRCA dataset when models are evaluated only on the PAM50 subset of genes. Results are reported as mean $\pm$ std over 5-fold site-aware cross-validation. \textbf{Bold} indicates the best mean value.}\vspace{-2.25mm}    
    \label{tab:main_results_brca_pam50}
    \centering
	\resizebox{0.625\textwidth}{!}{%
		% \begin{tabular}{l@{\hspace{1.0cm}}c}
% \toprule
%  &mean Pearson ($\uparrow$)\\
% \midrule
% UNI - Direct - ABMIL           &0.562$\pm$0.020\\
% UNI - Direct - Patch-Level     &0.541$\pm$0.015\\
% UNI - Contrastive     &\textbf{0.564}$\pm$0.020\\
% UNI - kNN             &0.415$\pm$0.019\\
% \midrule
% Resnet-IN - Direct - ABMIL           &0.373$\pm$0.070\\
% Resnet-IN - Direct - Patch-Level     &0.306$\pm$0.028\\
% Resnet-IN - Contrastive     &\textbf{0.449}$\pm$0.046\\
% Resnet-IN - kNN             &0.267$\pm$0.026\\
% \midrule
% \midrule
% UNI - Direct - ABMIL - Trained only on PAM50, 1 model           &\textbf{0.576}$\pm$0.020\\
% UNI - Direct - ABMIL - Trained only on PAM50, 2 models           &0.575$\pm$0.021\\
% UNI - Direct - ABMIL - Trained only on PAM50, 5 models           &0.572$\pm$0.012\\
% UNI - Direct - ABMIL - Trained only on PAM50, 10 models           &0.569$\pm$0.016\\
% UNI - Direct - ABMIL - Trained only on PAM50, 25 models           &0.566$\pm$0.019\\
% UNI - Direct - ABMIL - Trained only on PAM50, 50 models           &0.560$\pm$0.020\\
% \bottomrule
% \end{tabular}
\begin{tabular}{l@{\hspace{1.0cm}}c}
\toprule
 &mean Pearson ($\uparrow$)\\
\midrule
UNI - Direct - ABMIL           &0.562$\pm$0.020\\
UNI - Direct - Patch-Level     &0.541$\pm$0.015\\
UNI - Contrastive     &\textbf{0.564}$\pm$0.020\\
UNI - kNN             &0.415$\pm$0.019\\
\midrule
UNI - Direct - ABMIL - Trained only on PAM50, 1 model           &\textbf{0.576}$\pm$0.020\\
UNI - Direct - ABMIL - Trained only on PAM50, 2 models           &0.575$\pm$0.021\\
UNI - Direct - ABMIL - Trained only on PAM50, 5 models           &0.572$\pm$0.012\\
UNI - Direct - ABMIL - Trained only on PAM50, 10 models           &0.569$\pm$0.016\\
UNI - Direct - ABMIL - Trained only on PAM50, 25 models           &0.566$\pm$0.019\\
UNI - Direct - ABMIL - Trained only on PAM50, 50 models           &0.560$\pm$0.020\\
\bottomrule
\end{tabular}
	}
\end{table}

\begin{table}[h]
    \caption{\textbf{Top-performing genes for \textit{UNI - Direct - ABMIL} on TCGA datasets.} Top 20 genes (out of $N = 20{,}530$) with the highest prediction accuracy across the four TCGA datasets. Results are reported as mean $\pm$ standard deviation over 5-fold site-aware cross-validation and sorted by decreasing mean Pearson correlation.}\vspace{-2.25mm}    
    \label{tab:main_results_top-genes_abmil}
    \centering
	\resizebox{1.0\textwidth}{!}{%
		\begin{tabular}{@{\hspace{0.55cm}}ccc@{\hspace{0.55cm}}@{\hspace{0.55cm}}ccc@{\hspace{0.55cm}}@{\hspace{0.55cm}}ccc@{\hspace{0.55cm}}@{\hspace{0.55cm}}ccc@{\hspace{0.55cm}}}
\toprule
\multicolumn{3}{c}{\textbf{TCGA-BRCA}} &\multicolumn{3}{c}{\textbf{TCGA-HNSC}} &\multicolumn{3}{c}{\textbf{TCGA-STAD}} &\multicolumn{3}{c}{\textbf{TCGA-BLCA}}\\
Rank &Gene &Pearson ($\uparrow$)  &Rank &Gene &Pearson ($\uparrow$)  &Rank &Gene &Pearson ($\uparrow$)  &Rank &Gene &Pearson ($\uparrow$)\\
\midrule
%            BRCA                                 HNSC                               STAD                   BLCA
1  &FOXA1   &0.731$\pm$0.027      &1  &SGEF      &0.742$\pm$0.037      &1  &PKNOX2  &0.633$\pm$0.030      &1  &UPK2         &0.699$\pm$0.059\\
2  &MLPH    &0.726$\pm$0.048      &2  &LOC730101 &0.726$\pm$0.055      &2  &JAM2    &0.620$\pm$0.038      &2  &WARS         &0.696$\pm$0.068\\
3  &TBC1D9  &0.721$\pm$0.021      &3  &GLS2      &0.725$\pm$0.051      &3  &C1QTNF7 &0.617$\pm$0.056      &3  &TOX3         &0.679$\pm$0.068\\
4  &AGR3    &0.720$\pm$0.023      &4  &MAP7D1    &0.698$\pm$0.056      &4  &SCN4B   &0.612$\pm$0.069      &4  &TAP2         &0.676$\pm$0.082\\
5  &THSD4   &0.713$\pm$0.023      &5  &KIAA1609  &0.692$\pm$0.038      &5  &FCER1A  &0.601$\pm$0.082      &5  &KSR2         &0.670$\pm$0.070\\
6  &CCNE1   &0.711$\pm$0.032      &6  &ACPL2     &0.691$\pm$0.025      &6  &CNRIP1  &0.600$\pm$0.079      &6  &KRT6B        &0.670$\pm$0.061\\
7  &ESR1    &0.709$\pm$0.018      &7  &MYB       &0.691$\pm$0.067      &7  &TPX2    &0.597$\pm$0.123      &7  &DUSP7        &0.669$\pm$0.075\\
8  &XBP1    &0.701$\pm$0.025      &8  &C3orf58   &0.687$\pm$0.016      &8  &DNMT3B  &0.597$\pm$0.098      &8  &TYMP         &0.667$\pm$0.054\\
9  &ORC6L   &0.701$\pm$0.035      &9  &KRT14     &0.686$\pm$0.034      &9  &BHMT2   &0.593$\pm$0.031      &9  &TRAK1        &0.665$\pm$0.088\\
10 &CENPA   &0.701$\pm$0.037      &10 &RGS20     &0.685$\pm$0.086      &10 &MAPK10  &0.593$\pm$0.058      &10 &SLC30A2      &0.662$\pm$0.064\\
11 &GATA3   &0.700$\pm$0.045      &11 &THSD1     &0.684$\pm$0.035      &11 &GYPC    &0.593$\pm$0.097      &11 &KRT6C        &0.662$\pm$0.053\\
12 &DNALI1  &0.693$\pm$0.015      &12 &TUBB6     &0.681$\pm$0.062      &12 &FHL1    &0.591$\pm$0.055      &12 &C17orf28     &0.661$\pm$0.076\\
13 &CDC25A  &0.692$\pm$0.034      &13 &MYO3A     &0.677$\pm$0.070      &13 &GSTM5   &0.590$\pm$0.065      &13 &LILRA6       &0.658$\pm$0.063\\
14 &NOSTRIN &0.691$\pm$0.045      &14 &MT2A      &0.671$\pm$0.058      &14 &TCEAL7  &0.588$\pm$0.116      &14 &PDCD1LG2     &0.657$\pm$0.040\\
15 &SCUBE2  &0.690$\pm$0.014      &15 &SLC31A2   &0.670$\pm$0.059      &15 &ABCA8   &0.584$\pm$0.043      &15 &KLHDC7A      &0.654$\pm$0.082\\
16 &SPDEF   &0.690$\pm$0.043      &16 &MRAP2     &0.668$\pm$0.027      &16 &FAM107A &0.581$\pm$0.059      &16 &SLC9A2       &0.653$\pm$0.069\\
17 &PSAT1   &0.688$\pm$0.027      &17 &SP110     &0.665$\pm$0.051      &17 &FXYD1   &0.580$\pm$0.114      &17 &BHMT         &0.649$\pm$0.058\\
18 &CENPN   &0.688$\pm$0.027      &18 &TMEM116   &0.662$\pm$0.050      &18 &HJURP   &0.579$\pm$0.092      &18 &FAM190A      &0.647$\pm$0.037\\
19 &C6orf97 &0.687$\pm$0.034      &19 &SAMD12    &0.661$\pm$0.023      &19 &FGF7    &0.577$\pm$0.097      &19 &LOC100188947 &0.647$\pm$0.083\\
20 &SLC44A4 &0.686$\pm$0.035      &20 &CAV1      &0.661$\pm$0.044      &20 &TOP2A   &0.577$\pm$0.117      &20 &FCGR3A       &0.645$\pm$0.049\\
\bottomrule
\end{tabular}
	}
\end{table}

\begin{table}[h]
	\caption{\textbf{Top-performing genes for \textit{UNI - Contrastive} on TCGA datasets.} Top 20 genes (out of $N = 20{,}530$) with the highest prediction accuracy across the four TCGA datasets. Results are reported as mean $\pm$ standard deviation over 5-fold site-aware cross-validation and sorted by decreasing mean Pearson correlation. Compared to the results for the \textit{UNI - Direct - ABMIL} model in Table~\ref{tab:main_results_top-genes_abmil}, $17$ of the top $20$ genes are shared on TCGA-BRCA, $14$/$20$ genes on TCGA-HNSC, $13$/$20$ genes on TCGA-STAD, and $14$/$20$ genes on TCGA-BLCA.}\vspace{-2.25mm}
    \label{tab:main_results_top-genes_contrastive}
    \centering
	\resizebox{1.0\textwidth}{!}{%
		\begin{tabular}{@{\hspace{0.55cm}}ccc@{\hspace{0.55cm}}@{\hspace{0.55cm}}ccc@{\hspace{0.55cm}}@{\hspace{0.55cm}}ccc@{\hspace{0.55cm}}@{\hspace{0.55cm}}ccc@{\hspace{0.55cm}}}
\toprule
\multicolumn{3}{c}{\textbf{TCGA-BRCA}} &\multicolumn{3}{c}{\textbf{TCGA-HNSC}} &\multicolumn{3}{c}{\textbf{TCGA-STAD}} &\multicolumn{3}{c}{\textbf{TCGA-BLCA}}\\
Rank &Gene &Pearson ($\uparrow$)  &Rank &Gene &Pearson ($\uparrow$)  &Rank &Gene &Pearson ($\uparrow$)  &Rank &Gene &Pearson ($\uparrow$)\\
\midrule
%            BRCA                                  HNSC                                STAD                                 BLCA
1   &FOXA1   &0.731$\pm$0.030      &1   &LOC730101 &0.712$\pm$0.075      &1   &SCN4B   &0.616$\pm$0.105      &1   &UPK2     &0.695$\pm$0.040\\
2   &AGR3    &0.725$\pm$0.046      &2   &GLS2      &0.696$\pm$0.064      &2   &C1QTNF7 &0.616$\pm$0.073      &2   &WARS     &0.691$\pm$0.073\\
3   &MLPH    &0.722$\pm$0.049      &3   &SGEF      &0.691$\pm$0.019      &3   &PKNOX2  &0.610$\pm$0.081      &3   &KSR2     &0.686$\pm$0.065\\
4   &ESR1    &0.719$\pm$0.025      &4   &MAP7D1    &0.678$\pm$0.077      &4   &NEGR1   &0.599$\pm$0.080      &4   &C17orf28 &0.677$\pm$0.046\\
5   &THSD4   &0.717$\pm$0.025      &5   &CAV1      &0.676$\pm$0.050      &5   &JAM2    &0.597$\pm$0.069      &5   &TAP2     &0.668$\pm$0.082\\
6   &TBC1D9  &0.712$\pm$0.043      &6   &MYB       &0.672$\pm$0.083      &6   &RBMS3   &0.591$\pm$0.103      &6   &KLHDC7A  &0.663$\pm$0.046\\
7   &CCNE1   &0.709$\pm$0.027      &7   &KIAA1609  &0.672$\pm$0.028      &7   &BOC     &0.587$\pm$0.085      &7   &TOX3     &0.661$\pm$0.055\\
8   &C6orf97 &0.697$\pm$0.042      &8   &ACPL2     &0.664$\pm$0.043      &8   &JAM3    &0.586$\pm$0.071      &8   &DUSP7    &0.661$\pm$0.063\\
9   &DEGS2   &0.696$\pm$0.036      &9   &SLC31A2   &0.662$\pm$0.050      &9   &CCNA2   &0.585$\pm$0.098      &9   &SLC9A2   &0.657$\pm$0.082\\
10  &CENPA   &0.694$\pm$0.034      &10  &THSD1     &0.660$\pm$0.049      &10  &FHL1    &0.585$\pm$0.070      &10  &SLC30A2  &0.656$\pm$0.044\\
11  &XBP1    &0.694$\pm$0.036      &11  &RGS20     &0.657$\pm$0.093      &11  &MAPK10  &0.584$\pm$0.083      &11  &PDCD1LG2 &0.654$\pm$0.042\\
12  &SCUBE2  &0.694$\pm$0.015      &12  &KRT14     &0.655$\pm$0.042      &12  &HJURP   &0.583$\pm$0.087      &12  &TYMP     &0.653$\pm$0.078\\
13  &ORC6L   &0.692$\pm$0.042      &13  &TNFRSF12A &0.654$\pm$0.081      &13  &FCER1A  &0.582$\pm$0.078      &13  &TRAK1    &0.647$\pm$0.055\\
14  &CENPN   &0.692$\pm$0.037      &14  &SP110     &0.651$\pm$0.055      &14  &MFAP4   &0.582$\pm$0.067      &14  &RAB15    &0.646$\pm$0.089\\
15  &GATA3   &0.689$\pm$0.054      &15  &C3orf58   &0.648$\pm$0.014      &15  &BHMT2   &0.581$\pm$0.077      &15  &SNX31    &0.646$\pm$0.070\\
16  &SLC44A4 &0.688$\pm$0.051      &16  &RPS6KA4   &0.644$\pm$0.071      &16  &FXYD1   &0.580$\pm$0.102      &16  &RHOU     &0.645$\pm$0.093\\
17  &NOSTRIN &0.687$\pm$0.038      &17  &FHOD1     &0.643$\pm$0.064      &17  &DNMT3B  &0.579$\pm$0.075      &17  &UPK3A    &0.644$\pm$0.045\\
18  &LRRC17  &0.687$\pm$0.039      &18  &SBK1      &0.640$\pm$0.097      &18  &CNRIP1  &0.578$\pm$0.083      &18  &KRT6B    &0.641$\pm$0.078\\
19  &CA12    &0.687$\pm$0.030      &19  &FLRT3     &0.636$\pm$0.060      &19  &FAT4    &0.577$\pm$0.099      &19  &SERPINB1 &0.641$\pm$0.075\\
20  &SPDEF   &0.686$\pm$0.044      &20  &KRT6C     &0.636$\pm$0.061      &20  &TCEAL7  &0.576$\pm$0.116      &20  &UPK1A    &0.639$\pm$0.078\\
\bottomrule
\end{tabular}
	}
\end{table}

\clearpage
\begin{table}[h]
	\caption{\textbf{Prediction accuracy for all PAM50 genes on TCGA-BRCA for \textit{UNI – Direct - ABMIL}.} Results are reported as mean $\pm$ standard deviation over 5-fold site-aware cross-validation and sorted by decreasing mean Pearson correlation.}\vspace{-2.25mm}	
    \label{tab:main_results_brca_pam50_abmil_all_genes}    
    \centering
	\resizebox{1.0\textwidth}{!}{%
		\begin{tabular}{@{\hspace{0.55cm}}ccc@{\hspace{0.55cm}}@{\hspace{0.55cm}}ccc@{\hspace{0.55cm}}@{\hspace{0.55cm}}ccc@{\hspace{0.55cm}}@{\hspace{0.55cm}}ccc@{\hspace{0.55cm}}}
\toprule
Rank &Gene &Pearson ($\uparrow$)  &Rank &Gene &Pearson ($\uparrow$)  &Rank &Gene &Pearson ($\uparrow$)  &Rank &Gene &Pearson ($\uparrow$)\\
\midrule
1   &FOXA1  &0.731$\pm$0.027      &14  &BIRC5    &0.649$\pm$0.040      &27  &ANLN    &0.576$\pm$0.063      &40  &MMP11    &0.469$\pm$0.058\\
2   &MLPH   &0.726$\pm$0.048      &15  &NAT1     &0.635$\pm$0.042      &28  &CDC6    &0.571$\pm$0.053      &41  &KRT17    &0.465$\pm$0.036\\
3   &CCNE1  &0.711$\pm$0.032      &16  &GPR160   &0.634$\pm$0.054      &29  &MKI67   &0.569$\pm$0.053      &42  &TMEM45B  &0.460$\pm$0.045\\
4   &ESR1   &0.709$\pm$0.018      &17  &MAPT     &0.624$\pm$0.056      &30  &CXXC5   &0.568$\pm$0.039      &43  &MYC      &0.439$\pm$0.047\\
5   &ORC6L  &0.701$\pm$0.035      &18  &PGR      &0.619$\pm$0.034      &31  &TYMS    &0.558$\pm$0.044      &44  &KRT14    &0.417$\pm$0.040\\
6   &CDC20  &0.682$\pm$0.019      &19  &PTTG1    &0.616$\pm$0.023      &32  &SFRP1   &0.555$\pm$0.046      &45  &BAG1     &0.374$\pm$0.057\\
7   &KIF2C  &0.674$\pm$0.032      &20  &BCL2     &0.614$\pm$0.063      &33  &PHGDH   &0.552$\pm$0.034      &46  &BLVRA    &0.370$\pm$0.044\\
8   &MYBL2  &0.673$\pm$0.025      &21  &EXO1     &0.610$\pm$0.032      &34  &MIA     &0.542$\pm$0.052      &47  &ERBB2    &0.360$\pm$0.065\\
9   &MELK   &0.661$\pm$0.035      &22  &SLC39A6  &0.607$\pm$0.034      &35  &CDH3    &0.537$\pm$0.042      &48  &MDM2     &0.344$\pm$0.039\\
10  &FOXC1  &0.655$\pm$0.037      &23  &CCNB1    &0.600$\pm$0.060      &36  &CENPF   &0.524$\pm$0.046      &49  &FGFR4    &0.318$\pm$0.041\\
11  &CEP55  &0.654$\pm$0.052      &24  &NUF2     &0.600$\pm$0.028      &37  &EGFR    &0.513$\pm$0.039      &50  &GRB7     &0.181$\pm$0.107\\
12  &NDC80  &0.654$\pm$0.049      &25  &RRM2     &0.597$\pm$0.039      &38  &ACTR3B  &0.493$\pm$0.080      &  &  &\\
13  &UBE2C  &0.650$\pm$0.029      &26  &UBE2T    &0.593$\pm$0.041      &39  &KRT5    &0.477$\pm$0.037      &  &  &\\
\bottomrule
\end{tabular}
	}
\end{table}

\begin{table}[h]
	\caption{\textbf{Prediction accuracy for all PAM50 genes on TCGA-BRCA for \textit{UNI – Contrastive}.} Results are reported as mean $\pm$ standard deviation over 5-fold site-aware cross-validation and sorted by decreasing mean Pearson correlation.}\vspace{-2.25mm}	
    \label{tab:main_results_brca_pam50_contrastive_all_genes}
    \centering
	\resizebox{1.0\textwidth}{!}{%
		\begin{tabular}{@{\hspace{0.55cm}}ccc@{\hspace{0.55cm}}@{\hspace{0.55cm}}ccc@{\hspace{0.55cm}}@{\hspace{0.55cm}}ccc@{\hspace{0.55cm}}@{\hspace{0.55cm}}ccc@{\hspace{0.55cm}}}
\toprule
Rank &Gene &Pearson ($\uparrow$)  &Rank &Gene &Pearson ($\uparrow$)  &Rank &Gene &Pearson ($\uparrow$)  &Rank &Gene &Pearson ($\uparrow$)\\
\midrule
1   &FOXA1  &0.731$\pm$0.030      &14  &NDC80   &0.647$\pm$0.061      &27  &UBE2T  &0.576$\pm$0.040      &40  &KRT17   &0.484$\pm$0.045\\
2   &MLPH   &0.722$\pm$0.049      &15  &NAT1    &0.639$\pm$0.038      &28  &SFRP1  &0.574$\pm$0.027      &41  &TMEM45B &0.472$\pm$0.046\\
3   &ESR1   &0.719$\pm$0.025      &16  &MAPT    &0.632$\pm$0.054      &29  &ANLN   &0.572$\pm$0.041      &42  &MMP11   &0.468$\pm$0.060\\
4   &CCNE1  &0.709$\pm$0.027      &17  &GPR160  &0.628$\pm$0.055      &30  &CDC6   &0.569$\pm$0.043      &43  &MYC     &0.443$\pm$0.038\\
5   &ORC6L  &0.692$\pm$0.042      &18  &BCL2    &0.627$\pm$0.058      &31  &MKI67  &0.561$\pm$0.027      &44  &KRT14   &0.440$\pm$0.043\\
6   &CDC20  &0.676$\pm$0.032      &19  &PGR     &0.613$\pm$0.038      &32  &CDH3   &0.559$\pm$0.016      &45  &ERBB2   &0.392$\pm$0.046\\
7   &KIF2C  &0.661$\pm$0.036      &20  &PTTG1   &0.612$\pm$0.031      &33  &PHGDH  &0.550$\pm$0.041      &46  &BAG1    &0.368$\pm$0.059\\
8   &MYBL2  &0.659$\pm$0.033      &21  &SLC39A6 &0.601$\pm$0.020      &34  &TYMS   &0.547$\pm$0.054      &47  &BLVRA   &0.366$\pm$0.068\\
9   &FOXC1  &0.656$\pm$0.035      &22  &RRM2    &0.593$\pm$0.031      &35  &MIA    &0.542$\pm$0.034      &48  &MDM2    &0.354$\pm$0.026\\
10  &BIRC5  &0.655$\pm$0.038      &23  &EXO1    &0.589$\pm$0.029      &36  &EGFR   &0.538$\pm$0.034      &49  &FGFR4   &0.333$\pm$0.054\\
11  &MELK   &0.652$\pm$0.035      &24  &CXXC5   &0.585$\pm$0.054      &37  &KRT5   &0.494$\pm$0.028      &50  &GRB7    &0.247$\pm$0.081\\
12  &CEP55  &0.650$\pm$0.052      &25  &CCNB1   &0.580$\pm$0.058      &38  &CENPF  &0.492$\pm$0.041      &  & &\\
13  &UBE2C  &0.647$\pm$0.036      &26  &NUF2    &0.578$\pm$0.021      &39  &ACTR3B &0.491$\pm$0.089      &  & &\\
\bottomrule
\end{tabular}
	}
\end{table}

\begin{table}[h]
    \caption{\textbf{Prediction accuracy for all PAM50 genes on SCAN-B-Lund.} Spearman and Pearson correlations between predicted and observed gene-expression values for all PAM50 genes, for the \textit{H-optimus-1 - Direct - ABMIL} ensemble trained on TCGA-BRCA and evaluated on the external SCAN-B-Lund dataset. Genes are listed in alphabetical order.}\vspace{-2.25mm}
    \label{tab:scan-b-lund_pam50_results}
    \centering
	\resizebox{1.0\textwidth}{!}{%
		\begin{tabular}{@{\hspace{0.55cm}}ccc@{\hspace{0.55cm}}@{\hspace{0.55cm}}ccc@{\hspace{0.55cm}}@{\hspace{0.55cm}}ccc@{\hspace{0.55cm}}@{\hspace{0.55cm}}ccc@{\hspace{0.55cm}}}
\toprule
Gene &Spearman ($\uparrow$) &Pearson ($\uparrow$)  &Gene &Spearman ($\uparrow$) &Pearson ($\uparrow$)  &Gene &Spearman ($\uparrow$) &Pearson ($\uparrow$)  &Gene &Spearman ($\uparrow$) &Pearson ($\uparrow$)\\
\midrule
ACTR3B   &0.240  &0.456      &CXXC5    &0.329   &0.365      &KRT17    &0.384   &0.314      &ORC6L    &0.654   &0.608\\
ANLN     &0.590  &0.614      &EGFR     &0.528   &0.181      &MAPT     &0.450   &0.401      &PGR      &0.417   &0.259\\
BAG1     &0.265  &0.259      &ERBB2    &0.158   &0.232      &MDM2     &0.225   &0.143      &PHGDH    &0.328   &0.218\\
BCL2     &0.500  &0.445      &ESR1     &0.509   &0.410      &MELK     &0.632   &0.646      &PTTG1    &0.616   &0.638\\
BIRC5    &0.659  &0.565      &EXO1     &0.611   &0.627      &MIA      &0.439   &0.426      &RRM2     &0.584   &0.560\\
BLVRA    &0.214  &0.196      &FGFR4    &0.270   &0.198      &MKI67    &0.598   &0.617      &SFRP1    &0.540   &0.389\\
CCNB1    &0.593  &0.612      &FOXA1    &0.472   &0.514      &MLPH     &0.401   &0.434      &SLC39A6  &0.531   &0.376\\
CCNE1    &0.589  &0.389      &FOXC1    &0.467   &0.543      &MMP11    &0.503   &0.402      &TMEM45B  &0.329   &0.254\\
CDC6     &0.603  &0.397      &GPR160   &0.352   &0.329      &MYBL2    &0.649   &0.620      &TYMS     &0.562   &0.616\\
CDC20    &0.583  &0.692      &GRB7     &0.114   &0.293      &MYC      &0.153   &0.148      &UBE2C    &0.648   &0.620\\
CDH3     &0.399  &0.512      &KIF2C    &0.648   &0.698      &NAT1     &0.501   &0.356      &UBE2T    &0.610   &0.634\\
CENPF    &0.541  &0.561      &KRT5     &0.418   &0.307      &NDC80    &0.619   &0.647      &         &        &     \\
CEP55    &0.635  &0.658      &KRT14    &0.422   &0.299      &NUF2     &0.573   &0.474      &         &        &     \\
\bottomrule
\end{tabular}
	}
\end{table}

\end{document}